# Elemental Abundances in the Local Group: Tracing the Formation History of the Great Andromeda Galaxy

Thesis by
Ivanna A. Escala

In Partial Fulfillment of the Requirements for the
Degree of
Doctor of Philosophy

Caltech

CALIFORNIA INSTITUTE OF TECHNOLOGY
Pasadena, California

Ivanna A. Escala
ORCID: 0000-0002-9933-9551

*Para mi familia, y para mi*

# ACKNOWLEDGEMENTS

I would like to thank the many people that have provided me with guidance throughout my thesis, and who have helped me reach the point in my life defined by this achievement. It would have been infinitely more challenging without you all, and I am grateful beyond words.

First and foremost, I would like to thank my thesis advisor, Evan N. Kirby, for always making his students a priority. He has been an invaluable mentor, teacher, collaborator, and friend. I am especially thankful that Evan had enough trust in my ability as a competent and independent scientist to be supportive of my extended visit to Princeton. I am privileged to have him as an advisor.

I would also like to thank my various mentors throughout my undergraduate education at the University of California, San Diego, to whom I am indebted: Adam J. Burgasser, who granted me my first opportunity to do research in astronomy, and who believed in me when I most needed it; Dušan Kereš, who helped me develop my skills as a nascent researcher with immeasurable patience and kindness; and Alison Coil, who provided me with much appreciated advice and support on the challenges of navigating a career in astronomy. I would additionally like to thank my beautiful home state of California for providing me with an attainable path to higher education, in part by covering the cost of my undergraduate tuition.

Furthermore, I would like to express my gratitude to the additional mentors I had throughout my graduate career: Karoline M. Gilbert, who generously shared her expertise on M31 with me, and who invested significantly in my mentorship despite being a coast apart; Andrew Wetzel, who helped me find my footing as a first year graduate student, and under whose mentorship that I published my first paper; Jenny Greene, who magnanimously hosted me at Princeton during my extended visit; and Robyn E. Sanderson, who worked together with me to create Caltech Women in Physics, Mathematics, and Astronomy, and who provided me with valuable advice. Thank you all for advocating on my behalf.

I would like to thank our small Elemental Abundances in M31 collaboration (Evan N. Kirby, Karoline M. Gilbert, and Jennifer Wojno) for providing a wonderful and friendly environment that enabled me to flourish as a researcher. I would like to express my thanks to Princeton Astrophysics, for working to accommodate my bicoastal marriage, and for welcoming me into their department as a visitor.

Thanks to my thesis committee, Chuck Steidel, Philip F. Hopkins, Jim Fuller, Raja GuhaThakurta, Karoline M. Gilbert (informally), and formerly Shri Kulkarni, for providing me with feedback on my thesis plans, and for convening virtually for my defense, despite the fact that I will not be able to provide them with donuts.

Of course, my time in graduate school would not have been the same without the fantastic community of astronomy graduate students at Caltech. My decision to accept Caltech's offer of admission was partially motivated by this tight-knight community, and to this day I am certain that I made the best decision for myself by choosing Caltech astronomy. In particular, I would like to thank my cohort, Anna Y.-Q. Ho, Kaew Tinyanont, Yuguang Chen, and formerly Bhawna Motwani, for all the years we spent together advancing through the program. I am honored to be graduating with some of you this spring, and I look forward to seeing our careers progress. I would also like to extend special thanks to Nicha Leethochawalit and Gina Duggan, my seniors in the Kirby group, Jacob Jencson, for being a joy to have in the department, Denise Schmitz, for being a friend within the Hopkins group, and Tim Miller, who helped to make my first year of graduate school hilarious, despite his short time as a visitor at Caltech.

These acknowledgments would not be complete without mentioning Matthew E. Orr, my best friend throughout graduate school. I am thankful to have great memories of our many runs through La Cañada, the Hahamongna Watershed Park, and the Angeles National Forest, our expensive lattes from Copa Vida, our equally expensive cocktails from various bars across Los Angeles, and of course, the infamous "Goodbye Forever Tim." I am indebted to Matt for many reasons, most of all that he helped show me how to not be a robot (although I am glad that I retained some of my robotic tendencies) and the meaning of the phrase "Treat yo self."

It is undeniable that I could not have gotten by without a little help from my friends. Moving across coasts, albeit temporarily, in the middle of graduate school was no small burden, and I could not have shouldered it without a few irreplaceable friends. I owe much to Matt Orr, Nina Tompkin and Dimitar Ho, and Anna Ho, for assisting me with the physical act of moving, for generously hosting me throughout my various visits to Caltech while I was living in Princeton, and especially for letting me couch-surf for weeks upon my return to Caltech, when I was looking for housing in Pasadena's impossible market. Nina and Dim are among my most favorite people in the world, and I am so grateful that they have always stood by my side. Anna is a scientific force of nature to behold, and she is one of the most moral and trustworthy people that I will ever know. I sincerely hope that I will have the opportunity to return the innumerable favors of my friends, and to go beyond merely returning them in the future.

I would like to extend my heartfelt thanks to Anita, Ramin, and Shayda Abidi, my second family. The Abidis opened their home in San Diego to me and provided me with an endlessly warm and welcoming environment to further my undergraduate education. In all her kindness, Anita looked after me as if I was her own daughter; Ramin served as a role model to me, and taught me how to be "happily dissatisfied" in order to move forward in life; and Shayda was always full of encouragement for me and my scientific pursuits, and will always be one of my best friends.

Above all, I have undertaken this thesis for myself. But I have also done it for my family, to take full advantage of the educational opportunities that they did not have, but for which they sacrificed. For my mother, Ivannia C. Rodriguez Escala, who left everything she knew in her home country of Costa Rica to move to the United States so that I could be born an American citizen. For my father, Richard E. Escala, who worked so hard to provide for his family that, to this day, he has never taken the time off to return to his home country of Argentina. My mother is a shining beacon of fathomless understanding, compassion, and joy in the face of even the darkest parts of the world. My father is the epitome of self-reliance, unbreakable will, and admirable work ethic, even in the face of crippling global crises. I owe who I am to them, and I hope to embody the best of the both of them.

I would also like to thank my in-laws, Randal Moseley and Michele Dearmijo Moseley, for their unwavering support of my scientific career, and for offering to assist with the logistical challenges of a bicoastal marriage. Randy and Michele have become a reliable foundation in my life, and it is very meaningful to me that they have continually asked after my doctoral degree progress.

Last, but farthest from least, I would like to thank the love of my life, my husband, Eric R. Moseley. Eric is undoubtedly my biggest fan—he is a bottomless well of support, affection, praise, and devotion. He has thought the world of me, and it has made all the difference. His willingness to move from Princeton to California to be together for the remainder of the global pandemic is the single greatest thing that anyone has done for me in my short life. I am eternally grateful that my time as a graduate student at Caltech brought us together[1], and I look forward to helping to shape the future of astronomy and astrophysics together.

[1]This is your fault, Evan.

# ABSTRACT

The Local Group (LG) is an environment accessible to detailed studies of galaxy formation, providing a complement to the early universe. In particular, spectroscopy of resolved stellar populations in the LG provides kinematical and chemical information for individual stars that can be used to infer the history of $L_\star$ galaxies like the Milky Way (MW) and Andromeda (M31).

The *Gaia* revolution in the MW, combined with spectroscopy from APOGEE and other surveys, has enabled comprehensive observational studies of the MW's formation history. In addition, comparisons to simulations can be leveraged to maximally utilize such observational data to probe the hierarchical assembly of galaxies. Toward this goal, I have analyzed simulations of chemical evolution in LG dwarf galaxies to assess their ability to match observations.

The exquisite detail in which the MW has been studied is currently not achievable in any other $L_\star$ galaxy. For this reason, the MW is a template for our understanding of galaxy formation. M31 is the only external galaxy that we can currently hope study in a level of detail approaching the MW. Studies of M31 have recently taken on greater significance, given the growing body of evidence that its formation history differs substantially from that of the MW.

In an era of limited information about elemental abundances in M31, I have developed a technique to apply spectral synthesis to low-resolution stellar spectroscopy in order to measure abundances for individual giant stars in distant LG galaxies. Through undertaking the largest deep, spectroscopic survey of M31 to date with my collaborators, this has resulted in the first measurements of the elemental abundances in the inner stellar halo and stellar disk of M31, and the largest homogeneous catalog of elemental abundances in M31. With this foundational work, we have opened the doors to detailed studies of the chemical composition of M31.

Now, we can begin to ask–and answer–what differences in the elemental abundances of the M31 and the MW imply for our knowledge of galaxy formation in the broader universe. At the cusp of next-generation observational facilities and theoretical simulations, we can only advance toward this goal.

# PUBLISHED CONTENT AND CONTRIBUTIONS

Escala, Ivanna, et al. 2020a. "Elemental Abundances in M31: A Comparative Analysis of Alpha and Iron Element Abundances in the Outer Disk, Giant Stellar Stream, and Inner Halo of M31". *ApJ* 889 (2): 177.
I.E. contributed to data acquisition, performed the analysis, and wrote the manuscript. doi:10.3847/1538-4357/ab6659. arXiv: 1909.00006 [astro-ph.GA].

Escala, Ivanna, et al. 2020b. "Elemental Abundances in M31: Properties of Inner Stellar Halo". *ApJ* [submitted].
I.E. contributed to data acquisition, performed the analysis, and wrote the manuscript.

Escala, Ivanna, et al. 2019. "Elemental Abundances in M31: Alpha and Iron Element Abundances from Low-resolution Resolved Stellar Spectroscopy in the Stellar Halo". *ApJ* 878 (1): 42.
I.E. vetted the linelist, generated the grids of synthetic spectra, developed the software, contributed to data acquisition, performed the analysis, and wrote the manuscript. doi:10.3847/1538-4357/ab1eac. arXiv: 1811.09279 [astro-ph.GA].

Escala, Ivanna, et al. 2018. "Modelling chemical abundance distributions for dwarf galaxies in the Local Group: the impact of turbulent metal diffusion". *MNRAS* 474 (2): 2194–2211.
I.E. performed the analysis and wrote the manuscript. doi:10.1093/mnras/stx2858. arXiv: 1710.06533 [astro-ph.GA].

# TABLE OF CONTENTS

# LIST OF ILLUSTRATIONS

# LIST OF TABLES

*1*

# INTRODUCTION

## 1.1 Near-Field Cosmology with the Local Group

The term "near-field cosmology" was coined in 2002 in an influential review on galaxy formation by Freeman and Bland-Hawthorn. In contrast to "far-field cosmology," which relies on high-redshift galaxies to provide a snapshot of structure formation in the distant past, near-field cosmology aims to use *nearby* galaxies in our Local Group (LG) to reconstruct galaxy formation history. An intrinsic assumption of near-field cosmology is that the galaxies of the LG—the Milky Way (MW), its sister galaxy Andromeda (M31), and over 50 satellite and isolated dwarf galaxies—provide an informative, if not representative, picture of galaxy formation in the broader universe. Although both the near- and far-field contain clues to the fundamental, physical principles driving galaxy formation, near-field cosmology has the advantage of enabling exquisitely detailed studies based on bright, resolved stellar populations.

Whereas near-field cosmology is the ultimate goal, "galactic archaeology" is the indispensable tool utilized in its achievement. The foundational principle of galactic archaeology is that stellar populations preserve kinematical and chemical signatures on timescales of many billions of years, opening a window into the early universe, the recent past, the present, and even the future formation path of a galaxy. In particular, the chemical composition of stars, in the form of elemental abundances measured from stellar spectra, can serve as a fossil record of formation environment. Long after the distinct spatial and kinematical signatures have disappeared, a stellar population retains the chemical signatures imprinted at the time of its birth. Thus, stellar populations with common origins are identifiable in principle by using "chemical tagging" to associate stars with one another based on their unique set of elemental abundances.

In addition to kinematical and chemical information, stellar ages can be inferred from resolved stellar populations using asteroseismology and gyrochronology for very nearby stars, and more broadly, using color-magnitude diagrams throughout the LG (Section 6.3). Although the focus of this thesis is stellar structure within galaxies, it is important to mention that galactic archaeology also provides a promising avenue to study the distribution of dark matter in the nearby universe (Section 6.2), which is intimately connected to cosmology.

In one of the most pivotal papers in the field, Eggen, Lynden-Bell, and Sandage (1962) illustrated both the feasibility and power of galactic archaeology as an instrument to uncover the nature of galaxy formation. Based on an observed anti-correlation between orbital eccentricity and metallicity in a sample of 221 Galactic dwarf stars, Eggen, Lynden-Bell, and Sandage (1962) concluded that the metal-poor stars inhabited a distinct halo-like component of the Galaxy formed via the rapid, dissipative collapse of a protogalactic cloud. This hypothesis was challenged by Searle and Zinn (1978), who presented a more inhomogeneous and complex picture of stellar halo formation. By measuring metallicity for 177 red giant stars across 19 Galactic globular clusters, Searle and Zinn (1978) observed that cluster metallicities were independent of Galactocentric distance beyond 8 kpc, leading them to conclude that the stellar halo must have formed at least in part from distinct protogalactic fragments. Although Searle and Zinn did not consider a scenario in which the protogalactic fragments were external in origin, their study heralded the inception of galactic archaeology as a means to probe the theoretical framework of the hierarchical assembly of galaxies (e.g., Press and Schechter 1974).

Since the beginnings of this area of study, simulations of galaxy formation and evolution have contributed significantly to development of the hierarchical assembly scenario within the $\Lambda$CDM paradigm (e.g., White and Rees 1978; Blumenthal et al. 1984). In our current picture the of the formation of MW- and M31-like galaxies, the importance of hierarchical accretion of external galactic systems is difficult to disentangle from that of *in-situ* formation in the classical "protogalactic cloud" scenario. For example, stellar halos are ideal for studying hierarchical formation, owing to long dynamical times that maintain the coherence of accreted substructure. However, stellar halos are predicted to preserve signatures of *in-situ* stellar formation (e.g., Zolotov et al. 2009; Font et al. 2011; Tissera et al. 2013; Cooper et al. 2015), in addition to probing accretion history (e.g., Bullock and Johnston 2005; Font et al. 2006a; Cooper et al. 2010). Furthermore, the stellar halo and stellar disks of MW-like and M31-like galaxies are connected through accretion events that not only build up the halo, but can impact the evolution of the disk (Abadi et al. 2003; Peñarrubia, McConnachie, and Babul 2006; Tissera, White, and Scannapieco 2012). In turn, stellar disks can contribute to the inner stellar halo via heating mechanisms which serve as a distinct *in-situ* formation channel (Purcell, Bullock, and Kazantzidis 2010;

McCarthy et al. 2012; Tissera et al. 2013; Cooper et al. 2015).

In order to conclusively settle the relative importance of these various formation channels for $L_\star$ galaxies, detailed chemical and kinematical information of the resolved stellar populations of nearby galaxies, such as the MW and M31, are necessary.

**A Brief History of Near-Field Cosmology with the MW**

By definition, the galaxy in which we reside is the nearest, and thus the most accessible for studies of near-field cosmology. Over the last few decades, astronomers have tirelessly worked toward Freeman and Bland-Hawthorn's vision of the "New Galaxy," an exhaustive reconstruction of formation history for each stellar component of the MW.

The genesis of this objective can arguably be traced to the discovery of the tidally disrupting Sagittarius dwarf galaxy in 1994 by Ibata, Gilmore, and Irwin. At the time of its discovery, only the "classical" dwarf spheroidal galaxies and the Magellanic Clouds were known satellites of the MW. Although the Magellanic Clouds showed signs of tidal interaction with the MW via a stream of neutral gas, no such stellar streams were detected (e.g., Mathewson, Cleary, and Murray 1974; Putman et al. 1998). The findings of Ibata, Gilmore, and Irwin (1994) and Ibata, Gilmore, and Irwin (1995) provided direct evidence that at least some globular clusters and field stars in the MW's stellar halo originated from tidally disrupted, accreted dwarf galaxies, reinvigorating the hypothesis that minor mergers played a significant role in the formation of MW-like galaxies. Several years later, a plethora of substructure in the MW's stellar halo was detected in commissioning data from the Sloan Digital Sky Survey (SDSS; Yanny et al. 2000; Ivezić et al. 2000), much of which was attributed the newly discovered stream resulting from the cannibalization of Sagittarius (Ibata et al. 2001b). Despite early indications (e.g., Lynden-Bell and Lynden-Bell 1995; Majewski, Munn, and Hawley 1996; Helmi et al. 1999a), the degree of inhomogeneity and clustering observed in the stellar halo of the MW was surprising given that the survey had observed 1% of the sky, suggesting that substructure was likely a common feature of the MW's halo. The contemporaneous discovery of Andromeda's Giant Stellar Stream (GSS, Section 1.2; Ibata et al. 2001a) bolstered observational support for the predictions of galaxy formation theory (e.g., Cole et al. 1994; Johnston, Hernquist, and Bolte 1996; Klypin et al. 1999; Moore et al. 1999; Helmi and White 1999; Bullock, Kravtsov, and Weinberg 2000, 2001), providing the first indication that substructure of external origins may in fact be a common feature of stellar halos in general (see also Shang et al. 1998).

Notwithstanding the successes of the Searle and Zinn "fragment" hypothesis for the observed properties of the MW's stellar halo, a perceived challenge remained from a theoretical perspective. In 1996, Unavane, Wyse, and Gilmore attempted to quantify the number of Carina-like and Fornax-like galaxies expected to contribute to the accreted stellar halo of the MW based on its $B-V$ color and photometric metallicity, relying on the assumption that the accreted progenitor galaxies would have *the same* chemical properties as present-day dwarf spheroidal (dSph) satellites. With the development of 8-10 m class telescopes combined with efficient high-resolution spectrographs, direct measurements of the chemical abundances of MW dSphs emerged as a new field of study, enabling comparisons to the Galactic halo (e.g., Shetrone, Bolte, and Stetson 1998; Fulbright 2000; Shetrone, Côté, and Sargent 2001; Stephens and Boesgaard 2002; Fulbright 2002; Shetrone et al. 2003; Tolstoy et al. 2003; Venn et al. 2004). An unanimous conclusion among these works was that the chemical abundances of present-day dSphs disagree with the Galactic halo. Most notably, the dSphs have lower $\alpha$-element abundances at fixed metallicity compared to the MW. A proposed remedy to this issue in the hierarchical assembly paradigm was that the accreted "protogalactic fragments" were *not* chemically identical to present-day dSphs. For example, Venn et al. (2004) speculated that these progenitors could have been high-mass, although the distinct chemical signatures of Sagittarius relative to the MW presented a potential problem (Bonifacio et al. 2004). Alternatively, the progenitors could have been low-mass like the dSphs, but accreted early in the MW's formation history before they became chemically evolved.

The solution to this challenge for the hierarchical assembly paradigm resulted from the significant advance of coupling cosmological simulations (e.g., Wechsler et al. 2002; Bullock and Johnston 2005) of galaxy formation with chemical evolution modeling. Within this framework, Robertson et al. (2005) demonstrated that the observed differences in chemical abundance patterns between MW dSphs and the Galactic halo was a *natural* consequence of ΛCDM cosmology. The primary progenitors of stellar halos were predicted to be both massive ($\sim 5\times10^{10}$ $M_{\odot}$) and destroyed early ($\sim$10 Gyr ago). Thus, their star formation histories were truncated and dominated by the yields of core-collapse supernovae, resulting in metal-poor and $\alpha$-enhanced stars as observed in the Galactic halo (see also Font et al. 2006a; Johnston et al. 2008; Cooper et al. 2010).

While this debate on the relationship between the presumably accreted Galactic halo and MW dSphs occurred, early studies that focused on the MW's halo alone noticed stellar populations that appeared to be chemically and/or kinematically distinct (e.g., Nissen and Schuster 1997 and Chiba and Beers 2000, with even earlier indications by Majewski, Munn, and Hawley 1996 and Carney et al. 1996). Such observations inspired the concept of the "dual halo" (Norris 1994; Carollo et al. 2007; Zolotov et al. 2009; Carollo et al. 2010; Beers et al. 2012), which consisted of both a dissipative *in-situ* component (à la Eggen, Lynden-Bell, and Sandage 1962) and a dissipationless accreted component (à la Searle and Zinn 1978). In particular, Carollo et al. (2007) used SDSS data (York et al. 2000) to associate the metal-rich ([Fe/H] $\sim -1.6$), prograde inner halo (10–15 kpc) of the MW with an *in-situ* component, and metal-poor ([Fe/H] $\sim -2.2$), retrograde outer halo (15–20 kpc) with an accreted component. However, the perceived duality of the stellar halo was contested by Schönrich, Asplund, and Casagrande (2011) and Schönrich, Asplund, and Casagrande (2014), who argued that SDSS did not provide robust evidence for a two-component halo, but rather a single-component halo rife with substructure. Contemporaneously, Nissen and Schuster (2010), Nissen and Schuster (2011), and Schuster et al. (2012) identified two chemically and kinematically distinct stellar populations at fixed metallicity in the halo, characterized by prograde (retrograde) rotation and high (low) [$\alpha$/Fe] presumably corresponding to an *in-situ* (accreted) component.

As chemical abundance studies advanced our understanding of stellar halo formation, they presented new challenges for our comprehension of disk formation. Bensby, Feltzing, and Lundström (2003), Reddy et al. (2003), and Reddy, Lambert, and Allende Prieto (2006) identified two tracks in $\alpha$-element abundance at fixed metallicity in the MW's disk for subsolar metallicities, suggesting that the thin and thick disks have different origins. This dichotomy in the chemical properties of the MW's disk consistently appeared in subsequent surveys (e.g., Adibekyan et al. 2013; Bensby, Feltzing, and Oey 2014), as well as different scale lengths for the chemically thin and thick disks (e.g., Bensby et al. 2011; Cheng et al. 2012a). In order to explain these interesting abundance patterns, studies largely invoked internal mechanisms such as two-infall scenarios for disk formation (Brook et al. 2012; Minchev, Chiappini, and Martig 2014; Nidever et al. 2014).

With the dawn of industrial-scale spectroscopic (e.g., APOGEE, GALAH; Majewski et al. 2017; De Silva et al. 2015) and astrometric surveys (*Gaia*; Gaia Collaboration et al. 2016a; Gaia Collaboration et al. 2018) of the MW, our understanding of the MW's formation history has undergone a revolution. By cross-matching complementary data across surveys, it has become apparent that the properties of the inner halo are dominated by a single accretion event (Gaia-Enceladus-Sausage, or GES; Helmi et al. 2018; Belokurov et al. 2018; Haywood et al. 2018), which is intimately connected to the evolution of the disk and the emergence of the *in-situ* halo (Mackereth et al. 2018; Gallart et al. 2019; Mackereth et al. 2019b; Di Matteo et al. 2019; Belokurov et al. 2020; Bonaca et al. 2020). The "dual halo" consists of metal-poor, slightly retrograde tidal debris from the GES merger and a metal-rich, prograde, proto-disk heated to halo-like kinematics as a result of the merger.[1] Furthermore, the distinct chemical tracks in the MW's disk (further confirmed by APOGEE; Haywood et al. 2013; Hayden et al. 2014; Nidever et al. 2014) may directly result from this early (~10 Gyr), massive (4:1 merger ratio) GES event (Mackereth et al. 2018; Bonaca et al. 2020). Undoubtedly, future spectroscopic surveys of the MW (Section 6.2) will unveil new insights into the formation of its bulge, outer disk, and outermost halo.

**The Milky Way versus Andromeda**

The MW is the core of near-field cosmology. As a consequence, the MW has historically served as a template for galaxy formation and as the context in which we interpret studies of other $L_\star$ galaxies. However, this is potentially problematic given that the MW's sister galaxy, M31, appears to be more representative of a typical spiral galaxy, based on comparisons to local disk galaxies beyond the Local Group, in terms of rotational velocity, absolute K-band luminosity, and stellar halo metallicity (Hammer et al. 2007). In contrast, the MW seems to be systematically offset by $\sim 1\sigma$ in each of these parameters. Thus, studies of M31's history are essential to place our understanding of galaxy formation in a cosmological context.

[1] See Helmi (2020) for a discussion of the tension between the details of the dual halo picture proposed by Carollo et al. (2007) and Carollo et al. (2010) and that revealed by *Gaia* DR2.

Table 1.1: Properties of the Milky Way vs. M31

| Property | MW | M31 | References |
|---|---|---|---|
| Total Mass | (0.5–1.5) × $10^{12}$ $M_\odot$ | (1–2) × $10^{12}$ $M_\odot$ | 41, 42 |
| Virial Radius | 250–300 kpc | 290 kpc | 36, 47 |
| Stellar Mass | (5–7) ×$10^{10}$ $\mathrm{M}_\odot$ | 1 × $10^{11}$ $M_\odot$ | 40, 43 |
| V-band Luminosity | (3–4) × $10^{10}$ $L_{\odot,V}$ | 3 × $10^{10}$ $L_{\odot,V}$ | 4, 15 |
| Stellar Disk Mass | (3–4) × $10^{10}$ $M_\odot$ | (7–11) × $10^{10}$ $\mathrm{M}_\odot$ | 4, 46 |
| Rotational Velocity | 220 km $\mathrm{s}^{-1}$ | 250 km $\mathrm{s}^{-1}$ | 7, 47 |
| Disk Scale Length | 2 kpc | 5–6 kpc | 5, 11, 20 |
| Disk Scale Height | 0.6–0.7 kpc | 0.5–1 kpc | 5, 13 |
| Thin Disk? | Yes | Unclear | 9, 21, 35 |
| Disk Age-Dispersion | $\sigma_R \sim$ 30 km $\mathrm{s}^{-1}$ (4 Gyr) | $\sigma_v \sim$ 90 km $\mathrm{s}^{-1}$ (4 Gyr) | 21, 38 |
| Disk [Fe/H] Gradient | −(0.07–0.09) dex $\mathrm{kpc}^{-1}$ | −0.02 dex $\mathrm{kpc}^{-1}$ | 8, 30, 32 |
| Disk [$\alpha$/Fe] Tracks? | Yes | Unknown | 3, 37, 22 |
| Stellar Halo Mass | (1–2) × $10^9$ $M_\odot$ | 1 × $10^{10}$ $M_\odot$ | 17, 34, 39 |
| Stellar Halo Luminosity | (6–10) × $10^8$ $L_\odot$ | 2 × $10^9 L_\odot$ | 17, 34 |
| Stellar Halo [Fe/H] | −1.6 (−1.2?) | −1.1 | 6, 10, 14, 23 |
| Stellar Halo [$\alpha$/Fe] | 0.3 | 0.4 | 23, 44 |
| Halo [Fe/H] Gradient | Weak | −(0.01–0.02) dex $\mathrm{kpc}^{-1}$ | 10, 14, 23, 26 |
| Halo [$\alpha$/Fe] Gradient | Weak | Negligible | 23, 26 |
| Density Profile Slope | −(2–4) | −2 | 16, 28, 34, 39, 45 |
| Density Profile Break? | Yes | No | ... |
| In-Situ Stellar Halo? | Yes | Maybe | 2, 18, 19, 20, 27 |
| Last Major Accretion Event | GES (10 Gya) | GSS (1-4 Gya) | 1, 12, 24, 25, 31, 33 |
| Progenitor Stellar Mass | $10^{8.5-9}$ $M_\odot$ | $10^{9-10}$ $M_\odot$ | ... |

References.—(1) Belokurov et al. (2018), (2) Belokurov et al. (2020), (3) Bensby et al. (2011), (4) Bland-Hawthorn and Gerhard (2016), (5) Bovy et al. (2012), (6) Carollo et al. (2010), (7) Chemin, Carignan, and Foster (2009), (8) Cheng et al. (2012b), (9) Collins et al. (2011), (10) Conroy et al. (2019), (11) Courteau et al. (2011), (12) D'Souza and Bell (2018a), (13) Dalcanton et al. (2015), (14) de Jong et al. (2010), (15) de Vaucouleurs et al. (1991), (16) Deason, Belokurov, and Evans (2011), (17) Deason, Belokurov, and Sanders (2019), (18) Di Matteo et al. (2019), (19) Dorman et al. (2012), (20) Dorman et al. (2013), (21) Dorman et al. (2015), (22) Escala et al. (2020a), (23) Escala et al. (2020b), (24) Fardal et al. (2007), (25) Fardal et al. (2008), (26) Fernández-Alvar et al. (2017), (27) Gallart et al. (2019), (28) Gilbert et al. (2012), (29) Gilbert et al. (2014), (30) Gregersen et al. (2015), (31) Hammer et al. (2018), (32) Hayden et al. (2014), (33) Helmi et al. (2018), (34) Ibata et al. (2014), (35) Jurić et al. (2008), (36) Klypin, Zhao, and Somerville (2002), (37) Nidever et al. (2014), (38) Mackereth et al. (2019a), (39) Mackereth and Bovy (2020), (40) McMillan (2011), (41) Peñarrubia et al. (2014), (42) Patel, Besla, and Mandel (2017), (43) Seigar, Barth, and Bullock (2008), (44) Venn et al. (2004), (45) Walker et al. (2009), (46) Williams et al. (2017), (47) Xue et al. (2008)

In Table 1.1, I summarize the similarities and differences between the most relevant properties of the MW and M31 for this thesis. Most notably, there is growing evidence that the MW has a quiescent merger history that sets it apart from the active, recent merger history of M31 (Section 1.2). The MW exhibits a broken power law in its stellar halo density profile (e.g., Watkins et al. 2009; Deason et al. 2013), where the break radius is likely defined by the build-up of accreted stars from a single, ancient dwarf galaxy (Deason et al. 2018). M31 does not show any break in its density profile throughout its entire stellar halo (Gilbert et al. 2012), which indicates that M31 has likely had multiple and continuous contributions to its stellar halo from accreted galaxies.

In particular, the wealth of stellar structure visible in M31's stellar halo (Section 1.2) may partially result from a merger with an external galaxy of $M_\star \sim 10^{10} M_\odot$ (Hammer et al. 2018; D'Souza and Bell 2018a), whereas a $M_\star \sim 10^{8.5-9} M_\odot$ merger likely dominates the stellar halo of the MW (Helmi et al. 2018; Belokurov et al. 2018). Simulations predict that 70% of MW-mass and M31-mass galaxies should have experienced a major merger (Stewart et al. 2008). If so, a major merger in M31 would be in accordance with the fact that it seems to better represent spiral galaxies than the MW. Even if M31 materializes as a less representative galaxy than previously thought, detailed studies of M31 are still extremely valuable—both as a contrast to the MW, and in its own right as a unique example of $L_\star$ galaxy formation.

## 1.2 The History of Resolved Stellar Populations in Andromeda

> *"It no doubt appears strange that the magnificent Andromeda Spiral, which under a transparent sky is so evident to the naked eye, should be so faint spectrographically."*
>
> – V. M. Slipher, 1913

Around the year 964 CE, Andromeda was first described by the Persian astronomer Abd al-Rahman al-Sufi in the *Book of Fixed Stars* as *al-Latkha al-Sahabiya*, a "nebulous smear" or "little cloud" (Hafez 2010). Later, in 1785, William Herschel remarked that Andromeda was "undoubtedly the nearest of all the great nebulae" based on the "faint red colour" of its brightest regions (Herschel 1785). Although characterized as a nebula, as were all external galaxies, there were early suggestions that Andromeda—or Messier 31—was indeed a galaxy, or "island universe" (a term attributed to Immanuel Kant in 1755). In 1864, a spectrum of Andromeda's inner spheroid taken by William Huggins revealed stellar absorption features, where "the light appeared to cease very abruptly in the orange," instead of showing signatures of "the bright lines" characteristic of gaseous nebula (Huggins and Miller 1864). The first measurement of M31's line-of-sight velocity of $-300$ km s$^{-1}$ was the "greatest hitherto observed" of any astronomical object, suggesting that the "spirals as a class may have higher velocities than do the stars" (Slipher 1913). Observations of this nature in Andromeda culminated in The Great Debate of 1920, on the subject of whether "these spirals [were] in fact congeries of vast numbers of stars, like our own Galaxy" (Curtis 1988). Several years later, Edwin Hubble conclusively settled this debate by identifying Cepheid variable stars in photographic plates (taken at Mt. Wilson observatory) to infer that M31 was at least 275 kpc away (Hubble 1929). Verily, studies of M31 have served as a foundation of near-field cosmology for well over a century.

The history of resolved stellar population studies in M31 began in 1944 with Walter Baade. He used red-sensitive plates to show that resolved stars in M31's central regions had properties similar to Galactic globular clusters (Baade 1944). In this fundamental study, Baade developed the broadly influential framework of Population I and II stars through comparisons between red, old stars in M31's bulge and blue, young stars in M31's disk. The next advance in studies of resolved stellar populations in M31 did not occur until Mould and Kristian obtained CCD-based photometry for the galaxy and its companion, M33, in 1986 to discover that M31 had a comparatively high mean photometric metallicity ($[\mathrm{Fe/H}]_{\mathrm{phot}} \sim -0.6$ at 7 kpc). The age of modern M31 studies would not begin until the mid-1990s, ushered in by the launch of the *Hubble Space Telescope* in 1990 and the commissioning of the Keck telescopes in 1993 (Keck I) and 1996 (Keck II).

Early studies of resolved stars in M31's inner spheroid found that its surface brightness profile within 20-30 projected kpc followed a de Vaucouleurs profile similar to that of M31's bulge (Pritchet and van den Bergh 1994; Durrell, Harris, and Pritchet 2004; Irwin et al. 2005), in contrast to the power-law density profile found for the inner MW (e.g., Yanny et al. 2000). Furthermore, M31's stellar density was found to be $\sim$10 times higher than the MW at an analogous projected distance (Reitzel, Guhathakurta, and Gould 1998). Based on color-magnitude diagrams, early studies also established that M31's inner spheroid was primarily composed of red, photometrically metal-rich stellar populations with broad red giant branches indicative of a spread in metallicity and/or stellar age (Mould and Kristian 1986; Davidge 1993; Durrell, Harris, and Pritchet 1994; Holland, Fahlman, and Richer 1996; Rich et al. 1996; Reitzel, Guhathakurta, and Gould 1998; Durrell, Harris, and Pritchet 2001; Bellazzini et al. 2003; Durrell, Harris, and Pritchet 2004). The studies that investigated the photometric metallicity distribution as a function of radius did not find any evidence for a gradient within 30 projected kpc (Durrell, Harris, and Pritchet 2001; Bellazzini et al. 2003; Durrell, Harris, and Pritchet 2004).

An implicit assumption in the aforementioned studies that relied solely on photometry (e.g., Mould and Kristian 1986; Durrell, Harris, and Pritchet 1994; Holland, Fahlman, and Richer 1996) was that the spheroid of M31 corresponded to a halo-like component, except with bulge-like properties. By targeting the innermost regions of the spheroid ($\lesssim$10 projected kpc), such studies largely circumvented the issue of sample contamination by distant, unresolved background galaxies and foreground Galactic dwarf stars owing to M31's high stellar density. A separate question is what *type* of M31 giant stars were probed in these regions—genuine halo stars, or interloping disk giants—since a detailed structural decomposition of M31's inner spheroid had not been performed. As noted by Reitzel, Guhathakurta, and Gould (1998), *stellar spectroscopy* was necessary in order to kinematically identify a clean sample of M31 RGB stars in the then putative stellar halo. Reitzel and Guhathakurta (2002) presented first results from a Keck/LRIS spectroscopic survey, isolating likely M31 RGB stars (with line-of-sight velocities $< -220$ km s$^{-1}$) in a field at 19 kpc along the minor axis, to confirm the existence of metal-poor ([Fe/H] $< -1$) stars in M31 that had been suggested by previous HST imaging (e.g., Holland, Fahlman, and Richer 1996; Durrell, Harris, and Pritchet 2001). Using the new Keck/DEIMOS instrument (Faber et al. 2003) to develop spectroscopic (and photometric) diagnostics to isolate M31 RGB stars (Gilbert et al. 2006), Guhathakurta et al. (2005) discovered M31's previously elusive, extended, metal-poor halo component. In contrast to the bulge-like inner spheroid, this robustly identified halo had a power-law surface brightness profile (see also Irwin et al. 2005) extending out to 165 projected kpc, and was at least 3 times larger than any previously mapped stellar component in M31.

Subsequent studies supported the detection of a metal-poor stellar halo (with [Fe/H] as low as $\sim -2$) in M31 and aimed to describe its global metallicity properties (Kalirai et al. 2006b; Chapman et al. 2006; Koch et al. 2008). Both Kalirai et al. (2006a) and Koch et al. (2008) found evidence of a metallicity gradient in M31's stellar halo along the minor axis from photometric and calcium-triplet based metallicity estimates, respectively. Although Chapman et al. (2006) detected a significant metal-poor population from their co-added spectra taken primarily along major axis fields, they did not find any metallicity variation as a function of radius. In a more detailed view of the stellar halo, greater spectroscopic coverage painted a portrait of an increasingly complex structure. Larger spectroscopic samples of individual RGB stars from the Spectroscopic and Photometric Landscape of Andromeda's Stellar Halo (SPLASH) survey revealed the presence of kinematically distinct substructure too faint to detect with photometry alone, while enabling the characterization of the spatial, kinematic, and chemical properties of the prominent GSS tidal feature and related structures (Guhathakurta et al. 2006; Kalirai et al. 2006a; Gilbert et al. 2007; Gilbert et al. 2009; Fardal et al. 2012).

As spectroscopic studies of M31's halo advanced, so did photometric studies using *HST*, the Isaac Newton Telescope, and the Canada-France-Hawaii telescope. These censuses revealed a wealth of substructure and significant inhomogeneity in M31's stellar halo (Ferguson et al. 2002; McConnachie et al. 2003; Ferguson et al. 2005; Ibata et al. 2007; Richardson et al. 2009), most notably the GSS (Ibata et al. 2001a), and a kinematically detected massive, metal-rich, extended disk (Ibata et al. 2005). Taken together, these structures contributed to the emerging picture that the accretion of galactic systems was instrumental in building M31's halo. Analyses of *HST* color-magnitude diagrams consistently provided evidence in favor of intermediate age (~8-10 Gyr) stellar populations in M31's halo, outer disk, and GSS, and a possible connection between stellar populations in the apparently "smooth" stellar halo and GSS (Ferguson and Johnson 2001; Brown et al. 2003; Brown et al. 2006; Brown et al. 2007; Brown et al. 2008; Richardson et al. 2008). Models of the GSS formation showed that many properties of M31's inner halo could be explained by tidal debris related to the GSS merger event (Fardal et al. 2006; Font et al. 2006b; Fardal et al. 2007; Mori and Rich 2008; Sadoun, Mohayaee, and Colin 2014).

Motivated by the rich substructure content of M31's halo, a previous imaging survey (Ibata et al. 2007) was extended to form the Pan-Andromeda Archaeological Survey (PAndAS; McConnachie et al. 2009), which provides CFHT/MegaCam coverage in all four quadrants of M31's halo out to ~150 kpc. Both PAndAS and a concurrent minor-axis photometric survey using Subaru/HSC (Tanaka et al. 2010) aimed to constrain M31's formation history by translating the amount of detected substructure into a mass function of accreted progenitor galaxies. For example, McConnachie et al. (2018) estimated that the various distinct substructures in M31's stellar halo (as seen from PAndAS) were produced by at least 5 separate accretion events within the last ~4 Gyr.

In the era of large spectroscopic (SPLASH) and photometric (PAndAS) surveys of M31's stellar halo, such questions could be addressed using statistically significant samples of M31 RGB stars for the first time. Studies of the global properties of M31's halo revealed that it in fact possessed a steep (photometric) metallicity gradient (Gilbert et al. 2014; Ibata et al. 2014), suggesting that the inner and outer halo of M31 had either distinct formation mechanisms or dominant progenitor(s). Modern determinations of M31's surface brightness profile (Courteau et al. 2011; Gilbert et al. 2012; Ibata et al. 2014) showed that it followed a continuous power-law to large radii (~90 kpc) without any apparent break, in contrast to the MW (e.g., Watkins et al. 2009). This implies that M31 experienced a comparatively prolonged accretion history (e.g., Deason et al. 2013). M31's velocity dispersion profile has also been measured out to about half its virial radius (Gilbert et al. 2018), with the intention of constraining its total mass.

Globally representative studies of M31's disk advanced along with that of the stellar halo. At first, imaging was localized to several pencil-beam *HST* pointings in both the southwestern and northeastern disk, which showed evidence for bursts of star formation ~2 Gyr ago and a dominant formation epoch of $z \lesssim 1$ with a median stellar age of ~7.5 Gyr (Bernard et al. 2012; Bernard et al. 2015a; Bernard et al. 2015b). Using *HST*, the Panchromatic Hubble Andromeda Treasury (PHAT; Dalcanton et al. 2012), a multi-cycle program, produced the first spatially resolved map of M31's northeastern disk, based on millions of individual stars spanning 0 to 20 projected kpc. This resulted in studies of both the recent (Lewis et al. 2015) and ancient (Williams et al. 2017) star formation history across M31's disk, and the significant detection of an unusual *global* burst of star formation ~2-4 Gyr (Williams et al. 2015). Using PHAT photometry to isolate SPLASH spectroscopic targets in M31's disk, Dorman et al. (2012) performed the first detailed structural decomposition of M31's inner regions to disentangle M31's inner spheroid from its disk. In a series of studies, Dorman et al. (2012) detected significant rotation of the spheroid, Dorman et al. (2013) found evidence of kicked-up disk stars in M31's inner halo, and Dorman et al. (2015) measured a surprisingly steep age-velocity dispersion correlation in M31's disk. Both this fact and the asymmetric drift as a function of stellar age (Quirk et al. 2019) in M31's disk pointed to continuous heating events affecting its evolution. PHAT also resulted in the first systematic study of the (photometric) metallicity of M31's disk (Gregersen et al. 2015), whereas previously the (calcium-triplet based) metallicity of M31's disk had only been studied from DEIMOS masks covering a small total area (Ibata et al. 2005; Collins et al. 2011).

The kinematical properties of M31's disk, its star formation history, and the nature of substructure of M31's inner halo could potentially be explained by a major merger. This scenario could result in (1) the formation of an extended disk, (2) an unusually high velocity dispersion in the inner disk, (3) a global burst of star formation across the disk, (4) the wealth of substructure observed in the inner halo, such as the GSS, and (5) the observed similarity between the star formation history of the GSS and other probed regions in the inner halo. Indeed, such a major merger may be the merger that formed the GSS. In this unfolding scenario, the origin of the GSS is a 4:1 merger occurring within the last ~4 Gyr (Fardal et al. 2008; Hammer et al. 2010; Hammer et al. 2018; D'Souza and Bell 2018a; Quirk and Patel 2020), as opposed to the previously assumed origin in a minor merger (e.g., Fardal et al. 2007).

If such a merger occurred, the implications for our understanding of the formation of $L_\star$ galaxies in the universe are significant: because M31 may be more representative of spiral galaxies than the MW (Section 1.1), this suggests that the majority of spiral galaxies may have experienced a late major merger (e.g., Stewart et al. 2008). Furthermore, the differences between M31 and the MW (Section 1.1) may indicate that the MW is not a representative template. To conclusively address this possibility, 6D phase space information (achieved with proper motions and precise distances), detailed simulations of the formation of M31-like galaxies (Section 6), and perhaps most crucially, *chemical abundance measurements beyond metallicity* are necessary. Thus, we begin our story of elemental abundances in M31 following the first measurements of [$\alpha$/Fe] for 4 individual RGB stars in M31's outer halo (Vargas et al. 2014).

## 1.3 Thesis Outline

In the following chapters, I use both state-of-the-art simulations (Chapter 2) and new observational techniques (Chapter 3) to probe the complex relationship between chemical abundances of galaxies and their formation histories. In particular, I focus on measuring [Fe/H] and [$\alpha$/Fe] for individual red giant branch stars in M31 (Chapters 4, 5).

In Chapter 2, I perform a comprehensive analysis of chemical evolution predictions of simulated LG dwarf galaxies against observations. I model [Fe/H] distribution functions and [$\alpha$/Fe] vs. [Fe/H] abundance patterns in high-resolution, cosmological, zoom-in simulations to assess the impact of a sub-grid model for the turbulent diffusion of metals in gas on the realism of the simulated abundance distributions.

In Chapter 3, I develop a technique to apply spectral synthesis to low-resolution stellar spectroscopy ($R \sim 2500$) to measure [Fe/H] and [$\alpha$/Fe] for individual red giant branch stars in distant LG galaxies. Using this technique, I reliably make the first measurements of [Fe/H] from spectral synthesis and [$\alpha$/Fe] in the inner stellar halo of M31.

In Chapter 4, I apply the technique to measure [Fe/H] and [$\alpha$/Fe] for a sample of 70 individual red giant branch stars across the inner stellar halo, GSS, and outer disk of M31. These are the first measurements of [$\alpha$/Fe] in M31's disk. I also perform a comparative analysis of chemical abundances between the various probed stellar structures.

In Chapter 5, I further expand upon this sample to construct the largest homogeneous set of chemical abundances in M31 to date. Prior to the work of my collaborators and me, there were only 4 measurements of [$\alpha$/Fe] in M31, whereas now there are 230 measurements. I compile the available measurements of [Fe/H] and [$\alpha$/Fe] to analyze the global chemical properties of M31's inner stellar halo.

Finally, in Chapter 6, I present a view toward the future of near-field cosmology in the LG in the era of the 2020s and 2030s. I provide an overview of future instruments and telescopes and briefly discuss next steps for simulations of galaxy formation.

*2*

# MODELING CHEMICAL ABUNDANCE DISTRIBUTIONS FOR DWARF GALAXIES IN THE LOCAL GROUP: THE IMPACT OF TURBULENT METAL DIFFUSION

Ivanna Escala[1], Andrew Wetzel[2,3,4], Evan N. Kirby[1], Philip F. Hopkins[2], Xiangcheng Ma[2], Coral Wheeler[2], Dušan Kereš[5], Claude-André Faucher-Giguère[6], Eliot Quataert[7]

[1]Department of Astronomy, California Institute of Technology, 1200 E California Blvd, Pasadena, CA, 91125, USA
[2]TAPIR, California Institute of Technology, 1200 E California Blvd, Pasadena, CA, 91125, USA
[3]The Observatories of the Carnegie Institution for Science, 813 Santa Barbara St, Pasadena, CA, 91106, USA
[4]Department of Physics, University of California, Davis, 1 Shields Ave, Davis, CA 95616, USA
[5]Department of Physics, Center for Astrophysics and Space Science, University of California, San Diego, La Jolla, CA, 92093, USA
[6]Department of Physics and Astronomy and CIERA, Northwestern University, 2145 Sheridan Rd, Evanston, IL, 60208, USA
[7]Department of Astronomy and Theoretical Astrophysics Center, University of California, Campbell Hall, Berkeley, CA, 94720, USA

**Abstract**

We investigate stellar metallicity distribution functions (MDFs), including Fe and $\alpha$-element abundances, in dwarf galaxies from the Feedback in Realistic Environments (FIRE) project. We examine both isolated dwarf galaxies and those that are satellites of a Milky Way-mass galaxy. In particular, we study the effects of including a sub-grid turbulent model for the diffusion of metals in gas. Simulations that include diffusion have narrower MDFs and abundance ratio distributions, because diffusion drives individual gas and star particles toward the average metallicity. This effect provides significantly better agreement with observed abundance distributions in dwarf galaxies in the Local Group, including small intrinsic scatter in [$\alpha$/Fe] vs. [Fe/H] of $\lesssim 0.1$ dex. This small intrinsic scatter arises in our simulations because the interstellar medium in dwarf galaxies is well-mixed at nearly all cosmic times, such that stars that form at a given time have similar abundances to $\lesssim 0.1$ dex. Thus, most of the scatter in abundances at $z = 0$ arises from redshift evolution and not from instantaneous scatter in the ISM. We find similar MDF widths and intrinsic scatter for satellite and isolated dwarf galaxies, which suggests that environmental effects play a minor role compared with internal chemical evolution in our simulations. Overall, with the inclusion of metal diffusion, our simulations reproduce abundance distribution widths of observed low-mass galaxies, enabling detailed studies of chemical evolution in galaxy formation.

## 2.1 Introduction

Dwarf galaxies, which probe the low-mass ($M_* \lesssim 10^9$ $M_\odot$) end of the galaxy mass spectrum, serve as an environment to test models of galaxy formation. Dwarf galaxies are extremely sensitive to feedback effects from supernovae (SNe) and stellar winds, owing to their shallow gravitational potential wells. This results in significant mass loss (e.g., Dekel and Silk 1986), and thus metal loss, which is relevant for studies of galactic chemical evolution.

Owing to their small sizes, low masses, and relatively inefficient star formation, dwarf galaxies can be challenging to simulate accurately. However, given that they form out of a small volume, dwarf galaxies are ideal for targeted zoom-in simulations. Hydrodynamical simulations of galaxy evolution have achieved increasingly high baryonic particle mass resolution, such that only a few SNe occur per star particle, where each star particle represents a single stellar population. The limited mass-sampling of chemical enrichment histories thus becomes important for simulating dwarf galaxies in the given stellar mass range. Consequently, the predicted chemical evolution of dwarf galaxies will be impacted by the specific feedback implementation used in such simulations. For example, the predicted abundances are subject to stochastic sampling of nucleosynthetic events (e.g., the "enrichment sampling problem" of Wiersma et al. 2009).

A majority of the studies investigating the detailed properties of chemical evolution have been based on one-zone numerical models of galactic chemical evolution with instantaneous mixing (Lanfranchi and Matteucci 2003; Lanfranchi and Matteucci 2007, 2010; Lanfranchi, Matteucci, and Cescutti 2006; Kirby et al. 2011a). Only recently have hydrodynamical simulations of cosmological isolated dwarf galaxies attempted to go beyond accurately reproducing the stellar mass-metallicity relation (Ma et al. 2016) to include more detailed chemical evolution properties (Marcolini et al. 2008; Revaz et al. 2009; Sawala et al. 2010; Revaz and Jablonka 2012). For example, Sawala et al. (2010) simulated the metallicity distribution functions (MDFs) and $\alpha$-element abundance ratios for isolated dwarf galaxies, yet the MDF is broad compared to observations (Kirby et al. 2011a) and contains a pronounced, unobserved low-metallicity tail, whereas the scatter in the $\alpha$-element abundances is large, particularly at low-metallicity (Tolstoy, Hill, and Tosi; Kirby et al. 2011b; Frebel and Norris 2015). However, both MDFs and abundance ratios can serve as tests of the metal injection scheme, energy and momentum injection from stellar feedback, yields, and microphysics in the interstellar medium (ISM).

Until recently (e.g., Shen, Wadsley, and Stinson 2010; Shen et al. 2013; Pilkington et al. 2012; Brook et al. 2014), sub-grid turbulent diffusion has been neglected in many astrophysical simulations based on Lagrangian methods, such as smoothed particle hydrodynamics (SPH) or mesh-free methods. This is despite the fact that supersonic, compressible flows result in a turbulent cascade that transports momentum to small scales, where viscous forces begin to dominate (Wadsley, Veeravalli, and Couchman 2008). Furthermore, "standard" SPH methods are known to suppress mixing on small scales, in contrast to Eulerian codes (Agertz et al. 2007) and other finite-volume methods (Hopkins 2015), which include intrinsic numerical diffusion (e.g., Recchi, Matteucci, and D'Ercole 2001). In contrast to most prior galaxy evolution simulations, the Feedback in Realistic Environments (FIRE)[1] project (Hopkins et al. 2014; Hopkins et al. 2018) has recently implemented (Su et al. 2017) a model for turbulent metal diffusion (TMD) due to unresolved, small-scale eddies.

Su et al. (2017) considered the impact of sub-grid metal diffusion in the context of a realistic, multi-phase ISM influenced by strong stellar feedback processes. They found that sub-grid metal diffusion does not significantly impact cooling physics, having no systematic effect on galactic star formation rates. This is in contrast to the findings of Shen, Wadsley, and Stinson (2010) and Pilkington et al. (2012) on the effects of including fluid microphysics. For example, Shen, Wadsley, and Stinson (2010) concluded that simulations without sub-grid metal diffusion produce slightly fewer stars, since fewer gas particles experience gas cooling and subsequently turn into stars. Comparatively low spatial resolution, such that the turbulent driving scales are not resolved, could potentially explain the discrepancy (Su et al. 2017). Although sub-grid metal diffusion does not significantly impact cooling rates or star formation rates in the FIRE simulations, sub-grid metal diffusion is expected to strongly impact chemical evolution.

[1] The FIRE project website is `http://fire.northwestern.edu`.

Motivated by previous studies (e.g., Aguirre et al. 2005) that have shown that metals are too inhomogeneous in simulations, the introduction of a diffusive term on sub-grid scales has recently been explored as a promising solution to the problem of reproducing realistic MDFs and abundances. Shen, Wadsley, and Stinson (2010) implemented a turbulence-induced mixing model in SPH simulations based on velocity shear, as opposed to the velocity-dispersion based model of Greif et al. (2009). Using the latter model, Jeon, Besla, and Bromm (2017) incorporated metal diffusion into a fully cosmological study of the chemical abundances of ultra-faint dwarf galaxies. Williamson, Martel, and Kawata (2016) investigated sub-grid metal mixing in non-cosmological, isolated dwarf galaxies, and found that the metallicity of stars is not strongly dependent on how the diffusivity is calculated from the the velocity distribution. In addition, they observed a reduction of scatter in stellar abundances and the suppression of low-metallicity star formation. Pilkington et al. (2012) found that sub-grid metal diffusion reduced the overproduction of extremely metal poor stars, except for M33-like spiral galaxies, as opposed to dwarf galaxies.

Hirai and Saitoh (2017) explored the efficiency of sub-grid metal mixing in non-cosmological isolated dwarf galaxy simulations, focusing on reproducing the scatter in barium inferred from extremely metal-poor stars (Suda et al. 2008). They concluded that the timescale for metal mixing necessary to reproduce observations of barium is $\lesssim 40$ Myr, which is shorter than the typical dynamical timescale of dwarf galaxies ($\sim 100$ Myr). Kawata et al. (2014) investigated the impact of strong stellar feedback in a simulation of a WLM-like, non-cosmological dwarf disk galaxy, finding that including sub-grid diffusion maintained low-metallicity in star-forming regions owing to efficient metal mixing in the ISM. Comparing their results on the dispersion in the abundances of newly formed stars to observations, they concluded that their sub-grid diffusion was likely too strong. Revaz et al. (2016) showed that metal diffusion can reproduce the low scatter in $\alpha$-elements at low metallicity in particle-based simulations. However, they concluded that a "smoothed metallicity scheme" (Wiersma et al. 2009), in which metallicity is treated as a smoothly varying function and involves no explicit redistribution of metals, is preferable over sub-grid metal diffusion to reproduce the observed dispersion in abundances of dwarf galaxies.

In this chapter, we use the high-resolution, cosmological zoom-in simulations of the FIRE project to analyze the impact of turbulent metal diffusion on observables

Table 2.1: FIRE Simulation Properties

| Simulation[a] | $M_*$[b] ($10^6$ $M_\odot$) | $\langle$[Fe/H]$\rangle$[c] (dex) | $\sigma$[d] (dex) |
|---|---|---|---|
| **m10q** | 1.7 | −2.16 | 0.55 |
| **m10q.md** | 2.0 | −2.12 | 0.41 |
| **m10v**[2] | 1.1 | −1.82 | 0.52 |
| **m10v.md** | 1.5 | −1.54 | 0.34 |

Note. — All quantities are determined at $z = 0$. For all isolated dwarf galaxy simulations, the star particle spatial resolution $h$ is 1.4 pc and the mass resolution is 250 $M_\odot$. For more details on the methods used to simulate the cosmological, isolated dwarf galaxies, see Oñorbe et al. (2014), Hopkins et al. (2014), and Hopkins et al. (2018).

[a] The simulation naming convention reflects the halo mass, e.g., m10 $\implies M_{\rm halo} \sim 10^{10}$ $M_\odot$ at $z = 0$. The designations "q" and "v" reflect initial conditions that distinguish between halos dominated by early- and late-time star formation respectively. The addition ".md" indicates that the simulation was run with sub-grid turbulent metal diffusion (Section 2.2).

[b] Defined as the mass within the radius that contains 90% of the stellar mass, $r_{90}$.

[c] The mass-weighted average metallicity of star particles within $r_{90}$ (Eq 2.4) and,

[d] the associated standard deviation, or intrinsic spread in the metallicity distribution.

related to chemical evolution for simulated dwarf galaxies in the mass range $M_*(z = 0) \sim 7 \times 10^5 - 2 \times 10^8$ $M_\odot$. First, we study a small sample of cosmological field dwarf galaxies simulated at very high resolution, then expand our analysis to include satellite and isolated dwarf galaxies of a Milky Way (MW) mass halo (Wetzel et al. 2016). We find that the inclusion of a physically-motivated, sub-grid turbulent diffusion model produces MDFs and abundance ratios consistent with observations of Local Group (LG) dwarf galaxies. We confirm the necessity of including sub-grid metal mixing in Lagrangian hydrodynamical codes, while taking into account a multi-phase ISM, explicit stellar and radiative feedback, the impact of cosmological accretion, and environmental effects.

## 2.2 Simulations

[2]The galaxy labeled **m10v** analyzed in this chapter (and the version of the run with sub-grid metal diffusion, **m10v.md**) is not the usual **m10v** included in previous FIRE papers. Instead, this **m10v** is a significantly contaminated galaxy in the same cosmological volume with a larger stellar mass ($1.1 \times 10^6$ $M_\odot$, as compared to $\sim 10^5$ $M_\odot$) at the outskirts of the high-resolution region. We emphasize that the comparisons between our **m10v** and **m10v.md** are internally consistent, since they both suffer from low-resolution dark matter contamination, and that despite this, they still produce realistic dwarf galaxies. Additionally, the majority of our conclusions are based on the properties of the collective simulated dwarf galaxy sample, particularly **m10q** and the satellite and isolated dwarf

We present a generalized summary of the relevant features of the FIRE simulations (Hopkins et al. 2014; Hopkins et al. 2018). The simulations analyzed here (e.g., Table 2.1) were run with the "FIRE-2" (Hopkins et al. 2018) rerun of GIZMO[3] in its Meshless Finite Mass (MFM) mode (Hopkins 2015). All feedback quantities are based on stellar evolution models that are identical between "FIRE-1" (Hopkins et al. 2014) and "FIRE-2", such that galaxy-scale properties do not qualitatively change between versions of the code (Hopkins et al. 2018).

The MFM method combines advantages from both SPH and grid-based codes. FIRE cosmological simulations of dwarf galaxies have reproduced several key observations. These include the bursty star formation and outflows generated by low-mass galaxies at high redshift (Muratov et al. 2015), the stellar mass-halo mass relation at both $z = 0$ (Oñorbe et al. 2015) and at high redshift (Ma et al. 2018), the dark matter halo profile in dwarf galaxies (Chan et al. 2015), the size evolution and age gradients of dwarf galaxies (El-Badry et al. 2016), the stellar mass-metallicity relation (Ma et al. 2016), and the stellar kinematics of dwarf galaxies (Wheeler et al. 2017), all without calibration to match observations at $z = 0$.

### Gas Cooling, Star Formation, and Feedback

Gas follows a cooling curve from 10 – $10^{10}$ K, with cooling at low temperatures due to molecular transitions and metal-line fine structure transitions, and primordial and metal line cooling at higher temperatures ($\geq 10^4$ K). A uniform, redshift-dependent photoionizing background (Faucher-Giguère et al. 2009) is taken into account at each timestep when determining the cooling rates.

Star formation occurs only in dense, molecular, self-gravitating regions with $n_{\rm crit} > 1000\ {\rm cm}^{-3}$. When these conditions are met, stars form at 100% efficiency per local free-fall time, although stellar feedback rapidly regulates the global star formation efficiency to a few percent on the scales of giant molecular clouds (Orr et al. 2018). The newly formed star particle inherits its metallicity from its progenitor gas particle. 11 total chemical species (H, He, C, N, O, Ne, Mg, Si, S, Ca, Fe), including $\alpha$- and Fe peak elements, which are particularly relevant to constraining star formation history (SFH), are tracked in addition to the total metallicity (Wiersma et al. 2009). Each star particle is treated as a single stellar population with a Kroupa (2002) initial mass function (IMF), with known age, mass, and metallicity. Feedback quantities

galaxies around **m12i** (Section 2.5)

[3]The public version of GIZMO is available at http://www.tapir.caltech.edu/~phopkins/Site/GIZMO.html

such as luminosity, Type II SN rates, and the rates of mass and metal loss are calculated based on stellar population models (STARBURST99; Leitherer et al. 1999). SN explosions occur discretely, as opposed to modeling their collective effects. Metal yields for Type Ia SNe are adopted from Iwamoto et al. (1999), where the rates follow Mannucci, Della Valle, and Panagia (2006), including both prompt and delayed populations. Metal yields for Type II SNe (Nomoto et al. 2006) and stellar winds (AGB & O-stars) are also included, as well as their contributions to ejecta energy, momentum, and mass. All feedback quantities are deposited directly into the ISM (gas particles) surrounding a given star particle, where mass, energy, and momentum are conserved. Radiative feedback from local photo-ionization, photo-electric heating, and radiation pressure are also included.

**Turbulent Metal Diffusion**

Although metals are diffused via turbulence in a realistic ISM, this has yet to be taken into account in many galaxy evolution and formation simulations (Wadsley, Veeravalli, and Couchman 2008). Lagrangian codes, such as SPH and MFM, follow fluid elements of fixed mass. Particles conserve metallicity unless injected with metals or metal loss occurs owing to SNe/stellar winds. However, SPH, or any Lagrangian methods (MFM), do not account for additional mixing that occurs via sub-grid Kelvin-Helmholtz instabilities, Rayleigh-Taylor instabilities, and turbulent eddies between gas particles. That is, without sub-grid metal diffusion, the metals assigned to a given gas particle are locked to that particle for all time. Consequently, gas particles may never become enriched, resulting in artificial noise in the MDF. Other sources of noise that may impact the appearance of the MDF are addressed in Appendices 2.7 and 2.7. Moreover, even enriched particles contribute to an unrealistic spread in metallicity in the absence of sub-grid mixing. To account for such unresolved mixing processes, some of the simulations include an explicit metal diffusion term between particles, following the prescription investigated by Shen, Wadsley, and Stinson (2010) based on the Smagorinsky (1963) model,

$$\begin{aligned} \frac{\partial \mathbf{M}_i}{\partial t} + \nabla \cdot (D \nabla \mathbf{M}_i) = 0, \\ D = C_0 ||\mathbf{S}||_f \mathbf{h}^2, \end{aligned} \tag{2.1}$$

where $\mathbf{h}$ is the resolution scale, and $C_0$ is proportional to Smagorinsky-Lilly constant calibrated from direct numerical simulations (Su et al. 2017; Hopkins et al. 2018). For a discussion of the coefficient calibration, see Appendix 2.7. We adopt a value of $C_0 \approx 0.003$. The symmetric traceless shear tensor is given by

$$\mathbf{S} = \frac{1}{2}\left(\nabla\mathbf{v} + (\nabla\mathbf{v})^T\right) - \frac{1}{3}Tr\left(\nabla\mathbf{v}\right), \tag{2.2}$$

where $\mathbf{v}$ is the associated shear velocity.

More simplistically, $D \sim \lambda_{\rm eddy} v_{\rm eddy}$, where the largest unresolved eddies dominate the sub-grid diffusivity, i.e., $\lambda_{\rm eddy} \sim \mathbf{h}$. The only effect of a sub-grid prescription is to smooth variations in metallicity between fluid elements. However, since the shear tensor (Eq 2.2) can be artificially triggered by bulk motion such as rotation, the above model for turbulent diffusion likely over-estimates the true diffusivity. We further address the possibility of over-mixing and illustrate the robustness of our results with respect to the diffusion coefficient in Appendix 2.7.

## 2.3 Metallicity Distribution Functions

In this section, we define stellar metallicity in the simulations and analyze the resulting metallicity distribution functions in relation to observations of Local Group dwarf galaxies.

### Metallicity Definitions

We analyze the stellar-mass weighted[4] metallicity[5] distribution functions of the simulations to quantify the impact of metal diffusion. The abundances from the simulation are calculated from the absolute metal mass fractions per element of a star particle,

$$\begin{aligned}
[X/Y] &= \log_{10}\left(\frac{m_Y M_X}{m_X M_Y}\right) - (\log\epsilon_{X,\odot} - \log\epsilon_{Y,\odot}), \\
\log\epsilon_X &= \log_{10}\left(N_X/N_H\right) + 12,
\end{aligned} \tag{2.3}$$

where $X$ and $Y$ represent chemical species, $m_X$ is the atomic mass for a given species, $M_X$ is the metal mass fraction, and $\epsilon_{X,\odot}$ is the abundance relative to solar (Anders and Grevesse 1989; Sneden et al. 1992), observationally determined from $N_X$, the number density of the species.

[4]We weight the MDFs and mean metallicities by stellar mass, though mass-weighting does not significantly impact these quantities. The FIRE simulations include standard particle splitting and merging such that no particle ever deviates from the median particle mass by more than a factor of 3. That is, 99.99% of all star particles are within 0.2 dex of the median particle mass (Hopkins et al. 2018).

[5]We adopt the observational convention, where metallicity refers to stellar iron abundance ([Fe/H])

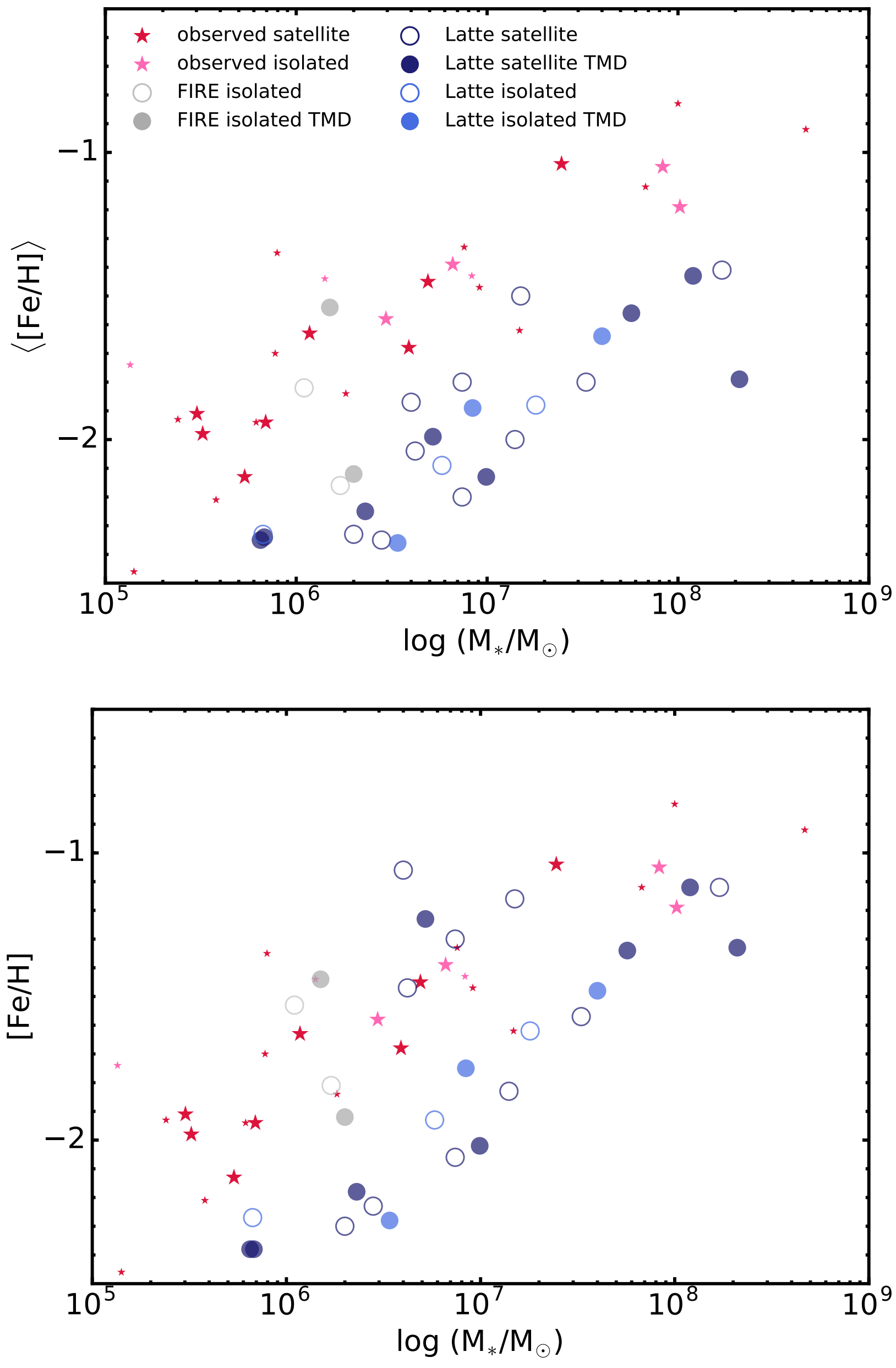

Figure 2.1: A comparison between the stellar mass-metallicity relation (MZR) at $z = 0$ for different definitions of the mean metallicity of simulated galaxies (Section 2.3). All data analysed in this chapter are shown, including isolated dwarfs galaxies from FIRE simulations (grey circles) (Table 2.1), dwarf galaxies from the Latte simulation (blue circles) (Section 2.5), and observed LG dwarf galaxies (red stars) (Table 2.2). The label TMD indicates the presence of sub-grid turbulent metal diffusion (filled circles). Additional observations of LG dIrrs and M31 dSphs are included (small red stars) (Kirby et al. 2013). We do not show observational errors for clarity. (*Top*) The MZR for ⟨[Fe/H]⟩, defined analogously to the observational data (Eq. 2.4). (*Bottom*) The MZR for the definition of mean metallicity used in Eq. 2.5, based on mean metal mass fractions of entire simulated galaxies. Note that the latter definition shows some overlap with the observations for $M_* > 10^6$ $M_\odot$. However, the definition motivated by observational methods is systematically offset from the observed MZR by $\sim 0.3$ dex.

We adopt the following definition of mean metallicity, $\langle[\mathrm{Fe/H}]\rangle$ (Table 2.1), motivated by observational measurements of Local Group dwarf galaxies (Kirby et al. 2010; Kirby et al. 2013; Table 2.2),

$$\langle[\mathrm{Fe/H}]\rangle = \frac{\sum_i^N [\mathrm{Fe/H}]_i \mathrm{m}_i}{\sum_i^N \mathrm{m}_i}, \tag{2.4}$$

where $[\mathrm{Fe/H}]_i$ is the metallicity of an individual star particle, calculated according to Eq. 2.3, $\mathrm{m}_i$ is the mass of the star particle, and $N$ is the total number of star particles in a given simulation. This is in contrast to the definition of mean metallicity based on mass-averaged metal mass fractions used previously in FIRE papers (Ma et al. 2016; Wetzel et al. 2016),

$$\begin{aligned} [\mathrm{Fe/H}] &= \log_{10}\left(\frac{\bar{\mathrm{f}}_{\mathrm{Fe}}}{\bar{\mathrm{f}}_{\mathrm{H}}}\right) - \log_{10}\left(\frac{\mathrm{f}_{\mathrm{Fe},\odot}}{\mathrm{f}_{\mathrm{H},\odot}}\right), \\ \bar{\mathrm{f}}_{\mathrm{Fe}} &= \frac{\sum_i^N \mathrm{f}_{\mathrm{Fe},i}\mathrm{m}_i}{\sum_i^N \mathrm{m}_i}, \end{aligned} \tag{2.5}$$

where $\mathrm{f}_{\mathrm{Fe},i}$ is the absolute iron mass fraction of a star particle. Eq. 2.4 is the mean of the logarithm, whereas Eq. 2.5 is the logarithm of the mean.

The Eq. 2.5 definition is appropriate for more distant galaxies, where the stellar metallicity is determined from galaxy-integrated spectra. In this case, stellar population synthesis models are used to measure Fe and Mg, which dominate the absorption features in stellar atmospheres (e.g., Gallazzi et al. 2005). However, Eq. 2.5 is not consistent with mean metallicity measurements based on spectra of resolved stellar populations, i.e., LG dwarf galaxies.

In the Eq. 2.5 mean metallicity definition, metal-rich stars are weighted heavily, particularly for galaxies with skewed MDFs, such that the scatter in the stellar mass-metallicity relation and the mean metallicity for such galaxies increases. As shown in the right panel of Figure 2.1, this results in a $\sim 0.3$ dex discrepancy in the FIRE stellar mass-metallicity relation between definitions of mean metallicity. Adopting the observationally motivated definition (Eq. 2.4) similarly results in a $\sim 0.3$ dex offset relative to the observed mass-metallicity relation for LG dwarf galaxies, whereas the alternate definition (Eq. 2.5) shows some overlap with observations.

The offset in the FIRE stellar mass-metallicity relation relative to observations of low-mass galaxies is likely caused by systematic uncertainties in the SNe Ia delay time distribution, and potentially the yields. The systematic offset ($\sim 0.03$ dex) from adopting the solar abundances of, e.g., Asplund et al. (2009), is negligible. Assuming a SNe Ia rate with prompt and delayed components (Mannucci, Della Valle, and Panagia 2006), as opposed to a power-law rate (e.g., Maoz and Graur 2017), can result in a factor $\sim 2$ reduction in the number of SNe Ia for a fixed stellar population over 10 Gyr, given the same minimum age for the onset of SNe Ia. Adopting a SNe Ia delay time distribution with a larger integrated number of events could therefore sufficiently increase the FIRE mean metallicity (Eq. 2.4) of simulated dwarf galaxies to result in better agreement with observations. Conclusively resolving this discrepancy is beyond the scope of this chapter, and will be addressed in future work.

Although the offset between the FIRE stellar mass-metallicity relation and observations is $\sim 0.3$ dex for low-mass ($M_* \lesssim 10^9$ $M_\odot$) galaxies, it only impacts the metallicity normalization, as opposed to comparisons of the overall MDF shape, the width of the MDF, and the intrinsic scatter in [$\alpha$/Fe] vs. [Fe/H]. In what follows, [Fe/H] refers to the Eq. 2.4 definition.

**Narrowing Effect of Turbulent Metal Diffusion**

Compared to simulations without sub-grid diffusion, we observe a narrowing of the characteristic width of the MDF when including sub-grid diffusion[6] (Figure 2.2). For **m10q**, the standard deviation of the MDF is reduced from 0.55 dex to 0.41 dex (a factor of $\sim 1.4$), whereas for **m10v** it is reduced from 0.52 dex to 0.34 dex (a factor of $\sim 1.5$, Table 2.1). This is in better agreement with the MDF width for a majority of the LG dwarf galaxies (Table 2.2), particularly for those within the mass range spanned by the simulations.

Although this narrowing effect of the MDF when including diffusion may initially seem counterintuitive, it is in accordance with expectations, given that individual particles are being driven toward the average metallicity as a result of sub-grid mixing. Star particles are no longer born on the low-metallicity tail of the distribution ([Fe/H] $\lesssim -3$ dex), which corresponds to extreme, improbably low metallicities, or the high-metallicity tail ([Fe/H] $\gtrsim -0.5$ dex), which corresponds to metallicities that are not observed in most LG dwarf galaxies.

[6]We acknowledge the potential of run-to-run variations in stellar mass and SFH due to stochasticity caused by random system perturbations (Su et al. 2017). Any individual detailed feature in the MDFs could be due to stochastic fluctuations, but general MDF properties such as the reduction of the MDF width and behavior at the tails are retained in statistical populations of simulated dwarf galaxies (Section 2.5). Stochastic effects are generally small in magnitude compared to the magnitude of systematic effects we observe in the MDF and $\alpha$-element abundance ratio distributions.

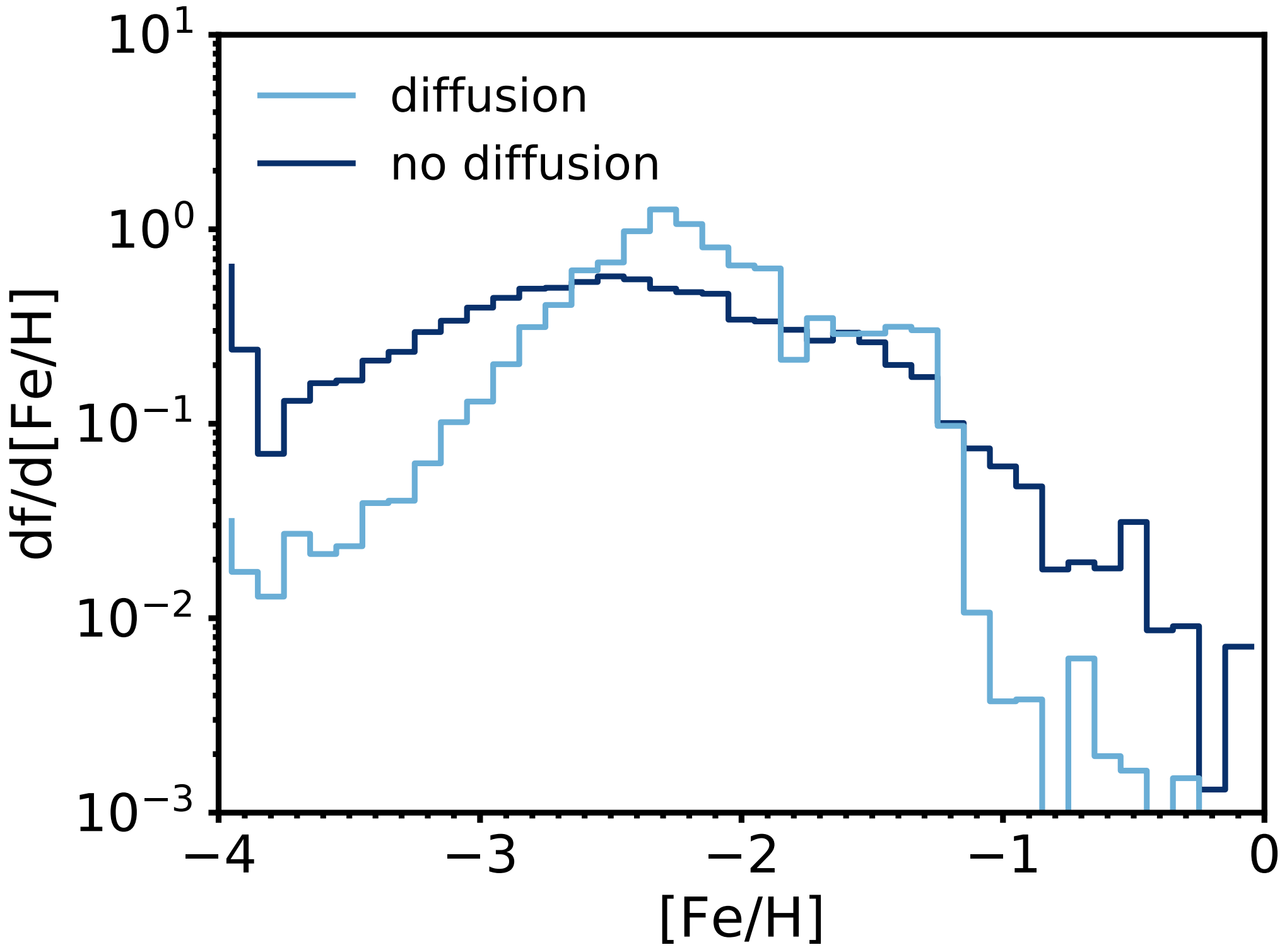

Figure 2.2: An example of the comparison between the stellar mass-weighted metallicity distribution functions of stars in the galaxy at $z = 0$ with and without turbulent metal diffusion for **m10q**. The metallicity distribution functions are plotted in terms of probability density on a log scale to emphasize the behavior at the tail of the distributions. The narrowing effect of diffusion is clear, resulting in a reduction in the width of the MDF by 0.14 dex, or a factor of 1.4 (40%). Similar behavior occurs in the case of **m10v**.

The Milky-Way mass FIRE simulation **m12i** (§5) (Wetzel et al. 2016) ($M_* \sim 6.5 \times 10^{10}$ $M_\odot$), including sub-grid metal diffusion, also exhibits a narrowing of the MDF. Considering that the effects of metal mixing microphysics on galaxy dynamics, as well as other global galaxy properties, are negligible (Su et al. 2017), our results are likely applicable to dwarf galaxy simulations within a broad mass range.

**Comparison to Observed MDFs**

Next, we investigate whether the narrowed theoretical MDFs are in better agreement with observations via comparison to those of MW satellite dwarf spheroidal (dSph) galaxies and LG isolated dwarf irregular (dIrr) galaxies (Table 2.2), for which we have metallicity measurements of $\gtrsim$ 100 red giants per galaxy (Kirby et al. 2010;

Table 2.2: Properties of Local Group Dwarf Galaxies

| Galaxy | D (kpc)[a] | $\log(M_*/M_\odot)$[b] | $\langle$[Fe/H]$\rangle$[c] (dex) | $\sigma$[d] (dex) | $N_{[Fe/H]}$[e] |
|---|---|---|---|---|---|
| **MW dSphs** | | | | | |
| Canes VenI | 217 ± 23 | 5.48 ± 0.09 | −1.91 | 0.44 (0.39) | 151 |
| Draco | 75 ± 5 | 5.51 ± 0.10 | −1.98 | 0.42 (0.35) | 333 |
| Ursa Minor | 75 ± 3 | 5.73 ± 0.20 | −2.13 | 0.43 (0.34) | 670 |
| Sextans | 85 ± 3 | 5.84 ± 0.20 | −1.94 | 0.47 (0.38) | 96 |
| Leo II | 233 ± 13 | 6.07 ± 0.13 | −1.63 | 0.40 (0.36) | 256 |
| Sculptor | 85 ± 4 | 6.59 ± 0.21 | −1.68 | 0.46 (0.44) | 365 |
| Leo I | 253 ± 15 | 6.69 ± 0.13 | −1.45 | 0.32 (0.28) | 774 |
| Fornax | 147 ± 9 | 7.39 ± 0.14 | −1.04 | 0.33 (0.29) | 665 |
| **dIrrs** | | | | | |
| Leo A | 787 ± 29 | 6.47 ± 0.09 | −1.58 | 0.42 (0.36) | 146 |
| Peg dIrr | 920 ± 29 | 6.82 ± 0.08 | −1.39 | 0.56 (0.54) | 99 |
| NGC 6822 | 459 ± 8 | 7.92 ± 0.09 | −1.05 | 0.49 (0.47) | 298 |
| IC 1613 | 758 ± 4 | 8.01 ± 0.06 | −1.19 | 0.37 (0.32) | 132 |

[a] Distance from the Milky Way (Kirby et al. 2014 and references therein).
[b] Stellar masses determined by Woo, Courteau, and Dekel (2008), with the exception of Canes Venatici I (Martin, de Jong, and Rix 2008).
[c] Error-weighted mean metallicity, determined analogously to Eq. 2.4. All galaxies have a standard error of the mean of 0.01 dex, except Leo A, with a standard error of the mean of 0.02 dex (Kirby et al. 2013).
[d] The MDF width (error-corrected in parenthesis), calculated by Kirby et al. 2013.
[e] The number of confirmed radial velocity members with [Fe/H] > −3 dex and measurement uncertainty ≤ 0.5 dex.

Kirby et al. 2013).[7]

To determine the similarity between the observed and simulated MDFs, we quantify the likelihood that the observed stars could have been drawn from the simulated MDF. The log-likelihood is given by $\ln L$,

$$L = \prod_i^n L_i, \tag{2.6}$$

[7] We do not anticipate any bias due to "mass-weighting" of red giants ($M_* \sim 0.8\ M_\odot$) in the mean metallicity or MDF for dSphs, which have uniformly old stellar populations. However, dIrrs contain intermediate age stellar populations, with higher metallicity and longer lifetimes on the red giant branch. For this reason, the mean metallicity of dIrrs may be slightly biased toward higher metallicity, and the MDF width may also be affected (Kirby et al. 2017; Manning and Cole 2017)

$$
\begin{aligned}
L_i = \int \mathrm{d[Fe/H]} \frac{dP}{d\mathrm{[Fe/H]}} \\
\times \left[ \frac{1}{\sqrt{2\pi}\sigma(\mathrm{[Fe/H]})_i} \exp\left( \frac{-(\mathrm{[Fe/H]} - \mathrm{[Fe/H]}_i)^2}{2\sigma(\mathrm{[Fe/H]})_i^2} \right) \right], \\
\frac{dP}{d\mathrm{[Fe/H]}} = \frac{1}{m} \sum_j^m \delta(\mathrm{[Fe/H]})_j
\end{aligned}
\tag{2.7}
$$

where $n$ is the number of measurements for a given observed dwarf galaxy, $i$ corresponds to an individual measurement (i.e. red giant), $m$ is the number of star particles in a given simulation, and $j$ corresponds to an individual star particle. $\sigma$([Fe/H]) is the observed measurement uncertainty in metallicity, and $\delta$([Fe/H]) is a delta function for a star particle with a given metallicity.

We approximate the probability distribution for the theoretical metallicity as a sum of a delta functions with no associated error. When computing the likelihood between each pair of simulated and observed galaxies, we exclude measurements with $\sigma$[Fe/H] > 0.5 dex from the observational data and [Fe/H] < −3 dex from both the observational and simulated data sets. At such low metallicity, it is more likely for the metal content of star particles to be dominated by a single SN event in the simulations. It is also possible for the star particles to be dominated by relics of Population III stars, owing to the initialization of particles at [Fe/H] = −4 in the absence of explicit modeling of the transition to Population II stars. The fraction of stars in the relevant stellar mass range for dwarf galaxies ($10^6$ $M_\odot \lesssim M_* \lesssim 10^8$ $M_\odot$) with [Fe/H] < −3 is $\lesssim$ 4.5% for the high-resolution simulations, such that it impacts the detailed shape of the low-metallicity tail of the distribution, as opposed to the MDF width. Thus, we exclude the potentially unphysical, extremely metal-poor stars present in the simulations that are not seen in observational data.

Figure 2.3 shows the results of the likelihood estimation method of comparison, where the theoretical MDFs are mass-weighted and smoothed to reproduce the effects of observational uncertainty. As an example, we consider the highest likelihood case for each simulated dwarf galaxy, where we compared to 12 LG dwarf galaxies with a variety of stellar masses and SFHs. We emphasize that the effects of including TMD, in relation to observations as discussed below, are generalizable to average populations of simulated dwarf galaxies (Section 2.5).

MW dSphs Ursa Minor (UMi) and Leo I ($M_* \sim 5.4 \times 10^5$ $M_\odot$ and $4.9 \times 10^6$ $M_\odot$ respectively; Woo, Courteau, and Dekel 2008) have the most statistically similar

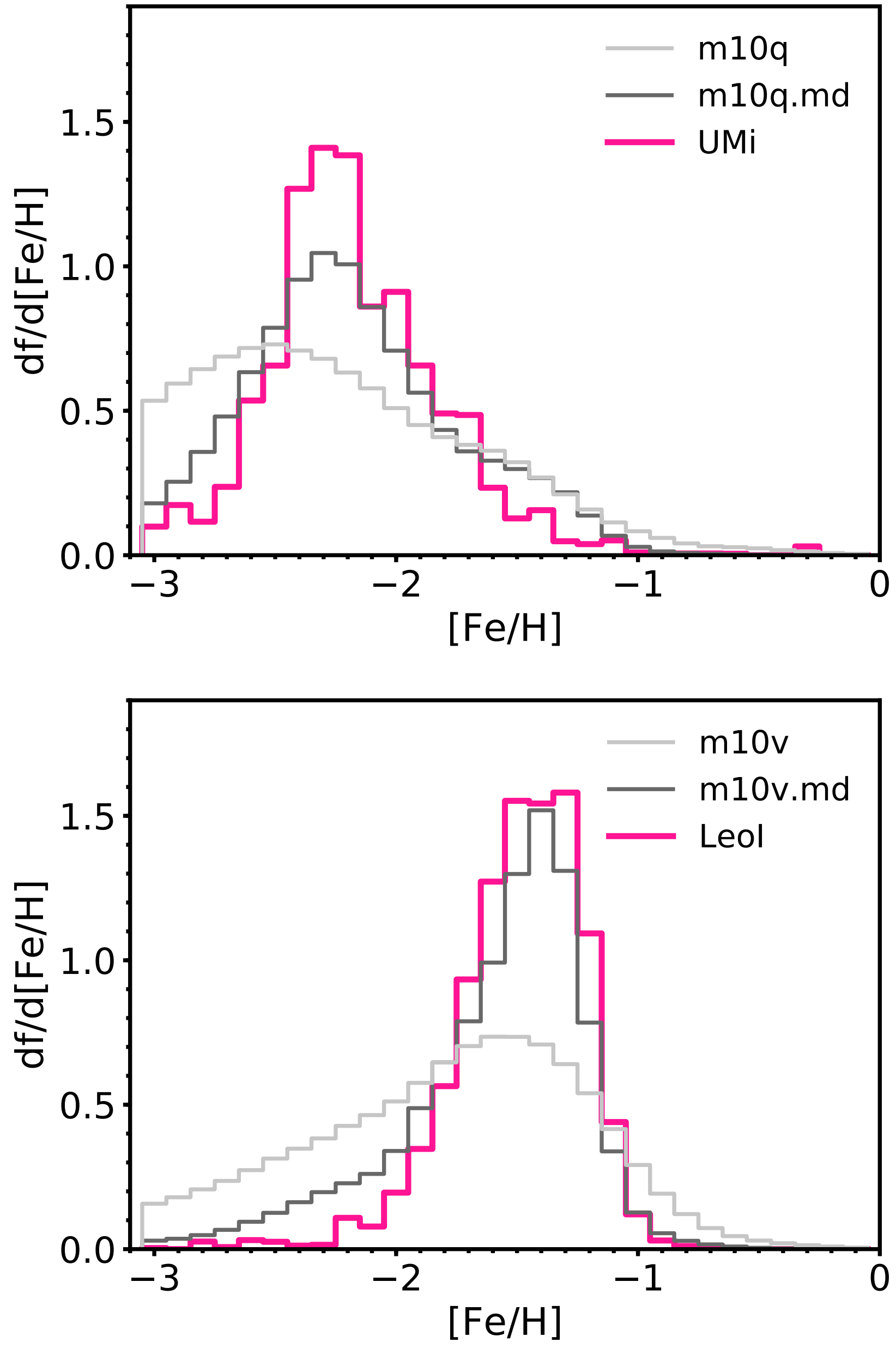

Figure 2.3: A comparison between the MDFs for **m10q** (left) and **m10v** (right) at $z$ = 0 ($M_* \sim 1$ - $2 \times 10^6$ $M_\odot$) with and without TMD plotted against observational data (Kirby et al. 2010) of MW dSph satellite galaxies (Ursa Minor, Leo I: $M_* \sim 5.4 \times 10^5$, $4.9 \times 10^6$ $M_\odot$ respectively; Woo, Courteau, and Dekel 2008). The two observed dSph MDFs exhibit the highest likelihood of being drawn from these two simulations. The theoretical MDFs are weighted by stellar mass and have been smoothed to reproduce the effects of observational uncertainty. By comparison with the MDFs without sub-grid metal mixing, TMD appears to better reproduce the narrowness, and overall shape, of the dwarf galaxy MDFs.

MDFs, according to Eq. 2.7, as compared to both simulations **m10q/m10q.md** and **m10v/m10v.md** respectively.[8] The **m10q.md**/UMi and **m10v.md**/Leo I pairs also have the highest average[9] likelihoods compared to all **m10q**/LG dwarf galaxy and **m10v**/LG dwarf galaxy pairs. The **m10q**/UMi pair has the highest probability of all LG dwarf galaxies of being drawn from that particular simulation. For **m10v**, comparisons to Leo II and Leo I yield similar likelihoods, such that one simulation is not strongly favoured over the other. In the following discussion, we compare **m10v** to Leo I given that it is strongly favoured by **m10v.md**.

Each highest-likelihood pair of simulated and observed dwarf galaxies have similarly shaped SFHs (Weisz et al. 2014), dominated by either an early burst of star formation (**m10q.md**/UMi) or rising late time star formation (**m10v.md**/Leo I). For **m10q** vs. UMi, $\ln L = -437$, whereas for **m10q.md** vs. UMi, $\ln L = -343$. For **m10v** vs. Leo I, $\ln L = -425$, whereas for **m10v.md** vs. Leo I, $\ln L = -129$. It is clear both from the increase in the values of $\ln L$ between cases with and without TMD and Figure 2.3 that TMD improves the ability of the simulations to match observations in terms of MDF shape. In general, the observed MDFs are narrower than the simulated MDFs without TMD (Table 2.1, 2.2). The ability of TMD to reproduce this effect results in the increased likelihood.

With the introduction of metal diffusion, it becomes possible to construct simulated and observed MDFs that are nearly indistinguishable. For both **m10q.md** and **m10v.md**, however, the simulations have a larger population of stars at low-metallicity as compared to observations. The lack of stars at the low-metallicity tail of the observed MDF may be caused by selection effects. Observational bias, which does not significantly affect the mean metallicity, may result in the preferential exclusion of rare, extremely metal-poor stars that tend to inhabit galaxy outskirts. These stars also may have been tidally stripped, now absent from satellite dwarf galaxy stellar populations. However, this only impacts the detailed shape of the metal-poor portion of the MDF (Kirby et al. 2013).

---

[8] In the case of Ursa Minor and **m10q**, we observe a 0.6 dex discrepancy between the simulated and observed galaxy stellar masses for similar mean metallicity. This results in an offset in the FIRE stellar mass-metallicity relation relative to observations. This offset is likely a consequence of our choice of Type Ia SNe delay time distribution, as discussed in Section 2.3. We emphasize that the offset in the FIRE stellar mass-metallicity relation at low stellar masses does not alter any of our conclusions regarding the MDF shape, or scatter in [$\alpha$/Fe] at fixed [Fe/H] (Section 2.4).

[9] The average log-likelihood is defined as $\hat{L} = \frac{1}{n} \ln L$, where $n$ is the number of measurements for a given observed dwarf galaxy. $\hat{L}$ estimates the expected log-likelihood of a single observation, given the "model" simulation. This estimator enables comparisons between different sample sizes, i.e., different LG dwarf galaxies, for a given simulation.

Nonetheless, the tails of the distribution are significantly reduced compared to the case without sub-grid metal mixing (Figure 2.3). The mean metallicities of the distributions approximately coincide, although we note the offset in the normalization of the mean metallicity (Section 2.3). Most significantly, the shapes of the distributions in terms of skewness and kurtosis are consistent with TMD runs, but inconsistent with non-TMD runs. This indicates that metal diffusion may be necessary to bring theoretical predictions into agreement with observations in realistic simulations of the formation and evolution of low-mass galaxies.

Although we have thus far considered only field galaxy simulations, they most resemble the classical dSphs in the sample, as opposed to LG dIrrs such as NGC 6822 and IC 1613, or even more massive dSphs such as Fornax. This is due to the limited mass sampling of our isolated dwarf galaxy simulations, which have masses in the range corresponding to the observed classical dSphs. The likelihood estimation method described above is, to first order, sensitive to mean metallicity, which is dictated by the stellar mass of the galaxy. By including the isolated dwarf galaxies that form in the zoom-in region well beyond the MW-mass host halo in the Latte simulation (Section 2.5), we expand the current isolated galaxy simulation suite with metal diffusion to contain more massive, metal-rich dwarf galaxies ([Fe/H] $\gtrsim$ 1.4) that may provide better analogues to more massive LG dwarf galaxies.

Overall, it appears that TMD results in a better match to observations. However, it is not immediately clear if metal diffusion is the only process that can narrow the MDF sufficiently to match observed galaxies. Alternatively, environmental effects, such as ram-pressure stripping (Lin and Faber 1983; Marcolini, Brighenti, and D'Ercole 2003), may have similar impacts on the appearance of the MDF as sub-grid metal diffusion. We explore this possibility in Section 2.5 by analysing simulated dwarf galaxies from the Latte simulation (Wetzel et al. 2016).

## 2.4 Alpha Element Distributions

In this section, we analyzed the relationship between $\alpha$-element abundances and metallicity in the simulations and performed comparisons to observations of Local Group dwarf galaxies.

### Narrowing of the Alpha-Element Abundance Ratio Distributions

Here, we represent the scatter in [$\alpha$/Fe] using [Si/Fe] as a proxy, owing to the lack of a theoretical analogue to measurements of $\alpha$-enhancement (Section 2.4). Figure 2.4 illustrates that, for the simulations, we observe the same reduction in scatter as in the MDF width for the $\alpha$-element abundance ratio distributions. There is also an apparent reduction in the envelopes corresponding to enrichment events of a single type. The envelopes originate from the initialization of the star particles at [Fe/H] = −4 and [$\alpha$/Fe] = 0, where enrichment by only Type II (Type Ia) SNe results in the upper (lower) envelope. The faint lower envelope disappears entirely with the inclusion of TMD. Sub-grid metal mixing reduces the probability that any given progenitor gas particle, where the star particle inherits the gas particles' metallicity, will only contain elements yielded from a single type of enrichment event.

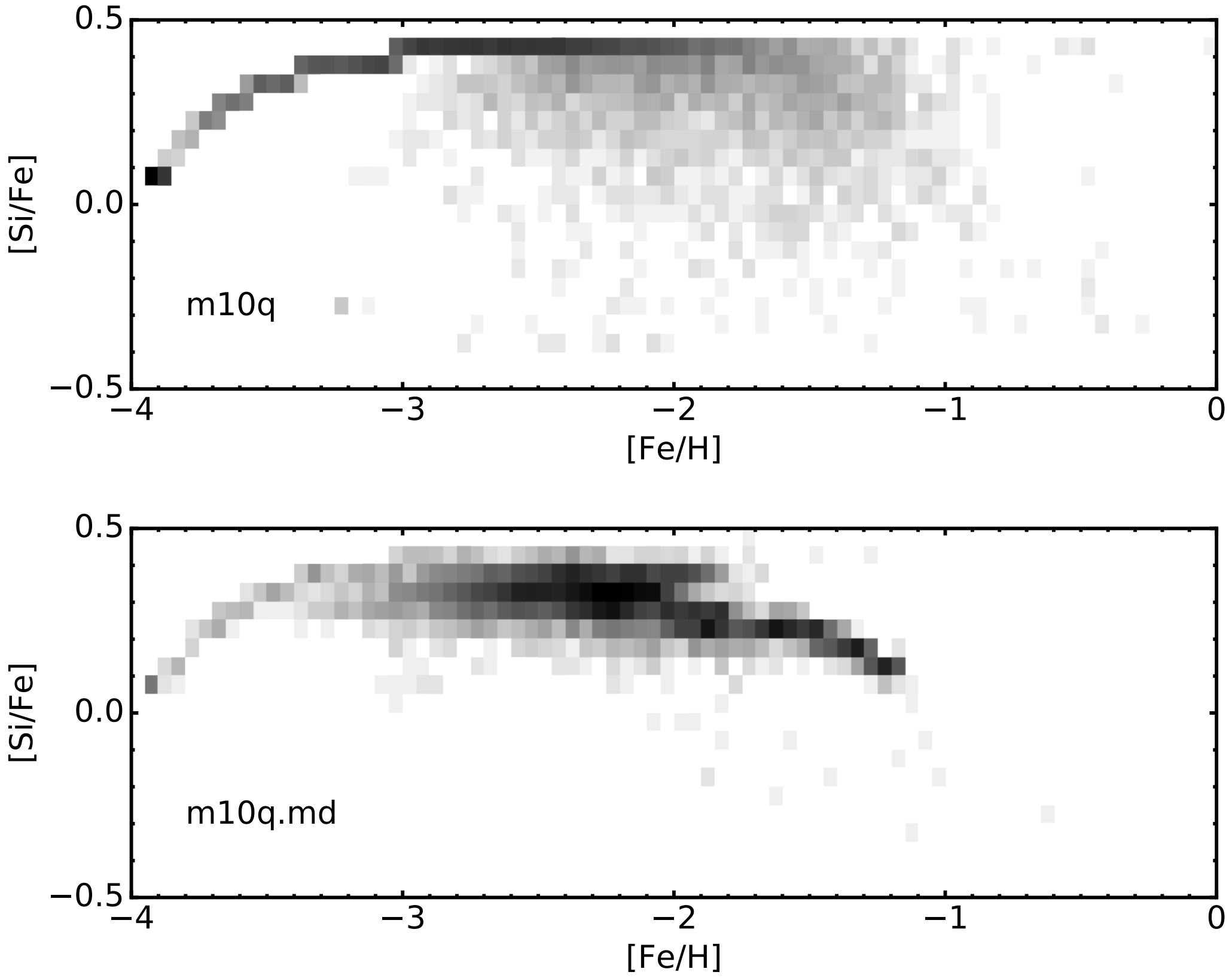

Figure 2.4: Silicon abundance versus iron abundance for **m10q** at $z = 0$ excluding (*top*) and including (*bottom*) sub-grid turbulent metal diffusion. The simulated data is represented as the relative number of star particles in 0.05 dex bins, with the darkest pixels corresponding to $\gtrsim 100$ particles. The reduction in scatter of [Si/Fe] as a function of [Fe/H] is clear (0.14 dex to 0.06 dex with diffusion). Note that the upper (lower) envelope in the case without diffusion corresponds to enrichment by only Type II (Type Ia) SNe. The inclusion of diffusion significantly reduces the envelope for Type II SNe, whereas it vanishes completely for Type Ia SNe. The abundances show less dispersion at a given metallicity with diffusion turned on. This is consistent with observations, which we show in Section 2.4 to have near-zero intrinsic scatter in [Si/Fe] at fixed [Fe/H].

In an analogous fashion to the MDF widths, we quantify the reduction in dispersion in the $\alpha$-element abundance ratio distributions caused by sub-grid turbulent metal diffusion. The intrinsic scatter in [$\alpha$/Fe] as a function of [Fe/H] has physical implications for metal mixing and chemical evolution. [Fe/H] correlates with stellar age, and so can be used as a proxy for time. Thus, a quantification of the intrinsic scatter in $\alpha$-enhancement contains information about the homogeneity of the ISM on typical galaxy evolution timescales.

We analyze the intrinsic scatter, $\Sigma$, in the $\alpha$-element abundance ratios as a function of metallicity for the simulations with and without sub-grid metal diffusion. We calculate $\Sigma$ by fitting a cubic spline (error-weighted for observations) in the range $-3 <$ [Fe/H] $< -0.5$ to the abundance ratios as a function of metallicity for both the simulated and observed LG dwarf galaxies. We assume that [Si/Fe] either remains constant or monotonically decreases with [Fe/H]. To calculate the intrinsic scatter at fixed [Fe/H], we first determine the distance in [Si/Fe] from the curve for each data point (Figure 2.5). The distribution of distances provides a standard deviation, which is the intrinsic scatter for the simulated data. Table 2.3 contains the results for the simulations.

The scatter for **m10q** reduces from $\Sigma = 0.14$ dex to 0.06 dex with sub-grid diffusion, whereas for **m10v**, which is dominated by late-time star formation, $\Sigma = 0.13$ to 0.03 dex. This corresponds to a reduction in the width by a factor of $\sim 1.2$ and 1.3 respectively, which is comparable to the narrowing factor of the MDF width (Section 2.3). The near zero value of $\Sigma$ in the case of TMD implies that metals in the cool gas of the ISM are well-mixed at any given [Fe/H], or time in the galaxy's evolutionary history. We show this explicitly for the simulations in Section 2.4.

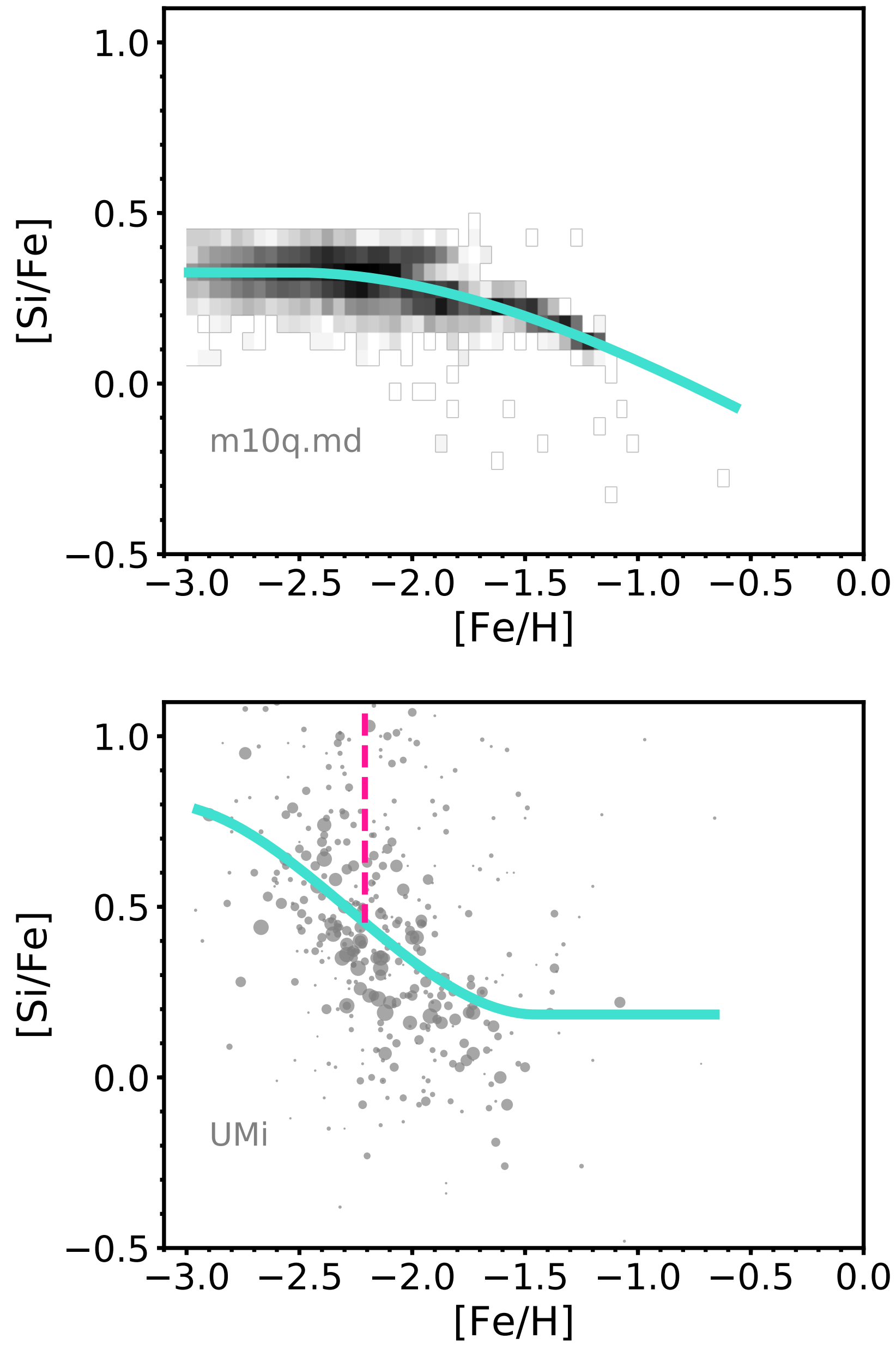

Figure 2.5: An illustration of the intrinsic scatter calculation for **m10q.md** (*top*) at $z = 0$ and Ursa Minor (*bottom*). The simulated data is colour-coded according to the relative number of star particles in 0.05 dex bins. The size of the data points for Ursa Minor is proportional to the inverse-squared measurement uncertainty in [Si/Fe]. The turquoise curve is the best-fit cubic spline, error-weighted for the observational data, assuming that [Si/Fe] either remains constant or monotonically decreases with [Fe/H]. The dotted magenta line is the difference between a data point and the curve approximating $\langle$[Si/Fe]$\rangle$, where this difference is used to determine $\sigma$ (Eq. 2.8).

**The Intrinsic Scatter at Fixed Time**

To identify the origin of the low intrinsic scatter in $\alpha$-elements at $z = 0$ in the simulations, we calculate the scatter at a fixed time in a galaxy's history. Using the formation times of star particles in the galaxy (within $r_{90}$; Table 2.1) at $z = 0$, we determine the dispersion in [Si/Fe], [Si/H], and [Fe/H] of star particles formed in 100 Myr time bins, a timescale comparable to the typical dynamical time of dwarf galaxies.

We base our analysis on $z = 0$ star particles partly because it is analogous to observational methods. Additionally, we expect a negligible contribution from mergers to the $z = 0$ stellar population in dwarf galaxies. As a check, we followed the evolution of simulated dwarf galaxy progenitors to high-redshift. Using a definition of in-situ formation within 10 kpc of the galactic centre, we found that $\gtrsim$ 98% of star particles present in the galaxy at $z = 0$ formed in-situ. Anglés-Alcázar et al. (2017) and Fitts et al. (2018) similarly found in a detailed study that ex-situ star formation contributes negligibly to the stellar mass growth of isolated dwarf galaxies. In principle, all star particles from these mergers could form a distinct [Si/Fe] vs. [Fe/H] track. In this way, all star particles brought in by mergers could contribute to, or even dominate the outliers of, the scatter in abundances at a given age. Despite this, we assume that a significant majority of stars present in simulated dwarf galaxies at $z = 0$ formed in-situ, with a negligible contribution from mergers.

Figure 2.6 illustrates the scatter in [Si/Fe], presented as a standard deviation, $\sigma$([Si/Fe]), with respect to lookback time for FIRE isolated dwarfs with and without sub-grid metal diffusion. We describe the scatter in terms of [Si/Fe] for consistency with the intrinsic scatter calculation (Section 2.4). We overplot the star formation rates, where **m10q/m10q.md** have SFHs dominated by early bursts, whereas late-time star formation dominates in **m10v/m10v.md**. The scatter in [Si/Fe] as a function of age clearly reduces with the inclusion of TMD. At any given time, the typical scatter in all quantities ([Si/Fe], as well as [Si/H] and [Fe/H]) ranges from $\sim$ 0.05 - 0.1 dex. The trend between scatter and age is characterized by mostly near-zero scatter, punctuated by periods of a relatively high rate of star formation that cause the ISM to become inhomogeneous and increase the scatter. We conclude that most of the scatter in [Si/Fe] (and [Si/H] and [Fe/H] by extension) at fixed time results from starbursts. However, the persistent near-zero scatter in the abundances of newly formed stars across cosmic time implies that most of the scatter present in stellar abundance distributions at $z = 0$ is caused by time evolution, as opposed to significant scatter in the abundances of newly formed stars at any fixed time.

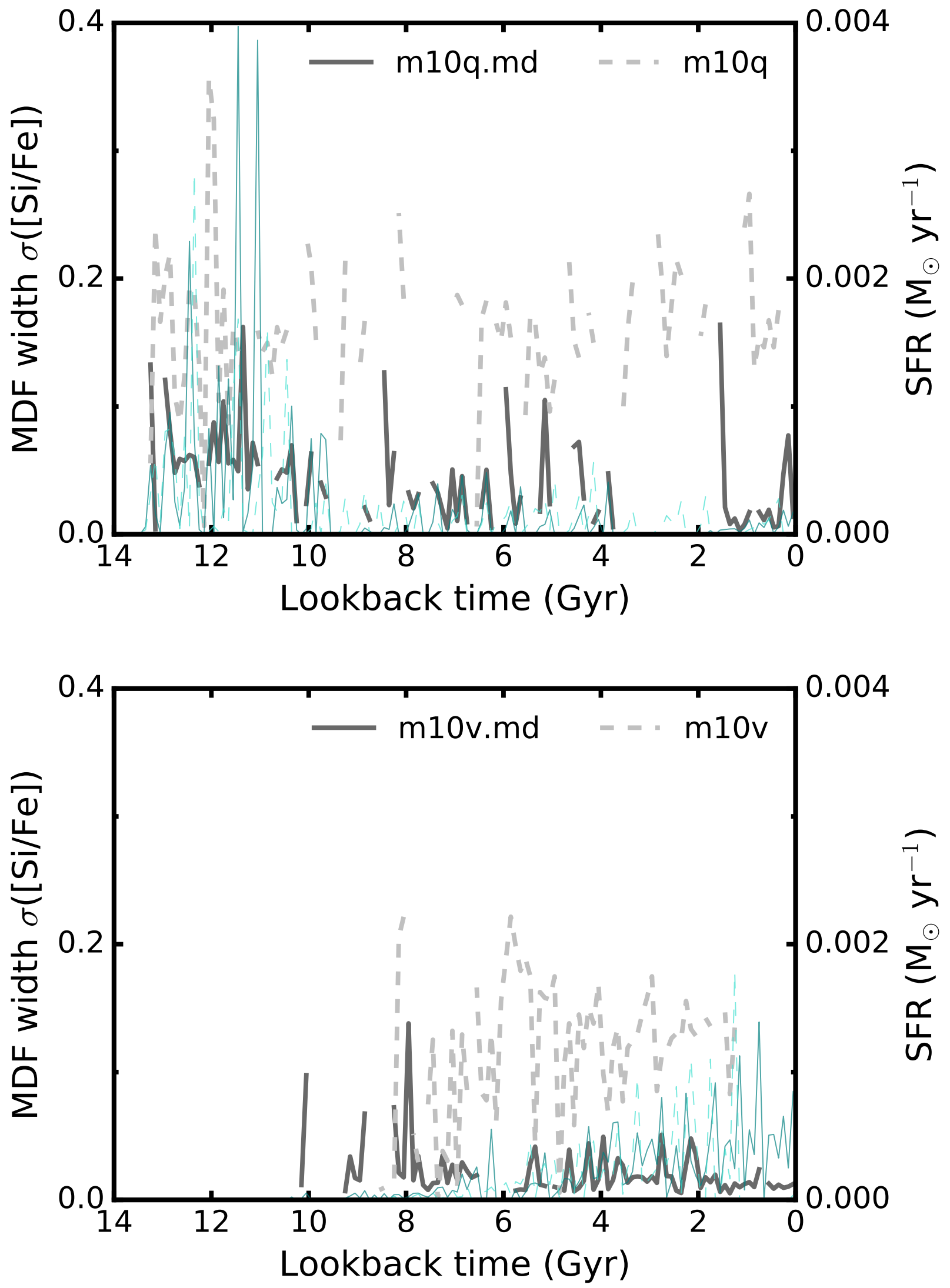

Figure 2.6: The standard deviation of the distribution of [Si/Fe], $\sigma$([Si/Fe]) (grey), for star particles formed in 100 Myr time windows versus lookback time in Gyr. Star formation rates (turquoise) are overplotted for comparison. Both **m10q** (*top*) and **m10v** (*bottom*) are shown for cases with (solid lines) and without (dashed lines) sub-grid metal diffusion. The typical scatter in [Si/Fe] at a fixed time is ~ 0.05 - 0.1 dex, where most of the scatter in [Si/Fe] at a fixed time is caused during short bursts of star formation. This implies that a majority of the $z$ = 0 scatter in the [Si/Fe] distribution (0.06 and 0.03 for **m10q.md** and **m10v.md** respectively) is due to time evolution of the quantity, as opposed to significant scatter in the ISM at a fixed time.

The fact that [Si/Fe] shows small scatter at any given time, where the naive expectation is that it should depend on both [Si/H] and [Fe/H], implies that Si enrichment correlates strongly with Fe enrichment. This is likely a result of the Type Ia and II yields (Iwamoto et al. 1999; Nomoto et al. 2006) and SN rates (Mannucci, Della Valle, and Panagia 2006; Leitherer et al. 1999) employed in FIRE. Type Ia SNe produce a Fe yield an order of magnitude larger than that of Type II SNe, whereas the total Si yield integrated over several Gyr differs only by a factor of 2 between the SNe II and Ia. The cumulative number of core-collapse events per star particle at ages of $\sim$ 100 Myr - 1 Gyr (corresponding to the delay time of Ias in FIRE; Hopkins et al. 2018) is 10 - 20 times larger than the integral number of SNe Ia explosions. As a consequence, the total amount of Fe produced is comparable between SNe II and Ia channels, whereas core-collapse singly dominates Si production. Thus the correlation between Si and Fe enrichment.

We thus conclude that (1) the most likely value of the intrinsic scatter in the abundances of stars forming at any given time across a simulated galaxy's evolutionary history is near-zero, (2) deviations from near-zero scatter at a given time result from starbursts, and (3) Si enrichment correlates strongly with Fe enrichment in FIRE. The ISM is well-mixed at all times in the simulations when taking into account TMD, excepting brief starburst periods. If the intrinsic scatter in [Si/Fe] vs. [Fe/H], where [Fe/H] approximates age, is also small for observations of LG dwarf galaxies, then observed dwarf galaxies have a nearly homogeneous ISM at a given time.

**Observational Intrinsic Scatter**

To determine if TMD produces results consistent with observations, we calculate the observational intrinsic (error-corrected) scatter. Despite having measured [$\alpha$/Fe] directly using spectral synthesis, simultaneously fitting all $\alpha$-element lines present in the spectra, we choose to represent the intrinsic scatter in the observations by a single $\alpha$-element. This is a consequence of the difficulty in interpreting a fit based on multiple chemical species and constructing a theoretical abundance ratio counterpart. [Mg/Fe] tends to have large uncertainties, resulting in fewer data points and a less precise determination of the scatter. Although [Ca/Fe] can be measured with higher precision, the theoretical [Ca/Fe] abundances do not agree with simulations for very low metallicity (Section 2.4). Thus, we adopt [Si/Fe] as a proxy for determining the intrinsic dispersion in $\alpha$-elements as a function of [Fe/H].

Table 2.3: Intrinsic Scatter in [$\alpha$/Fe] as a Function of [Fe/H]

| Galaxy | $\sigma^{2\text{a (dex)}}$ | $\Sigma^{\text{b}}$ (dex) | $s^{\text{c}}$ (dex) |
|---|---|---|---|
| **Simulations** | | | |
| **m10q** | ... | 0.144 | ... |
| **m10q.md** | ... | 0.058 | ... |
| **m10v** | ... | 0.127 | ... |
| **m10v.md** | ... | 0.033 | ... |
| **MW dSphs** | | | |
| Canes Venatici I | 0.901 | $\leq 0.073$ | 0.152 |
| Draco | 1.182 | $0.097^{+0.036}$ | 0.098 |
| Ursa Minor | 0.989 | $\leq 0.061$ | 0.069 |
| Sextans | 1.003 | $0.013^{+0.103}$ | 0.181 |
| Leo II | 1.061 | $0.060^{+0.046}$ | 0.104 |
| Sculptor | 1.173 | $0.078^{+0.021}_{-0.023}$ | 0.079 |
| Leo I | 0.991 | $\leq 0.055$ | 0.056 |
| Fornax | 1.417 | $0.140^{+0.016}_{-0.015}$ | 0.059 |
| **dIrrs** | | | |
| Leo A | 0.962 | $\leq 0.159$ | 0.171 |
| Peg dIrr | 0.667 | ... | 0.174 |
| NGC 6822 | 1.045 | $0.052^{+0.046}$ | 0.092 |
| IC 1613 | 1.083 | $0.089^{+0.082}$ | 0.160 |

Note. — All quantities calculated in terms of [Si/Fe] vs. [Fe/H] at $z = 0$. The ranges of $\Sigma$ represent the most likely values for observational data, although all calculations are consistent with zero.

[a] The initial calculation of the standard deviation of the differences between the data and the curve (Eq. 2.8), assuming zero intrinsic scatter.

[b] The intrinsic scatter, numerically solved for using Eq. 2.8 such that $\sigma^2 = 1$. Not all values of $\Sigma$ have both upper and lower limits, owing to limits on the range of the functional relationship between $\Sigma$ and $\sigma^2$. In some cases, where $\sigma^2 \sim 1$ (e.g., Ursa Minor), only an upper limit is possible to determine.

[c] The standard deviation of the sample variance for observational data (Eq. 2.9).

In the following calculation, we only consider the measurement uncertainty in [Si/Fe], as opposed to simultaneously taking into account the uncertainty in [Fe/H], $\delta$[Fe/H]. In comparison to the range of the data and the scales over which the slope of the relationship between [Si/Fe] and [Fe/H] varies, $\delta$[Fe/H] is insignificant. The typical value of d[Si/Fe]/d[Fe/H] varies between $\sim$ 0.1 - 0.2 and $\sim$ 0.2 - 0.8 for the simulations and observations respectively, which corresponds to a variation in [Si/Fe] of $\sim$ 0.01 - 0.02 and $\sim$ 0.02 - 0.08 dex for a typical $\delta$[Fe/H] $\sim$ 0.1 dex. This is small compared to the variation in [Si/Fe] given a typical $\delta$[Si/Fe] of $\sim$ 0.2 dex. We therefore conclude that the impact on our calculation of the intrinsic scatter owing to measurement uncertainty in [Fe/H] is negligible, such that the uncertainty in [Fe/H] can be reasonably neglected.

For the observational case, we first calculate $\sigma^2$, the variance of the distribution of distances for each galaxy (Figure 2.5), normalizing to the measurement uncertainty, i.e.,

$$\sigma^2 = \mathrm{var}\left[\frac{[\mathrm{Si/Fe}]_i - \mathrm{spline}([\mathrm{Fe/H}]_i)}{\left(\delta[\mathrm{Si/Fe}]_i^2 + \Sigma^2\right)^{1/2}}\right], \tag{2.8}$$

where $i$ is the index for a given red giant with measurements of both [Fe/H] and [Si/Fe], $\delta$[Si/Fe] is the measurement uncertainty in [Si/Fe], and $\Sigma$ is the intrinsic scatter in [Si/Fe] at fixed [Fe/H]. Eq. 2.8 follows a reduced chi-squared distribution with an expectation value of unity. We first calculate $\sigma^2$ assuming zero intrinsic scatter, then enforce the condition $\sigma^2 = 1$ to numerically solve for the most likely value of the intrinsic component, $\Sigma$.

We calculate the uncertainty of each measurement of $\Sigma$ using an unbiased estimator for the standard deviation of the sample variance,

$$s = \sqrt{\frac{2}{N-1}}. \tag{2.9}$$

Eq. 2.9 is dependent on the number of [Si/Fe] measurements, $N$, available for each dwarf galaxy, where in general $N < N_{[\mathrm{Fe/H}]}$ (Table 2.2). We then use $s$ to numerically solve for the corresponding upper and lower limits on $\Sigma$. We present our results for observations in Table 2.3.

If $\sigma^2 = 1$, $\Sigma = 0$, whereas if $\sigma^2 < 1$, the most likely value of $\Sigma$ cannot be properly determined. As long as $\sigma^2 > 1 + s$, we cannot reliably constrain the lower limit for $\Sigma$, as in the case for Draco, Fornax, and Sculptor. In these cases, $\Sigma$ is consistent with zero within one standard deviation. However, we still present the most likely range of $\Sigma$ for these dwarf galaxies in Table 2.3. Even if $\sigma^2 > 1$, not all values have both upper and lower limits, given the functional relationship between $\sigma^2$ and $\Sigma$. If $\sigma^2 < 1 - s$, as in the case of Peg dIrr, overestimated observational errors result in the lack of a well-determined $\Sigma$.

Accordingly, the intrinsic scatter is highly dependent on the magnitude of $\delta$[Si/Fe]. If $\delta$[Si/Fe] is increased (decreased), $\Sigma$ decreases (increases). Thus, the measurement uncertainty in [Si/Fe] must be both accurate and precise to produce a reliable estimate of the intrinsic scatter. The uncertainties of the $\alpha$-enhancement of the observed dwarf galaxies, determined using the Kirby, Guhathakurta, and Sneden (2008) method of spectral synthesis of medium-resolution spectroscopy, have been validated in terms of both precision and accuracy. Kirby, Guhathakurta, and Sneden (2008) showed that errors in [$\alpha$/Fe] determined from medium-resolution spectroscopy remain below 0.25 dex for spectra with sufficiently high signal-to-noise ($\gtrsim$ 20 Å). Based on duplicate observations of red giants in dwarf galaxies and the comparison to measurements of error on [$\alpha$/Fe] from high-resolution spectroscopy, Kirby et al. (2010) showed that the estimated uncertainties on $\alpha$-element abundance ratios are accurate, and that the uncertainties have not been underestimated, even on an absolute scale. This indicates that our analysis sets upper limits on the true value of the intrinsic scatter for observed LG dwarf galaxies.

In most cases, we find that the most likely values of $\Sigma$ range from 0 - 0.1 dex, except Fornax ($\Sigma = 0.14 \pm 0.02$), with upper limits of $\sim$ 0.1 - 0.17 dex. All values are consistent with zero intrinsic scatter. We do not take into account the sampling bias associated with the observations, which primarily target the central, denser, and more metal-rich regions of the dwarf galaxies (Kirby et al. 2011b). However, the calculations of both the likelihood statistic, $\hat{L}$ (Section 2.4, Eq. 2.11) and the intrinsic scatter, $\Sigma$, take into account the limited observational sample size and observational uncertainties. For the most likely intrinsic scatter (Table 2.3), larger, non-zero values result from sample size (Eq. 2.9), excepting Fornax. Assuming that galaxies with larger sample sizes provide a more accurate measurement of the intrinsic scatter, we can conclude that it is near zero for a majority of LG dwarf galaxies.[10]

[10]We note that medium-resolution spectroscopy data sets necessarily have larger uncertainties than those from high-resolution spectroscopy. Although high-resolution data sets do not contain as many stars, the smaller measurement uncertainties would provide a preferable data set for quantifying

If the ISM is indeed homogeneous at a given time, then we should find that the intrinsic scatter at a given metallicity is nearly zero (Section 2.4). We conclude that, based on the near zero intrinsic scatter in [Si/Fe] at a given [Fe/H] in both simulated and observational data, including sub-grid turbulent metal diffusion in simulations produces results consistent with observations. Based on this agreement, we infer that the ISM in LG dwarf galaxies is well-mixed, to within the given scatter, throughout a galaxy's history.

**Comparisons to Observed Alpha-Element Abundance Patterns**

To determine whether the $\alpha$-element abundance ratio distributions with sub-grid metal diffusion improve statistical agreement with observations, we compare the simulations to LG dwarf galaxies. We compute a likelihood estimator, simultaneously using [Fe/H], [Mg/Fe], [Si/Fe], and [Ca/Fe], which are included in both simulated and observed data sets, as constraints.

For each red giant star in a given observed dwarf galaxy that has a measurement for all four abundance ratios, we define a 4D Gaussian per star,

$$g_i = \prod_{j=1}^{4} \frac{1}{\sigma_{[X/Y]_{ji}}\sqrt{2\pi}} e^{-([X/Y]-[X/Y]_{ji})^2/2\sigma^2_{[X/Y]_{ji}}}, \tag{2.10}$$

where $[X/Y]$ is an abundance ratio, $\sigma_{[X/Y]}$ is the associated error, $j$ is the index for the abundance, and $i$ is the index for a star. Again, we considered only measurements with errors below 0.5 dex and [Fe/H] > −3 dex. The 4D probability distribution for an observed dwarf galaxy with $n$ measurements is therefore,

$$f([X/Y]) = \frac{1}{n}\sum_{i=1}^{n} g_i([X/Y]), \tag{2.11}$$

where the prefactor is included such that the integral of the function over 4D space is unity. This can be used to estimate the log-likelihood,

$$\begin{aligned} L &= \prod_{k=1}^{m} f([X/Y]_k), \\ \hat{L} &= \frac{1}{m}\ln L, \end{aligned} \tag{2.12}$$

---

the intrinsic scatter. For example, high-resolution data sets for Sculptor (de Boer et al. 2012b; Hill et al. 2019) and Fornax (Letarte et al. 2010; de Boer et al. 2012a) could be used for this purpose.

where $m$ is the number of star particles in a given simulated galaxy and $k$ is the index of each particle. In this case, we cite the average log-likelihood (Section 2.3) to enable comparisons between different simulation/LG dwarf galaxy pairs.

We choose to include calcium abundances in the statistic determination despite a systematic offset in [Ca/Fe] in the simulations relative to observations. The simulation yields for Type II SNe result in a maximum of [Ca/Fe] = 0.1 dex at low metallicity, despite observations indicating that [Ca/Fe] $\sim$ +0.2 - +0.4 dex (Venn et al. 2004; Kirby et al. 2011b). This is likely a consequence of our assumption that there is no strong dependence of the yield on metallicity for [Fe/H] $< -2$ dex in GIZMO. This is in contrast to the predictions of Nomoto et al. (2006), which assumes metallicity dependence of the yields at low [Fe/H].

We computed the likelihoods both with (4D) and without (3D) including calcium in the product in Eq. 2.10. The results of the likelihood comparison for the abundance ratios do not differ substantially between the 3D (excluding calcium) and 4D cases in terms of the maximum likelihood matches, so we adopt the results from the 4D case, assuming that it includes additional information and subsequently provides a tighter constraint.

The comparison between the pairs of observed and simulated galaxies results in the highest likelihoods (compared to every simulation/LG dwarf galaxy pair) of $\hat{L} = -0.456$ and $\hat{L} = -0.542$ for **m10q.md**/UMi and **m10v.md**/Leo I respectively. In comparison to the simulations without sub-grid metal diffusion, **m10v**/UMi and **m10q**/UMi have the highest likelihoods of $\hat{L} = -1.798$. By including TMD, we gain an increase in the statistical likelihood that the simulated and observed galaxies are similar. Thus, we further establish (Section 2.3) that Ursa Minor and Leo I are the most statistically similar dwarf galaxies in our sample to the simulations with metal diffusion, using abundance ratios in addition to MDFs. Including sub-grid diffusion enables galaxy simulations that provide good statistical analogues to observed dwarf galaxies.

## 2.5 Environmental Effects

We apply the same methodology outlined in Section 2.3 and Section 2.4 to analyse dwarf galaxy simulations captured in the high resolution regions of the Latte simulations (Wetzel et al. 2016). Latte is run with GIZMO in MFM-mode (Hopkins 2015) and includes the standard FIRE-2 implementation of gas cooling, star formation, stellar feedback, and metal enrichment as summarized in Section 2.2 (Hopkins et al. 2018). We consider the **m12i** simulations, with baryonic mass resolution of 7070 $M_\odot$ and star particle spatial resolution of 4 pc, run with and without TMD (Section 2.2). The original **m12i** did not include TMD. We present a version of the simulation, rerun with the same initial conditions and physics, including sub-grid metal diffusion in this chapter. We consider (sub)halos uncontaminated by low-resolution dark matter particles in the stellar mass range $5.5 \times 10^5$ $M_\odot$ < $M_*$ < $9.9 \times 10^9$ $M_\odot$, where the lower limit is based on our metallicity convergence tests (Hopkins et al. 2018). We define "satellite" and "isolated" dwarf galaxies around the MW-like host ($M_* \sim 6.5 \times 10^{10}$ $M_\odot$) by distance, $d_{\rm host}$, with $d_{\rm host}$ < 300 kpc and 300 kpc < $d_{\rm host}$ < 1 Mpc respectively, considering only simulated dwarf galaxies within the distance range of the observed LG dwarf galaxies (Table 2.2). Using these criteria, we identify 10 satellite and 3 isolated dwarf galaxies for **m12i** with $M_* \sim 7 \times 10^5$ - $2 \times 10^8$ $M_\odot$, covering a majority of the mass range spanned by the observed dwarf galaxies.

The Latte **m12i** simulation with TMD (**m12i.md**) is statistically consistent with the satellite mass function and the stellar 1D velocity dispersion of **m12i** without TMD (Wetzel et al. 2016). **m12i.md** similarly falls between the mass functions for the Milky Way and M31 (excluding the LMC, M33, and Sagittarius). Although **m12i.md** produces fewer satellites at $M_* \sim 10^6$ $M_\odot$, this likely arises because of stochasticity in when satellites are disrupted by the host (Garrison-Kimmel et al. 2017): the mass functions of isolated dwarf galaxies ($d_{\rm host}$ > 300 kpc), which are not affected by disruption, are nearly identical. Qualitatively, the stellar mass-metallicity relation between runs is in broad agreement, where the simulated dwarf galaxies agree with observations for $M_* \gtrsim 10^6$ $M_\odot$ for a definition of the mean metallicity based on average mass fractions (Section 2.3). This is consistent with our previous results, where including TMD narrows the scatter in the metallicity distribution, but does not significantly change the galaxy-averaged mean metallicity.

### Satellite and Isolated MDFs

Including TMD, we observe the same narrowing effect in the widths of the MDFs for the Latte satellite and isolated dwarf galaxies as in the isolated FIRE dwarf galaxies. Figure 2.7 shows the average MDF for dwarf galaxies with $M_* > 10^6$ $M_\odot$, excluding and including TMD. We show the MDFs for both satellite and isolated dwarf galaxy populations, where we normalize each individual dwarf galaxy MDF to its mean metallicity before averaging. The reduction in the width of the MDF is apparent, narrowing from 0.50 dex to 0.42 dex (a factor of ~ 1.2) on average for combined satellite and isolated galaxy populations. This agrees with our higher-resolution isolated dwarf galaxy simulations in Section 2.3.

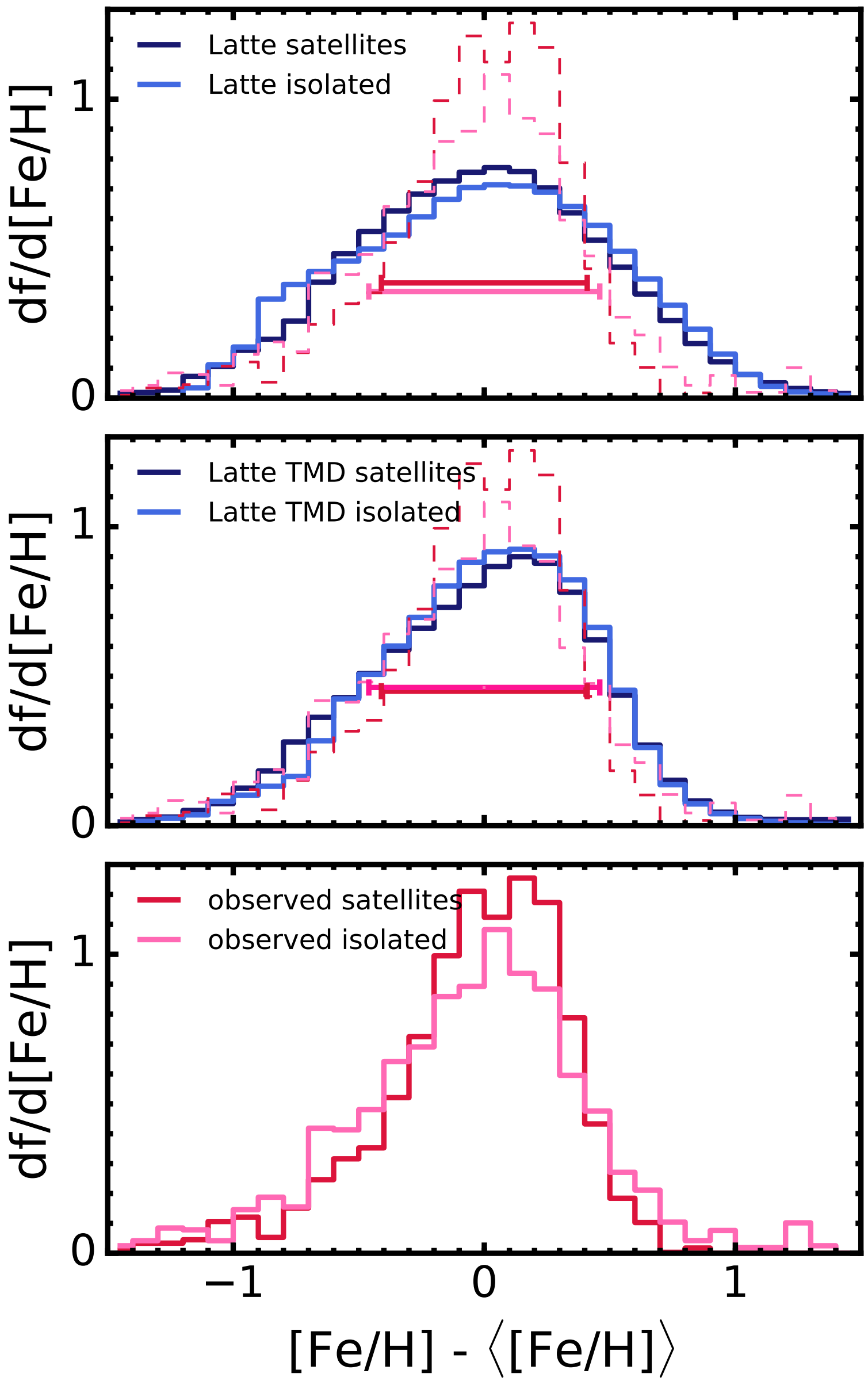

Figure 2.7: A comparison between average MDFs, relative to the mean metallicity for each dwarf galaxy, for both satellite and isolated dwarf galaxies with $M_* > 10^6$ $M_\odot$at $z = 0$. Data is shown for the Latte simulations without (*top*) and with (*middle*) TMD and observed LG dwarf galaxies (*bottom*). The simulated MDFs are mass-weighted and smoothed to reproduce the effects of measurement uncertainty, whereas the observational MDFs are error-weighted. For comparison, we overplot average width of the observed MDFs for both satellite (red) and isolated (pink) dwarf galaxies at half maximum, as well as the full observed MDFs (dotted), on the simulated data. TMD brings the MDFs into better agreement with the observations in terms of both width and shape of the distribution, where status as a satellite or isolated galaxy does not appear to impact MDF appearance for the simulations. A distinction between satellite and isolated galaxies may be present in the observed dwarf galaxies, although differences could be due to observational bias.

Figure 2.8 shows the MDF width, as a function of stellar mass, for all simulated and observed data sets. As also illustrated in Figure 2.7, the Latte dwarf galaxies with TMD have narrower MDF widths on average. The values for the Latte dwarfs are consistent with those from the FIRE isolated dwarfs, albeit potentially exhibiting more scatter. Figure 2.8 emphasizes that the simulations with TMD show reasonable agreement with observational, error-corrected MDF widths (Table 2.2), as opposed to runs without TMD. This is especially true for satellite dwarf galaxies. No apparent trend between MDF width and stellar mass exists for both simulated and observed dwarf galaxies. This implies that factors other than stellar mass, such as the star formation history, dictate the MDF width. In addition, it suggests that the MDF width converges in the simulations (Appendix 2.7), since there is no mass-dependent behavior that may result from secondary resolution effects (Section 2.5).

For the simulations, both Figures 2.7 and 2.8 do not show any systematic differences between MDF width for satellite versus isolated galaxies. The differences between the average individual MDF widths (mass-weighted and un-smoothed) across the entire stellar mass range of simulated satellite (0.51 to 0.43 dex, or a factor of $\sim$ 1.2) and isolated (0.47 dex to 0.39 dex, or a factor of $\sim$ 1.2) galaxies without and with TMD are comparable. For both **m12i** with and without TMD, the average MDF width of isolated dwarf galaxies is narrower than the corresponding average for satellite dwarf galaxies for $M_* > 10^6$ $M_\odot$ and $d_{\rm host} < 1$ Mpc. Within these constraints, there are only a few isolated dwarf galaxies in each run of **m12i**, as compared to the more numerous satellite dwarf galaxies. Expanding the sample size by incorporating a couple of uncontaminated isolated dwarf galaxies ($d_{\rm host} >$ 1 Mpc and $10^7$ $M_\odot < M_* < 10^9$ $M_\odot$) from the **m12i** simulated volume brings the average individual MDF width of isolated dwarf galaxies up to 0.50 dex and 0.42 dex, without and with TMD respectively. Thus, the average individual MDF widths between the simulated satellite and isolated dwarf galaxies are similar.

By including sub-grid metal diffusion, the average simulated MDF widths better approximate the averages of the error-corrected MDF width of individual observed galaxies (Table 2.2) across the entire stellar mass range for each galaxy population. The observed average satellite MDF width is 0.35 dex, whereas the observed average isolated MDF width is 0.42 dex, in comparison to 0.43 dex and 0.42 dex respectively for **m12i.md**. The observed average satellite MDF width is $\sim$ 80% that of the observed isolated galaxies, at odds with predictions from our simulations.

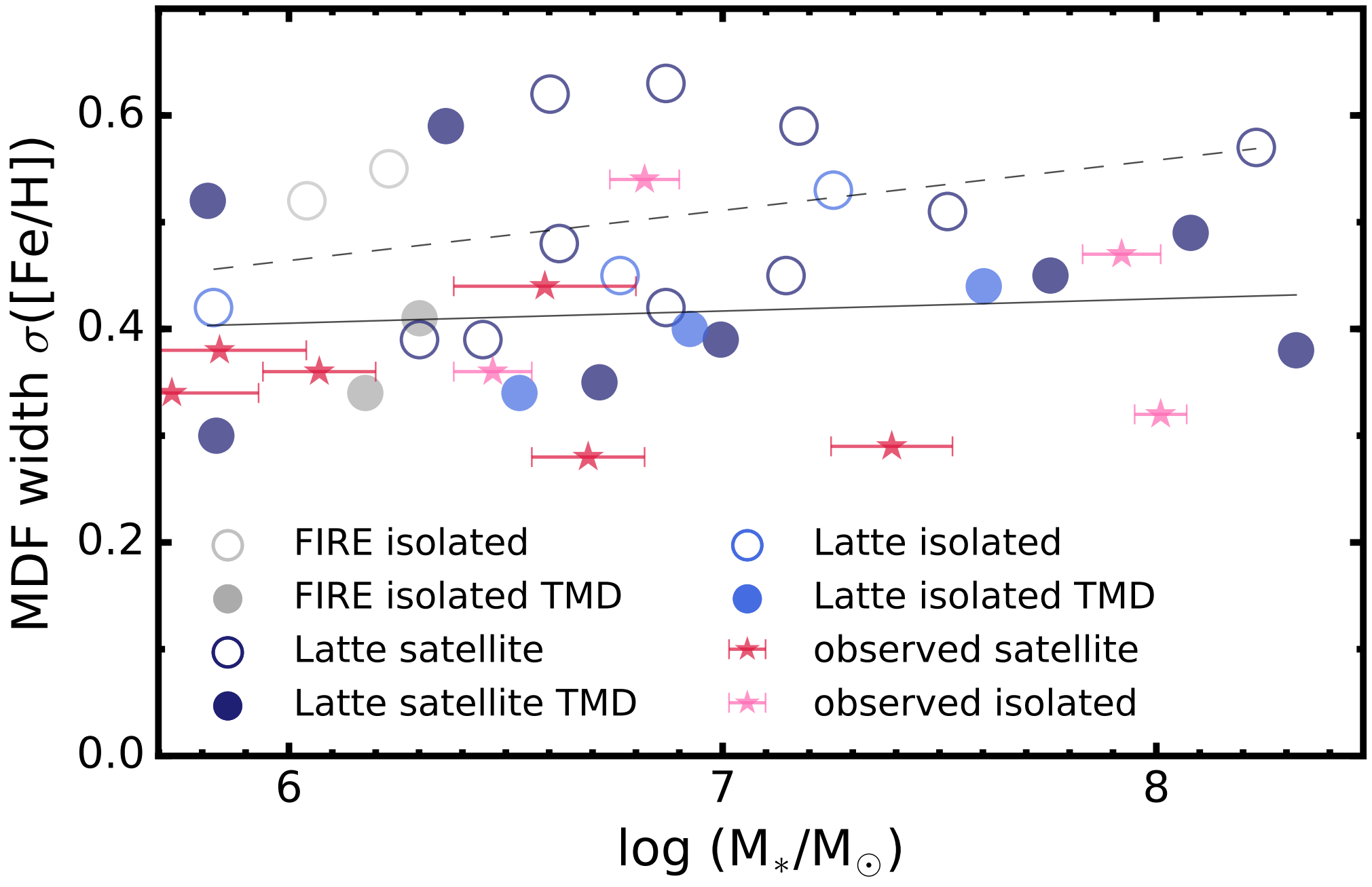

Figure 2.8: The MDF width as a function of stellar mass at $z = 0$. Best-fit lines for the trends in the simulations with stellar mass are shown for cases with (solid) and without (dashed) TMD. Latte satellite (dark blue circles) and isolated dwarf galaxies (light blue circles) with sub-grid metal diffusion, on average, have narrower MDFs across the entire stellar mass range, as compared to the run without sub-grid metal diffusion (open circles). For comparison, the error-corrected MDF width (Table 2.2) for observed satellite (red stars) and isolated (pink stars) dwarf galaxies are shown. No apparent trend of the MDF width with stellar mass is present for all simulated dwarf galaxies with sub-grid metal diffusion. Simulations including sub-grid diffusion agree reasonably with observations, and values for the MDF width between the field FIRE and Latte dwarf galaxies are consistent with each other.

Figure 2.7 shows the analogous observational average MDFs, separated according to satellite and isolated dwarf galaxy populations. In contrast to the simulations, the average satellite MDF is more sharply peaked and narrow (0.41 dex) than the broader average isolated MDF (0.46 dex), including scatter due to observational uncertainty. Kirby et al. (2013) found a less than 0.02 % likelihood that the distributions originate from the same parent distribution, attributing the disparity to differences between star formation histories (truncated vs. extended) for satellite and isolated dwarf galaxies (Mateo 1998; Orban et al. 2008; Weisz et al. 2014). Despite the similarity of the simulated satellite and isolated MDFs, both the Latte simulations and observations show no systematic difference in ⟨[Fe/H]⟩ at a given stellar mass for satellite and isolated dwarf galaxies. Further investigation of this discrepancy, such as quantifying impact of observational bias, is beyond the scope of this chapter.

However, the similarity between satellite and isolated galaxy MDFs for the same stellar mass range in the simulations indicates that TMD ultimately produces better agreement with observations. The alternate hypothesis, in which TMD mimics the impact of environmental effects (Section 2.3), such as ram-pressure stripping, on the appearance of satellite MDFs, is thus excluded for the FIRE simulations. This is further supported by a likelihood estimation comparison between the observed MDFs and the Latte simulations, in which the most similar simulated galaxies to observations were dictated solely by MDF shape, i.e., stellar mass range and star formation history, as opposed to any innate separation between isolated and dwarf galaxy MDFs. In general, the simulation suite including TMD provides better matches to the observed MDF shapes compared the case without TMD.

Figure 2.7 illustrates that TMD ultimately better reproduces the narrowness of the observed MDFs, the truncation of stars at high metallicity, and the skew toward high metallicity. We do not observe the latter effect in the average (or any individual) MDFs for the Latte simulations without sub-grid diffusion, for both isolated and satellite galaxies. The characteristic cutoff at high metallicity in MDFs can be attributed to ram-pressure stripping and the subsequent quenching of star formation (Bosler, Smecker-Hane, and Stetson 2007). Despite this, we see similar cutoffs at high metallicity for both satellite and isolated galaxies, where isolated galaxies remain star-forming to $z = 0$. It is unclear to what extent the cutoff is driven by environmental effects versus internal chemical evolution. Nonetheless, the cutoff appears in the simulations for $M_* \gtrsim 10^7$ $M_\odot$ *only* in the case of sub-grid metal mixing. This suggests that turbulent metal diffusion is the primary source of agreement between simulations and observations, in contrast to any significant role played by interaction with the host galaxy in the simulations.

**Intrinsic Scatter in the Latte Simulations**

We calculate the intrinsic scatter for Latte dwarf galaxies, with and without TMD. The average reduction in the intrinsic scatter for galaxies with $M_* > 10^6$ $M_\odot$ is from 0.16 to 0.12 dex, or a factor of $\sim$ 1.1. This is comparable to the reduction in the width of the MDF and the average reduction factor for the FIRE isolated dwarfs ($\sim$ 1.2). However, the reduction in intrinsic scatter including TMD at a given stellar mass is less pronounced (Figure 2.9) in the lower-resolution Latte simulations, as compared to the higher-resolution isolated FIRE dwarf galaxies. In addition, the Latte dwarf galaxies have larger intrinsic scatter overall as compared to the FIRE isolated dwarf galaxies at $M_* \lesssim 10^{6.5}$ $M_\odot$.

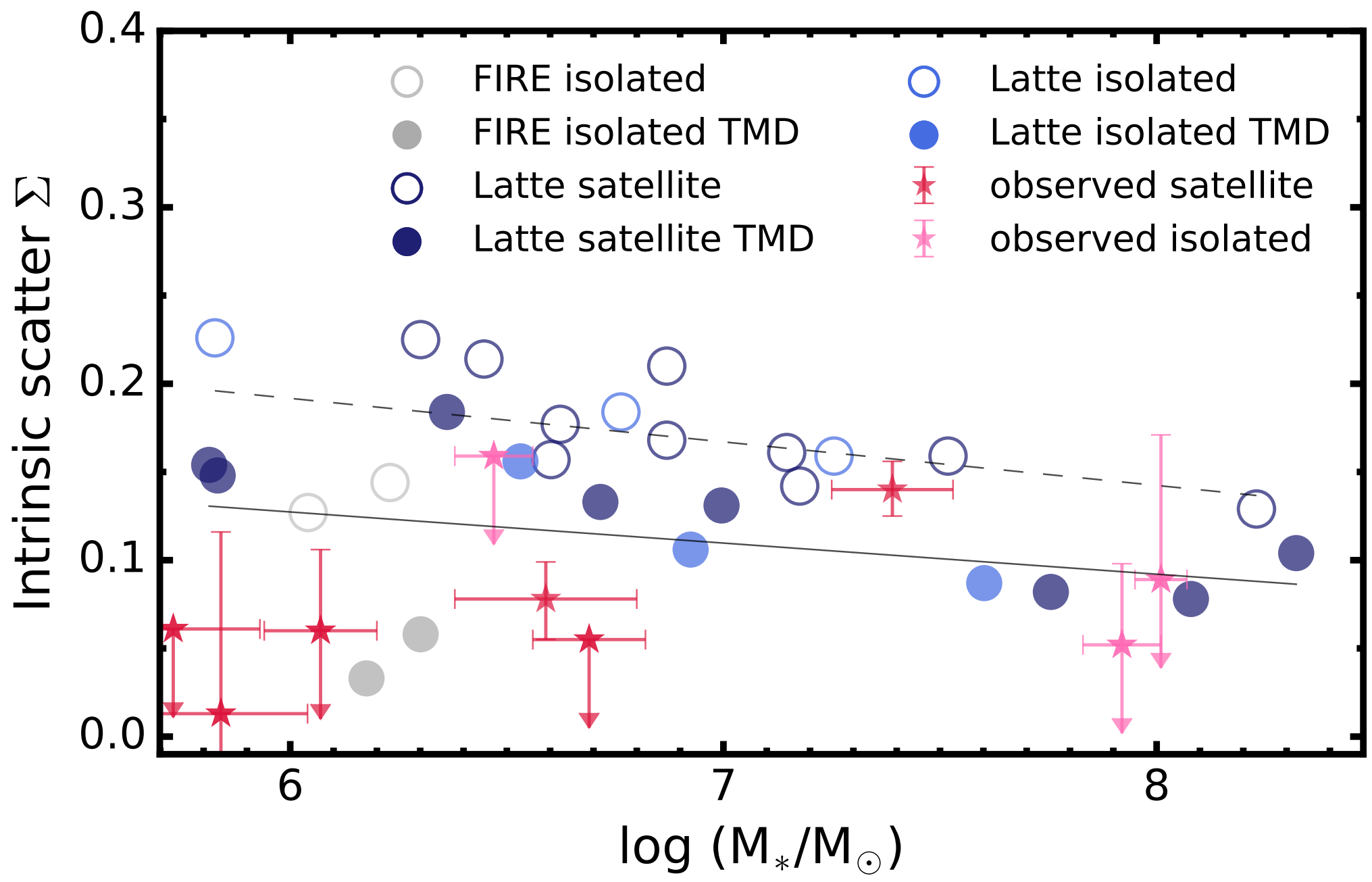

Figure 2.9: The intrinsic scatter in [Si/Fe] at fixed [Fe/H] as a function of stellar mass at $z = 0$. Best-fit lines for the trends in the simulations with stellar mass are shown for cases with (solid) and without (dashed) TMD. Latte satellite (dark blue circles) and isolated dwarf galaxies (light blue circles) with metal diffusion have lower intrinsic scatter across the stellar mass range $M_* \sim 10^6$ - $10^8$ $M_\odot$ as compared to the runs without sub-grid metal diffusion (open circles). The reduction in the intrinsic scatter upon including TMD is less pronounced for the Latte simulations as compared to the FIRE isolated dwarfs (grey circles) at a given stellar mass for $M_* \lesssim 10^{6.5}$ $M_\odot$, owing to resolution effects. For comparison, upper limits of the the most likely values of the intrinsic scatter for the for observed satellite (red stars) and isolated (pink stars) dwarf galaxies are shown. The higher-resolution isolated dwarf galaxy FIRE simulations, for which the effects of TMD are more important, are in better agreement with MW dSphs, which tend to have better measurements of $\alpha$-enhancement than LG dIrrs.

This is likely caused by secondary resolution effects (Appendix 2.7). Numerically enhanced burstiness (Hopkins et al. 2018) results in stronger periods of inhomogeneity for the less well-resolved low-mass ($M_* \lesssim 10^{6.5}$ $M_\odot$) Latte dwarf galaxies. By definition, the intrinsic scatter is more sensitive to inhomogeneities than the MDF width, which is more sensitive to the long-term star formation history, artificially increasing the dispersion in the $\alpha$-element abundances at fixed metallicity. For this reason, low-mass Latte dwarf galaxies do not agree with most observations of the MW dSphs with near-zero intrinsic scatter ($\lesssim 0.1$ dex), which have more reliable measurements of $\alpha$-enhancement than LG dIrrs (Figure 2.9). The impact of turbulent metal diffusion is more important for higher-resolution simulations, in which the effects of numerical noise become more pronounced (Appendix 2.7, 2.7), in terms of bringing simulations of dwarf galaxies into agreement with observations.

## 2.6 Summary & Discussion

We have examined the metallicity distribution functions and enrichment histories, including [Fe/H] $\alpha$-element abundances, of dwarf galaxies using the FIRE-2 cosmological simulations of dwarf galaxies, to investigate the chemical enrichment histories of dwarf galaxies. We have shown that turbulent metal diffusion, at levels suggested by converged simulations, is necessary to include in Lagrangian, hydrodynamical simulations of the formation and evolution of dwarf galaxies to obtain realistic predictions of chemical evolution.

Contrary to a majority past studies of the chemical properties of simulated dwarf galaxies, we have successfully modeled full abundance distributions, in addition to global properties such as the stellar mass-metallicity relation. As a caveat, we note the presence of an offset in the normalization of our stellar mass-metallicity relation compared with observations of low-mass dwarf galaxies. This is likely caused by systematic effects, such as our choice of the Type Ia SNe delay time distribution. However, this does not impact properties such as the MDF width and the intrinsic scatter in [$\alpha$/Fe] at fixed [Fe/H].

Through statistical comparison of FIRE simulations to observations of LG dwarf galaxies, we have demonstrated that simulations including TMD are in agreement with the width of the MDF and the intrinsic scatter in [$\alpha$/Fe] vs. [Fe/H]. For both the MDF and the $\alpha$-element abundance ratios, a reduction in the scatter occurs compared to simulations without TMD, as well as a reduction in numerical artifacts such as star particles with [Fe/H] < −3 and envelopes corresponding to Type Ia/Type II SNe yields.

The same effects are present in both satellite and isolated dwarf galaxies from the Latte simulation, albeit the reduction in [$\alpha$/Fe] vs. [Fe/H] scatter is subdued owing to resolution effects. Most significantly, we find that a distinction between satellite and isolated dwarf galaxies does not factor into our conclusions on the agreement between simulations and observations in terms of the shape and width of the MDF and dispersion in the $\alpha$-element abundance ratio distributions. Just as simulated and isolated dwarf galaxies are similar in these quantities, all dwarf galaxies, both observed and simulated, lie on the same stellar mass-metallicity relation regardless of environment or star formation history (Skillman, Kennicutt, and Hodge 1989; Kirby et al. 2013). In addition, both satellite and isolated dwarf galaxies primarily form as dispersion-dominated systems regardless of current proximity to the host (Wheeler et al. 2017; Kirby et al. 2017). This also poses a challenge to the traditional separation between dwarf galaxy populations. Our work suggests that galactic chemical evolution depends predominantly on the stellar mass of dwarf galaxies.

We analyze realistic galaxy evolution and formation simulations, taking into account a multi-phase ISM, explicit stellar feedback, and the impact of cosmological accretion. Our analysis serves as a robust confirmation of previous work done in the case of idealized, non-cosmological simulations. We have illustrated that a sub-grid turbulent diffusion model, owing to the physically-motivated nature of the implementation and its ability to match observations, is a valid alternative to methods such as a smoothed-metallicity scheme (Wiersma et al. 2009; Revaz et al. 2016). Similar to Williamson, Martel, and Kawata (2016), we find that the strength of mixing due to turbulent diffusion is stable against variations in the diffusion coefficient within an order of magnitude above a minimum diffusion strength (Appendix 2.7).

In addition, for the first time, we have presented an explicit calculation of the intrinsic scatter from medium-resolution spectroscopy of 12 LG dwarf galaxies. Previous studies of metal poor Galactic stars (Carretta et al. 2002; Cayrel et al. 2004; Arnone et al. 2005) similarly found near-zero intrinsic scatter for $\alpha$-elements at fixed metallicity. Based on $\alpha$-element abundance ratios (see Hirai and Saitoh 2017 for an analysis based on barium abundances), we conclude that the timescale for metal mixing is shorter than the typical dynamical timescale for dwarf galaxies. This is evidenced by the homogeneity of the ISM, as implied by the near zero intrinsic scatter for LG dwarf galaxies.

The implication of a well-mixed ISM for one-zone chemical evolution models (e.g., Lanfranchi and Matteucci 2003; Lanfranchi and Matteucci 2007, 2010; Lanfranchi, Matteucci, and Cescutti 2006) is that 3D hydrodynamical models (Mori, Ferrara, and Madau 2002; Revaz et al. 2009; Sawala et al. 2010) may not be necessary to relax the instantaneous mixing approximation, since the cool gas of the ISM becomes homogeneous within approximately a dynamical time for dwarf galaxies. This is in contrast to previous studies (Marcolini et al. 2008) that investigated the effects of inhomogeneous pollution by SNe on chemical properties in 3D hydrodynamical simulations of isolated dSphs. However, Marcolini et al. (2008) did not include a prescription for sub-grid metal mixing, which washes out temporary inhomogeneities in the ISM. Based on our analysis, one-zone approximations in chemical evolution models may be appropriate for dwarf galaxies.

Turbulent metal diffusion is important for accuracy in modelling the ISM as well as processes relevant for chemical evolution. The inclusion or exclusion of TMD will therefore influence predictions drawn from simulated chemical abundances. For example, Bonaca et al. (2017) justified the use of the Latte **m12i** primary halo as a Milky Way analogue for comparison to data from *Gaia* Data Release 1 (Gaia Collaboration et al. 2016c; Gaia Collaboration et al. 2016b) using the width of the MDF including TMD. The authors then drew inferences on the hierarchical formation of the Galaxy and its halo structure based on the simulation. Including turbulent metal diffusion can thus enable the use of simulations in detailed, chemical abundance based investigations of galaxy formation and evolution.

IE would like to thank the anonymous referee, in addition to Shea Garrison-Kimmel, Matthew Orr, and Denise Schmitz, for helpful comments that improved this chapter. Numerical calculations were run on the Caltech computing cluster “Zwicky” (NSF MRI award #PHY-0960291) and allocation TG-AST130039 granted by the Extreme Science and Engineering Discovery Environment (XSEDE) supported by the NSF. IE was supported by Caltech funds, in part through the Caltech Earle C. Anthony Fellowship, and a Ford Foundation Predoctoral Fellowship. AW was supported by a Caltech-Carnegie Fellowship, in part through the Moore Center for Theoretical Cosmology and Physics at Caltech, and by NASA through grants HST-GO-14734 and HST-AR-15057 from STScI. ENK was supported by NSF Grant AST-1614081. Support for PFH was provided by an Alfred P. Sloan Research Fellowship, NASA ATP Grant NNX14AH35G, NSF Collaborative Research Grant #1411920, and CAREER grant #1455342. CW was supported by the Lee A. DuBridge Postdoctoral Scholarship in Astrophysics. DK was supported by NSF grant AST-1715101 and the Cottrell Scholar Award from the Research Corporation for Science Advancement. CAFG was supported by NSF through grants AST-1412836 and AST-1517491, by NASA through grant NNX15AB22G, and by STScI through grant HST-AR-14293.001-A. EQ was supported in part by a Simons Investigator Award from the Simons Foundation and NSF grant AST-1715070.

## 2.7 Appendix: Numerical Effects

In this section, we explored the impact of numerical effects in the simulations on the conclusions of this chapter.

### Stochastic IMF Sampling

Here, we note that additional scatter can be introduced into the MDFs and abundances as a consequence of stochastic IMF sampling at sufficiently high resolutions, such that a star particle no longer approximates a single stellar population. Estimates from Revaz et al. (2016) for SPH methods suggest that the single stellar population approximation no longer holds for star particles with $M_* < 1000\ M_\odot$, which includes the standard FIRE dwarf galaxies with mass resolution of $\sim 250\ M_\odot$.

We do indeed expect that the IMF is not being individually well-sampled at these masses. However, we anticipate that this is more of an issue in ultra-faint dwarf galaxies, which have sufficiently low stellar masses such that the star formation history of the galaxy is substantially impacted by a single SN. In contrast, a majority of the FIRE dwarf galaxies have $M_* \gtrsim 10^6\ M_\odot$, such that the effects of individual SNe on galaxy-scale properties are negligible, or weak at most (Su et al. 2018).

We expect that the additional numerical scatter introduced by IMF sampling is offset by the inclusion of IMF-averaged yields (Iwamoto et al. 1999; Nomoto et al. 2006). Ultimately, we are concerned with total metallicity distributions and the behaviour of $\alpha$-element abundance ratios as a function of metallicity, as opposed to detailed abundance patterns. In the latter case, the effects of IMF sampling would be more pronounced, and would require proper quantification. Particularly in the case of metal diffusion, we do not anticipate that stochastic IMF sampling will have a significant effect on the MDFs and abundance ratios, since metal diffusion drives the metallicity of a star particle toward the average metallicity of the galaxy, and so can reduce artificial noise from a number of different sources.

Sub-grid metal mixing, which we have argued is necessary to include in Lagrangian simulations of galactic chemical evolution, has interesting implications for modeling individual abundance patterns. That is, the presence of sub-grid diffusion or the lack thereof dictates the amount of information that can be extracted from detailed abundance patterns. With the inclusion of metal diffusion, only the first few SN explosions individually affect a galaxy, after which metals are quickly homogenized.

Overall, this study will be informative for addressing how much scatter can be attributed to IMF sampling, and to what degree numerical artifacts can be reduced via sub-grid metal mixing.

**Metal Deposition**

Altering the metal deposition algorithm could potentially impact the MDF and abundances by introducing an additional source of numerical scatter. For example, gas particles with different distances from the "exploding" star particle may receive varying amounts of metals, a fixed number of particles may be injected with metals, or a fixed volume surrounding such an exploding star particle may be enriched. A more realistic metal deposition prescription, or one in which the metals are dispersed more uniformly, will intuitively introduce less scatter into chemical evolution observables as compared to, e.g., the injection of all metals from a SN into a single gas particle.

To illustrate the robustness of our results with respect to the details of mechanical feedback when including diffusion, we compare the unphysical, extreme case of single-particle metal injection to that of the standard FIRE implementation. In the standard method, an effective neighbor number is determined based on a kernel function and search radius within a sphere defined by these quantities, such that the resulting distribution of ejecta is isotropic, including regions with highly disordered gas particle positions (Hopkins et al. 2018).

Figure 2.10 shows the logarithmic MDFs of four simulations run at a comparatively lower mass resolution of 2100 $M_\odot$ for **m10q**. As expected, the MDF for the case of single-particle injection, without TMD, exhibits an unrealistically large scatter characterized by a series of disconnected peaks. In this prescription, some star particles are never enriched, whereas others are enriched to high values of metallicity ([Fe/H] $> -0.5$) that are not observed in LG dwarf galaxies.

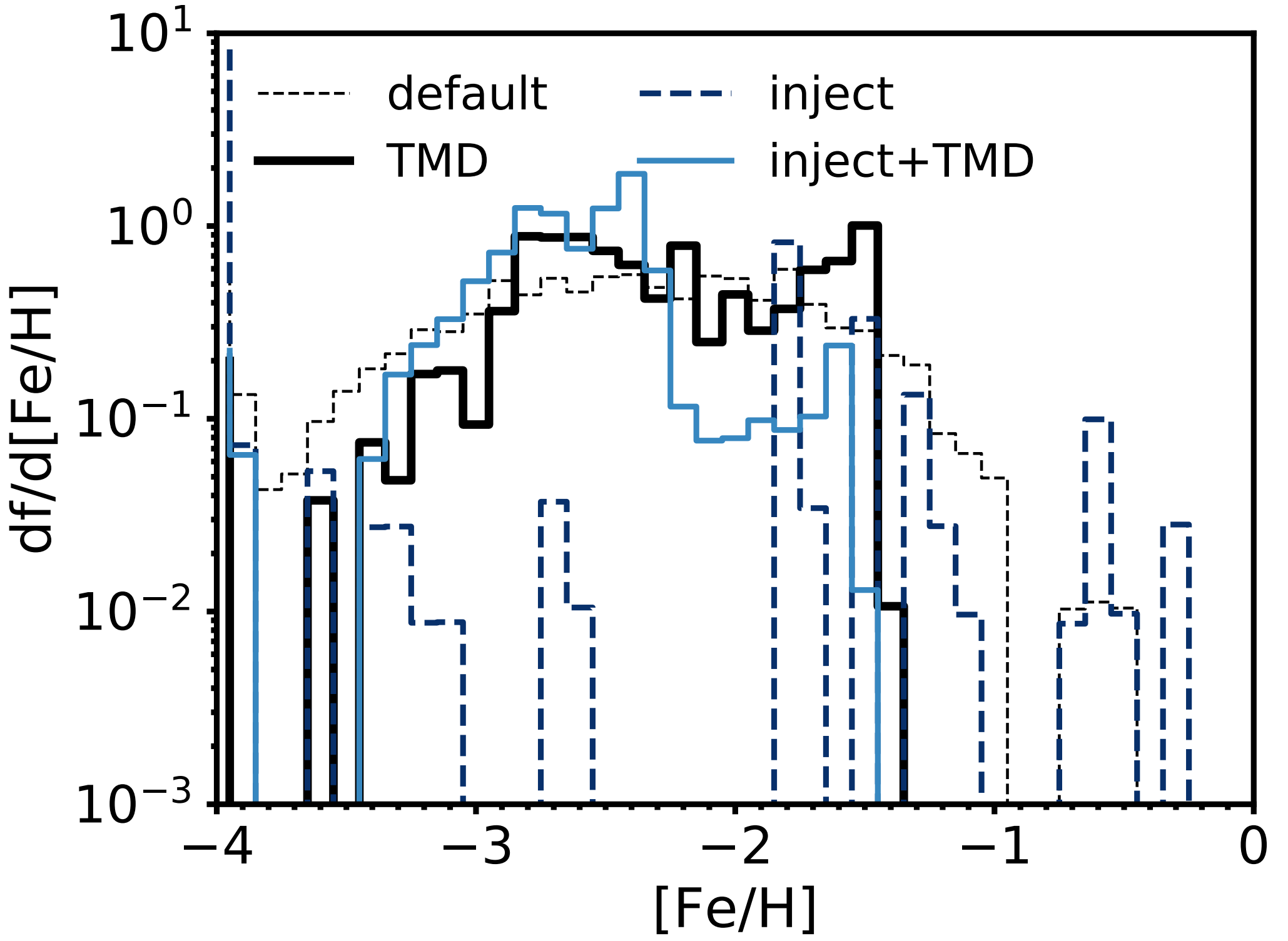

Figure 2.10: Stellar-mass weighted metallicity distribution functions at $z = 0$ with differing combinations of metal injection methods and including (solid lines) and excluding (dashed lines) turbulent metal diffusion. The simulations are run at a comparatively lower mass resolution of 2100 $M_{\odot}$, with the same initial conditions as **m10q**. The identifier "inject" indicates that all metals are injected into a single particle per enrichment event (blue lines), as opposed to the standard metal injection scheme (black lines), whereas "TMD" indicates the presence of turbulent metal diffusion. The metallicity distribution functions are plotted on a log scale to emphasize the behavior at the tail of the distributions. The inclusion of diffusion offsets the effects of the unphysical metal injection scheme, implying that diffusion stabilizes results against differences in the specific numerical implementation of metal deposition.

The inclusion of TMD brings the single-particle injection method into better approximate agreement with standard FIRE MDFs with and without sub-grid diffusion in terms of the shape of the distribution (Figure 2.10). All simulations presented in Figure 2.10 have approximately the same stellar mass at z = 0 ($M_* \sim 1.2$ - $1.7 \times 10^6$ $M_\odot$). The same narrowing effect of the MDF is exhibited, with a reduction in the width from $\sigma$[Fe/H] = 0.41 dex to 0.30 dex. This accompanies a significant shift of the average metallicity from -1.51 dex to -2.49 dex (as compared to average metallicity values of [Fe/H] = -2.18 and -2.15 for the standard metal deposition scheme with and without sub-grid diffusion). The narrowing effect is not as pronounced in the runs with standard metal deposition, with $\sigma$[Fe/H] = 0.50 and 0.46 dex with and without diffusion respectively, most likely owing to lower mass resolution (2100 $M_\odot$ as compared to 250 $M_\odot$).

Thus, we conclude that the presence of diffusion stabilizes the MDFs and abundances against differences in the specific numerical implementation of metal deposition.

**Diffusion Coefficient Calibration**

While the implementation of turbulent metal diffusion is more physical, given the nature of turbulence in the ISM, the diffusion coefficient must be calibrated independently for different numerical methods, owing to the different "effective resolution scale" of the turbulent cascade associated with each simulation. For example, the relevant scale at standard FIRE dwarf resolution ($\sim$ 250 $M_\odot$) will differ compared to other simulations, since a majority of the mixing is resolved. Colbrook et al. (2017) tested the FIRE implementation of turbulent metal diffusion in idealized, converged turbulent box simulations, verifying that the prescription is valid with $C \approx 0.05$, where $C_0 = \sqrt{2}C^2$, given our definitions of **h** and **S**. This is further confirmed by a more comprehensive study by Rennehan et al. (2019), which similarly indicates that $C \approx 0.03$ - $0.05$. Thus, a proper calibration of the coefficient is necessary to ensure that sub-grid metal diffusion approximates a physical process at a given resolution-scale. We emphasize through the numerical tests shown in Section 2.7 that the method is not especially sensitive to the exact value of the coefficient for a given calibration.

**Robustness of Results with Respect to the Diffusion Coefficient**

A potential drawback of the sub-grid turbulent metal diffusion implementation is possibly unphysical amounts of diffusion. To address the possibility of significant over-mixing, we varied the diffusion coefficient, $C_0$ (Section 2.2), to test the impact

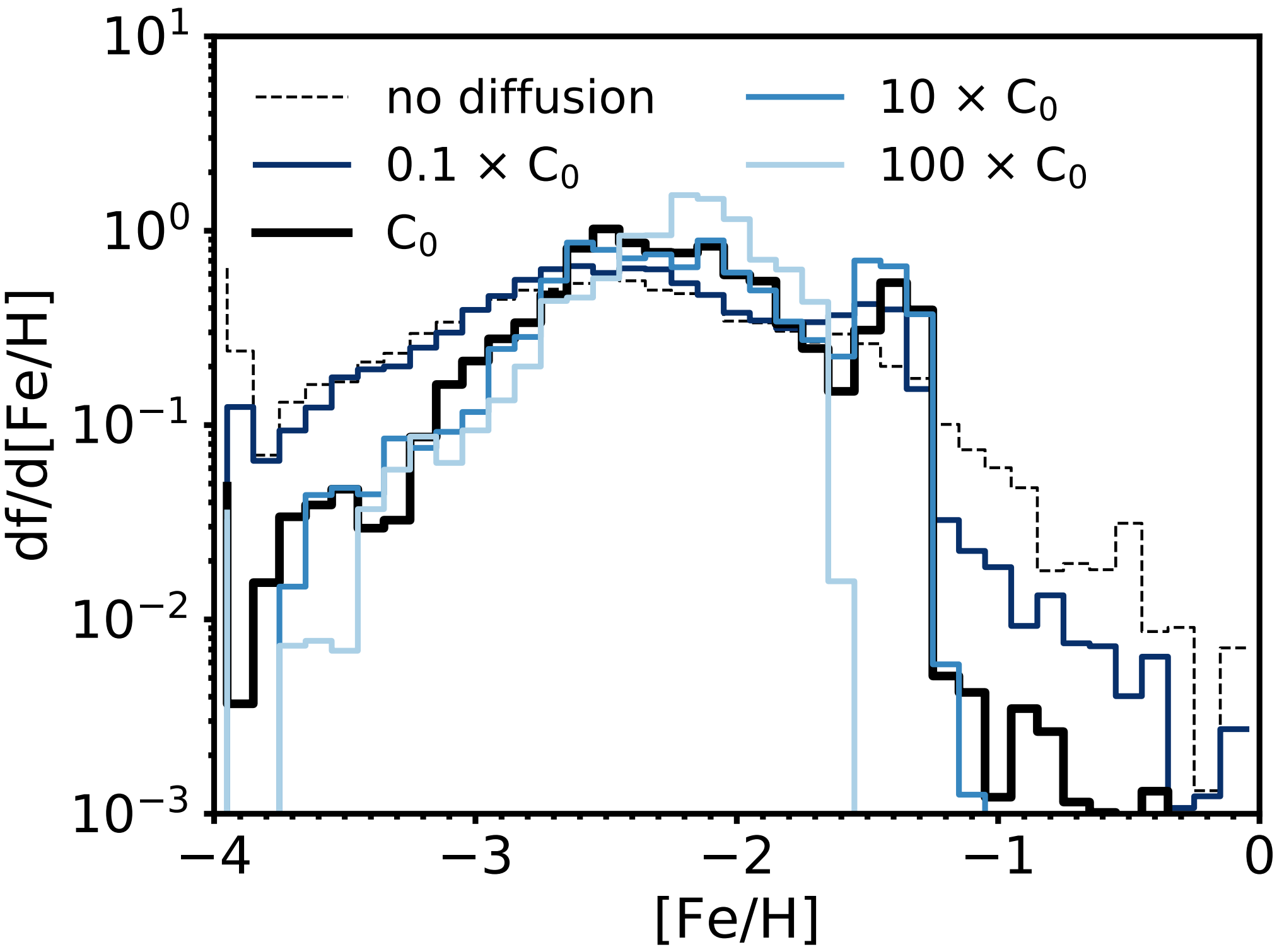

Figure 2.11: Stellar mass-weighted metallicity distribution functions at z = 0 with varied diffusion coefficients (blue lines), altering the overall diffusion strength, for runs of **m10q**. **m10q** without sub-grid metal diffusion is shown for reference (dashed black line). The metallicity distribution functions are plotted on a log scale to emphasize the behaviour at the tails of the distribution. $10 \times C_0$ represents a diffusion strength a factor of ten larger than the default value of $C_0 \approx 0.003$ (thick black line; Section 2.7), and so on. $C_0$ represents the minimum diffusion strength for significant sub-grid metal mixing to occur. The lack of a significant difference between diffusion effects for different values of the coefficient beyond our fiducial range suggests that MDF predictions are not extremely sensitive to the exact adopted value of the free parameter above the minimum diffusivity. However, in the case of the a diffusion strength larger by two orders of magnitude ($100 \times C_0$), the effects of over-mixing become apparent.

on the strength of metal diffusion on the MDF.

As illustrated in Figure 2.11, an increase in the coefficient by an order of magnitude for runs of **m10q** does not significantly change the appearance of the MDF above a minimum diffusivity. A reduction of the diffusion strength by an order of magnitude ($0.1 \times \mathrm{C}_0$), relative to the standard adopted diffusion strength ($C_0 = 0.003$), results in a broader MDF (0.50 dex). Although the detailed distribution at the high-metallicity tail changes for a diffusion strength of $0.1 \times \mathrm{C}_0$, the low-metallicity tail and main body of the MDF are comparable to the case without any sub-grid metal diffusion. We calibrate the value of $C_0$ such that it corresponds to the minimum diffusion strength that results in significant sub-grid metal mixing. Above $C_0 = 0.003$, increasing the diffusion strength results in some reduction in the number of stars present in both the high- and low-metallicity tails. Despite the increase in the diffusion coefficient, the widths of the distributions remain approximately constant at 0.44 dex and 0.46 dex for diffusion coefficients of $\mathrm{C}_0$ and $10 \times \mathrm{C}_0$ respectively. However, in the case of an increase in the coefficient by two orders of magnitude ($100 \times \mathrm{C}_0$), the effects of over-mixing become apparent with a reduction in the width to 0.29 dex, as well as in the complete absence of the tails of the distribution.

This suggests that MDF predictions are not extremely sensitive to the exact diffusion coefficients above a minimum diffusivity within an order of magnitude, as long as some degree of sub-grid turbulent mixing is present. Together, the insensitivity of the MDF to the diffusion coefficient and the narrowing of the MDF relative to runs without TMD (Section 2.3) imply that the timescale for metal mixing is shorter than a dynamical time for dwarf galaxies. As we show explicitly in Section 2.4, with any given burst of star formation, we can assume that the ISM is well-mixed and nearly homogeneous, resulting in a stellar population born with approximately the same metallicity.

Previous work by Williamson, Martel, and Kawata (2016) involving idealized, non-cosmological simulations of dwarf galaxies came to a similar conclusion concerning the robustness of the diffusion strength relative to the diffusion coefficient. However, this work does not include the infall of pristine gas, galaxy interactions, galaxy evolution through cosmological time, the initialization of star particles at low-metallicity at early times (to approximate the Population III to Population II transition), and self-regulated star formation, which are essential for an accurate model of the chemodynamical evolution of a galaxy. Thus, we confirm that the addition of diffusion still contributes to the robustness of results in the case of more realistic galaxy evolution simulations.

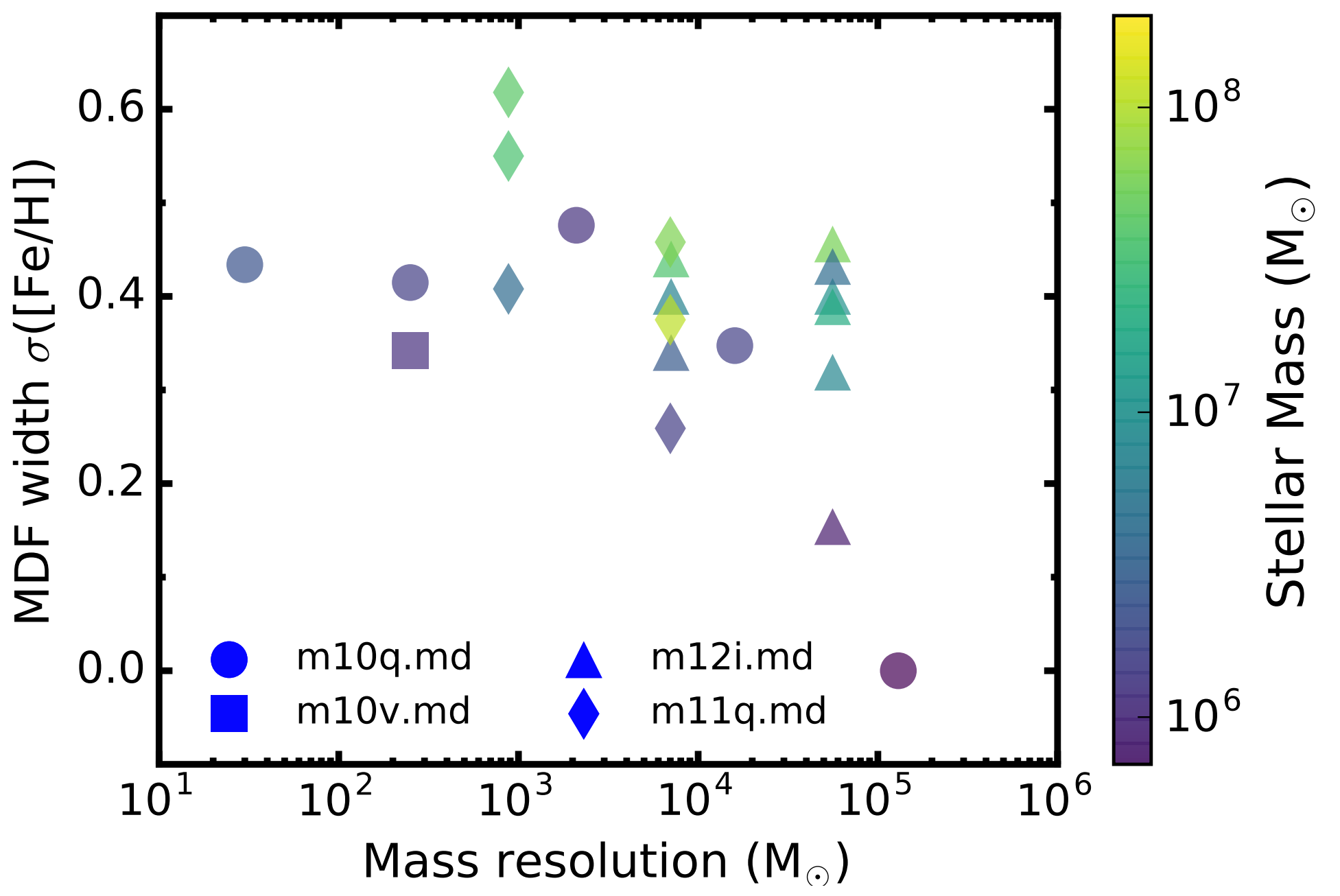

Figure 2.12: MDF width as a function of mass resolution at $z = 0$ for the FIRE isolated dwarfs **m10v** and **m10q**. Isolated dwarf galaxy subhalos of both **m11q**, a primary host with LMC mass, and **m12i** are also shown. We consider runs with TMD, since runs without TMD have star particles with more improbable metallicities and have comparatively enhanced numerical noise in the MDF. The data is colour coded according to the stellar mass of the galaxy, to account for the potential of associated variation in the MDF width. No systematic trend of the MDF width exits with mass resolution for baryonic particle masses $\lesssim 10^4$ $M_\odot$.

**Mass Resolution**

We consider the impact of mass resolution on the width of the MDF and the intrinsic scatter in [$\alpha$/Fe] vs. [Fe/H]. The star particle mass resolution for the isolated FIRE dwarfs is 250 $M_\odot$ whereas the Latte dwarf galaxies are at lower mass resolution (7070 $M_\odot$), and thus resolution effects may factor into our comparisons between isolated FIRE dwarf galaxies and dwarf galaxies from the Latte simulations.

First, we establish that most results, such as the stellar mass and star formation rates, converge with mass resolution after the Toomre scale (i.e. the largest self-gravitating structures) is resolved (Hopkins, Quataert, and Murray 2011; Hopkins et al. 2018). The metallicity and burstiness of dwarf galaxy star formation histories converge to within ~ 10 % (to ~ 20% maximum) differences for (1) $N \gtrsim 100$, where $N$ is the number of star particles in a galaxy, such that the self-enrichment history is well-sampled, and (2) star particle mass $\lesssim 10^4$ $M_\odot$ for $M_{halo} \sim 10^{10}$ $M_\odot$, such that numerically enhanced burstiness does not occur (Hopkins et al. 2018).

Figure 2.12 illustrates the results of mass resolution tests for the MDF width. Data from the FIRE isolated dwarfs, including an ultra high-resolution (30 $M_\odot$) run of **m10q** (Wheeler et al. 2019), in addition to isolated dwarf galaxies in the zoom-in region of both **m11q**, a LMC mass halo, and **m12i** are shown. We consider only isolated dwarfs for the resolution test to account for differences in the disruption of satellites by the host (Section 2.5). All runs include TMD, where runs without TMD contain star particles with more improbable metallicites and have comparatively enhanced numerical noise in the MDF.

Although the MDF width exhibits some variation for the isolated dwarf galaxies in simulations with a primary host, it can be attributed to the typical scatter (standard deviation $\lesssim 0.1$ dex) at a given stellar mass expected from stochastic effects. Nonetheless, the mean MDF width appears to systematically decrease for mass resolution $\gtrsim 10^4$ $M_\odot$. We note that the lowest resolution run, at baryonic particle mass of 160,000 $M_\odot$, has a near zero MDF width likely owing to a truncated SFH caused by numerically enhanced burstiness, a secondary resolution effect. Excluding the lowest resolution run and taking into account the typical scatter at a given stellar mass, we conclude that the MDF width of the FIRE isolated dwarfs and Latte satellite and isolated dwarf galaxies are converged for mass resolution $\lesssim 10^5$ $M_\odot$.

Similarly, Figure 2.13 shows mass resolution tests for the intrinsic scatter in [Si/Fe] at fixed [Fe/H]. In contrast to the MDF width, the intrinsic scatter increases with mass resolution $\gtrsim 10^3$ $M_\odot$ for the lowest mass simulated dwarf galaxies ($M_* \lesssim 10^{6.5}$ $M_\odot$). This is due to secondary resolution effects resulting in numerically enhanced burstiness, as discussed in Section 2.5. We quantify the degree of burstiness as a the fraction of the total $z = 0$ stellar mass formed when the star formation rate averaged over short timescales ($\sim$ 10 Myr) exceeds (by a factor of 1.5) the average star formation rate averaged over longer timescales ($\sim$ 100 Myr). We find that burstiness converges for baryonic particle mass resolution $\lesssim 10^3$ $M_\odot$. For mass resolution $\gtrsim 10^3$, burstiness in dwarf galaxies increases with decreasing mass resolution, while simultaneously exhibiting mass-dependent behaviour, such that lower-mass dwarf galaxies are burstier.

Hence, the intrinsic scatter in Latte dwarf galaxies with baryonic particle mass 7070 $M_\odot$increases for less well-resolved, lower-mass dwarf galaxies. The intrinsic scatter converges at higher mass resolution than the MDF width, because the burstiness is more sensitive to mass resolution than the long-term ($\sim$ 10 Gyr) star formation history. Since we cannot separate resolution effects from stochastic variation in the intrinsic scatter at a given stellar mass for these dwarf galaxies, we do not quantify the typical variation expected in the intrinsic scatter, $\Sigma$, as in the case of the MDF width, $\sigma$([Fe/H]). Considering only the better resolved, high-mass dwarf galaxies ($M_* \gtrsim 10^{6.5}$ $M_\odot$), no trend exists in the intrinsic scatter with baryonic particle mass resolution for $\lesssim 10^4$ $M_\odot$ for both the FIRE isolated and Latte dwarf galaxies.

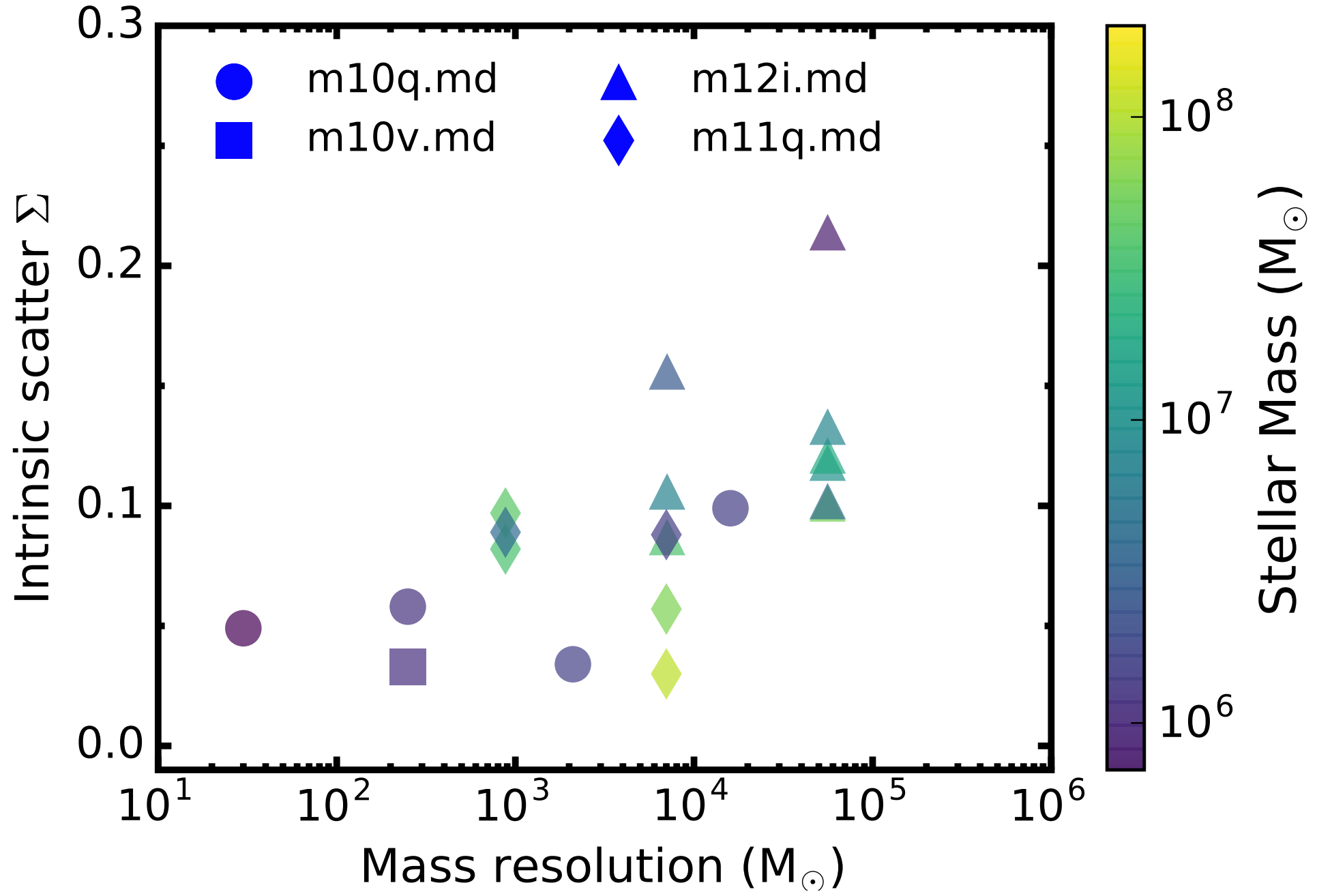

Figure 2.13: Intrinsic scatter in [Si/Fe] at fixed [Fe/H] as a function of mass resolution at $z = 0$. All other aspects of the figure are identical to Figure 2.12. In contrast to the MDF width, the intrinsic scatter, $\Sigma$, appears to converge for baryonic particle masses $\lesssim 10^3$ $M_\odot$, taking into account the full range of stellar masses. Considering only high-mass (better resolved) dwarf galaxies, the intrinsic scatter converges for mass resolution $\lesssim 10^4$ $M_\odot$. Thus, we expect $\Sigma$ to be larger on average for the Latte simulations as compared to the higher-resolution FIRE isolated dwarf galaxies at low stellar masses ($M_* \lesssim 10^{6.5}$ $M_\odot$.

*3*

# ELEMENTAL ABUNDANCES IN M31: ALPHA AND IRON ELEMENT ABUNDANCES FROM LOW-RESOLUTION RESOLVED STELLAR SPECTROSCOPY IN THE STELLAR HALO

Ivanna Escala[1,2], Evan N. Kirby[1], Karoline M. Gilbert[3,4], Emily C. Cunningham[5], Jennifer Wojno[4]

[1]Department of Astronomy, California Institute of Technology, 1200 E California Blvd, Pasadena, CA, 91125, USA
[2]Department of Astrophysical Sciences, Princeton University, 4 Ivy Lane, Princeton, NJ, 08544, USA
[3]Space Telescope Science Institute, 3700 San Martin Dr., Baltimore, MD 21218 USA
[4]Department of Physics & Astronomy, Bloomberg Center for Physics and Astronomy, John Hopkins University, 3400 N. Charles St, Baltimore, MD 21218, USA
[5]Department of Astronomy and Astrophysics, University of California, Santa Cruz, 1156 High St, Santa Cruz, CA, 95064, USA

**Abstract**

Measurements of [Fe/H] and [$\alpha$/Fe] can probe the minor merging history of a galaxy, providing a direct way to test the hierarchical assembly paradigm. While measurements of [$\alpha$/Fe] have been made in the stellar halo of the Milky Way, little is known about detailed chemical abundances in the stellar halo of M31. To make progress with existing telescopes, we apply spectral synthesis to low-resolution DEIMOS spectroscopy (R ~ 2500 at 7000 Å) across a wide spectral range (4500 Å $< \lambda <$ 9100 Å). By applying our technique to low-resolution spectra of 170 giant stars in 5 MW globular clusters, we demonstrate that our technique reproduces previous measurements from higher resolution spectroscopy. Based on the intrinsic dispersion in [Fe/H] and [$\alpha$/Fe] of individual stars in our combined cluster sample, we estimate systematic uncertainties of ~0.11 dex and ~0.09 dex in [Fe/H] and [$\alpha$/Fe], respectively. We apply our method to deep, low-resolution spectra of 11 red giant branch stars in the smooth halo of M31, resulting in higher signal-to-noise per spectral resolution element compared to DEIMOS medium-resolution spectroscopy, given the same exposure time and conditions. We find $\langle[\alpha/\mathrm{Fe}]\rangle = 0.49 \pm 0.29$ dex and $\langle[\mathrm{Fe/H}]\rangle = -1.59 \pm 0.56$ dex for our sample. This implies that—much like the Milky Way—the smooth halo field of M31 is likely composed of disrupted dwarf galaxies with truncated star formation histories that were accreted early in the halo's formation.

## 3.1 Introduction

Stellar chemical abundances are a key component in determining the origins of stellar halos of Milky Way (MW) like galaxies, providing insight into the formation of galaxy-scale structure. The long dynamical times of stellar halos allow tidal features to remain identifiable in phase space, in terms of kinematics and chemical abundances, for Gyr timescales. Stellar chemical abundances of stars retain information about star formation history and accretion times of progenitor satellite galaxies, even when substructures can no longer be detected by kinematics alone. In particular, measurements of metallicity[1] and $\alpha$-element abundances provide a way of directly testing the hierarchical assembly paradigm central to $\Lambda$CDM cosmology, providing a fossil record of the formation environment of stars accreted onto the halo.

The $[\alpha/\mathrm{Fe}]$ ratio serves as a useful diagnostic of formation history, given that it traces the star formation timescales of a galaxy (e.g., Gilmore and Wyse 1998). Type II supernovae (SNe) produce abundant $\alpha$-elements (O, Ne, Mg, Si, S, Ar, Ca, and Ti), increasing $[\alpha/\mathrm{Fe}]$, whereas Type Ia SNe produce Fe-rich ejecta, reducing $[\alpha/\mathrm{Fe}]$. While measurements of $[\alpha/\mathrm{Fe}]$ have been made in the stellar halo of the MW, little is known about the detailed chemical abundances of the stellar halo of M31. A comparable understanding of the properties of the MW and M31 stellar halos is required to verify basic assumptions about how the MW evolved, where such assumptions are used to extrapolate MW-based results to studies of galaxies beyond the Local Group.

Although high-resolution ($R \gtrsim 15{,}000$), high signal-to-noise (S/N) spectra enables simultaneous measurements of a star's temperature, surface gravity, and individual element abundances based on individual lines, it is impractical to achieve high enough S/N for traditional high-resolution spectroscopic abundance analysis (e.g., Kirby and Cohen 2012) for red giant branch (RGB) stars at the distance of M31 (783 kpc; Stanek and Garnavich 1998).

[1] We define metallicity in terms of stellar iron abundance, [Fe/H], where $[\mathrm{Fe/H}] = \log_{10}(n_{\mathrm{Fe}}/n_{\mathrm{H}}) - \log_{10}(n_{\mathrm{Fe}}/n_{\mathrm{H}})_\odot$

It is possible to obtain spectroscopic metallicity measurements of M31 RGB stars from medium-resolution spectra ($R \sim 6000$) using spectral synthesis (e.g., Kirby, Guhathakurta, and Sneden 2008). This method leverages the entire spectrum's metallicity information simultaneously, enabling measurements of abundances from relatively low S/N spectra. Kirby et al. (2008), Kirby et al. (2010), and Kirby et al. (2013) successfully measured [Fe/H] and [$\alpha$/Fe] in MW globular clusters (GCs), MW dwarf spheroidal (dSph) satellite galaxies, and Local Group dwarf irregular galaxies, showing that abundances can be measured to a precision of $\sim$ 0.2 dex from spectra with S/N $\sim$ 15 Å$^{-1}$.

Only in 2014 has spectral synthesis been applied to individual RGB stars in the M31 system for the first time (Vargas, Geha, and Tollerud 2014; Vargas et al. 2014). Existing spectroscopic chemical abundance measurements in M31 are primarily based on metallicity estimates from the strength of the calcium triplet (Chapman et al. 2006; Koch et al. 2008; Kalirai et al. 2009; Richardson et al. 2009; Tanaka et al. 2010; Ibata et al. 2014; Gilbert et al. 2014; Ho et al. 2015). Vargas, Geha, and Tollerud (2014) measured [$\alpha$/Fe] and [Fe/H] for a total of 226 red giants in 9 M31 satellite galaxies. Although Vargas, Geha, and Tollerud (2014) measured [$\alpha$/Fe] for 9 M31 dSphs, only a single dSph, And V, shows a clear chemical abundance pattern, where [$\alpha$/Fe] declines with [Fe/H]. However, the present spectroscopic sample size and measurement uncertainties of the And V data enable only qualitative conclusions about the chemical evolution of the dSph. Obtaining more quantitative descriptions of the chemical enrichment and star formation histories of the M31 system requires higher S/N spectroscopic data, which results in smaller uncertainties on abundance measurements. Only then can one-zone numerical chemical evolution models (Lanfranchi and Matteucci 2003; Lanfranchi and Matteucci 2007, 2010; Lanfranchi, Matteucci, and Cescutti 2006; Kirby et al. 2011a) be reliably applied to measurements to derive star formation and mass assembly histories.

Although Vargas, Geha, and Tollerud (2014) and Vargas et al. (2014) demonstrated the feasibility of measuring [Fe/H] and [$\alpha$/Fe] at the distance of M31, measuring [Fe/H] and [$\alpha$/Fe] more precisely requires deep ($\sim$6 hour) observations with DEIMOS using the 600 line mm$^{-1}$ grating to yield higher S/N for the same exposure time and observing conditions. For magnitudes fainter than $I_0 \sim 21$ (0.5 magnitudes below the tip of M31's RGB), sky line subtraction at $\lambda > 7000$ Å becomes the dominant source of noise in DEIMOS spectra observed with the 1200 line mm$^{-1}$ grating. Given the access to blue optical wavelengths granted by the 600 line mm$^{-1}$ grating, its spectra are less susceptible to the effects of sky noise. Additionally, using the 600 line mm$^{-1}$ grating achieves higher S/N per pixel for stars as faint as $I_0 \sim 21.8$.

Although using the 600 line mm$^{-1}$ grating with DEIMOS results in a gain in S/N and wavelength coverage, it corresponds to a decrease in spectral resolution (~2.8 Å FWHM, or R ~ 2500 at 7000 Å, compared to ~1.3 Å and R ~ 5400 for 1200 line mm$^{-1}$). Increasing the spectral range compensates for the decrease in spectral resolution, given the increase in the amount of available abundance information contained in the spectrum resulting from the higher density of absorption features at bluer optical wavelengths.

Here, we expand upon the technique first presented by Kirby, Guhathakurta, and Sneden (2008), applying spectral synthesis to low-resolution spectroscopy (LRS; R ~ 2500) across a wide spectral range ($\lambda \sim 4500$ - 9100 Å). In Section 3.2, we describe our data reduction and GC observations. Section 3.3 and Section 3.4 detail our preparations to the observed spectrum and the subsequent abundance analysis. This includes a presentation of our new line list and grid of synthetic spectra. In Section 3.5, we illustrate the efficacy of our technique applied to MW GCs and compare our results to chemical abundances from high-resolution spectroscopy in Section 3.6. We quantify the associated systematic uncertainties in Section 3.7. We conclude by measuring [$\alpha$/Fe] and [Fe/H] in a M31 stellar halo field in Section 3.8 and summarize in Section 3.9.

## 3.2 Observations

We utilize observations of Galactic GCs, MW dwarf spheroidal (dSph) galaxies, and MW halo stars (Table 3.1) taken using Keck/DEIMOS (Faber et al. 2003) to validate our LRS method of spectral synthesis. For our science configuration for all observations, including M31 observations (Section 3.8), we used the GG455 filter with a central wavelength of 7200 Å, in combination with the 600ZD grating and 0.7” slitwidths. When targeting individual stars, such as the MW halo stars in Table 3.1, we utilized the longslit, as opposed to a slitmask intended to target multiple stars simultaneously.

The spectral resolution for the 600 line mm$^{-1}$ grating is approximately ~2.8 Å FWHM, compared to ~1.3 Å FWHM for the 1200 line mm$^{-1}$ grating used in prior observations (Kirby et al. 2010; Kirby et al. 2013). The wavelength range for each spectrum obtained with the 600ZD grating is within 4100 Å − 1 $\mu$m, where we generally omit $\lambda \lesssim 4500$ Å, owing to poor S/N in this regime and the presence of the G band. We also omit $\lambda > 9100$ Å, which extends beyond the wavelength coverage of our grid of synthetic spectra (Section 3.4).

Table 3.1: MW and dSph DEIMOS Observations

| Object | Slitmask | $\alpha_{J2000}$ | $\delta_{J2000}$ | Date | Seeing (") | Airmass | $t_{exp}$[a] (s) | $N_{target}$ | $N_{member}$[b] |
|---|---|---|---|---|---|---|---|---|---|
| MW Globular Clusters | | | | | | | | | |
| NGC 2419 | n2419c | 07:38:09.67 | +38:51:15.0 | 2015 Oct 9 | 0.6 | 1.23 | $2 \times 1380$ | 92 | 61 |
| NGC 1904 (M79) | 1904l2 | 05:24:15.37 | −24:31:31.3 | 2015 Oct 8 | 0.8 | 1.40 | $2 \times 1260$ | 96 | 18 |
| NGC 6864 (M75) | n6864a | 20:06:14.03 | −21:55:16.4 | 2015 May 19 | 0.9 | 1.56 | $3 \times 1080$ | 86 | 35 |
| NGC 6341 (M92) | n6341b | 17:17:23.68 | +43:06:49.4 | 2018 Oct 11 | 0.6 | 1.52 | $6 \times 300$ | 146 | 33 |
| NGC 7078 (M15) | 7078l1 | 21:29:48.03 | +12:10:23.0 | 2015 May 19 | 0.9 | 1.16 | $2 \times 1140$ | 169 | 48 |
| dSphs | | | | | | | | | |
| Draco | dra11 | 17:19:46.87 | +57:57:21.1 | 2019 Mar 10 | 1.6 | 1.39 | 4280 | 138 | ... |
| Canes Venatici I | CVnIa | 13:28:02.47 | +33:32:49.5 | 2018 May 20 | 1.0 | 1.3 | $6 \times 1200$ | 122 | ... |
| Ursa Minor | bumib | 15:09:28.75 | +67:13:07.1 | 2018 May 20 | 1.0 | 1.68 | $7 \times 1200$ | 124 | ... |
| MW Halo | | | | | | | | | |
| HD20512 | LVMslitC | 03:18:27.14 | +15:10:38.29 | 2019 Mar 10 | 1.2 | 1.35 | 10 | ... | ... |
| HD21581 | LVMslitC | 03:28:54.48 | −00:25:03.10 | 2019 Mar 10 | 1.2 | 1.45 | 171 | ... | ... |
| HD88609 | LVMslitC | 10:14:28.98 | +53:33:39.34 | 2019 Mar 10 | 1.2 | 1.43 | 44 | ... | ... |
| SAO134948 | LVMslitC | 06:46:03.69 | +67:13:45.75 | 2019 Mar 10 | 1.2 | 1.49 | 180 | ... | ... |
| BD+80245 | LVMslitC | 08:11:06.23 | +79:54:29.55 | 2019 Mar 10 | 1.2 | 1.99 | $3 \times 300$ | ... | ... |

[a] In the case of multiple exposures with unequal exposure time, we indicate the total exposure time.

[b] Number of RGB and AGB members (Kirby et al. 2016) per slitmask. For dSph slitmasks, we do not fully evaluate membership since we utilized these observations only for comparison to the HRS literature (Section 3.6).

To extract one-dimensional spectra from the raw DEIMOS data, we used a modification of version 1.1.4 of the data reduction pipeline developed by the DEEP2 Galaxy Redshift Survey (Cooper et al. 2012; Newman et al. 2013). Guhathakurta et al. (2006) provides a detailed description of the data reduction process. Modifications to the software include those of Simon and Geha (2007), where the pipeline was re-purposed for bright unresolved stellar sources (as opposed to faint, resolved galaxies). In addition, we include custom modifications to correct for atmospheric refraction in the two-dimensional raw spectra, which affects bluer optical wavelengths, and to identify lines in separate arc lamp spectra, as opposed to a single stacked arc lamp spectrum.

## 3.3 Preparing the Spectrum for Abundance Measurement

Here, we detail our preparations to the observed spectrum for the subsequent abundance analysis.

### Telluric Absorption Correction

Unlike the red side of the optical (6300 - 9100 Å), there is no strong telluric absorption in the bluer regions (4500 - 6300 Å). As such, we do not make any corrections to the observed stellar spectra to take into account absorption from Earth's atmosphere in this wavelength range.

For the red (6100 - 9100 Å), we correct for the absorption of Earth's atmosphere using the procedure described in Kirby, Guhathakurta, and Sneden (2008). We adopt HD066665 (B1V), observed on April 23, 2012 with an airmass of 1.081, using a long slit in the same science configuration (Section 3.2) as our data, as our spectrophotometric standard.

### Spectral Resolution Determination

In contrast to Kirby, Guhathakurta, and Sneden (2008), who determined the spectral resolution as a function of wavelength based on the Gaussian widths of hundreds of sky lines, we assume a constant resolution, expressed as the typical FWHM of an absorption line, across the entire observed spectrum (~4500–9100 Å). Owing to the dearth of sky lines at bluer wavelengths, the number of available sky lines is insufficient to reliably determine the resolution as a function of wavelength. As an alternative, we introduce an additional parameter, $\Delta\lambda$, or the spectral resolution, into our chi-squared minimization, which determines the best-fit synthetic spectrum for each observed spectrum (Section 3.4).

**Continuum Normalization**

It is necessary to normalize the observed flux by its slowly varying stellar continuum in order to meaningfully compare to synthetic spectra for the abundance determination (Section 3.4). To obtain reliable abundances from spectral synthesis of low- and medium-resolution spectra dominated by weak absorption features, the continuum determination must be accurate (Shetrone et al. 2009; Kirby et al. 2009). This is particularly important for bluer wavelengths, where absorption lines are so numerous and dense that we cannot define "continuum regions" (Kirby, Guhathakurta, and Sneden 2008) in the blue. Instead, we utilize the entire observed spectrum, excluding regions with strong telluric absorption and bad pixels, to determine the continuum for 4500 − 9100 Å. In contrast to Kirby, Guhathakurta, and Sneden (2008), we do not utilize continuum regions at redder wavelengths (6300 - 9100 Å), despite the fact that they can be reliably defined, to maintain consistency in the continuum normalization method between each wavelength region of the observed spectrum.

We determined the initial continuum fit to the raw observed spectrum, which we shift into the rest frame, using a third-order B-spline with a breakpoint spacing of 200 pixels, excluding 5 pixels around the chip gap and at the start and stop wavelengths of the observed spectrum. In all steps, we weighted the spline fit by the inverse variance of each pixel in the observed spectrum. We performed sigma clipping, such that pixels that deviate by more than $5\sigma$ ($0.1\sigma$) above (below) the fit are excluded from the subsequent continuum determination, where $\sigma$ is the inverse square root of the inverse variance array. We did not perform the fit iteratively beyond the above steps, given that our stringent criterion to prevent the numerous absorption lines from offsetting our continuum determination can eliminate a significant fraction of the pixels from subsequent iterations of the fit.

To further refine our continuum determination, we recalculate the continuum fit iteratively in the initial step of the abundance analysis (Section 3.4). Once we have found a best-fit synthetic spectrum, we divide the continuum-normalized observed spectrum by the best-fit synthetic spectrum to construct a "flat noise" spectrum, which captures the higher order terms in the observed spectrum not represented in the fit. We fit a third-order B-spline with a breakpoint spacing of 100 pixels to the flat noise spectrum, excluding $3\sigma$ deviant (above and below the fit) pixels, dividing the continuum-normalized observed spectrum by this fit. The modified continuum-normalized spectrum is then used in the next iteration of the continuum refinement until convergence is achieved (Section 3.4).

Table 3.2: Spectral Features (4100 - 6300 Å)

| Feature | Wavelength(s) (Å) |
|---|---|
| H$\delta$ | 4101.734 |
| Ca I | 4226.730 |
| G band (CH absorption) | 4300-4315 |
| H$\gamma$ | 4340.462 (4335-4345)[a] |
| H$\beta$ | 4861.35 (4856-4866) |
| Mg I (b4) | 5167.322 |
| Mg I (b2) | 5172.684 |
| Mg I (b1) | 5183.604 |
| Mg H | 4845,5622 |
| Na D1,D2 | 5895.924,5889.951 (5885-5905) |

[a] Wavelength regions indicated in parenthesis indicate regions that are omitted from the spectral fit.

**Pixel Masks**

In addition to wavelength masks corresponding to a particular abundance (Section 3.4), we constructed a pixel mask for each analyzed observed spectrum. Typically excluded regions include 5 pixels on either side of the chip gap between the blue and red sides of the CCD, areas with improper sky line subtraction, the region around the Na D1 and D2 lines (5585 - 5905 Å), and other apparent instrumental artifacts. Table 3.2 includes a summary of prominent spectral features in DEIMOS spectra between 4100 - 6300 Å, where wavelength ranges given in parenthesis indicate regions that are masked. For example, we excluded 10 Å regions around H$\gamma$ (4335 - 4345 Å) and H$\beta$ (4856 - 4866 Å). Given that MOOG (Sneden 1973), the spectral synthesis software utilized to generate our grid of synthetic spectra (Section 3.4), does not incorporate the effects of non-local thermodynamic equilibrium, it cannot properly model the strong Balmer lines. If necessary, we also masked regions where the initial continuum fit failed, most often owing to degrading signal-to-noise as a function of wavelength at bluer wavelengths ($\lesssim$ 4500 Å). As for the red (6300 - 9100 Å), we adopted the pixel mask from Kirby, Guhathakurta, and Sneden (2008), which excludes spectral features such as the Ca II triplet, H$\alpha$, and regions with strong telluric absorption.

### Signal-to-Noise Estimation

We estimate S/N per Angstrom for objects observed with the 600 line mm$^{-1}$ grating from wavelength regions of the spectrum utilized in the initial continuum determination (Section 3.3). Given that we cannot define continuum regions for wavelengths blueward of 6300 Å, we calculate the S/N after the continuum refinement process (Section 3.4). We estimate the noise as the deviation between the continuum-refined observed spectrum and the best-fit synthetic spectrum and the signal as the best-fit synthetic spectrum itself. The S/N estimate per pixel is the median of the S/N as a function of wavelength calculated from these quantities, where we exclude pixels that exceed the average noise threshold by more than $3\sigma$. To convert to units of per Angstrom, we multiply this quantity by the inverse square root of the pixel scale ($\sim$ 0.64 Å for the 600 line mm$^{-1}$ grating).

## 3.4 Chemical Abundance Analysis

Here, we present a new library of synthetic spectra in the range 4100 - 6300 Å. In this section, we describe our procedure for spectral synthesis in the blue, where we use our new grid in conjunction with the red grid of Kirby, Guhathakurta, and Sneden (2008) to measure abundances across an expanded optical range (4100 - 9100 Å).

### Line List

We constructed a line list of wavelengths, excitation potentials (EPs), and oscillator strengths (log $gf$) for atomic and molecular transitions in the spectral range covering 4100 - 6300 Å for stars in our stellar parameter range ($T_{\mathrm{eff}}$ > 4000 K). We queried the Vienna Atomic Line Database (VALD; Kupka et al. 1999) and the National Institute of Standards and Technology (NIST) Atomic Spectra Database (Kramida, Ralchenko, and Reader 2016) for all transitions of neutral or singly ionized atoms with EP < 10 eV and log $gf$ > -5, supplementing the line list with molecular (Kurucz 1992) and hyperfine transitions (Kurucz 1993). All Fe I line oscillator strengths from Fuhr and Wiese (2006) are included in the NIST database.

Next, we compared synthetic spectra (Section 3.4) of the Sun and Arcturus, generated from our line list and model stellar atmospheres, to high resolution spectra (Hinkle et al. 2000) of the respective stars. We adopted $T_{\mathrm{eff}}$ = 5780 K, log $g$ = 4.44 dex, [Fe/H] = 0 dex, and [$\alpha$/Fe] = 0 dex for the Sun. For Arcturus, we adopted $T_{\rm eff}$ = 4300 K, log $g$ = 1.50 dex, and [Fe/H] = -0.50 dex, and [$\alpha$/Fe] = 0 dex (Peterson, Dalle Ore, and Kurucz 1993).

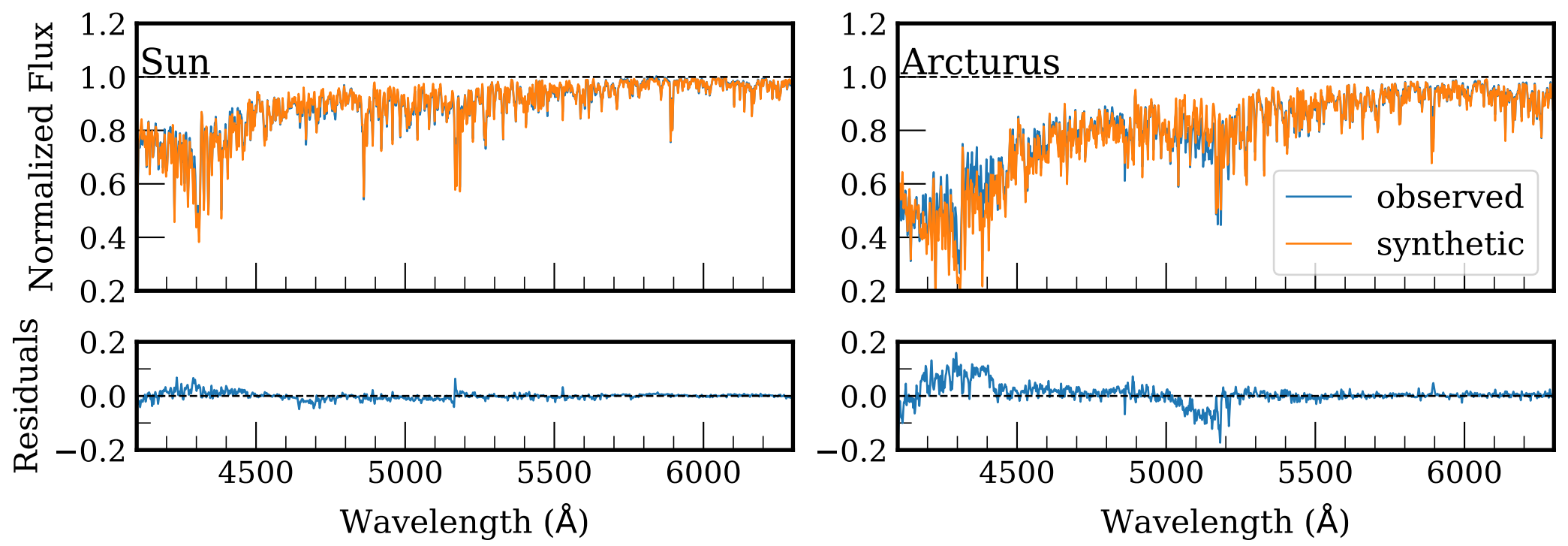

Figure 3.1: (*Top*) A comparison between high-resolution spectra (Hinkle et al. 2000) (*blue*) of the Sun (*left*) and Arcturus (*right*) and synthetic spectra (*orange*) generated using the blue line list (Section 3.4). Both spectra are smoothed to the expected resolution of the DEIMOS 600ZD grating ($\sim 2.8$ Å). For a description of synthetic spectrum generation, see Section 3.4. (*Bottom*) The difference between the observed and synthetic spectra for the Sun and Arcturus. To improve the agreement between the synthetic and observed spectra, we have manually vetted the line list, adjusting the oscillator strengths of discrepant atomic transitions as necessary. In this process, we have favored the Sun over Arcturus, thus the larger residuals between the observed and synthetic spectra for the latter star (which has a lower effective temperature).

To produce agreement between the synthetic and observed spectra, we vetted the line list by manually adjusting the oscillator strengths of aberrant atomic lines as necessary. We preferred the Sun over Arcturus in this process, given that Arcturus is a cool K-giant star with stronger molecular absorption features (e.g., the G band) that are more difficult to match. For features absent from the line list, which could not be resolved by considering lines with log $gf < -5$, we included Fe I transitions with EPs and log $gf$ to match the observed strength in both the Sun and Arcturus. We present the final blue line list in a format compatible with MOOG in Table 3.3. The line list contains 132 chemical species (atomic, molecular, neutral, and ionized), including 74 unique elements and 2 molecules (CN and CH). In total, the line list contains 53,164 atomic line transitions and 58,062 molecular transitions.

Figure 3.1 illustrates a comparison between the Hinkle spectra and their syntheses for the Sun and Arcturus. At the expected resolution of the DEIMOS 600ZD grating ($\sim 2.8$ Å), the mean absolute deviation of the residuals between the observed spectra and their syntheses across the wavelength range of the line list are $8.3 \times 10^{-3}$ and

Table 3.3: Blue Line List (4100 - 6300 Å)

| Wavelength (Å) | Species[a] | EP[b] (eV) | log $gf$[c] |
|---|---|---|---|
| 5183.409 | 57.1 | 0.403 | -0.6 |
| 5183.414 | 69.1 | 4.744 | -2.65 |
| 5183.436 | 24.1 | 6.282 | -3.172 |
| 5183.465 | 26.0 | 3.111 | -5.06 |
| 5183.466 | 27.0 | 4.113 | -1.187 |
| 5183.493 | 106.00112 | 3.244 | -2.848 |
| 5183.506 | 106.00113 | 1.569 | -3.974 |
| 5183.518 | 607.0 | 1.085 | -4.211 |
| 5183.544 | 26.0 | 5.064 | -3.886 |
| 5183.55 | 58.1 | 1.706 | -2.27 |
| 5183.565 | 106.00112 | 3.55 | -2.811 |
| 5183.578 | 607.0 | 1.204 | -2.653 |
| 5183.598 | 23.1 | 6.901 | -3.568 |
| 5183.604 | 12.0 | 2.717 | -0.167 |
| 5183.615 | 607.0 | 1.085 | -3.003 |
| 5183.683 | 607.0 | 1.205 | -4.499 |
| 5183.683 | 106.00113 | 1.29 | -4.921 |
| 5183.686 | 106.00113 | 2.371 | -3.586 |
| 5183.708 | 40.0 | 0.633 | -1.62 |
| 5183.709 | 22.1 | 1.892 | -2.535 |
| 5183.748 | 58.1 | 1.482 | -1.56 |
| 5183.794 | 106.00112 | 1.43 | -2.866 |
| 5183.803 | 24.0 | 5.277 | -3.52 |

Note. —The line list presented here is a subset of the entire line list, which spans 4100 - 6300 Å. The range of wavelengths presented here spans 0.4 Å around the strong Mg I line at 5183.604 Å. The line list is formatted to be compatible with MOOG.

[a] A unique code corresponding to chemical species. For example, 12.0 indicates Mg I, 22.1 indicates Ti II, 106.00112 indicates CN for carbon-12, and 106.0113 indicates CN for carbon-13.

[b] Excitation potential.

[c] Oscillator strength. Transitions modified in the vetting process have less than three decimal places.

$2.2 \times 10^{-2}$ for the Sun and Arcturus respectively.

**Synthetic Spectra**

We employed the ATLAS9 (Kurucz 1993) grid of model stellar atmospheres, with no convective overshooting (Castelli, Gratton, and Kurucz 1997). We based our grid on recomputed (Kirby et al. 2009 and references therein) ATLAS9 model atmospheres with updated opacity distribution functions, available for [$\alpha$/Fe] = 0.0 and +0.4 (Castelli and Kurucz 2003). We adopted the solar composition of Anders and Grevesse 1989, except for Fe (Sneden et al. 1992). The elements considered to be $\alpha$-elements are O, Ne, Mg, Si, S, Ar, Ca, and Ti.

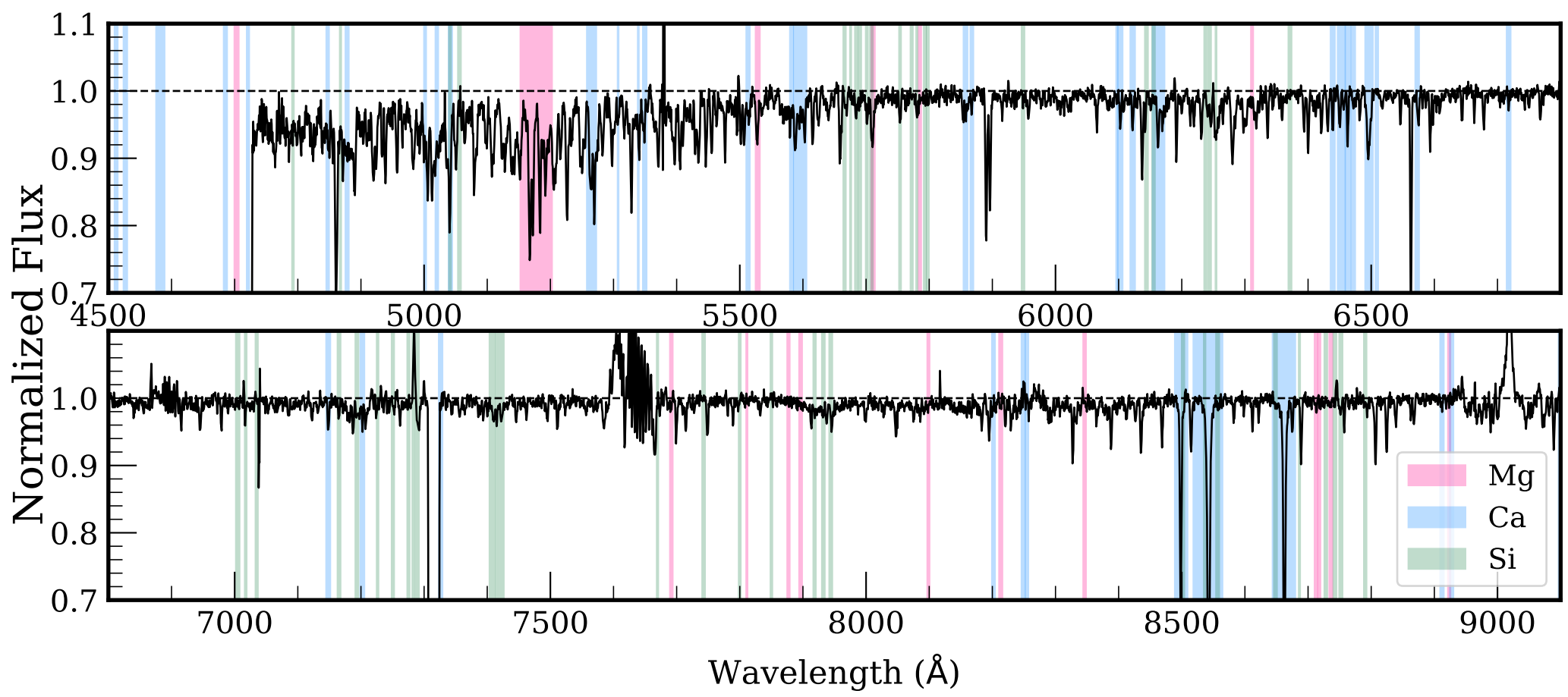

Figure 3.2: Wavelength regions sensitive to changes in [$\alpha$/Fe] for the spectral resolution of the DEIMOS 600ZD grating (~2.8 Å). We show an example spectrum (*black*) over the wavelength range 4500−9100 Å, where we corrected the spectrum for telluric absorption (Section 3.3) and performed an initial continuum normalization (Section 3.3). We do not show spectral regions with wavelengths below 4500 Å, since low S/N generally prevents utilization of the observed spectrum in this wavelength range. The spectrum is for a star in the globular cluster NGC 2419. Spectral regions sensitive to Mg, Ca, and Si are shown as highlighted ranges in magenta, blue, and green respectively. The atmospheric value of [$\alpha$/Fe] is measured using the union of the Mg, Ca, and Si spectral regions.

For stellar parameters between grid points, we linearly interpolated to generate model atmospheres within the ranges 3500 K < $T_{eff}$ < 8000 K, 0.0 < log $g$ < 5.0, −4.5 < [Fe/H] < 0.0, and -0.8 < [$\alpha$/Fe] < +1.2. A full description of the grid is presented in Table 3.4. Here, [$\alpha$/Fe] represents a *total* $\alpha$-element abundance for the atmosphere, which augments the abundances of individual $\alpha$-elements without distinguishing between their relative abundances. In total, the grid contains 316,848 synthetic spectra.

We generated the synthetic spectra using MOOG (Sneden 1973), an LTE spectral synthesis software. MOOG takes into account neutral hydrogen collisional line broadening (Barklem, Piskunov, and O'Mara 2000; Barklem and Aspelund-Johansson 2005), in addition to radiative and Stark broadening and van der Waals line damping. The most recent version (2017) includes an improved treatment of Rayleigh scattering in the continuum opacity (Alex Ji, private communication). The resolution of each generated synthetic spectrum is 0.02 Å.

Table 3.4: Parameter Ranges of Blue Grid (4100 - 6300 Å)

| Parameter | Minimum Value | Maximum Value | Step |
|---|---|---|---|
| $T_{\rm eff}$ (K) | 3500 | 5600 | 100 |
| | 5600 | 8000 | 200 |
| $\log g$ (cm s$^{-2}$) | 0.0 ($T_{\rm eff}$ < 7000 K) | 5.0 | 0.5 |
| | 0.5 ($T_{\rm eff}$ > 7000 K) | 5.0 | 0.5 |
| [Fe/H] | −4.5[a] ($T_{\rm eff} \leq$ 4100 K) | 0.0 | 0.1 |
| | −5 ($T_{\rm eff}$ > 4100 K) | 0.0 | 0.1 |
| [$\alpha$/Fe] | −0.8 | 1.2 | 0.1 |

[a] Below [Fe/H] < −4.5 for $T_{\rm eff} \leq$ 4100 K, certain stellar atmosphere models fail to converge when solving for molecular equilibrium in each atmospheric layer. Synthetic spectra with [Fe/H] < −4.5 exist for a majority of $T_{\rm eff}$-log $g$ pairs for $T_{\rm eff}$ $\leq$ 4100 K, but our grid is complete for all parameter combinations only above [Fe/H] = −4.5 in this regime.

**Photometric Constraints**

To reduce the dimensionality of parameter space and to optimize our ability to find the global chi-squared minimum in the parameter estimation (Section 3.4), we constrained the effective temperature and surface gravity of the synthetic spectra by available Johnson-Cousins $VI$ photometry for red giant stars in our sample. The photometric effective temperature is estimated using a combination of the Padova (Girardi et al. 2002), Victoria-Regina (VandenBerg, Bergbusch, and Dowler 2006), and Yonsei-Yale (Demarque et al. 2004) sets of isochrones, assuming an age of 14 Gyr and an $\alpha$-element abundance of 0.3 dex. If available, we also employed the Ramírez and Meléndez (2005) color temperature. We adopted a single effective temperature ($T_{\rm eff,phot}$) and associated uncertainty ($\sigma_{T_{\rm eff,phot}}$) from an average of the isochrone/color temperatures for each star.

We determined the photometric surface gravity in a similar fashion. However, no color-log $g$ relation exists, so we could not include this additional source for the photometric surface gravity. Unlike the effective temperature, we did not solve for log $g$ using spectral synthesis techniques, as the errors on the photometric surface gravity are negligible when the distance is known. Additionally, LRS and MRS spectra cannot effectively provide constraints on its value owing to the lack of ionized lines in the spectra. Thus, we held $\log g$ fixed in the abundance determination.

### [Fe/H] and [$\alpha$/Fe] Regions

In order to increase the sensitivity of the synthetic spectrum fit to a given abundance measurement, we constructed wavelength masks that highlight regions that are particularly responsive to changes in [Fe/H] and [$\alpha$/Fe]. We employed the same procedure as Kirby et al. (2009) to make the masks, starting with a base synthetic spectrum for each combination of $T_{\rm eff}$ (3500 - 8000 K in steps of 500 K) and log g (0.0 - 3.5 in steps of 0.5 dex). We assumed a bulk metallicity [Fe/H] = -1.5 and solar [$\alpha$/Fe] for the atmosphere. We then generated synthetic spectra with either enhanced or depleted values of individual element abundances (Fe, Mg, Si, Ca, and Ti) for each $T_{\rm eff}$-log $g$ pair and compare to the base synthetic spectra, identifying wavelength regions that differ by more than 0.5%. In the determination of the [Fe/H] and [$\alpha$/Fe] wavelength regions, we smoothed all synthetic spectra used to an approximation of the expected resolution of the 600ZD grating ($\sim$ 2.8 Å) across the entire spectrum (4100 - 9100 Å). We then compared the spectral regions for each element against the line list and high signal-to-noise (S/N > 100) spectra of cool ($T_{\rm eff}$ < 4200 K) globular cluster stars, eliminating any regions that do not have a corresponding transition in the line list or an absorption feature in the spectra.

Although our measurements reflect the atmospheric value of [$\alpha$/Fe], we constructed the associated wavelength mask from the regions sensitive to changes in the individual elements Mg, Si, and Ca. We excluded Ti from the [$\alpha$/Fe] mask owing to the prevalence of regions sensitive to Ti at bluer optical wavelengths, such that we cannot meaningfully isolate its elemental abundance. Figure 3.2 illustrates our [$\alpha$/Fe] mask across the entire usable spectral range (4500−9100 Å; Section 3.2). The [Fe/H] spectral regions cover 92% and 51% of the wavelength range in the blue and red respectively, whereas the [$\alpha$/Fe] regions span 15% and 12% of the same wavelength ranges. The overlap between the [Fe/H] regions and [$\alpha$/Fe] regions is 16% and 23% in the blue and red respectively. We emphasize that [Fe/H] and [$\alpha$/Fe] are measured separately and iteratively (Section 3.4).

### Parameter Determination from Spectral Synthesis

Here, we outline the steps involved in our measurement of atmospheric parameters and elemental abundances from spectral synthesis of low-resolution spectra. Figure 3.3 illustrates an example of a final continuum-normalized observed spectrum and best-fit synthetic spectrum resulting from our measurement procedure for a high S/N RGB star in a MW GC. Our method is nearly identical to that of Kirby et al. (2009), excepting our introduction of an additional free parameter, $\Delta\lambda$, the resolution of the observed spectrum. We use a Levenberg-Marquardt algorithm to perform each comparison between a given observed spectrum and a synthetic spectrum. We weight the comparison according to the inverse variance of the observed spectrum. In each step, the synthetic spectra utilized in the minimization are interpolated onto the observed wavelength array and smoothed to the fitted observed resolution, $\Delta\lambda$, prior to comparison with a given observed spectrum.

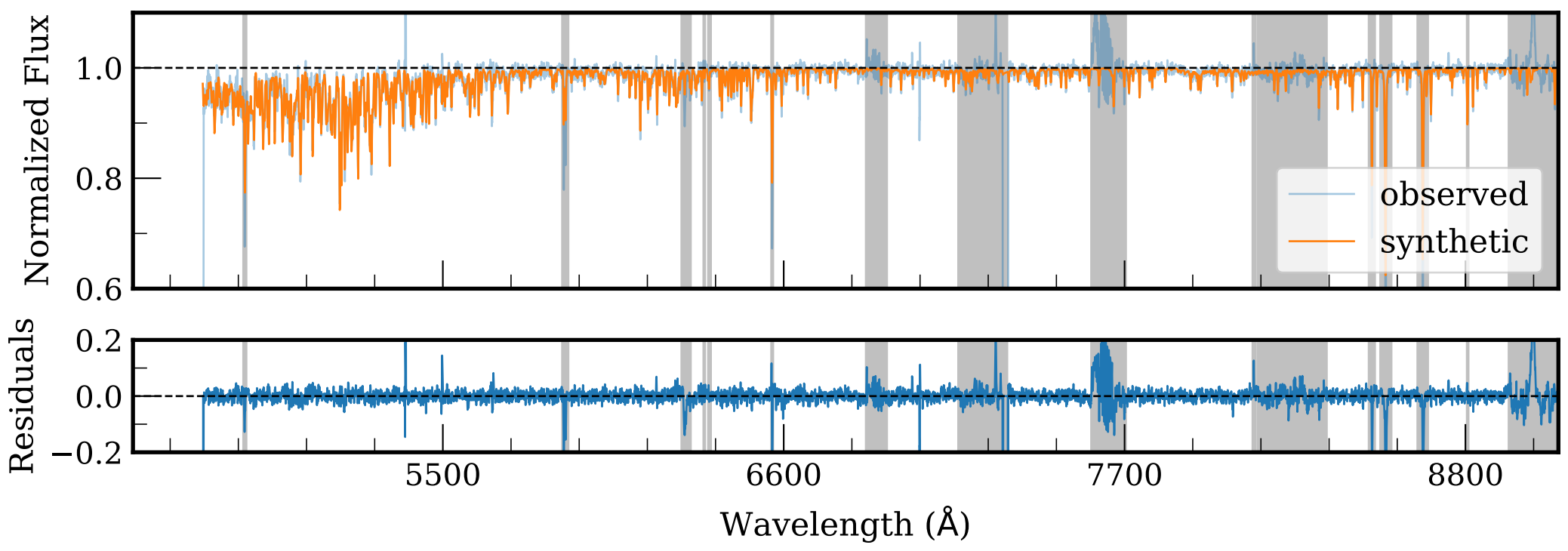

Figure 3.3: An example of a continuum-normalized observed spectrum (*light blue*) and its best-fit synthetic spectrum (*orange*). The observed spectrum corresponds to the same object in Figure 3.2. We show the the entire fitted wavelength range (~ 4500 - 9100 Å) for this object. The highlighted regions (*grey*) correspond to our standard mask (Section 3.3), which excludes lines such as H$\beta$, Na D1 and D2 from the fit. We adopt the parameters of the best-fit synthetic spectrum for the observed spectrum, $T_{\rm eff}$ = 4300 K, log $g$ = 0.71 dex, [Fe/H] = −2.07 dex, [$\alpha$/Fe] = 0.29 dex, and $\Delta\lambda$ = 2.60 Å FWHM. We measure $\chi^2_\nu$ = 1.63 for the quality of the fit across the full wavelength range, based on the regions of the spectrum used to measure [Fe/H] (Section 3.4). The normalized residuals (*dark blue*) between the continuum-refined observed spectrum and best-fit synthetic spectrum are also shown. The residuals have been scaled by the inverse variance of the observed spectrum and the degree of freedom of the fit, such that each pixel represents the direct contribution to $\chi^2_\nu$.

1. *$T_{\it eff}$, [Fe/H], and $\Delta\lambda$, first pass.* All three parameters are allowed to vary simultaneously in the fit. We use only regions sensitive to [Fe/H] (Section 3.4) in this measurement. We choose to measure $\Delta\lambda$ simultaneously with $T_{\rm eff}$ and [Fe/H] to prevent the chi-squared minimizer from under- or over-smoothing the synthetic spectrum to compensate for the initial guesses of $T_{\rm eff}$ and [Fe/H], which are offset from the final parameter values corresponding to the global $\chi^2$ minimum. The [Fe/H] regions cover almost the entire spectrum (92%) in the wavelength range 4100 - 6300 Å and a majority of the spectrum (51%) in the range 6300 - 9100 Å, such that using the entire spectrum to measure $\Delta\lambda$ does not change the results within the statistical uncertainties.

We assume a starting value of $T_{\rm eff,phot}$ (Section 3.4) for the spectroscopic effective temperature. $T_{\rm eff}$ is constrained by photometry using a Gaussian prior, such that $\chi^2$ increases if $T_{\rm eff}$ deviates substantially from $T_{\rm eff,phot}$, as defined by the associated error in the photometric effective temperature ($\sigma_{T_{\rm eff,phot}}$). As motivated in Section 3.4, log $g$ is fixed at the photometric value in all steps. We initialize [Fe/H] at $-2$ dex, where we performed tests to ensure that the final value of [Fe/H] does not depend on the initial guess. Similar to the approach for $T_{\rm eff}$, we enforce a Gaussian prior with a mean of 2.8 Å and standard deviation of 0.05 Å on $\Delta\lambda$, according to the expected spectral resolution for the 600ZD grating.[2] For the first iteration of the continuum refinement, [$\alpha$/Fe] remains fixed at solar. In subsequent iterations, [$\alpha$/Fe] is fixed at the value determined in step 2. The other parameters are allowed to vary until the best-fit synthetic spectrum is found.

2. *[α/Fe], first pass.* $T_{\rm eff}$, [Fe/H], and $\Delta\lambda$ are fixed at the values determined in step 1 while [$\alpha$/Fe] is allowed to vary, assuming a starting value of solar for the first iteration of the continuum refinement. Otherwise, the starting value is the value of [$\alpha$/Fe] determined in the last iteration of the continuum refinement. As in the case of [Fe/H], the final value of [$\alpha$/Fe] does not depend on the initial guess. In the determination of the best-fit synthetic spectrum, only wavelength ranges sensitive to variations in $\alpha$-element abundance are considered (Section 3.4).

3. *Iterative continuum refinement.* After a best-fit synthetic spectrum is determined according to steps 1 and 2, we refine the continuum normalization according to Section 3.3. We perform the continuum refinement iteratively, enforcing the convergence conditions that the difference in parameter values between the previous and current iteration cannot exceed 1 K, 0.001 dex, 0.001 dex, and 0.001 Å for $T_{\rm eff}$, [Fe/H], [$\alpha$/Fe], and $\Delta\lambda$ respectively. If these conditions are not met in a given iteration, the continuum-refined spectrum is used to repeat steps 1 and 2 until convergence is achieved. If the maximum number of iterations ($N_{\text{iter, max}}$ = 50) is exceeded, which occurs for a small fraction of observed spectra, we do not include the observed spectra in the subsequent analysis.

[2] Approximating the spectral resolution by constant value of $\Delta\lambda$ over the full spectrum does not impact the determination of either [Fe/H] or [$\alpha$/Fe]. Even if $\Delta\lambda$ over-smooths (under-smooths) the spectrum in the fitting procedure, it should not alter the identified $\chi^2$ minimum for [Fe/H] and [$\alpha$/Fe], given that the effect of over-smoothing (under-smoothing) the core and wings of an absorption feature should effectively negate each other. However, approximating the spectral resolution by $\Delta\lambda$ could increase $\chi^2_\nu$, consequently increasing the statistical uncertainties on the abundances.

4. *[Fe/H], second pass.* [Fe/H] is redetermined, where $T_{\text{eff}}$, $\Delta\lambda$, and [$\alpha$/Fe] are fixed at their converged values from step 3. We use the final continuum-refined observed spectrum determined in step 3 in this step and all remaining steps.

5. *[$\alpha$/Fe], second and final pass.* We repeat step 2, holding [Fe/H] fixed at the value determined in step 4.

6. *[Fe/H], third and final pass.* We repeat step 4, holding [$\alpha$/Fe] fixed at the value determined in step 5.

## 3.5 Globular Cluster Validation Tests

We demonstrate the robustness of our LRS technique by applying it to a set of MW GCs: NGC 2419, NGC 1904 (M79), NGC 6864 (M75), and NGC 6341 (M92) and NGC 7078 (M15).

NGC 2419 is a luminous outer halo GC located ~ 90 kpc away from the Galactic center (Harris et al. 1997) with multiple stellar populations, but no detected variation in [Fe/H] (Cohen and Kirby 2012). NGC 6864 also exhibits evidence for chemically distinct populations, including a marginal spread in [Fe/H] (~ 0.07 dex; Kacharov, Koch, and McWilliam 2013). It is a relatively young GC (Catelan et al. 2002) at a Galactocentric radius of ~ 15 kpc (Harris et al. 1997). NGC 1904 (~ 19 kpc; Harris et al. 1997) possesses an extended blue horizontal branch, but it is otherwise a typical cluster. NGC 6341 (~ 9 kpc; Harris et al. 1997) is notable primarily for being very metal-poor ([Fe/H] ~ −2.3 dex). NGC 7078 is similarly metal-poor, and has been observed to exhibit variations in $\alpha$-elements (Sneden, Pilachowski, and Kraft 2000; Carretta et al. 2009c). For a summary of observations used in our validation tests, see Table 3.1.

### Membership

For all subsequent analysis in this section, we utilize only stars that have been identified as RGB or AGB star members by Kirby et al. (2016). Membership is defined using both radial velocity and metallicity criteria based on MRS, such that any star whose measurement uncertainties are greater than $3\sigma$ from the mean of either radial velocity or metallicity is not considered a member. The colors and magnitudes of member stars must also conform to the cluster's giant branch.

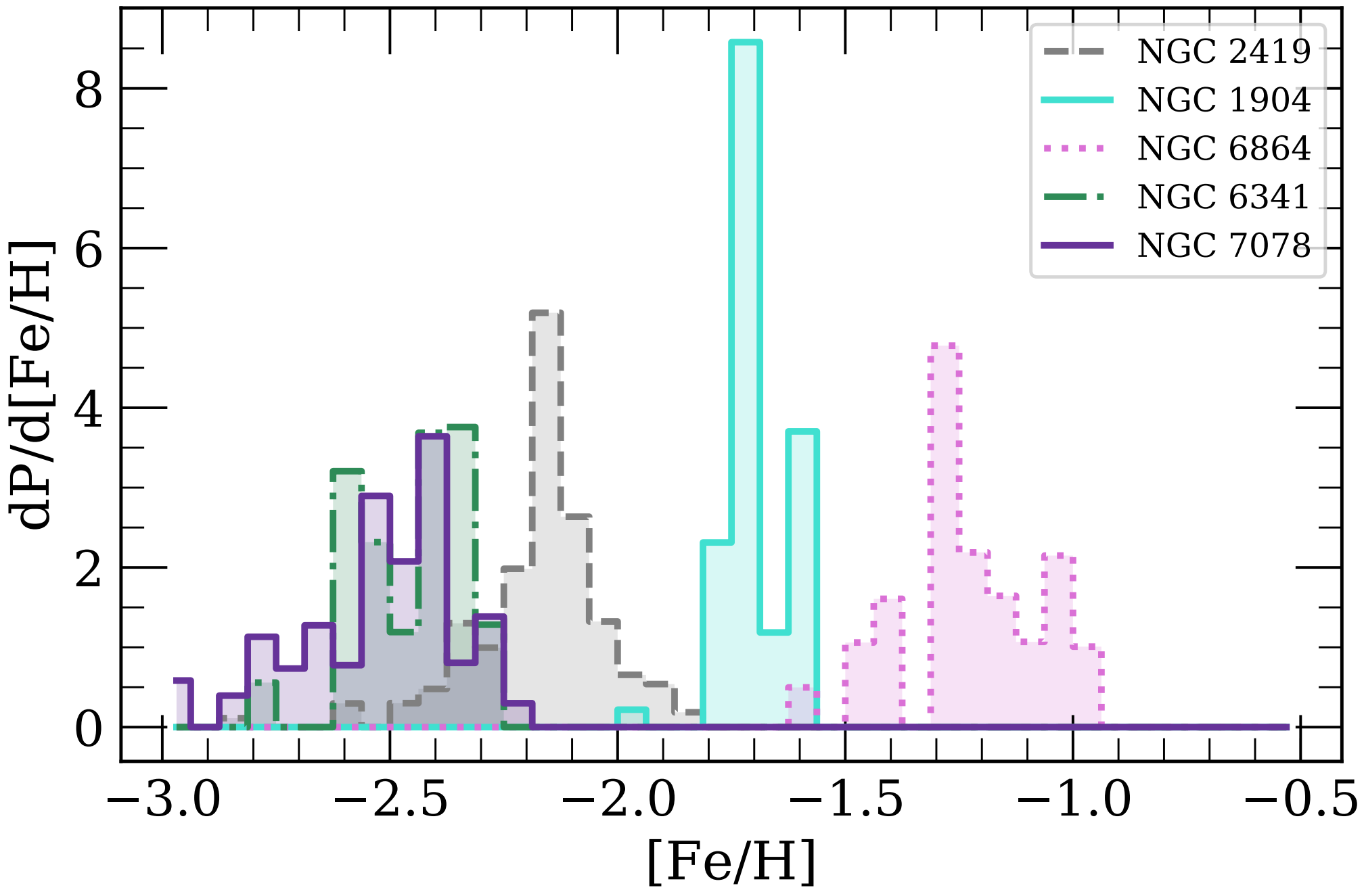

Figure 3.4: Error-weighted metallicity ([Fe/H]) distribution functions for RGB members of Galactic globular clusters NGC 2419 (*grey*), NGC 1904 (*cyan*), NGC 6864 (*magenta*), NGC 6341 (*green*), and NGC 7078 (*purple*). Only stars for which $\delta$[Fe/H] < 0.3 dex are shown. We find mean cluster metallicities of −2.14 dex, −1.70 dex, −1.22 dex, −2.45 dex, and −2.50 dex for the four respective clusters.

**Metallicity**

As described in Section 3.4, we measure metallicity from spectral regions sensitive to variations in [Fe/H]. In addition to membership criteria (Section 3.5), we further refine our sample by requiring that the $5\sigma$ contours in $T_{\rm eff}$, [Fe/H], and [$\alpha$/Fe] (Section 3.4) identify the minimum. This condition is effectively equivalent to requiring that a given star has sufficient S/N, a converged continuum iteration, and overall high enough quality fit ($\chi^2_\nu$) to produce a reliable abundance measurement.

We illustrate our results for [Fe/H] in the form of metallicity distribution functions (Figure 3.4) for NGC 2419, NGC 1904, NGC 6864, and NGC 6341, where we weight the distribution according to the *total* error in [Fe/H]. For a discussion of the measurement uncertainties, including systematic uncertainties, see Section 3.7. We find that $\langle$[Fe/H]$\rangle$ = −2.18 ± 0.15 dex, −1.70 ± 0.08 dex, −1.23 ± 0.15 dex, −2.45 ± 0.12 dex, and −2.53 ± 0.19 dex for NGC 2419, NGC 1904, NGC 6864, NGC 6341, and NGC 7078 respectively, where $\langle$[Fe/H]$\rangle$ is weighted according to the inverse variance of the total measurement uncertainty. These values approximately agree with the corresponding quantities from HRS: −2.12 ± 0.09 (Cohen and Kirby 2012), −1.58 ± 0.03 (Carretta et al. 2009c), −1.16 ± 0.07 (Kacharov, Koch, and McWilliam 2013), −2.34 dex (Sneden, Pilachowski, and Kraft 2000)[3], and −2.32 ± 0.07 dex respectively. In particular, we note that we find a spread in [Fe/H] for M15 that is likely not intrinsic (Carretta et al. 2009a), but rather a consequence of measurement uncertainty. Our estimate of the systematic uncertainty in [Fe/H] (Section 3.7) incorporates this dispersion in [Fe/H] measurements. We present a detailed comparison of our [Fe/H] measurements to HRS abundances in Section 3.6.

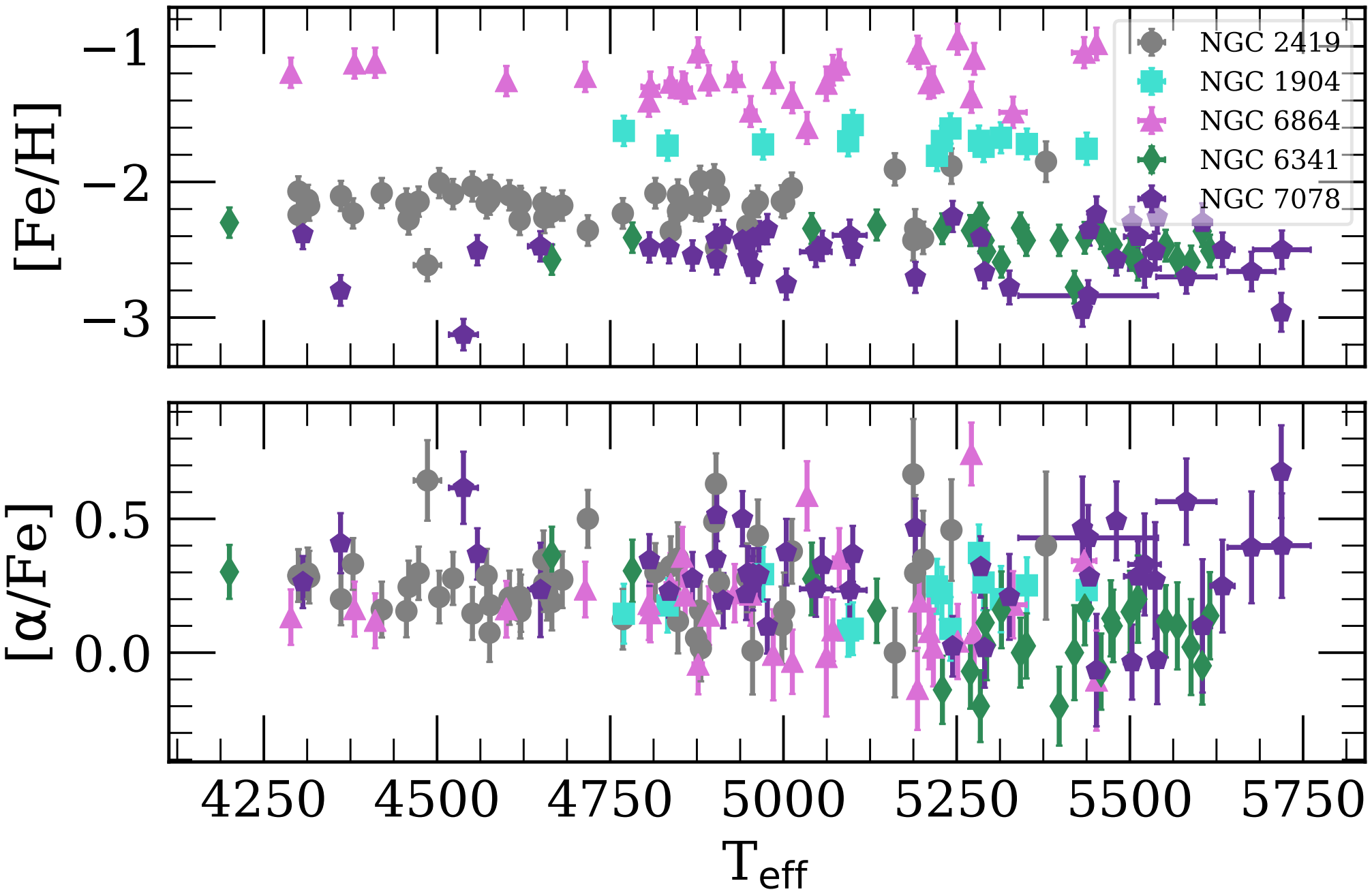

Figure 3.5: (*Top*) metallicity ([Fe/H]) and (*bottom*) [$\alpha$/Fe] vs. spectroscopic effective temperature ($T_{\rm eff}$). We show only stars with $\delta$[Fe/H] < 0.3 dex and $\delta$[$\alpha$/Fe] < 0.3 dex in each panel. The lack of a trend between both [Fe/H] and [$\alpha$/Fe] with respect to $T_{\rm eff}$ for each GC implies that our chemical abundance analysis is robust to systematic covariance in these parameters.

As another example of our ability to reliably recover abundances, we show [Fe/H] and [$\alpha$/Fe] vs. spectroscopically determined $T_{\rm eff}$ in Figure 3.5 for all GCs. In a nearly mono-metallic population like a GC, correlation of metallicity with other fitted parameters, such as $T_{\rm eff}$, would indicate the presence of systematic effects. Because $T_{\rm eff}$ is strongly covariant with [Fe/H], the fitting procedure might erroneously select a lower value of [Fe/H] and $T_{\mathrm{eff}}$ in order to match spectral features. Figure 3.5 presents evidence against any such correlation. The same argument can be extended to the $\alpha$-element abundance of a GC, given the assumption of chemical homogeneity. Similarly, we do not see any correlation between [$\alpha$/Fe] and $T_{\mathrm{eff}}$.

[3] Sneden, Pilachowski, and Kraft (2000) do not cite random uncertainties on their abundances. We represent their [Fe/H] and [$\alpha$/Fe] (Section 3.6) values as simple means.

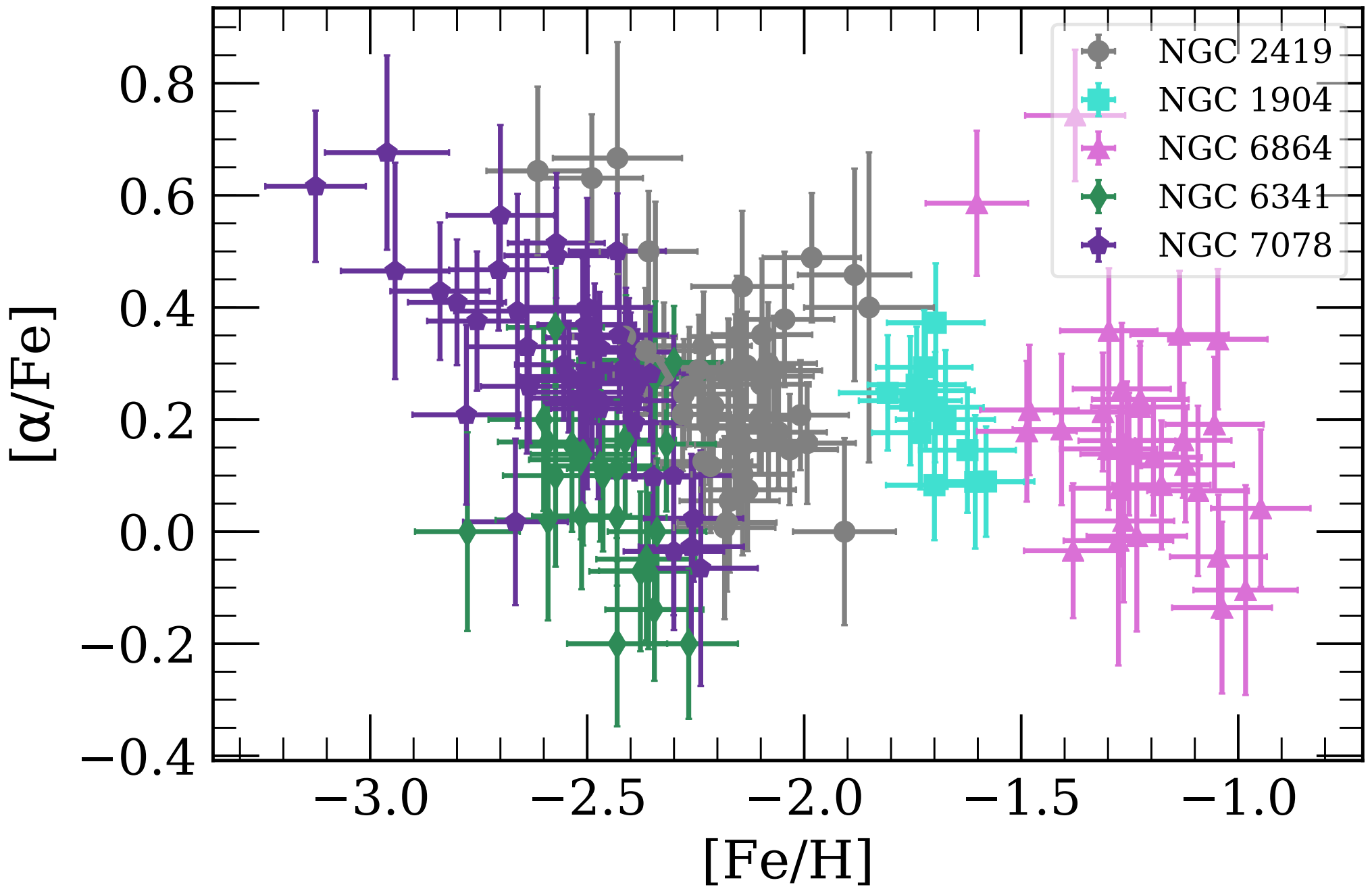

Figure 3.6: Atmospheric [$\alpha$/Fe] versus [Fe/H] for the same data set as Figure 3.4, where we also exclude points with $\delta$([$\alpha$/Fe]) > 0.3 dex. There is no apparent anticorrelation between [$\alpha$/Fe] and [Fe/H] within a GC, indicating that our method does not show any unphysical covariance between these two parameters.

**$\alpha$-element Abundance**

Similarly, we do not anticipate a correlation between [Fe/H] and [$\alpha$/Fe] within a GC. [$\alpha$/Fe] abundance impacts the determination of [Fe/H] via its contribution of $H^-$ opacity to the stellar atmosphere through electron donation. Thus, the abundance of [$\alpha$/Fe] alters stellar atmospheric structure, requiring a re-evaluation of [Fe/H] in the spectral fitting process. The presence of trends between [$\alpha$/Fe] with [Fe/H] (e.g., increasing [$\alpha$/Fe] with decreasing [Fe/H]) within a GC, which we expect to contain no such correlations, would indicate systematic effects in measuring abundances. As summarized in Figure 3.6, no such systematics are consistently present in our data for each GC. Even in the worst case scenario of M15, a massive, very metal-poor GC with known $\alpha$-element variation, any apparent anticorrelation is primarily driven by a few outliers in both [Fe/H] and [$\alpha$/Fe].

## 3.6 Comparison to High-Resolution Spectroscopy

Here, we compared our derived stellar parameters and elemental abundances from LRS to HRS literature measurements as a validation of our method.

### High-Resolution Data

Given the variety in approaches of HRS studies of the MW GCs, MW dSphs, and MW halo stars listed in Table 3.1, we provide a summary of the stellar parameter determination and abundance analysis in each case. For all systems, membership is determined based on radial velocities.

- *NGC 2419*: Using Keck/HIRES ($R$ ~ 34,000) spectroscopy, Cohen and Kirby (2012) measured [Fe/H], [Mg/Fe], [Si/Fe], and [Ca/Fe] for 13 RGB stars in NGC 2419. They used MOOG (Sneden 1973) in combination with Castelli and Kurucz (2003) atmospheric models to derive equivalent widths from neutral lines across the wavelength range 4500 - 8350 Å, including the Mg triplet. Stellar parameters were set to photometric values. Measurement uncertainties represent the dispersion of the mean abundance based on the various lines used in the abundance determination.

- *NGC 6864*: Kacharov, Koch, and McWilliam (2013) used Magellan/MIKE ($R$ ~ 30,000) to observe 16 RGB stars in NGC 6864 over a wavelength range of 3340 - 9150 Å. They measured [Fe/H], [Mg/H], [Si/H], and [Ca/H] via equivalent width measurements using MOOG and Castelli and Kurucz (2003) atmospheric models. Mg was measured from a single line (5711 Å). They determined $T_{\rm eff}$ from excitation equilibrium and surface gravities from $T_{\rm eff}$, extinction-corrected bolometric magnitude ($M_{\rm bol}$), and the known distance to the cluster. The measurement uncertainties are a combination of the random error (based on the number of lines used in the abundance analysis for a given element) and a component that reflects the error from adopted stellar atmosphere parameters. For the latter component, we adopt the larger, more conservative errors that reflect averages based on the entire GC sample of Kacharov, Koch, and McWilliam (2013).

- *NGC 1904*: From VLT/UVES ($R$ ~ 40,000), Carretta et al. (2009c) and Carretta et al. (2010) performed an abundance analysis based on equivalent width measurements for Fe, Mg, Si, and Ca, respectively, for a sample of 10 RGB stars, over the wavelenth range 4800 - 6800 Å. Following Carretta et al. (2009b), the authors adopted $T_{\rm eff}$ from calibrated $V$-$K$ colors and surface gravity from $T_{\rm eff}$, $M_{\rm bol}$, and distance moduli, and used Kurucz atmosphere models (with convective overshooting). To determine errors in the stellar atmospehric parameters, Carretta et al. repeated their analysis for each star, varying a single atmospheric parameter each time, to derive an average internal error, in addition to the rms error.

- *NGC 6341*: Based on WIYN/Hydra (R ~ 20,000, 5740 Å< $\lambda\lambda$ < 5980 Åspectra, Sneden, Pilachowski, and Kraft (2000) measured [Fe/H], [Si/Fe], and [Ca/Fe] abundances for RGB stars in NGC 6341 and NGC 7078. A single Si transition is used to determine [Si/Fe]. They adopted atmospheric parameters from $B - V$ photometry calibrations and employed MARCS atmosphere models in combination with MOOG. Given that Sneden, Pilachowski, and Kraft (2000) did not provide an estimate of abundance errors, we assume an uncertainty of 0.1 dex for all elemental abundances.

- *NGC 7078*: We utilize a compilation of data from Carretta et al. (2009c) and Carretta et al. (2010) and Sneden et al. (1997) and Sneden, Pilachowski, and Kraft (2000). The latter two studies employed similar techniques, where Sneden et al. (1997) used a mix of observations from both Hamilton and HIRES. In contrast to Sneden, Pilachowski, and Kraft (2000), Sneden et al. (1997) measured Mg abundances, based on a combination of equivalent width measurements $1-2$ strong Mg I lines and spectral synthesis of a slightly weaker line. We use [Fe I/H], as opposed to [Fe/H], which is given as the mean of [Fe I/H] and [Fe II/H] and assume 0.1 dex uncertainties on the HRS abundances.

- *MW dSphs*: For Canes Venatici I (CVnI) and Ursa Minor (UMi), we find a single star in common between each of our DEIMOS slitmasks and the HRS literature (François et al. 2016; Shetrone, Côté, and Sargent 2001). François et al. (2016) used VLT/X-shooter spectra ($\lambda$ = 300 nm $-$ 1 $\mu$m, $R$ = 7900 $-$ 12600) to measure [Fe/H], [Mg/Fe], and [Ca/Fe] for two stars in CVnI. They determined $T_{\rm eff}$ using a color-temperature relation and $V, I_C$ photometry, log $g$ from $M_{\rm bol}$, and abundances via spectral synthesis using OSMARCS (Gustafsson et al. 1975; Plez, Brett, and Nordlund 1992) atmosphere models. We estimated the HRS abundance errors based on the typical uncertainties of the published data. Based on Keck/HIRES spectroscopy (4540 Å$\lesssim$ $\lambda\lambda$ $\lesssim$ 7020Å), Shetrone, Côté, and Sargent (2001) measured [Fe/H], [Mg/Fe], and [Ca/Fe] for several stars in UMi, using MOOG and MARCS model atmospheres. They provided large upper limits on [Si/Fe], which we exclude from our analysis. Atmospheric parameters were determined simultaneously and iteratively using dereddened $(B-V)$ color-temperature relations, excitation equilibrium, and ionization balance. As for Draco, we use a mixture of data (Shetrone, Bolte, and Stetson 1998; Shetrone, Côté, and Sargent 2001; Fulbright, Rich, and Castro 2004; Cohen and Huang 2009). The methods of Shetrone, Bolte, and Stetson (1998) are nearly identical to those of Shetrone, Côté, and Sargent (2001), as is the case for Cohen and Huang (2009) in relation to Cohen and Kirby (2012). Aside from using Keck/HIRES spectroscopy and adopting photometric values for atmospheric parameters, the analysis of Fulbright, Rich, and Castro (2004) is similar to that of Fulbright (2000) (see discussion of MW halo HRS abundances). Additionally, the majority of Draco stars do not have a HRS measurement of Si abundance.

- *MW Halo*: We selected 5 MW halo stars from Fulbright (2000) with $\log g <$ 3.5 dex and $[\alpha/\mathrm{Fe}] \lesssim 0.15$ dex or $[\alpha/\mathrm{Fe}] \gtrsim 0.4$ dex. Using high-resolution ($R \sim$ 50,000), high S/N Lick/Hamilton, Fulbright (2000) measured Mg, Si, and Ca for 168 halo and disk stars. $T_{\mathrm{eff}}$ and $\log g$ were determined iteratively using Fe lines, with initial guesses determined from $V-K$ photometry. Abundances were determined from equivalent width measurements using MOOG and Kurucz model atmospheres with convective overshooting. We use [Fe I/H], as opposed to [Fe/H], which is given as the mean of [Fe I/H] and [Fe II/H]. The abundance errors reflect the rms uncertainty in each elemental abundance.

We emphasize that the HRS abundances do not include true estimates of systematic uncertainty, e.g., resulting from limitations from the selected grid of model atmospheres or line list. Additionally, we are comparing our homogeneous LRS abundances to an inhomogeneous HRS sample. As a result, some of the differences among the HRS studies can be attributed simply to different abundance measurement tools and techniques.

**Abundance Comparison**

We find good agreement in [Fe/H] for the 30 stars common between both data sets,where we have included abundance measurements of MW dSphs and MW halo stars, in addition to GCs, to expand our sample size with HRS overlap. In order to perform the comparison, we have shifted the HRS abundances (Cohen and Meléndez 2005; Gratton et al. 2003; Asplund et al. 2009; Grevesse and Sauval 1998; Anders and Grevesse 1989) to the same solar abundance scale as the LRS abundances. Figure 3.7 shows a strong correlation between the LRS and HRS measurements across a wide metallicity range ($\sim -3.0 - 0.0$ dex).

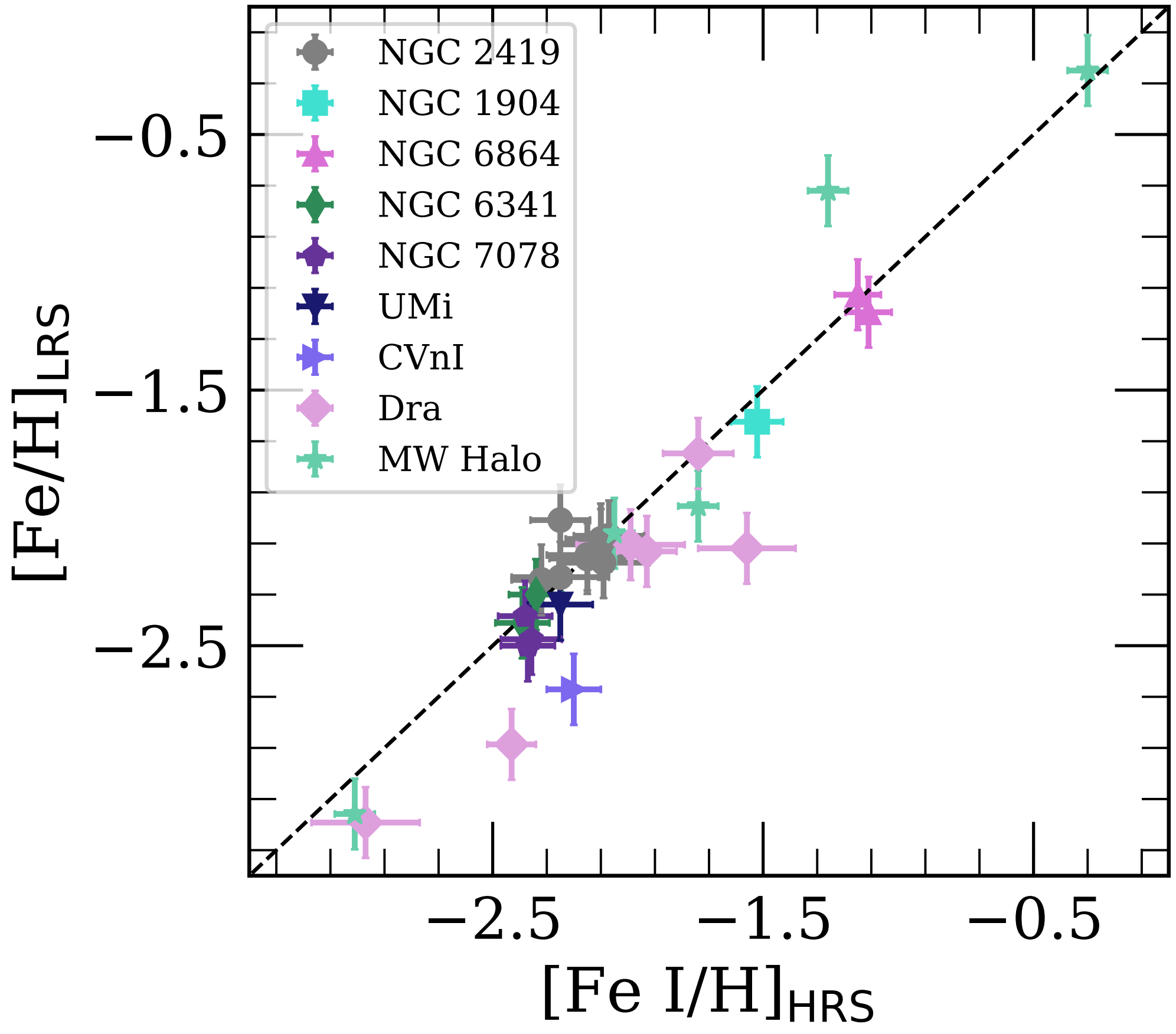

Figure 3.7: Metallicity measured from LRS ([Fe/H]) versus HRS ([Fe I/H]) for MW GCs, dSphs, and halo stars for NGC 2419, NGC 1904, and NGC 6864. Between LRS and HRS data sets, we have measurements in common for 30 stars. Although some scatter is present between data sets, $[\text{Fe I/H}]_{\text{HRS}}$ and [Fe/H] are strongly correlated at high significance across multiple orders of magnitude in metallicity.

In order to compare our $[\alpha/\text{Fe}]$ measurements to an analogous HRS quantity, we construct $[\alpha/\text{Fe}]_{\text{HRS}}$ based on a weighted sum of Mg, Si, and Ca elemental abundances. To derive the weights, we start with a reference synthethic spectrum defined by $T_{\text{eff}} = 4400$ K, $\log g = 1.0$ dex, $[\text{Fe/H}] = -1.8$ dex, which correspond to mean parameter values from HRS studies of NGC 2419, NGC 1904, NGC 6864, and NGC 6341, and $[\alpha/\text{Fe}] = 0$. We assume a spectral resolution of $\Delta\lambda = 2.8$ Å and interpolate

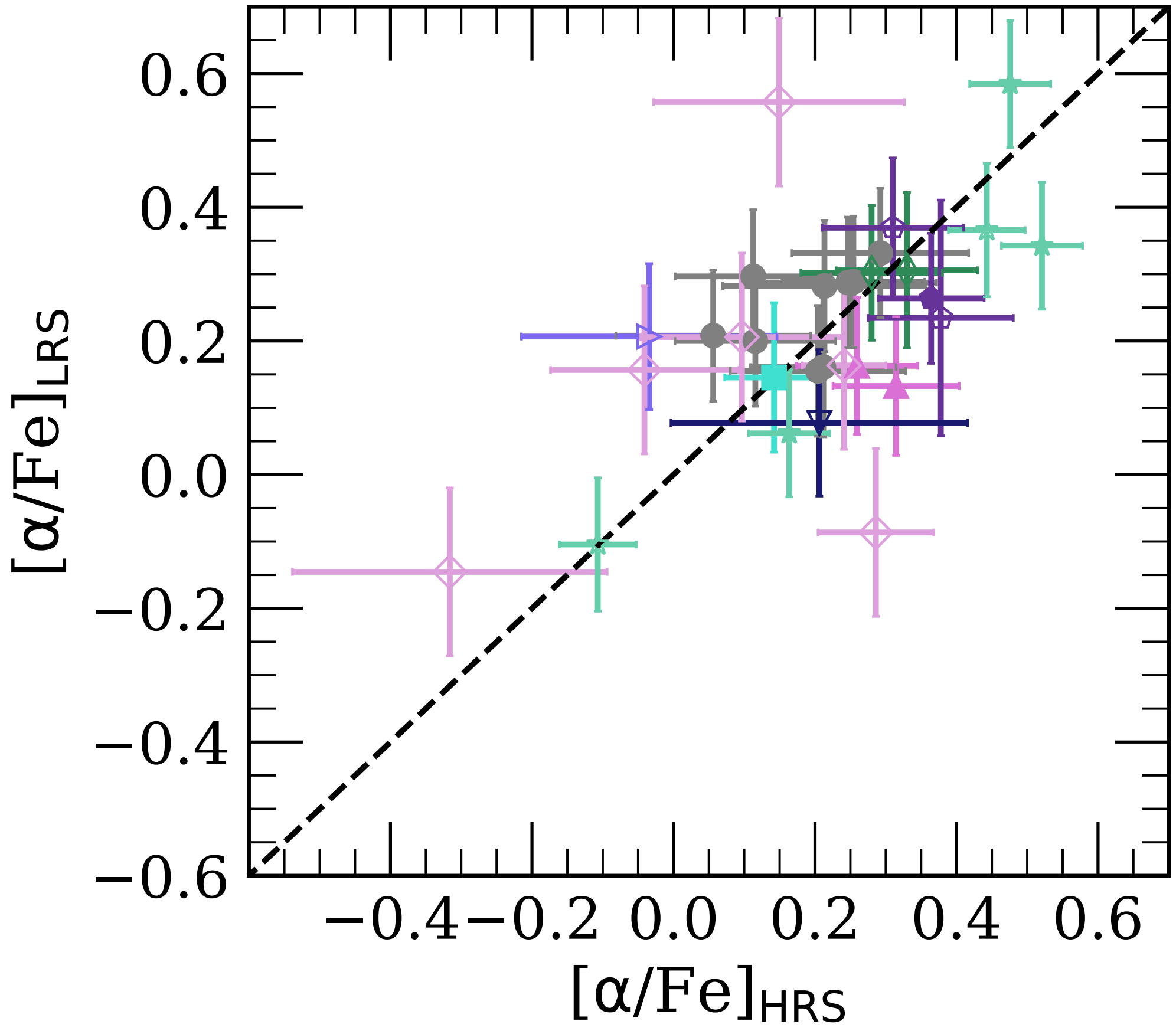

Figure 3.8: $[\alpha/\mathrm{Fe}]_{\mathrm{HRS}}$ versus $[\alpha/\mathrm{Fe}]_{\mathrm{LRS}}$ for the same data set as Fig. 3.7. We construct $[\alpha/\mathrm{Fe}]_{\mathrm{HRS}}$ based on a weighting (Section 3.6) of its individual $\alpha$-element abundances. Open symbols indicate values of $[\alpha/\mathrm{Fe}]_{\mathrm{HRS}}$ constructed from two or fewer individual $\alpha$-element abundance ratios. The LRS and HRS abundances exhibit broad agreement and show a clear correlation, despite the intrinsic differences between the sets of measurements.

the synthetic spectrum onto a wavelength array with spacing equal to the pixel scale of the 600 line $\mathrm{mm}^{-1}$ grating ($\sim 0.64$ Å). Next, we enhance/deplete the $\alpha$-element abundance by 0.1, 0.2, and 0.3 dex, calculating the sum of the absolute difference between the reference and enhanced/depleted synthetic spectrum in each case. For each $\alpha$-element, we utilize only the relevant wavelength regions (Section 3.4) and spectral coverage that corresponds to our data set (4500 - 9100 Å). Additionally, we exclude contributions from masked wavelength regions (Section 3.3). We adopt the normalized average value of the summed absolute flux differences as our final weight for a given element, i.e.,

Table 3.5: $\langle[\alpha/\text{Fe}]\rangle$ in MW GCs

| GC | $\langle[\alpha/\text{Fe}]_{\text{LRS}}\rangle$ (dex) | $\langle[\alpha/\text{Fe}]_{\text{HRS}}\rangle$ (dex) |
|---|---|---|
| NGC 2419 | 0.26 ± 0.13 | 0.21 ± 0.09 |
| NGC 1904 | 0.20 ± 0.09 | 0.28 ± 0.02 |
| NGC 6864 | 0.18 ± 0.18 | 0.28 ± 0.07 |
| NGC 6341 | 0.10 ± 0.16 | 0.37 |
| NGC 7078 | 0.30 ± 0.15 | 0.32 ± 0.06 |

Note.— The HRS references for NGC 2419, NGC 1904, NGC 6864, NGC 6341, are Cohen and Kirby (2012), Carretta et al. (2009c) and Carretta et al. (2010), Kacharov, Koch, and McWilliam (2013), and Sneden, Pilachowski, and Kraft (2000).

$$\begin{aligned}[\alpha/\text{Fe}]_{\text{HRS}} = 0.282 \times [\text{Mg/Fe}]_{\text{HRS}} + 0.136 \times [\text{Si/Fe}]_{\text{HRS}} \\ + 0.582 \times [\text{Ca/Fe}]_{\text{HRS}},\end{aligned} \tag{3.1}$$

$$\begin{aligned}\delta[\alpha/\text{Fe}]_{\text{HRS}} = \Big[\big(0.282 \times \delta[\text{Mg/Fe}]_{\text{HRS}}\big)^2 + \big(0.136 \times \delta[\text{Si/Fe}]_{\text{HRS}}\big)^2 \\ + \big(0.582 \times \delta[\text{Ca/Fe}]_{\text{HRS}}\big)^2\Big]^{1/2}.\end{aligned} \tag{3.2}$$

In Figure 3.8, we utilize Eqs. 3.1 and 3.2 to directly compare HRS and LRS $\alpha$-element abundances for the same sample as Figure 3.7. A clear positive correlation exists between $[\alpha/\text{Fe}]_{\text{LRS}}$ and $[\alpha/\text{Fe}]_{\text{HRS}}$, with some degree of scatter present. We emphasize that $[\alpha/\text{Fe}]_{\text{HRS}}$ represents only an approximation to the atmospheric value of $[\alpha/\text{Fe}]$, given the fundamental differences between the HRS and LRS methods (Section 3.6).

In Table 3.5, we summarize our findings for $\langle[\alpha/\text{Fe}]\rangle_{\text{LRS}}$ in MW GCs and compare to equivalent HRS measurements constructed using Eqs. 3.1 and 3.2.[4] In the case of NGC 6341, we construct $[\alpha/\text{Fe}]_{\text{HRS}}$ based only on [Ca/Fe] and [Si/Fe] because Sneden, Pilachowski, and Kraft (2000) did not measure [Mg/Fe]. In any instance of an incomplete set of Mg, Ca, and Si abundances for our HRS comparisons, we renormalized the weights in Eqs. 3.1 and 3.2 accordingly. The average LRS

[4]The HRS reference used to construct $[\alpha/\text{Fe}]_{\text{HRS}}$ for NGC 7078 is based on the most recent study of $\alpha$-enhancement in this cluster. The star-to-star HRS comparisons of Figures 3.7 and 3.8 are based on a compilation of values from Sneden et al. (1997) and Sneden, Pilachowski, and Kraft (2000) and Carretta et al. (2009c) and Carretta et al. (2010) for NGC 7078.

$\alpha$-element abundances are in some cases lower than the HRS measurements by ~0.1 dex, but overlap within the associated standard deviation on the measurement, excepting the case of NGC 6341. We find a significantly lower value of $\langle[\alpha/\mathrm{Fe}]\rangle$, although we note that it is particularly difficult to compare between LRS and HRS in this case. Given that NGC 6341 lacks HRS Mg abundances, we cannot perform a meaningful direct comparison of our constructed $\langle[\alpha/\mathrm{Fe}]_{\mathrm{HRS}}\rangle$ to our measurements of $\langle[\alpha/\mathrm{Fe}]_{\mathrm{LRS}}\rangle$, which include Mg.

This apparent offset in between the LRS and HRS measurements can be characterized in terms of a systematic uncertainty component. We find that a star-by-star comparison of [$\alpha$/Fe] and [Fe/H] shows that our LRS results are consistent with those from HRS within the systematic uncertainties (Section 3.7). We estimate the systematic error by calculating the additional error term that would be required to force the LRS and HRS measurement to agree within one standard deviation. The relevant equation is,

$$\frac{1}{N}\sum_{i}^{N}\frac{\left(\epsilon_{\mathrm{LRS,i}}-\epsilon_{\mathrm{HRS,i}}\right)^2}{\left(\delta\epsilon^2_{\mathrm{LRS,i}}+\delta\epsilon^2_{\mathrm{HRS,i}}+\delta\epsilon^2_{sys}\right)}=1, \tag{3.3}$$

where $\epsilon$ represents a given elemental abundance, such as [Fe/H] or [$\alpha$/Fe], $\delta\epsilon$ is the corresponding statistical uncertainty on the measurement, $i$ is an index representing a given star in common between both the HRS and LRS data sets, and $N$ is the total number of common stars.

For $\epsilon$ = [Fe/H] and $N$ = 30 stars, we numerically solve Eq. 3.3 to find $\delta\epsilon_{\mathrm{sys}}$ = 0.176 dex. A majority of this systematic term is driven by a single MW halo star with a discrepant value of $[\mathrm{Fe/H}]_{\mathrm{LRS}}$. Excluding it from the calculation, we find $\delta\epsilon_{\mathrm{sys}}$ = 0.143 dex for $N$ = 29 stars. This value is likely more representative of the true systematic uncertainty, given that it is not subject to extreme outliers and exhibits better agreement with an independent calculation of the systematic uncertainty from the intrinsic spread in GCs (Section 3.7, Table 3.6). In the case of $\epsilon$ = [$\alpha$/Fe] we find $\delta\epsilon_{\mathrm{sys}}$ = 0.058 dex and 0.039 dex, respectively, in the cases of $N$ = 30 and $N$ = 29. For consistency, we adopt the latter value to reflect the systematic uncertainty in $[\alpha/\mathrm{Fe}]_{\mathrm{LRS}}$ from HRS comparisons. Given the intrinsic heterogeneity of the HRS sample (Section 3.6) and the comparatively limited sample size ($N$ = 29 versus $N$ = 154), we chose not to adopt these values as our systematic uncertainty. Instead, we favor values calculated from the internal spread in GCs (Section 3.7).

Table 3.6: Systematic Uncertainty

| Parameter | $\delta_{\mathrm{sys,HRS}}$[a] (dex) | $N_{\mathrm{HRS}}$ | $\delta_{\mathrm{sys,gc}}$[b] (dex) | $N_{\mathrm{gc}}$[c] |
|---|---|---|---|---|
| [Fe/H] | 0.143 | 29 | 0.111 | 154 |
| [$\alpha$/Fe] | 0.039 | 29 | 0.094 | 68 |

[a] Number of stars with both LRS and HRS abundance measurements used to determine the systematic uncertainty from HRS
[b] The systematic uncertainty as calculated from the intrinsic spread in GCs (Section 3.7). We adopt these values over the systematic uncertainty determined from comparison to HRS, $\delta_{\mathrm{sys,HRS}}$ (Section 3.6), given the heterogeneity of the HRS sample (Section 3.6).
[c] Number of stars used to determine the systematic uncertainty from the intrinsic spread in GCs.

## 3.7 Systematic Uncertainty from Internal Spread in Globular Clusters

The total uncertainty on fitted parameters is composed of two components added in quadrature, the statistical (fit) uncertainty, $\delta_{\mathrm{fit}}$, and a systematic component, $\delta_{\mathrm{sys}}$. The fit uncertainty is calculated according to the reduced chi-squared value ($\chi^2_\nu$) and the diagonals of the covariance matrix of the fit ($\sigma_{ii}$), i.e., $\sigma_{ii}(\chi^2_\nu)^{1/2}$. We calculate $\chi^2_\nu$ using only the regions of the observed spectrum utilized in the fit, e.g., in the case of [Fe/H], we use the wavelength regions sensitive to [Fe/H] (Section 3.4) and not excluded by the pixel mask (Section 3.3). The systematic component encapsulates uncertainty intrinsic to our method, owing to sources such as the linelist (Section 3.4), assumptions involved in spectral synthesis (Section 3.4), details of our method, such as the continuum normalization (Section 3.3) and fitting procedure (Section 3.4), and covariance with other fitted parameters.[5]

### Metallicity

Because most GCs, including those in our sample, are nearly monometallic (Carretta et al. 2009a), we can derive an estimate of systematic uncertainty in [Fe/H], $\delta[\mathrm{Fe/H}]_{\mathrm{sys}}$, by enforcing the condition that the intrinsic dispersion in the GC is zero, i.e.,

$$\sigma^2 = \mathrm{var}\left[\frac{[\mathrm{Fe/H}]_i - \langle[\mathrm{Fe/H}]\rangle}{\left(\delta[\mathrm{Fe/H}]^2_{\mathrm{fit},i} + \delta[\mathrm{Fe/H}]^2_{\mathrm{sys}}\right)^{1/2}}\right] = 1, \tag{3.4}$$

[5] We do not completely characterize the systematic uncertainty on $T_{\mathrm{eff}}$ and $\Delta\lambda$ because our primary goal is to determine abundances. The systematic errors $\delta([\mathrm{Fe/H}])_{\mathrm{sys}}$ and $\delta([\alpha/\mathrm{Fe}])_{\mathrm{sys}}$ already account for errors propagated by inaccuracies in $T_{\mathrm{eff}}$ and $\Delta\lambda$. All of the of uncertainties in $T_{\mathrm{eff}}$ and $\Delta\lambda$ presented in this chapter reflect only the statistical uncertainty.

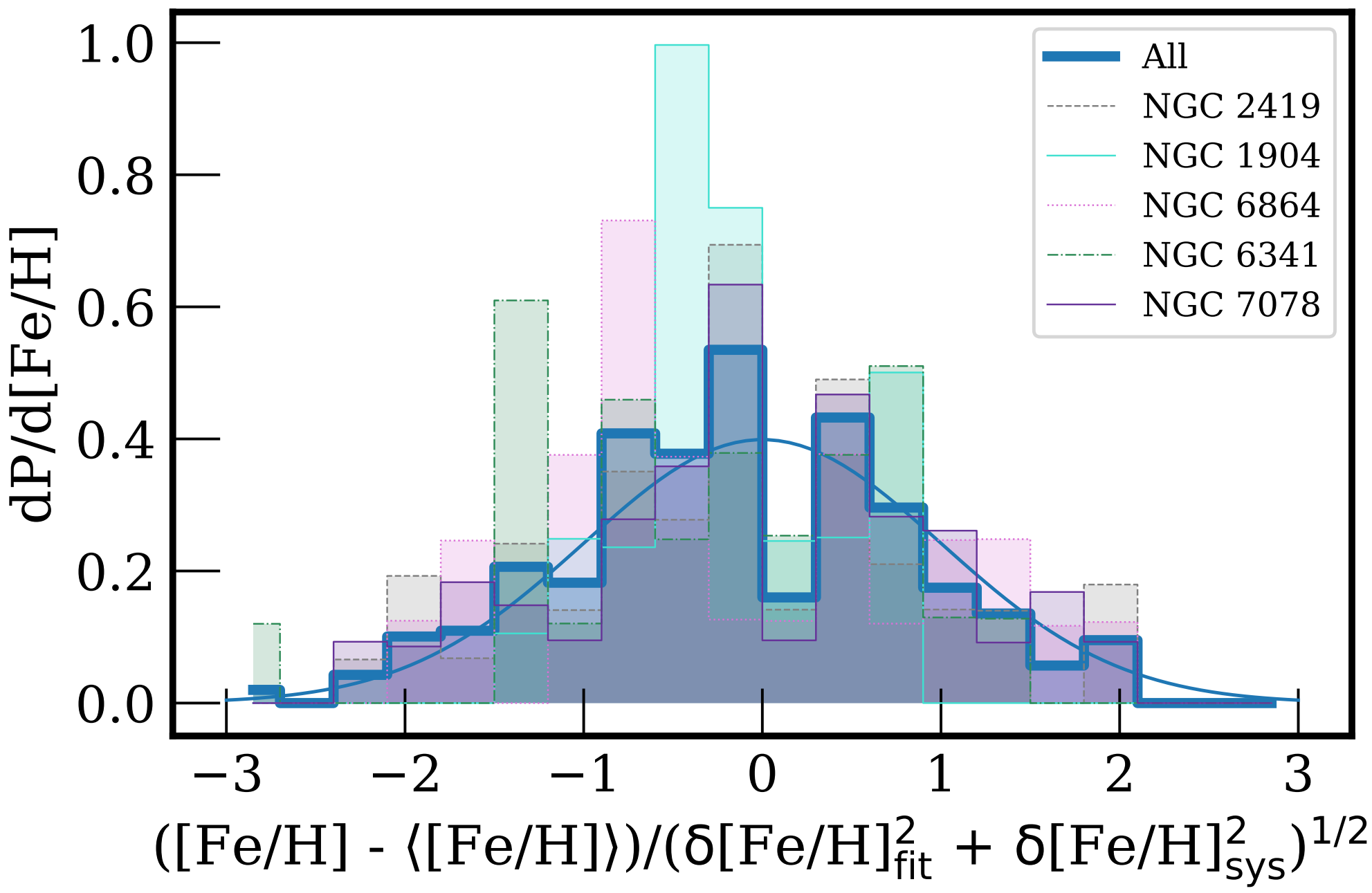

Figure 3.9: Probability distribution function of [Fe/H] normalized to the mean metallicity of a given cluster (⟨[Fe/H]⟩) and weighted by the total error in metallicity. We show the distributions for NGC 2419 (*grey*), NGC 1904 (*cyan*), NGC 6864 (*magenta*), NGC 6341 (*green*) , and all five clusters (*blue*) 154 stars). The total error is composed of the statistical uncertainty from the fit ($\delta$[Fe/H]$_{\rm fit}$) and the systematic uncertainty ($\delta$[Fe/H]$_{\rm sys}$). We determine the systematic uncertainty from the intrinsic dispersion in the combined distribution for all three clusters. The Gaussian defined by the systematic uncertainty ($\delta$([Fe/H])$_{\rm sys}$ = 0.111 dex) is overplotted.

where $i$ is the index for a star in the GC, $\delta$[Fe/H]$_{\rm fit}$ is the S/N-dependent statistical uncertainty in [Fe/H], and ⟨[Fe/H]⟩ is the mean metallicity of the GC, where the mean is weighted by the statistical uncertainty on each measurement of [Fe/H]. Eq. 3.4 follows a reduced chi-squared distribution with an expectation value of unity. Enforcing the condition $\sigma^2 = 1$, we can numerically solve for the most likely value of the systematic uncertainty.

First, we refine our sample by removing outliers in [Fe/H] for each GC. Following Kirby et al. (2016), we calculate the mean metallicity, ⟨[Fe/H]⟩, and standard deviation, $\sigma$([Fe/H]), for each cluster, and we remove stars that deviate by more than 2.58$\sigma$ (99% confidence level). Then, we re-compute ⟨[Fe/H]⟩ and $\sigma$([Fe/H]) from this refined sample, including stars from the full sample that fulfill the criteria $|$[Fe/H] $-$ ⟨ [Fe/H]⟩$|$ $-$ $\delta$([Fe/H])$_{\rm fit}$ $< 3\sigma$. The inclusion of $\delta$( [Fe/H])$_{\rm fit}$ in the inequality allows stars to be considered members even if they fall outside of the allowed metallicity range, as long as some part of their $1\sigma$ confidence intervals falls within the range.

After performing this sigma clipping for each individual cluster, we subtract the mean cluster metallicity from each star's measurement of [Fe/H], and we solve for the intrinsic dispersion based on the combined sample (Eq. 3.4). We obtain a systematic uncertainty in [Fe/H] of $\delta([\mathrm{Fe/H}])_{\mathrm{sys}} = 0.111$ dex (Table 3.6) based on 154 stars. We present an illustration of this method in Figure 3.9, where we show the probability distributions for the total-error-weighted metallicity of each cluster, in addition to the the combined GC sample. The fact that the combined distribution is well-approximated by a Gaussian with $\sigma = 1$ indicates that the calculated systematic uncertainty sufficiently accounts for the observed metallicity spread.

Thus, the total error is,

$$\delta([\mathrm{Fe/H}])_{\mathrm{tot}} = \sqrt{\delta([\mathrm{Fe/H}])^2_{\mathrm{fit}} + \delta([\mathrm{Fe/H}])^2_{sys}}. \quad (3.5)$$

In general, the statistical fit uncertainty for [Fe/H] is negligible compared to the systematic error for GCs. However, this is not be the case for M31, given the low value of the expected S/N.

**$\alpha$-element Abundance**

To determine the systematic uncertainty in [$\alpha$/Fe], $\delta([\alpha/\mathrm{Fe}])_{\mathrm{sys}}$, we calculate the intrinsic dispersion in the clusters, analogously to Eq. 3.4. Whereas it is generally reasonable to assume that GCs have negligible spread in [Fe/H], the assumption of zero intrinsic variation in [$\alpha$/Fe] must be evaluated individually for each cluster. For example, abundance analysis of HRS has detected a significant spread in Mg for NGC 2419, where a minority of the population is Mg-abnormal (Cohen, Huang, and Kirby 2011; Cohen and Kirby 2012). Large star-to-star variations in Mg have also been found for NGC 7078 from HRS studies (Sneden, Pilachowski, and Kraft 2000; Carretta et al. 2009c). Although NGC 6864 possesses chemically distinct populations, O is the only $\alpha$-element that exhibits significant variation within the cluster, as opposed to Mg, Si, or Ca (Kacharov, Koch, and McWilliam 2013). NGC 6341 is not known to possess $\alpha$-element variations (Sneden, Pilachowski, and Kraft 2000), with the caveat that no recent Mg abundances from HRS have been published to our knowledge.

We therefore construct our combined GC sample from NGC 1904, NGC 6864, and NGC 6341 to compute $\delta([\alpha/\mathrm{Fe}])_{\mathrm{sys}}$, obtaining a value of 0.094 dex (Table 3.6) from

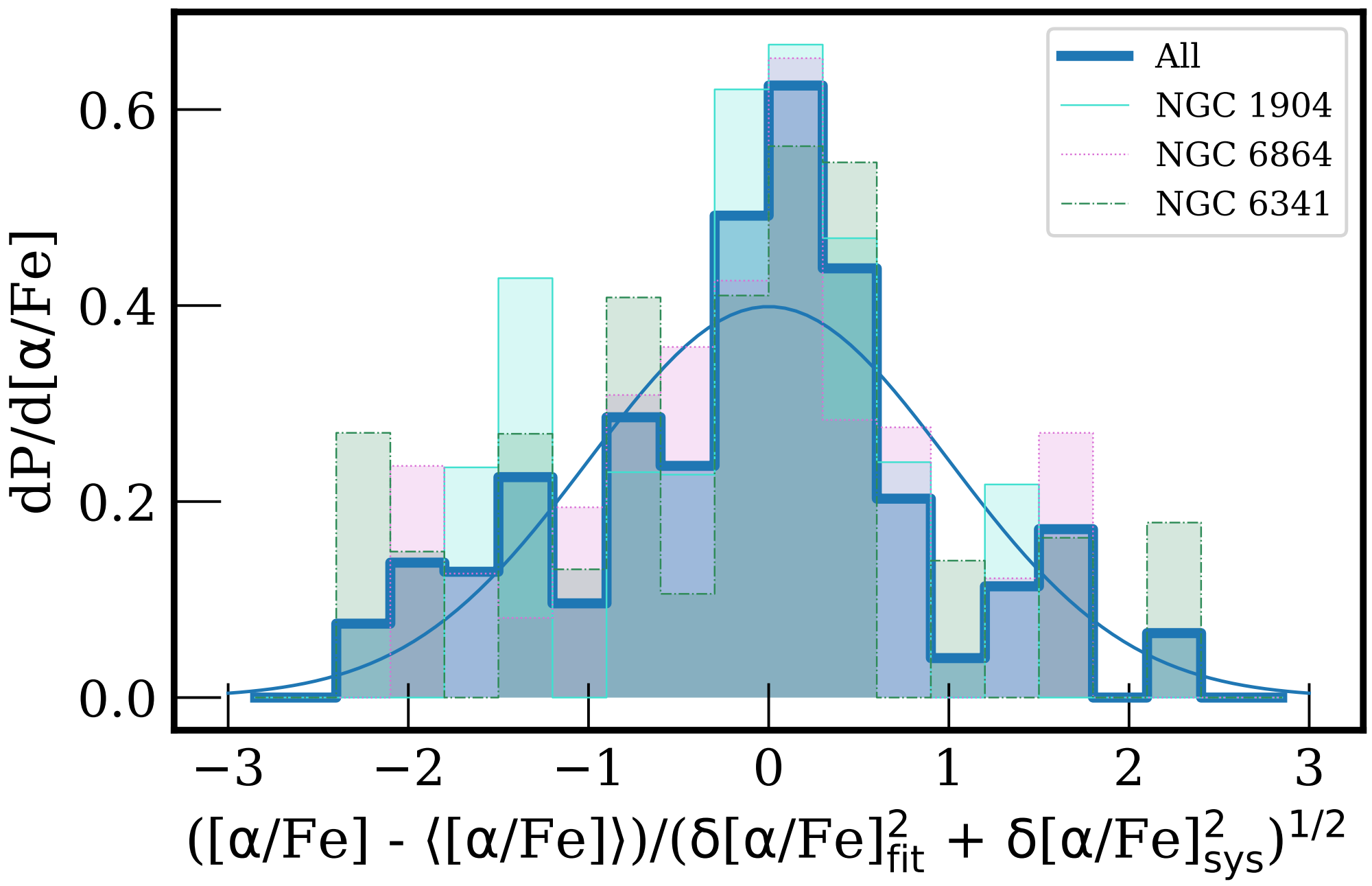

Figure 3.10: Probability distributions used to determine the systematic uncertainty, as in Figure 3.9, except for the case of [$\alpha$/Fe]. We show the distributions for NGC 1904 (*cyan*), NGC 6864 (*magenta*), NGC 6341 (*green*), and all three clusters (*blue*) (68 stars). We find that $\delta$([$\alpha$/Fe])$_{\rm sys}$ = 0.094 dex. (Section 3.7)

Table 3.7: M31 Stellar Halo Observations (f130_2)

| Object | Date | $\theta_s$ (”) | $\langle X \rangle$ | $t_{\rm exp}$ (s) | $N$ |
|---|---|---|---|---|---|
| f130_2a[a] | 2018 Jul 19 | 1.0 | 1.53 | 5639 | 37 |
| f130_2b | 2018 Jul 19 | 1.0 | 1.16 | 5758 | 37 |
| f130_2a | 2018 Aug 14 | 0.86 | 1.29 | 4140 | 37 |
| f130_2a | 2018 Oct 10 | 0.83 | 1.84 | 3000 | 37 |
| f130_2a | 2018 Oct 11 | 0.60 | 1.49 | 2400 | 37 |

[a] Slitmasks indicated “a” and “b” are identical, except that the slits are titled according to the median parallactic angle at the approximate time of observation.

68 stars. Figure 3.10 illustrates that the adopted error floor in [$\alpha$/Fe] describes the data well. We anticipate a smaller value of $\delta$([$\alpha$/Fe])$_{\rm sys}$ relative to $\delta$([Fe/H])$_{\rm sys}$, given that the systematic effects (uncertainties in the line list, atmospheric parameters, continuum normalization, etc.) that impact [Fe/H] tend to similarly affect [$\alpha$/H].

## 3.8 The Star Formation History of the Stellar Halo of M31

We apply our spectral synthesis technique to spectra of individual RGB stars in the stellar halo of M31. We select a field with no identified substructure (Gilbert et al. 2007) as an example. We will apply our method to additional M31 stellar halo fields in following chapters.

### Halo Field Observations

The field, f130_2, is located at 23 kpc in projected radius along the minor axis of M31, and was first observed and characterized by Gilbert et al. (2007) using the Keck II/DEIMOS 1200 line $mm^{-1}$ grating. We selected it owing to its proximity to the 21 kpc halo field of Brown et al. (2007), for which Brown et al. (2009) presented catalogs of deep optical photometry obtained using the Advanced Camera for Surveys (ACS) on the *Hubble Space Telescope*.

Table 3.7 summarizes our observations of the M31 stellar halo field, which we observed with the same configuration as described in Section 3.2. The total exposure time was 5.8 hours. Following Cunningham et al. (2016), we designed two separate slitmasks for the single field, with the same mask center, mask position angle, and target list, but with differing slit position angles. Switching slitmasks in the middle of the observation allows us to approximately track the change in parallactic angle over the course of the night. This technique mitigates flux losses due to differential atmospheric refraction (DAR), which disproportionately affects blue wavelengths. Thus, it is especially important to consider DAR when observing with the 600 line $mm^1$ grating, which covers a wider spectral range than any other DEIMOS grating.

### Sample Selection

The observed field, at a M31 galactocentric radius of 23 kpc, includes a non-negligible contamination fraction of Milky Way foreground dwarf stars. In order to identify both secure and marginal M31 members ($\langle L_{i,\text{no v}} \rangle > 0$; Gilbert et al. 2012), we used a likelihood-based method (Gilbert et al. 2006) that relies on three criteria to determine membership: the strength of the Na I $\lambda\lambda$8190 absorption line doublet, the *(V, I)* color-magnitude diagram location, and photometric versus spectroscopic (Ca II $\lambda\lambda$8500) metallicity estimates. Following Gilbert et al. (2007), we excluded radial velocity as a criterion to result in a more complete sample. In total, we identified 37 M31 stellar halo members ($20 \lesssim I_0 \lesssim 22.5$) in this field out of 106 targets.

We required that our abundance measurement technique determined the abundances reliably (Section 3.5): $\delta$([Fe/H]) < 0.5 and $\delta$([$\alpha$/Fe]) < 0.5. We also required that the $5\sigma$ $\chi^2$ contours in $T_{\rm eff}$, [Fe/H], and [$\alpha$/Fe] (Section 3.4) identify the minimum. Both of these criteria effectively mimic a S/N cut (S/N $\gtrsim$ 10 Å$^{-1}$). Lastly, we manually screened member stars for molecular TiO bands between 7055−7245 Å (Cenarro et al. 2001; Gilbert et al. 2006), where affected stars exhibit a distinctive pattern. Stars with strong TiO absorption tend to be more metal-rich ([Fe/H] $\gtrsim -1.5$), have red colors ($(V - I)_0 > 2.0$), and can also show unusual $\chi^2$ contours in [$\alpha$/Fe]. We omitted 7 M31 member stars that passed the aforementioned cuts, which meet the $(V - I)_0$ color criterion and show spectral evidence of strong TiO absorption. In total, this reduces the sample size to 11 stars (S/N $\sim$ 10−30 Å$^{-1}$), for which we present a summary of stellar parameters and chemical abundances in Table 3.8.

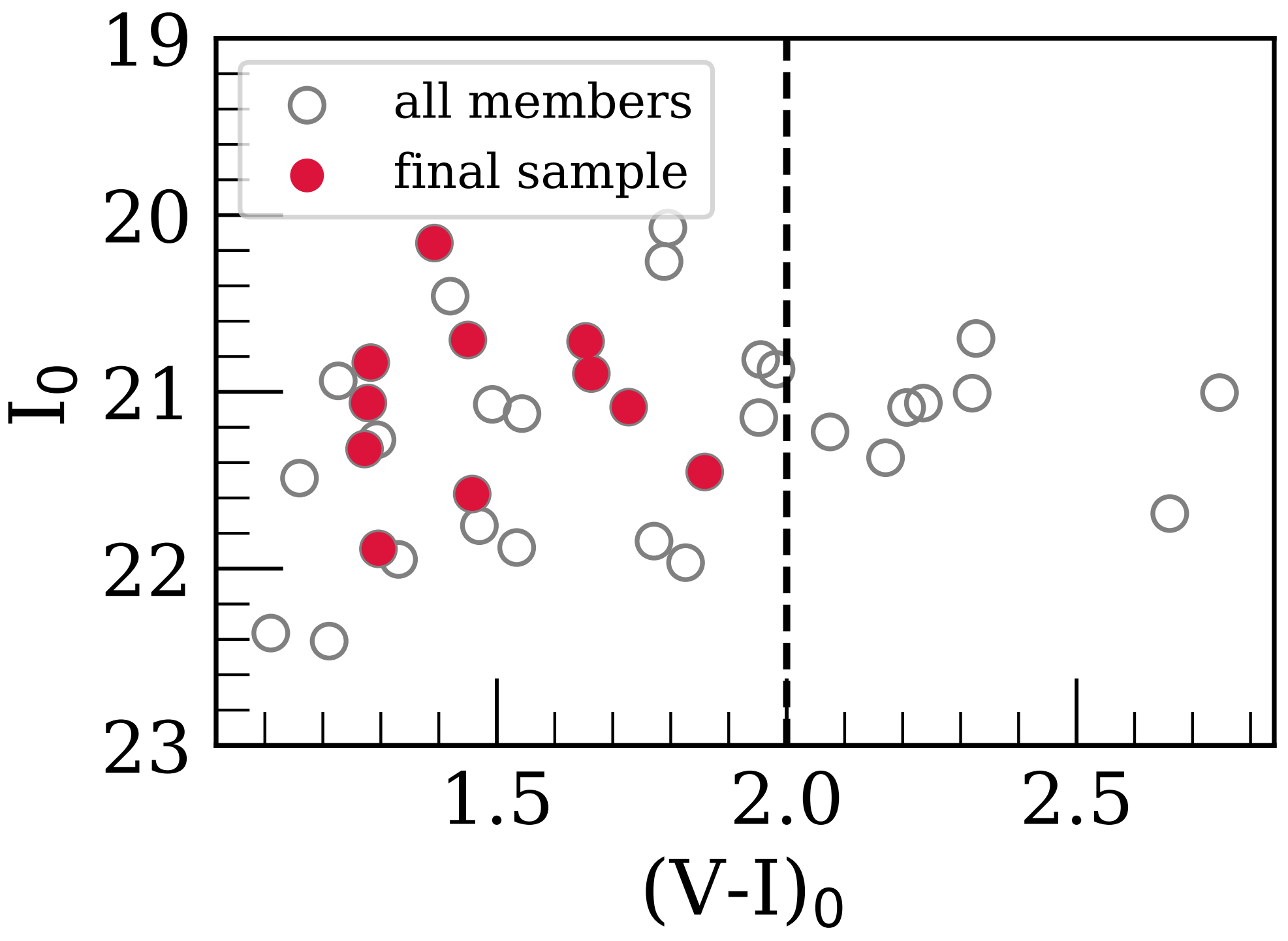

Figure 3.11: ($I$, $V - I$) color-magnitude diagram for M31 RGB stars from a 23 kpc field (f130_2) with no identified substructure. We show both stars contained in the final sample (*red filled circles*) (Section 3.8) and the full sample of member stars (*grey open circles*). The dashed vertical line represents a rough threshold in color above which stars are likely to show evidence for strong TiO absorption in their spectra ($(V - I)_0 > 2.0$). The final sample of stars shows significant overlap with the full distribution of M31 members, aside from known color biases that we have introduced in our sample selection.

In Figure 3.11, we show the $(I, V-I)$ color-magnitude diagram for all 37 M31 RGB stars in f130_2, highlighting the stars contained in our final sample. No stars in our final sample have $(V-I)_0 > 2.0$ due to the aforementioned exclusion of stars with TiO absorption. Excluding this known color bias, the final sample of RGB stars is well-sampled from the full color-magnitude distribution of M31 member stars for this field.

**Kinematics**

Given the proximity of our 23 kpc field to the various structures present in the inner halo of M31, we analyzed the kinematics of our finalized 11 star sample relative to that of the broader field. We adopt the heliocentric velocity measurements of Gilbert et al. (2007), which are based on ~1 hour observations obtained with the DEIMOS 1200 line mm$^{-1}$ grating. As discussed in detail by Gilbert et al. (2007), f130_2 contains no detectable substructure and is consistent with the kinematics of a hot stellar halo. Additionally, f130_2 is not a significant contributor to the nearby $-300$ km s$^{-1}$ kinematically cold component known as the Southeast shelf (Gilbert et al. 2007; Fardal et al. 2007). Neither is the field spatially coincident with this feature, given that it is located at a larger minor-axis distance (23 kpc) than its outermost extent in projected radius (18 kpc).

In order to confirm that our final sample is not biased in radial velocity, we present its velocity distribution compared to that of all 37 M31 RGB stars with successful radial velocity measurements in the field in Figure 3.12. We also show the velocity distribution for 128 M31 RGB stars from the more encompassing field f130 (Gilbert et al. 2007), composed of 3 distinct slitmasks, including f130_2. Based on a two-sided Kolmogorov-Smirov test, our final sample is consistent at the 99% level with the kinematics of a hot spheroid representing the velocity distribution of f130 ( $\bar{v} = -260$ km s$^{-1}$, $\sigma_v = 132$ km s$^{-1}$; Gilbert et al. 2007). We identify this kinematic component with the virialized stellar halo of M31.

We also investigate whether f130_2 contains any chemically distinct stellar populations. Figure 3.13 illustrates the relationship between [Fe/H] and radial velocity for our final sample. For a more complete representation of these two quantities, we identified stars in f130_2 that possessed well-constrained [Fe/H] measurements (Section 3.8), without enforcing any criteria on the quality of the [$\alpha$/Fe] measurements. We do not find compelling evidence for correlations between [Fe/H] and radial velocity for f130_2, such that we conclude that there are no kinematically or chemically distinguishable stellar populations within this field. The presence of a tidal feature is not necessary to explain the metallicity or velocity distribution in this field, which is fully consistent with a virialized, phase-mixed stellar population. This conclusion is supported by inspection of the color-magnitude diagram (Figure 3.11) and the velocity distribution for the broader field (Figure 3.12).

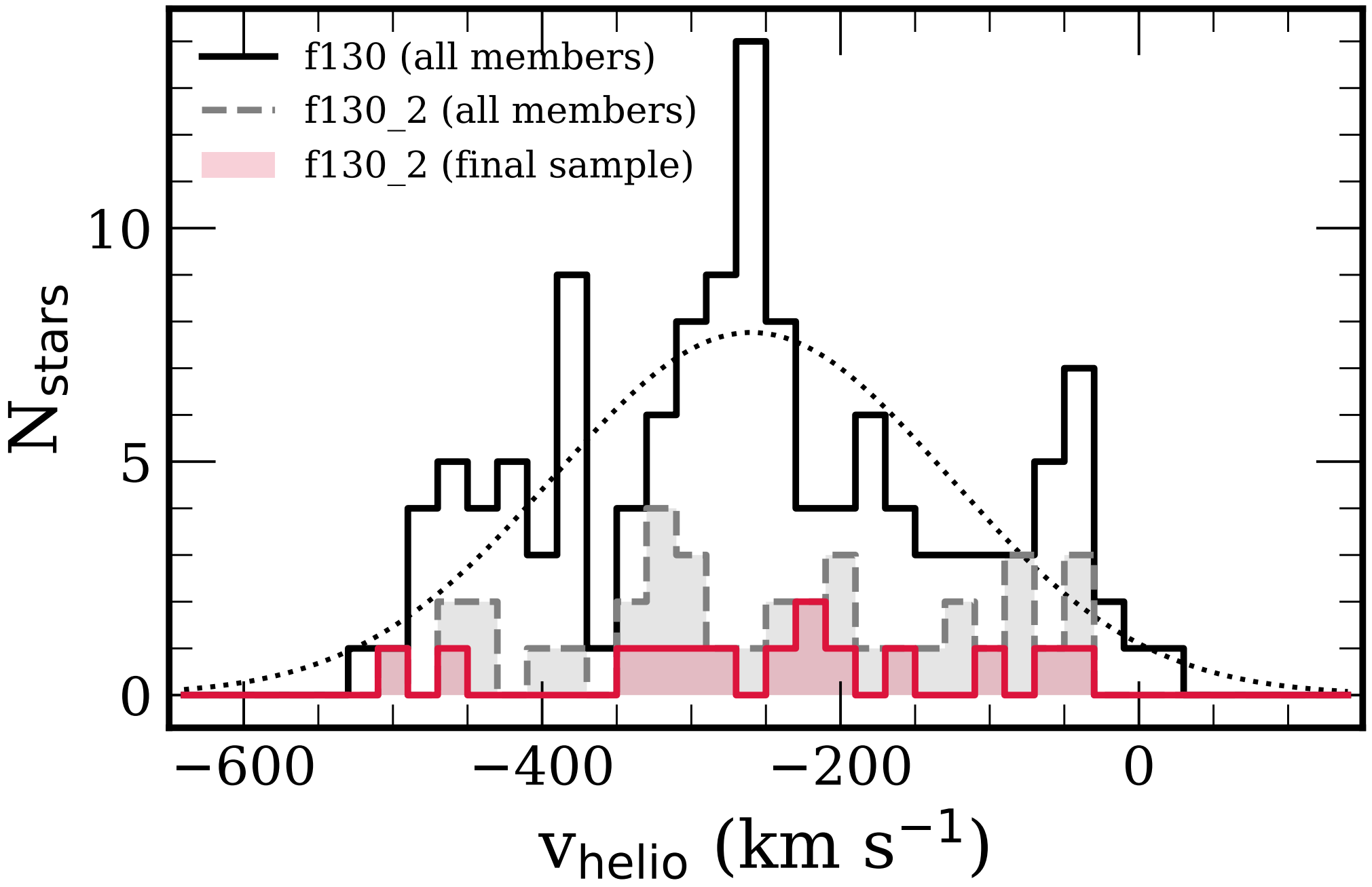

Figure 3.12: Heliocentric velocity histogram for the final 11 star sample drawn from f130_2 (*red*) compared to the distributions for all 37 M31 RGB stars in f130_2 (*grey*) with successful radial velocity measurements. We also show the velocity histogram for 128 M31 RGB stars (*black*) from the broader sample of nearby fields, including f130_2, known as f130 (Gilbert et al. 2007). The dotted line is the best-fit Gaussian ($\bar{v}$ = −260 km s$^{-1}$, $\sigma_v$ = 132 km s$^{-1}$; Gilbert et al. 2007) to f130, which corresponds to a kinematically hot spheroid component with no detected substructure (i.e., the smooth stellar halo of M31). We find that our final sample is consistent with the kinematics of the hot spheroid.

We acknowledge the possibility that kicked-up M31 disk stars, which are kinematically indistinguishable from halo stars (Dorman et al. 2013), could contribute to f130_2. However, given the distance of f130_2 along the minor axis, it is unlikely that this fraction exceeds ∼1% (Dorman et al. 2013). Based on our above kinematic analysis, we can rule out any significant contribution to f130_2 (∼100 kpc in the disk plane) by the extended disk of M31 (Ibata et al. 2005) (outermost extent ∼ 40 − 70 kpc in the disk plane). Thus, we conclude that our final sample of 11 M31 RGB

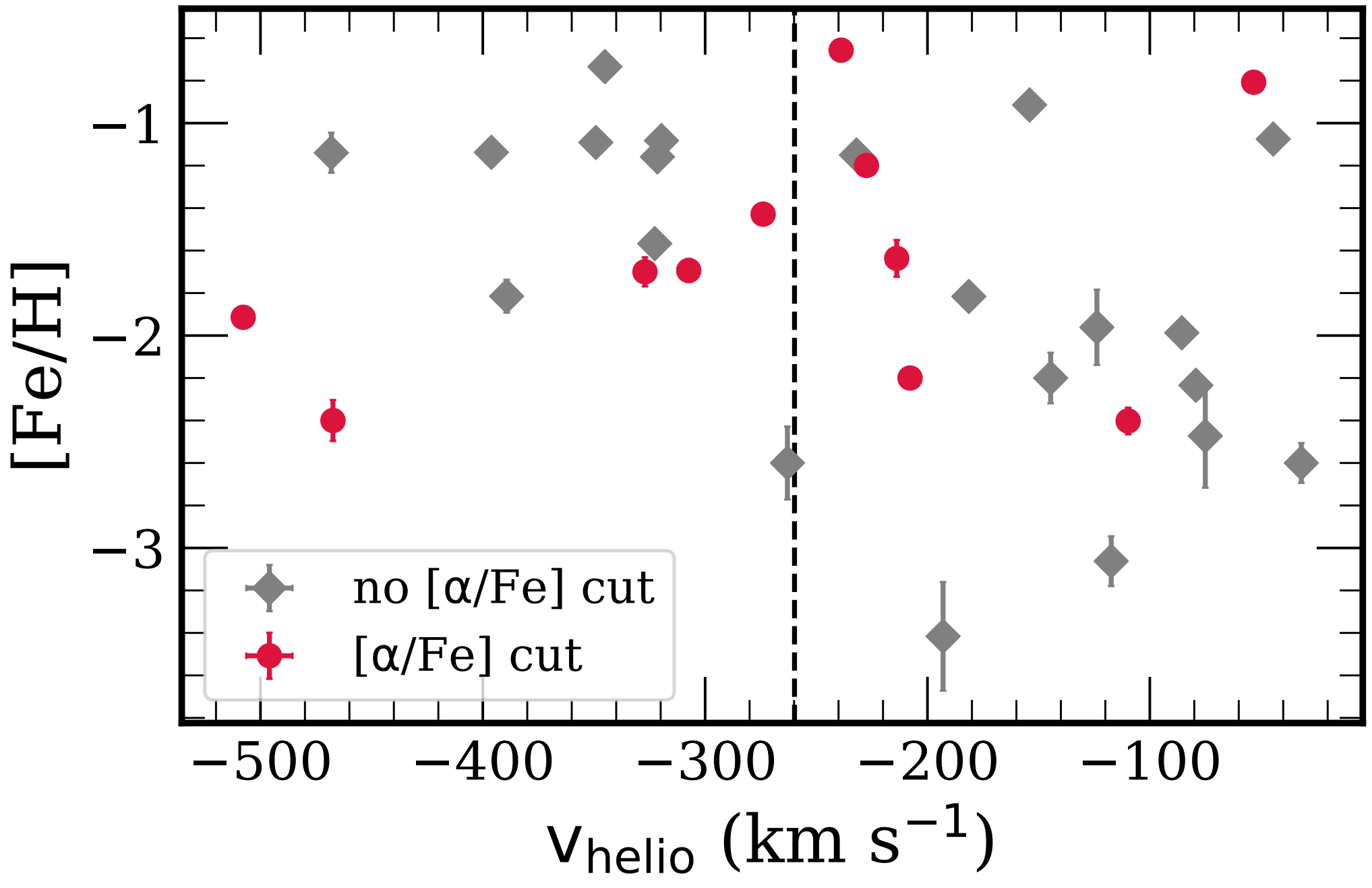

Figure 3.13: Spectroscopic metallicity ([Fe/H]) as a function of heliocentric velocity for M31 RGB stars in f130_2. The typical uncertainty in velocity (~3.5 km $s^{-1}$) is smaller than the size of the data points. In addition to our final sample (*red circles*), which contains only reliable [$\alpha$/Fe] measurements (§ 3.8), we show a broader sample (*grey diamonds*) that contains well-constrained [Fe/H] measurements, with no cuts on [$\alpha$/Fe]. No apparent correlation exists between radial velocity and metallicity in the field, where our final sample is representative of the broader sample.

stars accurately represents the properties of the smooth stellar halo of M31 in this region.

**Results and Interpretation**

Our 11 measurements increase the previous sample size for [$\alpha$/Fe] measurements in the stellar halo of M31 from 4 stars (Vargas et al. 2014). For our field, we find inverse-variance weighted values of $\langle$[Fe/H]$\rangle$ = −1.59 dex (for a comparison to previous work, see Appendix 3.10), $\sigma$([Fe/H]) =0.56 dex, $\langle$[$\alpha$/Fe]$\rangle$ = 0.49 dex, and $\sigma$([$\alpha$/Fe]) = 0.29 dex for our uniform, $\alpha$-enhanced halo field at 23 kpc.

In addition to our 11 measurements of [$\alpha$/Fe] and [Fe/H], Figure 3.14 includes the 4 outer halo stars from Vargas et al. (2014) for comparison. Vargas et al. (2014) utilized Gilbert et al.'s (2012) sample of M31 halo stars to identify stars within existing M31 dSph fields (Vargas, Geha, and Tollerud 2014) for deeper spectroscopic follow-up. They narrowed their sample by enforcing the criteria that the stars were high-likelihood M31 members with S/N sufficient to measure abundances from MRS (S/N $\gtrsim$ 15 Å$^{-1}$). Their finalized sample originates from the metal-poor outer halo of M31 between ~70 – 140 kpc. We re-compute the inverse-variance weighted average elemental abundances from their data, finding $\langle$[Fe/H]$\rangle$ = −1.70 dex, $\sigma$([Fe/H]) = 0.27 dex, $\langle$[$\alpha$/Fe]$\rangle$ = 0.28 dex, and $\sigma$([$\alpha$/Fe]) = 0.22 dex. In contrast to our work, Vargas et al. (2014) applied an empirical correction factor to convert between the measured, atmospheric value of [$\alpha$/Fe] and the average [$\alpha$/Fe] calculated from individual $\alpha$-element abundances.

Table 3.8: Parameters of 11 M31 RGB Stars

| Object | $T_{eff}$[a](K) | log $g$ (dex) | [Fe/H] (dex) | [$\alpha$/Fe] (dex) | $\Delta\lambda$[a](Å) | S/N (Å$^{-1}$) |
|---|---|---|---|---|---|---|
| 1282178 | 4339 ± 7 | 0.39 | −2.4 ± 0.17 | 0.4 ± 0.32 | 2.79 ± 0.04 | 26 |
| 1292468 | 3796 ± 4 | 0.67 | −0.66 ± 0.14 | 0.52 ± 0.37 | 2.75 ± 0.03 | 12 |
| 1292496 | 4368 ± 5 | 0.7 | −0.81 ± 0.14 | 0.0 ± 0.27 | 2.86 ± 0.02 | 24 |
| 1292507 | 3899 ± 5 | 0.46 | −1.69 ± 0.15 | 0.61 ± 0.35 | 2.72 ± 0.05 | 19 |
| 1302682 | 4075 ± 5 | 0.85 | −1.43 ± 0.14 | 0.83 ± 0.18 | 2.79 ± 0.02 | 19 |
| 1302710 | 4264 ± 9 | 1.07 | −1.64 ± 0.16 | 0.4 ± 0.4 | 2.83 ± 0.04 | 10 |
| 1302971 | 3858 ± 3 | 0.53 | −1.7 ± 0.15 | 0.75 ± 0.39 | 2.79 ± 0.04 | 15 |
| 1303039 | 4144 ± 4 | 0.52 | −2.2 ± 0.15 | −0.07 ± 0.25 | 2.83 ± 0.02 | 24 |
| 1303200 | 4337 ± 4 | 0.88 | −1.91 ± 0.15 | 0.07 ± 0.28 | 2.83 ± 0.02 | 26 |
| 1303382 | 4356 ± 3 | 0.78 | −2.4 ± 0.15 | 0.7 ± 0.19 | 2.79 ± 0.02 | 30 |
| 1303502 | 3914 ± 3 | 0.39 | −1.2 ± 0.14 | 0.53 ± 0.15 | 2.96 ± 0.02 | 28 |

[a] As discussed in Section 3.7, the errors presented for $T_{eff}$ (and $\Delta\lambda$) represent only the random component of the total uncertainty.

As expected for a smooth halo field, we do not find evidence for a trend of [$\alpha$/Fe] as a function of [Fe/H], in contrast to the expected abundance pattern (decreasing [$\alpha$/Fe] with [Fe/H]) for fields dominated by a single, recent accretion event (such as the Giant Southern Stream; Ibata et al. 2001a) or dwarf galaxies. Additionally, the fact that our [$\alpha$/Fe] measurements at 23 kpc are consistent with those at ~70−140 kpc (Figure 3.14) over the same metallicity range (−2.5 dex $\lesssim$ [Fe/H] $\lesssim$ −1.5 dex) suggests the lack of a significant radial trend with [$\alpha$/Fe] in M31 stellar halo fields absent of substructure. We also find that our 23 kpc field is on average 0.2 dex more metal rich than the outer halo Vargas et al. (2014) measurements (see Appendix 3.10 for a discussion of potential selection effects). In combination with the approximately constant value of [$\alpha$/Fe] with both [Fe/H] and radius, this may indicate that we are probing the same extended halo component, which is metal-poor, $\alpha$-enhanced, and underlies substructure at all radii (Chapman et al. 2006; Gilbert et al. 2012; Ibata et al. 2014).

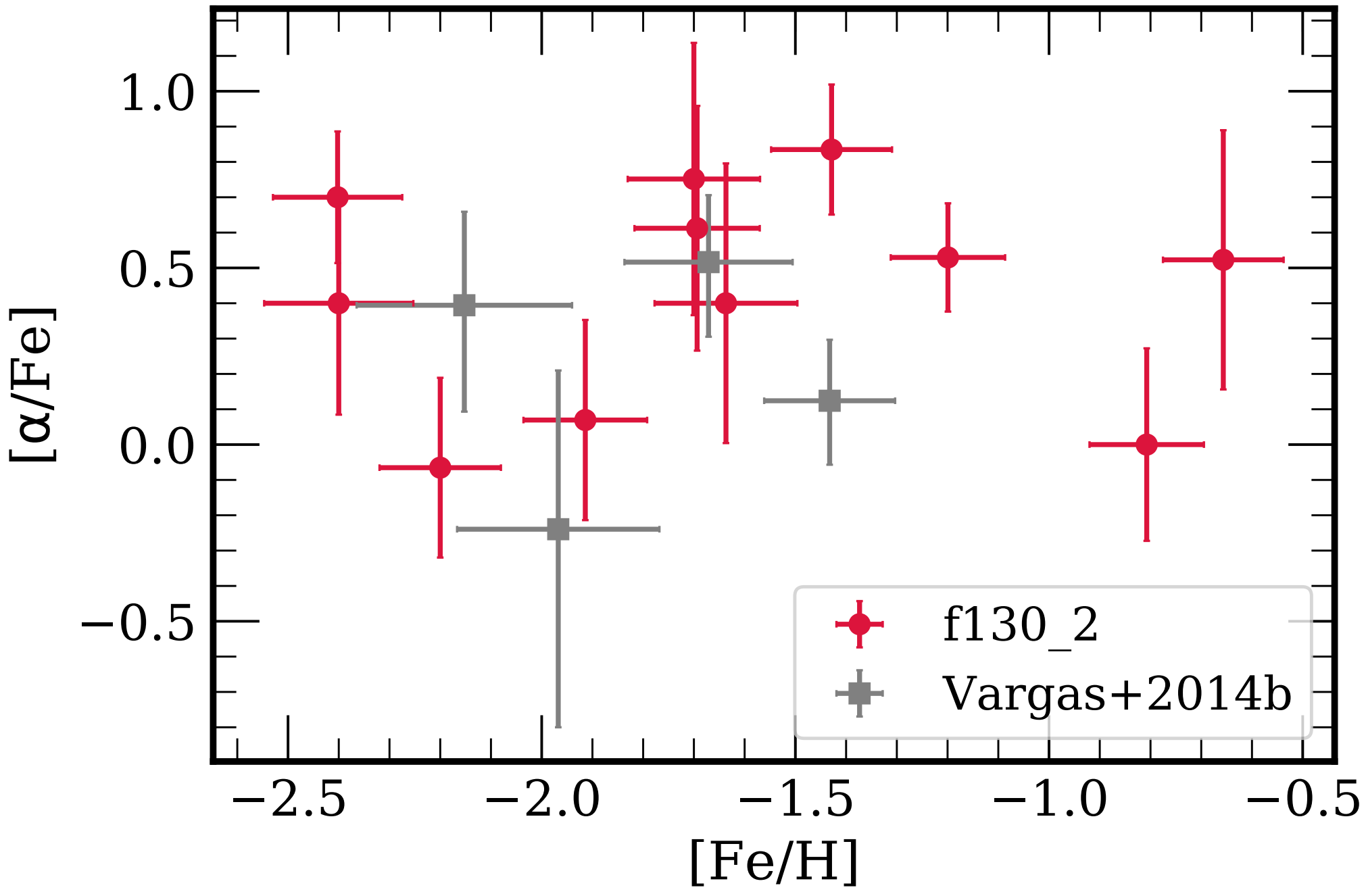

Figure 3.14: [$\alpha$/Fe] vs. [Fe/H], measured from LRS, for M31 RGB stars (*red circles*) in the 23 kpc field. We show a subset of our entire sample, with $\delta$([Fe/H]) < 0.5 dex and $\delta$([$\alpha$/Fe]) < 0.5 dex, considering only stars with reliable abundance measurements. In total, we present [$\alpha$/Fe] measurements for 11 M31 halo stars, increasing the previous sample size of 4 stars (Vargas et al. 2014). We plot the latter sample of metal-poor halo stars (*grey squares*) (S/N $\gtrsim$ 15 Å$^{-1}$) over our data set for comparison (S/N ~ 10–30 Å$^{-1}$).

Given the low luminosity of the smooth halo component (L ~ $1.9\times10^8$ $L_\odot$ for $[Fe/H]_{phot} < -1.1$ dex), Ibata et al. (2014) inferred that it would consist of many low luminosity structures accreted at early times. In terms of star formation history (SFH), high $\alpha$-element abundances indicate that the stellar population in f130_2 is characterized by rapid star formation and is dominated by the yields of Type II supernovae. Recognizing that the outer regions ($\gtrsim$ 20 kpc) of the stellar halo are most likely formed via accretion (Johnston et al. 2008; Cooper et al. 2010; Tissera, White, and Scannapieco 2012), we infer that the disrupted dwarf galaxies that were the progenitors of this field likely had short SFHs. Their SFHs could have been truncated by accretion onto M31.

Interestingly, the slightly lower average $\alpha$-element abundance (0.28 dex) of Vargas et al. (2014) could suggest that the outer halo is composed of progenitors with more extended chemical evolution as compared to the inner halo. If true, this would be in accordance with expectations from hierarchical buildup of the stellar halo (Johnston et al. 2008; Font et al. 2008). However, we cannot draw a robust conclusion on this matter given that the average $\alpha$-element abundances, similar to the case of [Fe/H], between Vargas et al.'s 2014 sample and our sample are consistent at the $1\sigma$ level, which is compounded by limited sample sizes.

Our inferred SFH for f130_2 qualitatively agrees with the trend derived from deep photometry in a nearby *HST*/ACS field located at 21 kpc along the minor axis. The mask centers of the fields are separated by 6.33 arcmin on the sky, or 1.44 kpc, assuming a distance to both fields of 783 kpc (Stanek and Garnavich 1998). Using the Brown et al. (2006) method of comparing theoretical isochrones to color-magnitude diagrams, Brown et al. (2007) derived a SFH for the ACS field, assuming [$\alpha$/Fe] = 0. They found a wide range of stellar ages and metallicities, providing support for an accretion origin, as opposed to early monolithic collapse. The field exhibits evidence for an extended SFH, with the majority of stellar ages between ~8−10 Gyr, with a small but non-negligible ($\lesssim$ 5%) population of stars with ages $\lesssim$ 8 Gyr. The wide range of metallicity (−2.5 < [Fe/H] < −0.5 dex) that we find in this chapter is consistent with a multiple progenitor hypothesis. If the nearby ACS field is representative of f130_2, this implies a composition for f130_2 of intermediate-age system(s) that had elevated star formation rates, quenched at latest $\lesssim$ 8 Gyr ago.

Comparing our average $\alpha$-element abundance to that of other systems, we find that, in general, they are similarly $\alpha$-enhanced. $\langle$[$\alpha$/Fe]$\rangle$ for the 23 kpc M31 halo field agrees with that of M31 GCs (0.37 ± 0.16 dex) within 20 kpc of the galactic center (Colucci et al. 2009). Additionally, the metal-poor MW halo possesses elevated $\alpha$-element abundance ratios of approximately +0.4 dex (Venn et al. 2004; Cayrel et al. 2004; Ishigaki, Chiba, and Aoki 2012; Bensby, Feltzing, and Oey 2014), which is comparable to our result.

Drawing comparisons to M31 dwarf galaxies is less straightforward, given that their average $\alpha$-element abundance varies from approximately solar to highly $\alpha$-enhanced (~0.5 dex) (Vargas, Geha, and Tollerud 2014). This may indicate a range of star formation timescales for these systems, where some are dominated by old stellar populations ($\gtrsim$ 10 Gyr ago) and others possess intermediate-age (~7-10 Gyr ago) stars, although the systematic uncertainties on their SFHs at early times are large (Weisz et al. 2014). Vargas, Geha, and Tollerud (2014) also found M31 dwarf galaxies to vary in terms of their internal [$\alpha$/Fe] vs. [Fe/H] abundance patterns, ranging from constant (e.g., And VII; Tollerud et al. 2012) to decreasing [$\alpha$/Fe] with respect to [Fe/H] (And V; Tollerud et al. 2012). The latter case is in accordance with abundance trends found in MW dwarf spheroidal galaxies (Shetrone, Côté, and Sargent 2001; Shetrone et al. 2003; Tolstoy et al. 2003; Venn et al. 2004; Kirby et al. 2009; Kirby et al. 2011b) and systems with more extended SFHs.

In terms of $\alpha$-enhancement and SFH, our field resembles old M31 dSphs, although it is possible that f130_2 contains intermediate age stars (Brown et al. 2007). Vargas, Geha, and Tollerud (2014) inferred that a present-day stellar halo constructed from M31 dwarf galaxies would be metal-rich, where $\langle$[Fe/H]$\rangle \sim -0.7$ dex ($-1.4$ dex) for their full sample (old dwarf galaxies only), with a distinct $\alpha$-element abundance pattern as compared to the MW halo. Given the similarly flat [$\alpha$/Fe]-[Fe/H] trend between f130_2 and And VII, and the similar $\langle$[Fe/H]$\rangle$ and [Fe/H] range between f130_2 and old M31 dSphs, it is possible that the progenitors of f130_2 were composed of systems similar to And VII. In order to meaningfully test if systems similar to present day M31 dwarf galaxies could have contributed to the smooth halo component, or whether the $\alpha$-element abundance pattern of the smooth halo of M31 differs from that of the MW, we would require larger sample sizes across more halo fields.

### 3.9 Summary

In an effort to increase the amount of available high-quality data in M31, we have developed a method of measuring [Fe/H] and [$\alpha$/Fe] from low-resolution spectroscopy of individual RGB stars. We applied our technique to a field in M31's smooth stellar halo component.

The primary advantages of utilizing low-resolution spectroscopy are (1) the substantial increase in wavelength coverage (from $\sim$ 2800 Å with MRS to $\sim$ 4600 Å with LRS) available to constrain the abundances and (2) the accompanying increase in S/N per pixel for the same exposure time and observing conditions. To make spectral synthesis of DEIMOS LRS a reality, we generated a new grid of synthetic spectra spanning 4100 – 6300 Å based on a line list we constructed for bluer optical wavelengths. We find the following results:

1. Testing our technique on Galactic GCs, we do not find evidence for any systematic covariance between fitted parameters, such as $T_{\rm eff}$ and [Fe/H]. In light of the the fundamental inhomogeniety of the various HRS samples compared to our LRS data set, our measurements broadly agree with HRS abundances.

2. Based on the intrinsic dispersion in [Fe/H] and [$\alpha$/Fe] of Galactic GCs with no known abundance variations in Fe, Mg, Ca, or Si, we estimate error floors of $\delta$([Fe/H])$_{sys}$ = 0.111 dex and $\delta$([$\alpha$/Fe])$_{sys}$ = 0.094 dex.

3. We present measurements for 11 RGB stars of [Fe/H] *and* [$\alpha$/Fe] in the stellar halo of M31, increasing the previous sample size of 4 stars. The field has no identified substructure and is located at 23 kpc in galactocentric projected radius. We find that $\langle$[Fe/H]$\rangle$ = −1.59 ± 0.56 dex and $\langle$[$\alpha$/Fe]$\rangle$ = 0.49 ± 0.29 dex for this field.

4. $\langle$[$\alpha$/Fe]$\rangle$ agrees with the value of the MW halo plateau (~0.4 dex), M31 GCs, and some $\alpha$-enhanced M31 dwarf galaxies. Our measurements exhibit overlap with previously published [$\alpha$/Fe] measurements for M31 halo RGB stars at larger projected radii (70−140 kpc), showing no evidence for a significant radial trend in [$\alpha$/Fe] in our limited sample.

5. Given its high $\alpha$-enhancement and low metallicity, we surmise that the smooth halo field is likely composed of disrupted dwarf galaxies with elevated star formation rates and truncated SFHs, accreted early in the formation history of M31.

In the following chapter, we will measure [Fe/H] and [$\alpha$/Fe] from ~6 hour observations of individual RGB stars in additional M31 halo and tidal stream fields with deep *HST* photometry (Brown et al. 2006).

The authors thank the anonymous reviewer for a careful reading of the published manuscript on which this chapter is based. We also thank Alis Deason for assistance in line list vetting, Gina Duggan for useful discussions on generating grids of synthetic spectra, Raja Guha Thakurta for help with observations and insightful conversations, and Luis Vargas and Marla Geha for sharing their data for M31 outer halo RGB stars. IE acknowledges support from a Ford Foundation Predoctoral Fellowship and the NSF Graduate Research Fellowship under Grant No. DGE-1745301, as well as the NSF under Grant No. AST-1614081, along with ENK. KMG and JW acknowledge support from NSF grant AST-1614569. ECC was supported by a NSF Graduate Research Fellowship as well as NSF Grant No. AST-1616540. The analysis pipeline used to reduce the DEIMOS data was developed at UC Berkeley with support from NSF grant AST-0071048.

## 3.10 Appendix: Comparison to Literature Mean Metallicity

In this chapter, we focused on the determination of [$\alpha$/Fe] in M31 stellar halo RGB stars. Given the limited sample size of previously existing equivalent measurements, we can only directly compare our [$\alpha$/Fe] measurements to the Vargas et al. (2014) sample. However, an extensive body of literature exists on [Fe/H] estimates in the stellar halo of M31, which we discuss in detail here in the context of our measurements.

As presented in Section 3.8, we find $\langle$[Fe/H]$\rangle$ = $-1.59$ dex and $\sigma$([Fe/H]) = 0.56 dex for f130_2. In contrast, Brown et al. (2007) estimated $\langle$[Fe/H]$\rangle_{\rm phot}$ = -0.87 dex for the nearby ACS field from color-magnitude diagram based SFHs, where their value is more metal-rich than our mean metallicity by 0.71 dex. In terms of both star counts and metallicity, Brown et al. (2007) characterized this field as straddling a transition region between the metal-rich inner halo and the metal-poor outer halo. Although the extended halo ($\gtrsim$ 60 kpc) is known to be metal-poor based on both photometric and Ca triplet metallicity indicators (Guhathakurta et al. 2006; Kalirai et al. 2006b; Chapman et al. 2006; Koch et al. 2008; Gilbert et al. 2014; Ibata et al. 2014), a majority of photometric studies find that the inner halo (20$-$30 kpc) is as metal-rich as $-0.7$ dex for fields unpolluted by Giant Southern Stream debris (Guhathakurta et al. 2006; Gilbert et al. 2014). Based on an imaging survey, Ibata et al. (2014) found [Fe/H]$_{\rm phot}$ = $-0.7$ dex at 30 kpc for [$\alpha$/Fe] = 0, where the mean metallicity does not decline to $-1.5$ dex until 150 kpc. Assuming [$\alpha$/Fe] = 0.3 dex, Kalirai et al. (2006b) found $\langle$[Fe/H]$\rangle_{\rm phot}$ = $-1.48$ dex and $\sigma$([Fe/H]$_{\rm phot}$) = 0.11 dex for the extended metal-poor halo ($\gtrsim$ 60 kpc). They based their measurements on photometry from fields with $\sim$1 hour DEIMOS spectroscopy, but they did not include f130_2 in their analysis of inner halo fields, for which they found $\langle$[Fe/H]$\rangle_{\rm phot}$ = $-0.94$ dex and $\sigma$([Fe/H]$_{\rm phot}$) = 0.60 dex around 30 kpc. Similarly, based on 397 stars between 20$-$40 kpc, Gilbert et al. (2014) found $\langle$[Fe/H]$\rangle_{\rm phot}$ = $-0.70$ dex and $\sigma$([Fe/H]$_{\rm phot}$) = 0.53 dex for [$\alpha$/Fe] = 0 in this region (including more metal-rich Giant Southern Stream debris).

Clearly, our value of $\langle$[Fe/H]$\rangle$ = $-1.59$ dex for f130_2 is discrepant with photometric studies of M31's inner halo. This could be a consequence of selection effects against metal-rich stars, given that we discarded stars with strong TiO absorption (Section 3.8). However, we also consider alternative explanations. There are indications that (1) a smooth, metal-poor halo component with no detected substructure is found at all radii, and (2) the photometric metallicities likely overestimate the degree to which the inner halo is metal-rich. Using Ca triplet equivalent width measurements from stacked DEIMOS spectra, Chapman et al. (2006) analyzed major axis fields (and one minor axis field) in M31's stellar halo, finding evidence for a metal-poor stellar halo component ($[Fe/H]_{CaT} = -1.4$ dex) detectable at all radii between 10−70 kpc with no apparent metallicity gradient. In an analysis of M31's surface brightness profile, Gilbert et al. (2012) confirmed the detection of this distinct halo component. Additionally, Ibata et al. (2014) found that the smooth halo is ~0.2 dex more metal-poor than fields dominated by substructure, where metallicities of $-2.5 < [Fe/H] < -1.1$ tend to characterize fields throughout the halo with little to no substructure. In contrast to Kalirai et al. (2006b), Koch et al. (2008) analyzed the same DEIMOS fields (including f130_2) using Ca triplet metallicities, finding values systematically more metal-poor in mean metallicity by ~0.75 dex. The large discrepancy likely results from differences in sample selection and metallicity measurement methodology (photometric vs. Ca triplet based).

Whether the methodology employed is photometric, Ca triplet based, or utilizes spectral synthesis can result in substantial differences in metallicity estimates for the same sample (e.g., Lianou, Grebel, and Koch 2011). Most relevantly, photometric studies often assume $[\alpha/Fe] = 0$, which can inflate metallicity estimates significantly compared to assuming an $\alpha$-enhanced field. Using VandenBerg, Bergbusch, and Dowler (2006) isochrones, assuming 10 Gyr old stellar populations (Brown et al. 2007), a distance modulus of $(m - M)_0 = 24.63 \pm 0.20$ (Clementini et al. 2011), and $[\alpha/Fe] = 0$ dex, we found $\langle[Fe/H]\rangle_{phot} = -1.40$ dex for our sample of 11 M31 RGB stars. If we instead assume $[\alpha/Fe] = 0.3$, we obtain $\langle[Fe/H]\rangle_{phot} = -1.60$ dex, corresponding to a decrease in the mean photometric metallicity of 0.19 dex. We find nearly identical results by repeating the calculation with a different set of isochrones (Demarque et al. 2004).

The assumptions intrinsic to photometric metallicities, combined with the large amount of tidal debris present in the inner halo of M31 that is included in many previously published measurements in this radial range, are sufficient to explain the large difference between our value of $\langle[Fe/H]\rangle$ for f130_2 and previous analyses in the inner halo of M31. A primary strength of our study is that we can determine both $[\alpha/Fe]$ and $[Fe/H]$ from spectroscopy, without prior assumptions on either parameter. We acknowledge that we may be preferentially sampling brighter, more metal-poor stars in this field, given that we are S/N-limited and select against stars with strong TiO absorption. However, given that we can measure both [Fe/H] and [$\alpha$/Fe] reliably from some of the highest quality spectra in M31's halo yet obtained, we conclude that $\langle$[Fe/H]$\rangle$ = $-1.59$ dex is likely an accurate representation of our final sample's mean metallicity. Thus, it is possible that our sample in f130_2 represents the metal-poor halo that underlies substructure (Chapman et al. 2006; Gilbert et al. 2014; Ibata et al. 2014) in the inner halo of M31.

*4*

# ELEMENTAL ABUNDANCES IN M31: A COMPARATIVE ANALYSIS OF ALPHA AND IRON ELEMENT ABUNDANCES IN THE OUTER DISK, GIANT STELLAR STREAM, AND INNER HALO

Ivanna Escala[1,2], Karoline M. Gilbert[3,4], Evan N. Kirby[1], Emily C. Cunningham[5], Jennifer Wojno[4] Puragra GuhaThakurta[5]

[1]Department of Astronomy, California Institute of Technology, 1200 E California Blvd, Pasadena, CA, 91125, USA

[2]Department of Astrophysical Sciences, Princeton University, 4 Ivy Lane, Princeton, NJ, 08544, USA

[3]Space Telescope Science Institute, 3700 San Martin Dr., Baltimore, MD 21218 USA

[4]Department of Physics & Astronomy, Bloomberg Center for Physics and Astronomy, John Hopkins University, 3400 N. Charles St, Baltimore, MD 21218, USA

[5]Department of Astronomy & Astrophysics, University of California, Santa Cruz, 1156 High St, Santa Cruz, CA, 95064, USA

[6]UCO/Lick Observatory, Department of Astronomy & Astrophysics, University of California Santa Cruz, 1156 High Street, Santa Cruz, California 95064, USA

*Abstract

We measured [Fe/H] and [$\alpha$/Fe] using spectral synthesis of low-resolution stellar spectroscopy for 70 individual red giant branch stars across four fields spanning the outer disk, Giant Stellar Stream (GSS), and inner halo of M31. Fields at M31-centric projected distances of 23 kpc in the halo, 12 kpc in the halo, 22 kpc in the GSS, and 26 kpc in the outer disk are $\alpha$-enhanced, with $\langle$[$\alpha$/Fe]$\rangle$ = 0.43, 0.50, 0.41, and 0.58, respectively. The 23 kpc and 12 kpc halo fields are relatively metal-poor, with $\langle$[Fe/H]$\rangle$ = −1.54 and −1.30, whereas the 22 kpc GSS and 26 kpc outer disk fields are relatively metal-rich with $\langle$[Fe/H]$\rangle$ = −0.84 and −0.92, respectively. For fields with substructure, we separated the stellar populations into kinematically hot stellar halo components and kinematically cold components. We did not find any evidence of a radial [$\alpha$/Fe] gradient along the high surface brightness core of the GSS between ∼17−22 kpc. However, we found tentative suggestions of a negative radial [$\alpha$/Fe] gradient in the stellar halo, which may indicate that different progenitor(s) or formation mechanisms contributed to the build up of the inner versus outer halo. Additionally, the [$\alpha$/Fe] distribution of the metal-rich ([Fe/H] > −1.5), smooth inner stellar halo ($r_{proj} \lesssim 26$ kpc) is inconsistent with having formed from the disruption of progenitor(s) similar to present-day M31 satellite galaxies. The 26 kpc outer disk is most likely associated with the extended disk of M31, where its high $\alpha$-enhancement provides support for an episode of rapid star formation in M31's disk possibly induced by a major merger.

## 4.1 Introduction

Stellar halos probe various stages of accretion history, as well as preserving signatures of *in-situ* stellar formation (Zolotov et al. 2009; Cooper et al. 2010; Font et al. 2008; Font et al. 2011; Tissera et al. 2013; Tissera et al. 2014). The stellar halo and stellar disk of $L_\star$ galaxies are connected through accretion events that not only build up the halo, but can impact the evolution of the disk (Abadi et al. 2003; Peñarrubia, McConnachie, and Babul 2006; Tissera, White, and Scannapieco 2012). Additionally, stellar disks can contribute to the inner stellar halo via heating mechanisms (Purcell, Bullock, and Kazantzidis 2010; McCarthy et al. 2012; Tissera et al. 2013). The formation history of these various structural components are imprinted in its stellar populations at the time of their formation via chemical abundances (Robertson et al. 2005; Bullock and Johnston 2005; Font et al. 2006a; Johnston et al. 2008; Zolotov et al. 2010; Tissera, White, and Scannapieco 2012). In particular, measurements of $\alpha$-element abundances (O, Ne, Mg, Si, S, Ar, Ca, and Ti) encode information concerning the relative timescales of Types Ia and core-collapse supernovae (e.g., Gilmore and Wyse 1998) and the epoch of accretion onto the host $L_\star$ galaxy, whereas [Fe/H] measurements provide information concerning the star formation duration of a stellar system.

The Andromeda galaxy (M31) is ideal for studies of stellar halos and stellar disks, given that it is viewed nearly edge-on (de Vaucouleurs 1958). In contrast to the Milky Way (MW), M31 appears to be more representative of a typical spiral galaxy (Hammer et al. 2007). Thus, M31 serves as a complement to the MW in studies of galaxy formation and evolution. Although much has been learned about the global properties of M31 and its tidal debris through photometry and shallow spectroscopy (e.g., Kalirai et al. 2006b; Ibata et al. 2005; Ibata et al. 2007; Ibata et al. 2014; Gilbert et al. 2007; Gilbert et al. 2009; Gilbert et al. 2012; Gilbert et al. 2014; Gilbert et al.

2018; Koch et al. 2008; McConnachie et al. 2009; McConnachie et al. 2018), the level of detail available in the MW to study its accretion history from resolved stellar populations (Haywood et al. 2018; Deason et al. 2018; Helmi et al. 2018; Gallart et al. 2019; Mackereth et al. 2019b) is currently not achievable in M31.

In particular, the distance to M31 (785 kpc; McConnachie et al. 2005) has historically precluded robust spectroscopic measurements of [$\alpha$/Fe] and [Fe/H] for individual stars. The majority of chemical information of individual RGB stars in M31 and its dwarf satellite galaxies originate from photometric metallicity estimates or spectroscopic metallicity estimates from the strength of the calcium triplet (Chapman et al. 2006; Kalirai et al. 2006b; Koch et al. 2008; Kalirai et al. 2009; Richardson et al. 2009; Collins et al. 2011; Gilbert et al. 2014; Ibata et al. 2014; Ho et al. 2015). However, the degree to which photometric and calcium triplet based metallicity estimates accurately measure iron abundance alone is uncertain (Battaglia et al. 2008a; Starkenburg et al. 2010; Lianou, Grebel, and Koch 2011; Da Costa 2016). It was only in 2014 that Vargas, Geha, and Tollerud presented the first spectroscopic chemical abundances in the M31 system based on spectral synthesis of medium-resolution (Kirby, Guhathakurta, and Sneden 2008; Kirby et al. 2009) spectroscopy.

Here, we present the third contribution of a deep spectroscopic survey of the stellar halo, tidal streams, disk, and present-day satellite galaxies of M31. The first work in this series (Escala et al. 2019, hereafter E19) applied a new technique of spectral synthesis of low-resolution ($R \sim 2500$) spectroscopy to individual RGB stars in the smooth, metal-poor halo of M31 at $r_{\rm proj}$ = 23 kpc. These were the first measurements of [$\alpha$/Fe] and [Fe/H] of individual stars in the inner halo of M31. Gilbert et al. (2019a), hereafter G19, presented the first [$\alpha$/Fe] and [Fe/H] measurements in the Giant Stellar Stream (GSS; Ibata et al. 2001a) of M31, located at $r_{\rm proj}$ = 17 kpc. In this chapter, we present [$\alpha$/Fe] and [Fe/H] measurements for three additional fields in the inner halo at $r_{\rm proj}$ = 12 kpc, the GSS at $r_{\rm proj}$ = 22 kpc, and outer disk of M31 at $r_{\rm proj}$ = 26 kpc. These three fields, in addition to the smooth halo field of E19, all overlap with *Hubble Space Telescope* (*HST*) Advanced Camera for Surveys (ACS) pointings with inferred color-magnitude diagram based star formation histories (Brown et al. 2006; Brown et al. 2007; Brown et al. 2009). The 26 kpc outer disk field represents the first abundances in the disk of M31.

Section 4.2 details our observations and summarizes the properties of relevant, nearby spectroscopic fields in M31. In Section 4.3, we describe the changes and

Table 4.1: M31 DEIMOS Observations[a]

| Object | Date | $\theta_s$ (") | $X$ | $t_{\rm exp}$ (s) | $N$ |
|---|---|---|---|---|---|
| | | 12 kpc Halo Field (H) | | | |
| H1 | 2014 Sep 29 | 0.90 | 1.67 | 1097 | 110 |
| H1 | 2014 Sep 30 | 0.90 | 2.16 | 5700 | ... |
| H1 | 2014 Oct 1 | 0.73 | 2.11 | 5700 | ... |
| H2[b] | 2014 Sep 29 | 0.9 | 1.29 | 2400 | 110 |
| H2 | 2014 Sep 30 | 0.80 | 1.39 | 4200 | ... |
| H2 | 2014 Oct 1 | 0.90 | 1.32 | 4320 | ... |
| | | 22 kpc GSS Field (S) | | | |
| S1 | 2014 Sep 30 | 0.70 | 1.12 | 4800 | 114 |
| S1 | 2014 Oct 1 | 0.75 | 1.11 | 3600 | ... |
| S2 | 2014 Sep 29 | 0.90 | 1.07 | 2400 | 114 |
| S2 | 2014 Sep 30 | 0.70 | 1.07 | 4261 | ... |
| S2 | 2014 Oct 1 | 0.75 | 1.07 | 4800 | ... |
| | | 26 kpc Disk Field (D) | | | |
| D1 | 2014 Sep 30 | 0.60 | 1.15 | 4200 | 126 |
| D1 | 2014 Oct 1 | 0.60 | 1.18 | 4320 | ... |
| D2 | 2014 Sep 29 | 0.70 | 1.43 | 3600 | 126 |
| D2 | 2014 Sep 30 | 0.70 | 1.41 | 4683 | ... |
| D2 | 2014 Oct 1 | 0.60 | 1.41 | 4320 | ... |

Note. —The columns of the table refer to slitmask name, date of observation, seeing in arcseconds, airmass, exposure time per slitmask in seconds, and number of stars targeted per slitmask.

[a] The observations for f130_2, which we further analyze in this chapter, were published by Escala et al. (2019).

[b] Slitmasks indicated "1" and "2" are identical, except that the slits on "2" are tilted according to the parallactic angle at the approximate time of observation.

improvements to our abundance measurement technique (E19) and discuss our abundance sample selection. We define our membership criteria for M31 RGB stars and model their velocity distributions in Section 4.4, with a focus on separating the stellar halo from substructure. Section 4.5 presents the full abundance distributions and separates them into kinematic components. We discuss our abundances in the context of the existing literature on M31 in Section 4.6.

## 4.2 Observations

In this section, we describe our spectroscopic observations in M31 and the known properties of each spectroscopic field.

Table 4.2: DEIMOS 600ZD Velocity Templates

| Object | Spec. Type | X | $t_{exp}$ (s) |
|---|---|---|---|
| HD 103095 | K1 V | 1.39 | 20 |
| HD 122563 | G8 III | 1.37 | 20 |
| HD 187111 | G8 III | 1.72 | 20 |
| HD 38230 | K0 V C | 1.08 | 720 |
| HR 4829 | A2 V C | 1.64 | 100 |
| HD 109995 | A0 V C | 1.54 | 99 |
| HD 151288 | K7.5 V | 1.04 | 20 |
| HD 345957 | G0 V | 1.62 | 200 |
| HD 88609 | G5 III C | 1.45 | 45 |
| HR 7346 | B9 V | 1.34 | 20 |

Note. — All templates were observed on 2019 Mar 10.

**Data**

We summarize our deep M31 observations for fields H, S, and D in Table 4.1. The slitmasks for H, S, and D were observed for a total of 6.5, 5.5, and 5.9 hours, respectively. The M31 stars in these fields were included as additional targets on the slitmasks first presented by Cunningham et al. (2016), which were intended to target MW foreground halo stars. We utilized the Keck/DEIMOS (Faber et al. 2003) 600 line mm$^{-1}$ (600ZD) grating with the GG455 order blocking filter, a central wavelength of 7200 Å, and 0.8" slitwidths. Two separate slitmasks were designed for each field, with the same mask center, mask position angle, and target list, but with differing slit position angles. This enabled us to approximately track the changes in parallatic angle throughout the night, minimizing flux losses due to differential atmospheric refraction at blue wavelengths. The spectral resolution is approximately ~2.8 Å FWHM. As discussed in E19, using a low-resolution grating (comparing to the medium-resolution DEIMOS 1200G grating, ~1.3 Å FWHM) provides the advantage of higher signal-to-noise per pixel for the same exposure time and observing conditions. The similarly deep (5.8 hours) observations for an additional field, f130_2, which we further analyze in this chapter, were published by E19. Additionally, we observed radial velocity templates (Section 4.3; Table 4.2) in our science configuration.

**Field Properties**

The fields H, S, and D are located at approximately 12 kpc, 22 kpc, and 26 kpc, respectively, away from the M31 galactic center in projected radius. The DEIMOS slitmasks were designed to target RGB stars near the well-studied halo21, halo11, stream, and disk fields presented in the catalog of Brown et al. (2009). The wide *HST*/ACS images were obtained in the broad $V$ and $I$ filters and reach ∼1.5 magnitudes fainter than the oldest main-sequence turn-off. Table 4.3 summarizes the positioning on the sky of all four 600ZD fields and the accompanying *HST*/ACS pointings. Figure 4.1 provides an illustration relative to the galactic center of M31 for these fields, including relevant 1200G fields (H11, H13s, H13d, and f207_1). We also include the 1200G fields f115_1, f116_1, and f123 (Gilbert et al. 2007) in Figure 4.1, given their proximity to field H. The 1200G fields are not analyzed in this chapter, but their known kinematics are useful for placing our 600ZD observations in context. The dimensions of each DEIMOS slitmask are approximately 16’×4’, whereas the ACS images are comparatively small, spanning 202”×202”.

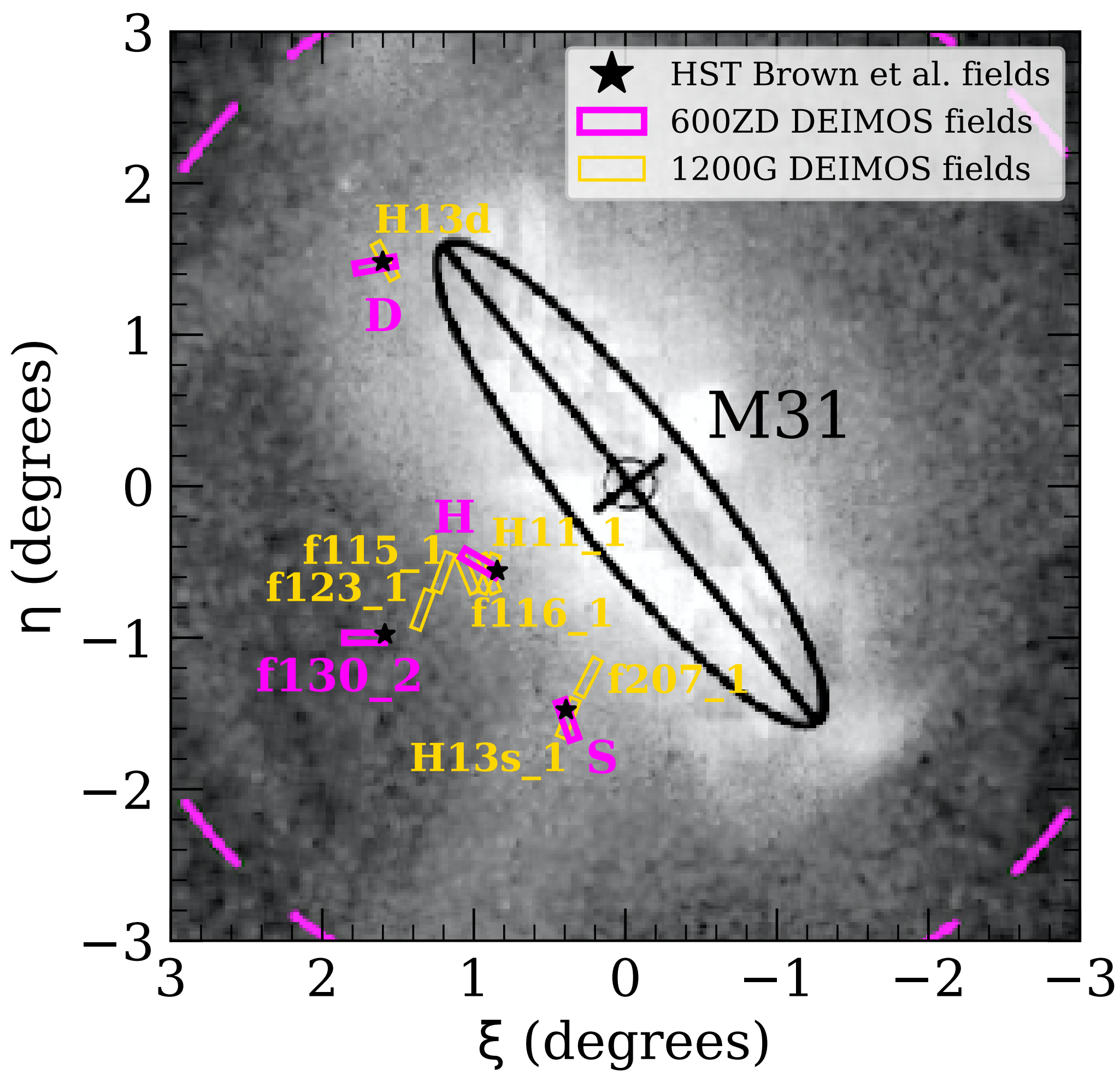

Figure 4.1: The location of M31 DEIMOS fields observed with the 600ZD grating (Section 4.2 of this chapter, E19; magenta rectangles), the 1200G grating (Kalirai et al. 2006a; Gilbert et al. 2007; Gilbert et al. 2009; yellow rectangles), and *HST*/ACS fields (Brown et al. 2009; black stars) in M31-centric coordinates, overlaid on the PAndAS star count map (McConnachie et al. 2018). The dashed magenta line corresponds to 50 projected kpc. The ACS fields are represented as points given their extent on the sky (202”×202”) relative to the DEIMOS masks. Our spectroscopic fields span M31’s outer disk at 26 kpc, Giant Stellar Stream at 22 kpc, and the inner halo at 12 kpc and 23 kpc.

Table 4.3: M31 Field Positions

| Field | $r$ (kpc)[a] | $\alpha_{J2000}$ | $\delta_{J2000}$ | P.A.[b] | ACS Field | $d_{ACS}$ (kpc)[c] |
|---|---|---|---|---|---|---|
| f130_2 | 23 | 00:49:37.49 | +40:16:07.0 | +90 | halo21 | 1.44 |
| H | 12 | 00:46:34.09 | +40:45:38.6 | −120 | halo11 | 1.35 |
| S | 22 | 00:44:15.98 | +39:43:31.5 | +20 | stream | 0.92 |
| D | 26 | 00:49:20.59 | +42:43:44.7 | +100 | disk | 0.58 |

[a] Projected radius of the mask center from M31 galactocenter.
[b] Slitmask position angle, in degrees east of north.
[c] Projected distance from DEIMOS mask center to pointing center of corresponding ACS field.

The field S is nearly identical to H13s_1, which was first observed for ∼1 hour using the 1200 line mm$^{-1}$ (1200G) grating on DEIMOS by Kalirai et al. (2006a) and later re-analyzed using an improved spectroscopic data reduction by Gilbert et al. (2009). Field S is located southeast of an additional Giant Stellar Stream field, f207_1 (Gilbert et al. 2009). f207_1 is located near the eastern edge of the highest surface brightness region of the GSS core, at a projected radius of ∼17 kpc. The DEIMOS 1200G fields H11 and H13d, which overlap with the southwestern and northwestern edges of the 600ZD fields H and D respectively, were also first observed by Kalirai et al. (2006a). Field H11 was subsequently re-analyzed by Gilbert et al. (2007) following improvements in the reduction technique. The field f130_2, which is located at 23 kpc in projected radius, has been previously studied by E19, for which shallow spectroscopy was first published by Gilbert et al. (2007).

Based on the nearby 1200G fields, we expect that the properties of fields H, S, and D will generally reflect the inner halo of M31, the GSS, and the outer northeastern disk of M31, respectively, although other components are present in these fields. In particular, field H is likely polluted by stars belonging to a substructure known as the Southeast shelf, which is associated with the GSS progenitor (Section 4.6; Gilbert et al. 2007; Fardal et al. 2007). Field S should contain a secondary kinematically cold component of unknown origin in addition to the GSS core (Kalirai et al. 2006a; Gilbert et al. 2009; Gilbert et al. 2019a). E19 showed that f130_2 is likely associated with the "smooth", metal-poor component of M31's stellar halo. We refer to fields H, S, D, and f130_2 interchangeably as the 12 kpc inner halo, 22 kpc GSS, 26 kpc outer disk, and 23 kpc smooth halo fields where appropriate to emphasize the physical properties of the M31 fields.

## 4.3 Abundance Determination

We use spectral synthesis of low-resolution stellar spectroscopy (E19) to measure stellar parameters and abundances from our deep observations of M31 RGB stars. In summary, we measure [Fe/H] and [$\alpha$/Fe] from regions of the spectrum sensitive to Fe and $\alpha$-elements (Mg, Si, Ca), respectively, by comparing to a grid of synthetic spectra degraded to the resolution of the DEIMOS 600ZD grating. We also measure the spectroscopic effective temperature, $T_{\rm eff}$, informed by photometric constraints, and fix the surface gravity, log $g$, to the photometric value. Measurements of [Fe/H] and [$\alpha$/Fe] using spectra obtained with the 600ZD grating are generally consistent with equivalent measurements (Kirby, Guhathakurta, and Sneden 2008) from 1200G spectra (Appendix 4.8). For a detailed description of the low-resolution spectral synthesis method, see E19. In the following subsections, we describe improvements and changes to our technique since E19.

### Photometry

We utilized wide-field (1 deg$^2$) $g'$ and $i'$ band photometry from the Pan-Andromeda Archaeological Survey (PAndAS) catalog (McConnachie et al. 2018) for the fields H, S, and D. The images were obtained from MegaCam on the 3.6 m Canada-France-Hawaii Telescope. We extinction-corrected the photometry assuming field-specific interstellar reddening values from the dust reddening maps of Schlegel, Finkbeiner, and Davis (1998), with the corrections defined by Schlafly and Finkbeiner (2011). We used the conversion between reddening and extinction adopted by Ibata et al. (2014). For stars present in the DEIMOS fields but absent from the PAndAS point source catalog (~20-30% of M31 RGB stars on a given slitmask), we sourced photometry from CFHT/MegaCam images obtained by Kalirai et al. (2006a) and reduced with the CFHT MegaPipe pipeline (Gwyn 2008). We cross-validated the MegaPipe photometry against that of PAndAS for common stars to verify that the photometry is accurate for the majority of stars.

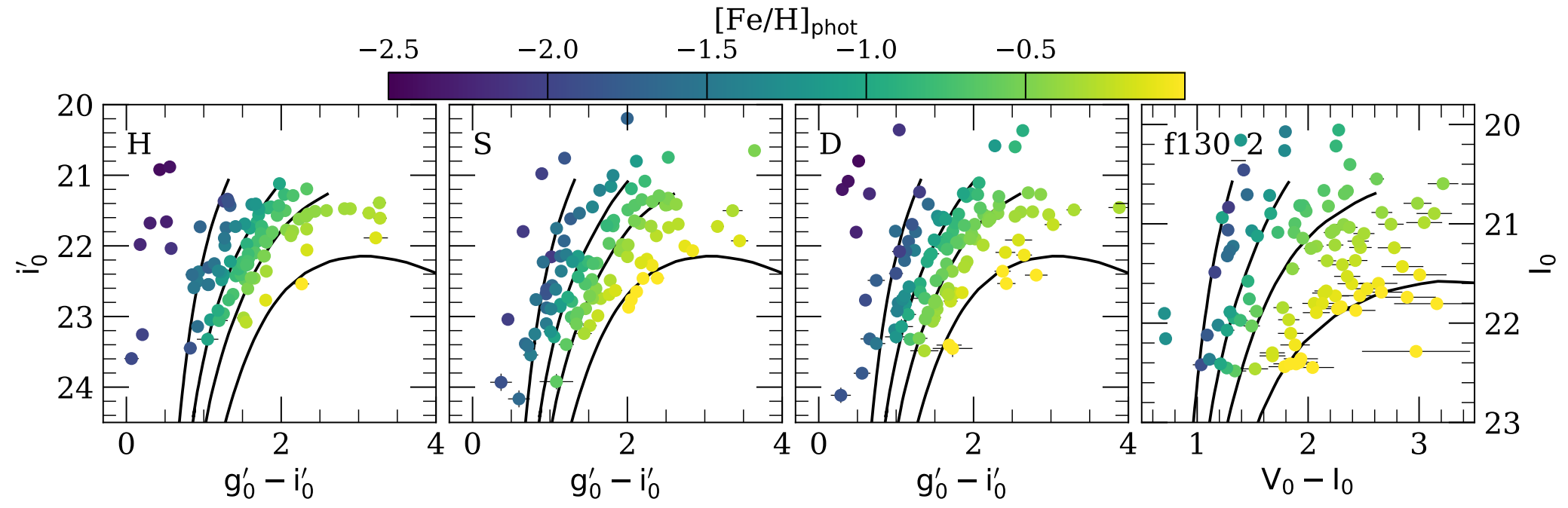

Figure 4.2: ($i'_0$, $g'_0 - i'_0$) and ($V_0 - I_0$, $I_0$) color-magnitude diagrams for all stars (M31 RGB stars and MW foreground dwarf stars) in the 12 kpc inner halo field (H), 22 kpc GSS field (S), the 26 kpc outer disk field (D), and the 23 kpc smooth halo field (f130_2). The points are color coded according to the photometric metallicity estimated for each star from the PARSEC (Marigo et al. 2017) isochrones ($-2.2 <$ [Fe/H] $< +0.2$) assuming an age of 9 Gyr for H, S, and D and 12 Gyr for f130_2 (Section 4.3). For reference, we overplot a few PARSEC isochrones (*black lines*) with, from left to right, [Fe/H] = −2.2, −1.1, −0.5, and +0.1. Stars that are bluer than the most metal-poor isochrone (taking into account photometric errors) are likely MW dwarf stars.

In contrast to E19, we did not use multiple isochrone sets to calculate photometrically-based quantities such as $T_{\rm eff,phot}$ and log $g$. We employed the most recent version of the PARSEC (Marigo et al. 2017) isochrones, which are available in the relevant filters for a wide range of stellar ages and metallicities between $-2.2 <$ [Fe/H] $<$ 0.2 and [$\alpha$/Fe] = 0. For stars positioned above the tip of the red giant branch[1], we linearly extrapolate to obtain estimates of $T_{\rm eff,phot}$, log $g$, and [Fe/H]$_{\rm phot}$. Similarly, we extrapolate blueward of the most metal-poor isochrone to determine $T_{\rm eff,phot}$ and log $g$ for these stars. We assumed a distance modulus relative to M31 of $m - M$ = 24.63 ± 0.20 (Clementini et al. 2011). We utilized the same Johnson-Cousins $V$, $I$ photometry for f130_2 as in E19, but determined photometric parameters using the PARSEC isochrones as described above. Figure 4.2 illustrates our usage of the PARSEC isochrones to determine photometrically-based quantities, where we have color-coded the color-magnitude diagrams (CMDs) according to the estimated photometric metallicity. We assumed ages of 9 Gyr for H, S, and D based on mean stellar ages of 9.7 Gyr, 8.8 Gyr, and 7.5 Gyr, respectively, in the corresponding ACS fields (Table 4.3) inferred from CMD-based star formation histories (Brown et al. 2006). For f130_2, we assumed an age of 12 Gyr, where it was inferred to have a mean stellar age of 11 Gyr (Brown et al. 2007).

[1] Stars that have magnitudes brighter than the tip of the red giant branch, according to the assumed isochrone set and distance modulus, are either a consequence of photometric errors or AGB stars. None of these stars are in our final abundance sample (Figure 4.3).

Although other isochrone sets (e.g., Dotter et al. 2007; Dotter et al. 2008) are also available in these filters, we based our selection in part on whether the isochrones contained contributions from molecular TiO (Section 4.3) in the stellar atmosphere models used to compute the evolutionary tracks.

**Spectral Resolution**

Previously, we approximated the spectral resolution as constant with respect to wavelength ($\Delta\lambda$). In E19, we determined $\Delta\lambda$ for each star by fitting the observed spectrum in a narrow range centered on the expected resolution for the 600ZD grating. In actuality, for our DEIMOS configuration, the spectral resolution is a slowly varying function of wavelength.

We employed this approximation to circumvent the problem of an insufficient number of sky lines at bluer wavelengths to empirically determine the spectral resolution as a function of wavelength ($\Delta\lambda(\lambda)$). Alternatively, including arc lines in the fitting procedure can provide constraints in this wavelength regime. Using a combination of Gaussian widths from both sky lines and arc lines, we utilized a maximum-likelihood approach (K. McKinnon et al., in preparation) to determine $\Delta\lambda(\lambda)$ for each star. In the few cases per slitmask where the spectral resolution determination fails (e.g., owing to an insufficient number of arc and sky lines), we assumed $\Delta\lambda$ = 2.8 Å, the expected resolution of the 600ZD grating (E19). For the case of multiple observations per star, we calculate $\Delta\lambda(\lambda)$ as the average of the individual measurements on different dates of observation for a given star.

In addition to $\Delta\lambda(\lambda)$, we determined a resolution scale parameter. This parameter accounts for the fact that the resolution as calculated from the sky lines and arc lines, which fill the entire slit, slightly overestimates $\Delta\lambda$ for the stellar spectrum, whose width depends on seeing. First, we included the resolution scale as a free parameter in our abundance determination, measuring its value, $f_i$, for each object on a given slitmask. However, given that each individual measurement is subject to noise, the final measurement, $f$, is the average of the individual measurements for the entire slitmask. The resolution scale parameter is primarily a function of seeing, and therefore should be constant for a single slitmask. In the final abundance determination, we fixed the spectral resolution at $f\Delta\lambda(\lambda)$.

Based on our globular cluster calibration sample from E19, we confirmed that utilizing $\Delta\lambda(\lambda)$, as opposed to the $\Delta\lambda$ approximation, alters our abundances within the $1\sigma$ uncertainties. We re-calculated the systematic error in [Fe/H] and [$\alpha$/Fe] from the internal spread in globular clusters, finding $\delta$([Fe/H])$_{\rm sys}$ = 0.130 and $\delta$([$\alpha$/Fe])$_{\rm sys}$ = 0.107. We repeated our chemical abundance analysis for f130_2 (E19), fixing $\Delta\lambda(\lambda)$ to its empirically derived value for each star, and present these abundances in Section 4.5.

**Radial Velocity**

We cross-correlated the observed spectrum with empirical templates of high signal-to-noise (S/N) stars (Cooper et al. 2012; Newman et al. 2013), which we observed with the 600ZD grating in our science configuration (Table 4.2). We shifted the templates to the rest frame based on their *Gaia* DR2 (Gaia Collaboration et al. 2016a; Gaia Collaboration et al. 2018) radial velocities, except for HD 109995 (Gontcharov 2006). The templates do not possess any A-band velocity offsets, as the template stars were trailed through the slit while observing. We utilized the full template spectrum ($\sim$ 4500 – 11000 Å) to shift the science spectrum into the rest frame. In cases where the full-spectrum radial velocity determination failed, we instead utilized the wavelength regions near the calcium triplet (8450 Å $< \lambda <$ 8700 Å). Additionally, we apply an A-band correction, which significantly impacts the determination of the heliocentric velocity. We determined random velocity errors from Monte Carlo resampling with $10^3$ trials.

Following an improvement in the spectroscopic data reduction, Gilbert et al. (2009) and Gilbert et al. (2018) found a typical velocity precision of $\sim$ 5 – 7 km s$^{-1}$ for low S/N ($\sim$ 10–12 Å$^{-1}$) M31 RGB stars observed with the 1200G grating, including a systematic component of $\sim$ 2 km s$^{-1}$ from repeat observations of stars (Simon and Geha 2007). For our entire sample (including MW dwarf stars) with successful radial velocity measurements, our median velocity uncertainty is 11.6 km s$^{-1}$, incorporating a systematic error term for the 600ZD grating based on repeat observations of over 300 stars (5.6 km s$^{-1}$; Collins et al. 2011). The reduced velocity precision for the 600ZD grating is a consequence of its lower spectral resolution.

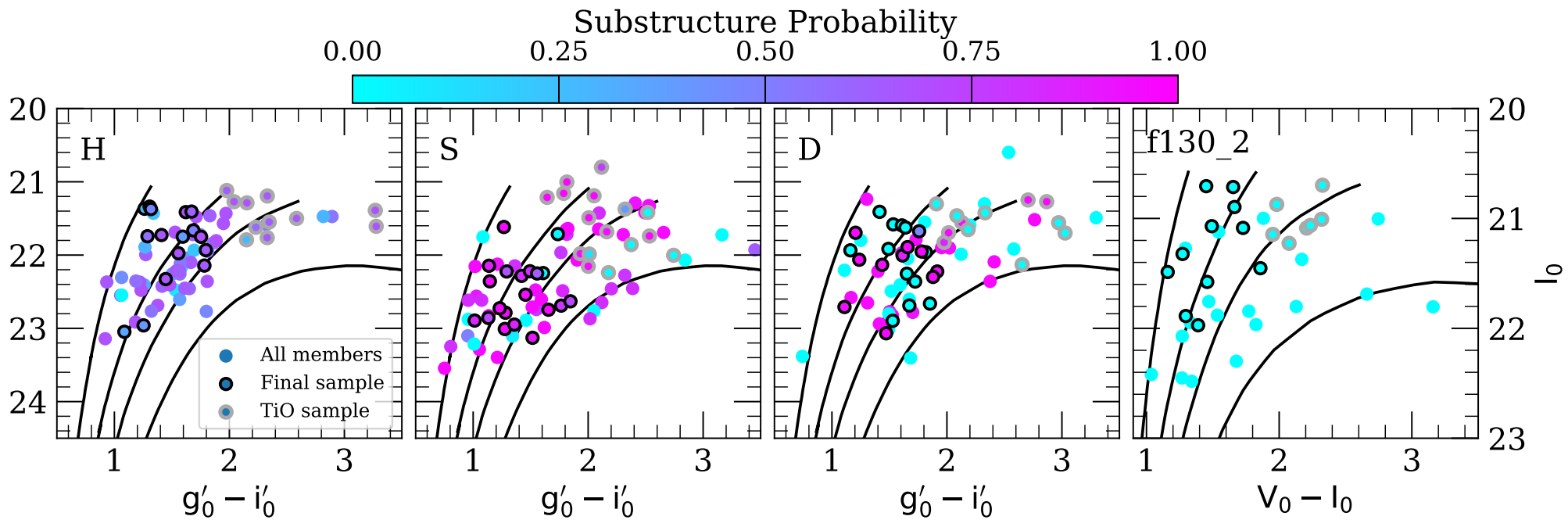

Figure 4.3: Color magnitude diagrams of M31 RGB stars (Section 4.4) reflecting selection effects in the 12 kpc inner halo field (H), 22 kpc GSS field (S), 26 kpc outer disk field (D), and 23 kpc smooth halo field (f130_2). Stars are color-coded according to probability of belonging to *any* substructure component in a given field (Section 4.4). Magenta (blue) points are likely (unlikely) to be associated with substructure. For each field, we show all M31 RGB stars, M31 RGB stars with spectroscopic abundance measurements constituting our final sample of 70 total stars (*black outlined circles*; Section 4.3), and M31 RGB stars with spectroscopic abundance measurements that otherwise pass our selection criteria, but show signatures of TiO absorption (*grey outlined circles*). We overplot PARSEC isochrones in the appropriate filter for each field with, from left to right, [Fe/H] = −2.2, −1.0, −0.5, and 0. On average, the omission of TiO stars from our final sample results in a bias against red stars ($g'_0 - i'_0 \gtrsim 2$ and $V_0 - I_0 \gtrsim 2$), which disproportionately affects the substructure components (relative to the stellar halo components) in the 12 kpc halo field and 22 kpc GSS field.

### Abundance Sample Selection

As in E19, we included only *reliable* measurements for M31 RGB stars (Section 4.4) in our final samples, i.e., $\delta$[Fe/H] < 0.5, $\delta[\alpha/\mathrm{Fe}] < 0.5$, and well-constrained parameter estimates based on the $5\sigma$ $\chi^2$ contours for all fitted parameters. Unreliable abundance measurements are often a consequence of insufficient S/N. In addition, we excluded spectra of stars with sufficient S/N for a reliable measurement that found a minimum at the cool end of the $T_{\mathrm{eff}}$ range (3500 K) spanned by our grid of synthetic spectra. We also manually screened member stars for evidence of strong molecular TiO absorption between 7055−7245 Å, finding that 41%, 44%, 34%, and 39% of the measurements for H, S, D, and f130_2 passing the reliability cuts were affected by TiO. We excluded these stars from the subsequent abundance analysis, given our uncertainty in our ability to accurately and precisely measure abundances for stars with TiO in the absence of a suitable calibration sample. We found that 16, 20, 23, and 11 of the measurements in H, S, D, and f130_2 can be considered reliable based on the above criteria, resulting in a final sample of 70 stars. Our sample of stars affected by TiO that otherwise pass our selection criteria is composed of 46 stars across all four fields.

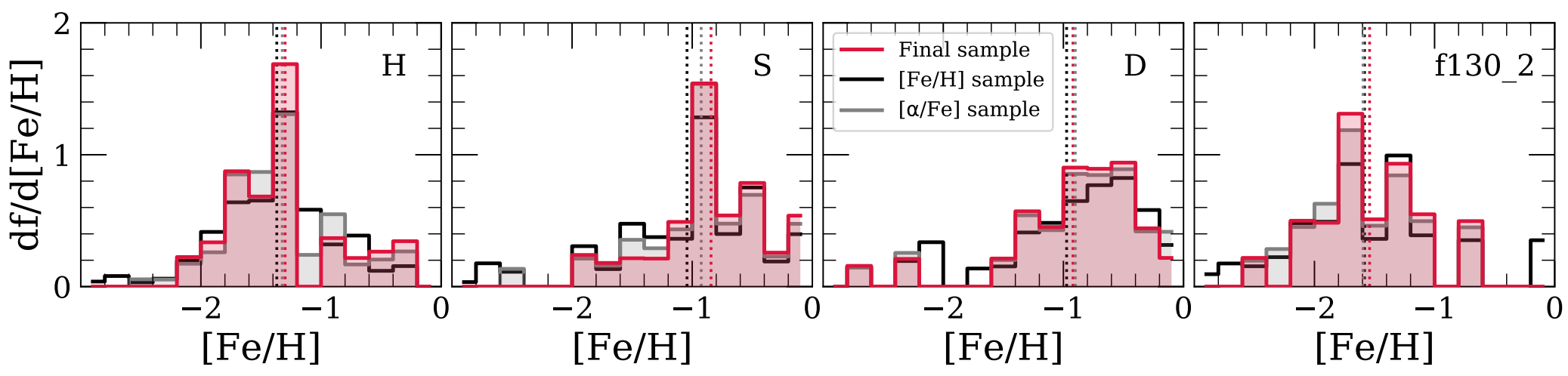

Figure 4.4: Metallicity distribution functions (MDFs), represented in terms of probability density, for subsets of our total sample of M31 RGB stars with abundance measurements (Section 4.3). The MDFs are weighted according to the inverse variance of the total measurement uncertainty in [Fe/H]. We have subdivided our total sample into M31 RGB stars with a [Fe/H] measurement (*black histogram*) and with both a [Fe/H] and [$\alpha$/Fe] measurement (*grey histogram*), regardless of the error on the measurement. The red histogram shows our final sample, which includes M31 RGB stars with both [Fe/H] and [$\alpha$/Fe] measured to within 0.5 dex. Stars with TiO absorption are excluded from all subsets. The inverse-variance weighted means of all samples are indicated as dashed verticle lines. The metallicity distributions of the finalized sample are weakly biased against metal-poor stars with low S/N spectra.

**Selection Effects on the Abundance Distributions**

Given that our spectroscopic abundance determination is S/N limited and it is unclear how the omission of TiO from our linelist (E19) impacts our abundance measurements for stars with strong TiO absorption, we investigated the impact of our selection criteria (Section 4.3) on the properties of our final sample. Figure 4.3 shows CMDs of all M31 RGB stars (Section 4.4) in each field, where we have highlighted our final sample. We also show the subset of stars with spectroscopic evidence of TiO absorption that otherwise pass our selection criteria. Excluding stars with TiO translates to an effective color bias of $g'_0 - i'_0 \lesssim 2$ and $V_0 - I_0 \lesssim 2$. Additionally, our final sample probes brighter magnitudes, particularly in H. Fainter stars tend to have lower S/N, which results in either an uncertain (e.g., $\delta$([$\alpha$/Fe]) > 0.5) or failed abundance measurement. In principle, this should not affect the metallicity distribution, so long as the final sample spans the majority of the color range of the CMD.

We quantitatively assessed the metallicity bias introduced by excluding M31 RGB stars with imprecise spectroscopic abundance measurements. We separated our sample of M31 RGB stars with spectroscopic [Fe/H] measurements into three subsets: (1) a sample with successful [Fe/H] determinations as dictated by the $5\sigma$ $\chi^2$ contours, with no restrictions on the errors, (2) a sample with both successful [Fe/H] and [$\alpha$/Fe] measurements, with no restrictions on the errors, and (3) our final sample, with successful [Fe/H] and [$\alpha$/Fe] measurements and $\delta$([Fe/H]) < 0.5 and $\delta$([$\alpha$/Fe]) < 0.5. All three subsets exclude TiO stars. As illustrated by Figure 4.4, the inverse-variance weighted metallicity distribution functions appear similar between the three subsets. The error-weighted mean metallicity for the most inclusive sample is more metal-poor than the final sample by ~0.04−0.07 dex for fields H, D, and f130_2. The difference between samples is 0.20 dex for field S, owing to very metal poor stars present in sample (1) that were omitted from sample (3). If we assume that sample (1) better represents the true spectroscopic metallicity of M31 RGB stars in the field, then we can conclude that S/N limitations, which increase measurement uncertainty, results in a weak bias in our final sample against metal-poor stars with low S/N spectra.

Regarding the known color bias introduced by excluding TiO stars, we analyzed the photometric metallicity distributions of each sample. The isochrone set we employed to calculate $T_{\rm eff}$, log $g$, and $[{\rm Fe/H}]_{\rm phot}$ (Section 4.3) for H, S, and D (PARSEC; Marigo et al. 2017) were generated using models that included molecular TiO absorption. This allowed us to estimate the metallicity of all M31 RGB stars, many of which are not included in our final sample. Assuming [$\alpha$/Fe] = 0, we found that our final sample is biased toward lower $[{\rm Fe/H}]_{\rm phot}$ relative to the full sample of M31 RGB stars. We found that $\langle[{\rm Fe/H}]\rangle_{\rm phot}$ = −0.89, −0.76, −0.69, and −0.76 for all M31 RGB stars in H, S, D, and f130_2, respectively. For our final sample, we found that $\langle[{\rm Fe/H}]\rangle_{\rm phot}$ = −1.17, −0.96, −0.87, and −1.15 for H, S, D, and f130_2. Thus, on average, our final sample is biased toward lower $[{\rm Fe/H}]_{\rm phot}$ by ~0.2−0.4 dex. Much of this effect is a consequence of the exclusion of TiO stars from the final sample. Including the subset of TiO stars, we obtain $\langle[{\rm Fe/H}]\rangle_{\rm phot}$ = −0.91 dex, −0.85 dex, −0.73 dex, and −0.86 dex, for H, S, D, and f130_2, reducing the bias in the final sample to ~0.02−0.10 dex more metal poor than the full sample. Based on this, we can conclude the primary source of bias against metal-rich stars originates from excluding TiO stars. However, the exact amount by which we might be biased in [Fe/H] is unclear, given that $[{\rm Fe/H}]_{\rm phot}$, which has no knowledge of [$\alpha$/Fe] and is degenerate with stellar age, cannot be translated into spectroscopic [Fe/H].

We do not anticipate that selection effects impacting the color distribution of our final sample incur a bias in [$\alpha$/Fe] relative to the full sample of M31 RGB stars. The width, or color range, of the RGB is largely dictated by [Fe/H], as opposed to $\alpha$-enhancement (Gallart, Zoccali, and Aparicio 2005). However, S/N limitations may affect the [$\alpha$/Fe] distribution of the final sample, resulting in a weak bias against $\alpha$-poor stars with low S/N spectra.[2]

## 4.4 Kinematic Analysis of the M31 Fields

In this section, we identified M31 RGB stars in each spectroscopic field and analyzed their velocity distributions.

### M31 Membership

Given that foreground MW dwarf stars and M31 stars are spatially coincident and exhibit significant overlap in both their velocity distributions and CMDs, identifying bona fide M31 RGB stars is nontrivial. In E19, we utilized the probabilistic method of Gilbert et al. (2006) to carefully assess the likelihood of membership for stars in our spectroscopic sample. This method incorporates up to four criteria to determine membership for a majority of M31 fields: the strength of the Na I $\lambda\lambda$8190 absorption line doublet, the *(V, I)* color-magnitude diagram location, photometric versus spectroscopic (Ca II $\lambda\lambda$8500) metallicity estimates, and the heliocentric radial velocity. However, we cannot use this exact approach for fields H, S, and D, owing to the diversity of utilized photometric filters.

Instead, we determine membership based on three criteria: (1) Na I $\lambda\lambda$8190 absorption strength, (2) CMD location, and (3) heliocentric radial velocity. Given that the strength of the Na I doublet depends on surface gravity, it can effectively separate M31 RGB stars from foreground MW M dwarfs (Schiavon et al. 1997). We excluded stars with clear signatures of the Na I doublet as nonmembers of M31. We classified stars as MW dwarf stars if they have colors bluer than the most metal-poor isochrone (Section 4.3) by an amount greater than their photometric error. Such stars are $\gtrsim$ 10 times more likely to be MW dwarf stars than M31 RGB stars (Gilbert et al. 2006). Lastly, we adopted a radial velocity cut of $v_{\rm helio} < -150$ km s$^{-1}$ for fields H, S, and f130_2 to select for M31 RGB stars. Using a sample of $\gtrsim$ 1000 probablistically identified M31 RGB stars, Gilbert et al. (2007) found that contamination from MW dwarf stars is largely constrained to $v_{\rm helio} > -150$ km s$^{-1}$. The estimated contamination fraction using this radial velocity cut, in combination with the additional membership diagnostics of Gilbert et al. (2006), is 2-5% across their entire sample, where contamination is defined as the fraction of bona fide MW dwarf stars classified as M31 RGB stars.

[2]Additionally, if our abundance measurements of TiO stars are indeed valid, we cannot eliminate the possibility that our final sample is biased toward lower [$\alpha$/Fe] by $\sim$ 0.1–0.2 dex (e.g., Figure 4.7).

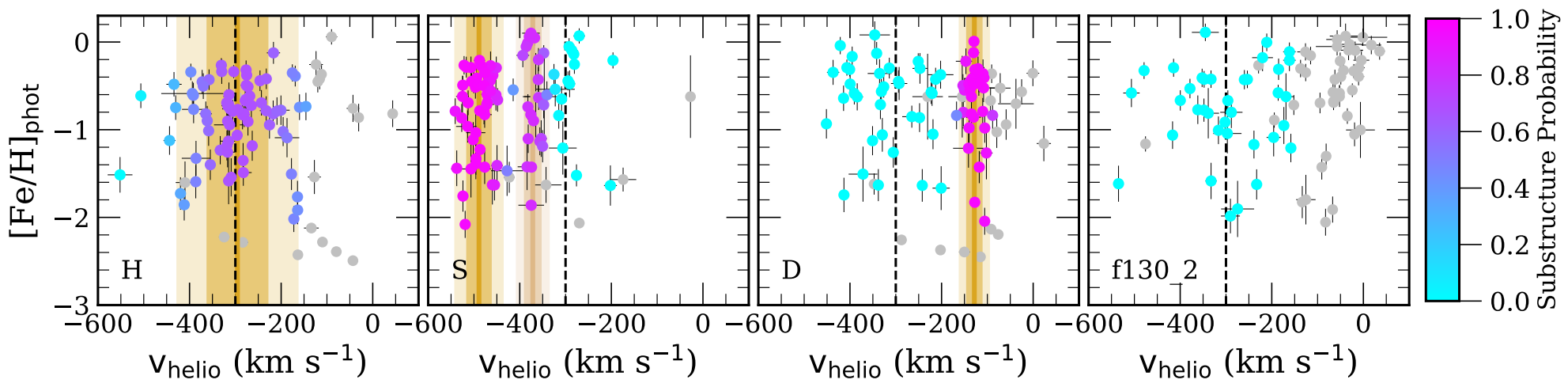

Figure 4.5: Heliocentric radial velocity versus photometric metallicity (Section 4.3) for stars with successful velocity measurements in the 12 kpc halo (H), 22 kpc GSS (S), 26 kpc disk (D), and 23 kpc halo (f130_2) M31 fields. The velocity errors represent only the random component of the uncertainty. M31's systemic velocity ($v_{\rm M31}$ = −300 km s$^{-1}$) is indicated as a dashed vertical line. Stars classified as MW foreground dwarfs (Section 4.4) are shown in grey, whereas stars classified as M31 RGB stars are color-coded according to its probability of belonging to substructure, as in Figure 4.3. The vertical bands represent the mean velocity and the velocity dispersion ($\mu$, $\mu \pm \sigma$, $\mu \pm 2\sigma$; Table 4.4) of the primary (*orange*) and secondary (*tan*) substructure components in each field. We identified 73, 84, 68, and 36 stars as M31 RGB stars in the 12 kpc, 22 kpc, 26 kpc, and 23 kpc fields, respectively.

To evaluate the performance of our binary membership determination, we compared our results to those of stars with Gilbert et al. (2006) membership probabilities. For fields H, S, and f130_2, 11%, 24%, and 74%, respectively, of our sample with successful radial velocity measurements have associated M31 membership probabilities. Assuming $m - M$ = 24.47 mag (Gilbert et al. 2006), we accurately recovered 87%, 98%, and 97% of both secure and marginal M31 members, including radial velocity as a membership diagnostic ($L_i > 0$; Gilbert et al. 2012), in H, S, and f130_2, respectively. The fraction of stars present in our M31 RGB samples that are classified as MW dwarf stars using the method of Gilbert et al. (2006) is 0% across all three fields. Given that we used similar membership criteria to Gilbert et al. (2006) and were able to reproduce their results to high confidence, we estimate that our true MW contamination fraction is ~2-5% across fields H, S, and f130_2.

Stars in field D do not possess previously determined membership probabilities to which we could compare. The rotation of M31's northeastern disk produces a redshift relative to M31's systemic velocity, such that the peak of the disk is located at $v_{\rm disk} \sim -130$ km s$^{-1}$ (Section 4.4). The presence of the disk invalidates the use of a $v_{\rm helio} < -150$ km s$^{-1}$ velocity cut as a diagnostic for M31 membership in field D. Instead, we employed a less conservative radial velocity cut of $v_{\rm helio} < -100$ km s$^{-1}$ to identify potential M31 RGB stars. This cut likely recovers the majority of M31 members in this field, but increases the MW contamination fraction (in the velocity range of $-150$ km s$^{-1}$ $< v_{\rm helio} < -100$ km s$^{-1}$). Gilbert et al. (2006) estimated that *only* using a radial velocity cut of $v_{\rm helio} < -100$ km s$^{-1}$ results in a 10% contamination fraction in inner halo fields. For disk fields covering an area of 240 arcmin$^2$ within their color-magnitude selection window, including outer disk fields in the northeastern quadrant, Ibata et al. (2005) used predictions from Galactic models (Robin et al. 2003) to argue that MW contamination in disk fields is negligible (~5%) for $v_{\rm helio} < -100$ km s$^{-1}$. Therefore, we expect that the MW contamination fraction in field D is ~5-10%, where the relatively high density of stars in the pronounced disk feature at $\sim -130$ km s$^{-1}$ should minimize contamination.

Figure 4.5 illustrates our membership determination for H, S, D, and f130_2 in terms of the relationship between $v_{\rm helio}$ and [Fe/H]$_{\rm phot}$. We identified 73, 84, 68, and 36 RGB stars as M31 members in fields H, S, D, and f130_2, respectively, out of 90, 89, 84, and 78 targets with successful radial velocity measurements. Using the same membership criteria as in fields H and S, we re-determined membership homogeneously for f130_2, resulting in a final 11 star sample with reliable abundances (Section 4.3) that is *not* identical to the 11 star sample presented in E19. We included some stars that were originally excluded in E19 as a consequence of lacking membership probabilities from shallow 1200G spectra (owing to failed radial velocity measurements). We excluded some stars that were originally included in E19 as a result of using radial velocity as a membership diagnostic, where we did not take radial velocity into account to determine membership in E19 to avoid kinematic bias.

**Kinematic Decomposition**

In Figure 4.6, we present the heliocentric radial velocity distributions for M31 RGB stars in all four fields. We also show the full velocity distributions for stars with successful radial velocity measurements, including MW contaminants, for a total of 105, 111, 124, and 64 stars in fields H, S, D, and f130_2. Field f130_2 was shown to have no detected substructure by Gilbert et al. (2007), which is consistent with our velocity distribution (see also E19). For fields H and S, velocity distributions have previously been analyzed in fields that contain partial overlap (Figure 4.1). The mean velocity of substructure along GSS fields (Gilbert et al. 2009) and the velocity dispersion of substructure near the 12 kpc inner halo field (Gilbert et al. 2007) are known to vary with radius. Thus, to compare abundances of different kinematic components within H, S, and D, it is necessary to characterize the velocity distributions of the current sample. In particular, fields S and D show clear evidence of substructure from inspection of Figure 4.6, such as the GSS ($\sim$ −500 km s$^{-1}$) and the kinematically cold component of unknown origin (Kalirai et al. 2006a; Gilbert et al. 2009) located at approximately −400 km s$^{-1}$ in field S, and M31's outer northeastern disk (−130 km s$^{-1}$) in field D. Although less clear, the velocity distribution of H is more strongly peaked at the systemic velocity of M31, $v_{\rm M31} = -300$ km s$^{-1}$, than expectations for a pure stellar halo component, which suggests the presence of substructure (Section 4.6).

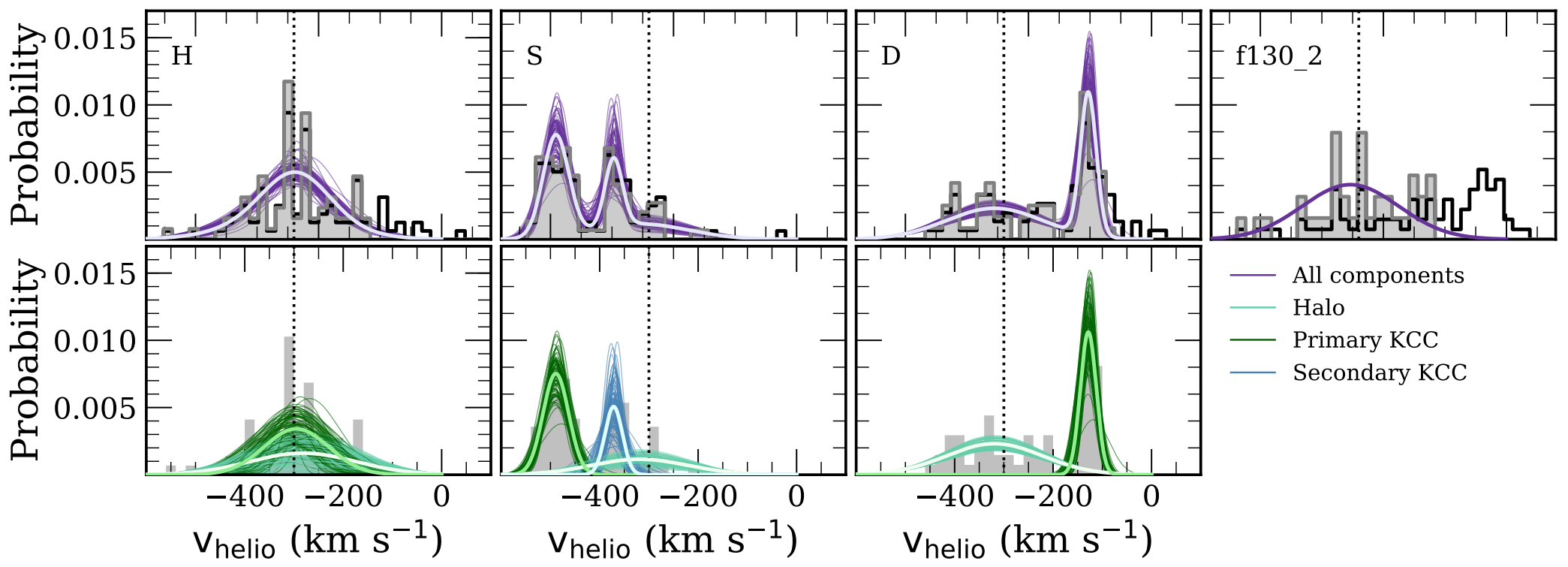

Figure 4.6: Heliocentric radial velocity distributions of stars with successful velocity measurements (Section 4.3, *top panels, black histograms*), including foreground MW dwarf stars (Section 4.4), and velocity distributions for M31 RGB stars (*grey filled histograms*) in the 12 kpc (H), 22 kpc (S), 26 kpc (D), and 23 kpc (f130_2) fields. We also show full velocity distribution models for M31 RGB stars (*top panels, purple lines*) and, for fields with substructure, models of the kinematic components (*bottom panels*). We include both stellar halo (*light green lines*) and kinematically cold components (KCCs; *dark green and blue lines*) (Section 4.4). M31's systemtic velocity is indicated as a dotted vertical line. For fields with substructure, we show 100 randomly sampled models from the converged portion of the MCMC chain to represent the uncertainty in fitting the velocity distribution. For each field, we also show the full velocity model (and its components, where applicable) as defined by the 50$^{\text{th}}$ percentile parameter values (*thick lines*; Table 4.4). All fields show evidence for M31 halo stars (distributed in a kinematically hot component around the systemic velocity). The GSS is located in the 22 kpc field at approximately −490 km s$^{-1}$, including the KCC of unknown origin at −370 km s$^{-1}$ (Kalirai et al. 2006a; Gilbert et al. 2009). The 12 kpc substructure likely corresponds to the Southeast shelf (Section 4.6; Fardal et al. 2007; Gilbert et al. 2007), a tidal feature originating from the GSS progenitor. M31's disk appears as the prominent feature centered at −130 km s$^{-1}$ in the 26 kpc field.

We separated fields with indications of substructure–H, S, and D–into kinematically cold components and the kinematically hot stellar halo by describing the velocity distributions as a Gaussian mixture, such that the log likelihood function is given by,

$$\ln \mathcal{L} = \sum_{i=1}^{n} \ln \left( \sum_{k=1}^{K} f_k \mathcal{N}(v_i | \mu_k, \sigma_k^2) \right), \tag{4.1}$$

where $i$ is the index representing a M31 RGB star, $v_i$ is its heliocentric radial velocity, and $n$ is the total number of M31 RGB stars in a field. $K$ is the *total* number of components in a field, including the stellar halo component, where $k$ represents the index for a given component, and $f_k$ represents the normalized fractional contribution of each component to the total distribution. Each component is described by a mean velocity, $\mu_k$, and velocity dispersion, $\sigma_k$.

Given our usage of radial velocity as a diagnostic for membership (Section 4.4), which excludes stars with MW-like velocities as nonmembers, the velocity distributions for M31 members in our fields are kinematically biased toward negative heliocentric velocities. As a consequence, the positive velocity tail of the stellar halo distribution in each field is truncated, such that we could not reliably fit for a halo component in each field (i.e., the velocity dispersion of the fitted halo component would likely be smaller than the true velocity dispersion of M31's stellar halo in a given region). Therefore, we fixed the stellar halo component in each field. Gilbert et al. (2018) measured global properties of the M31 stellar halo's velocity distribution as a function of radius using over 5000 M31 RGB stars across 50 fields. They used the likelihood of M31 membership (Section 4.4; without the use of radial velocity as a diagnostic) as a prior, simultaneously fitting for all M31 and MW components. This resulted in a kinematically unbiased estimation of parameters characterizing the M31 halo's stellar velocity distribution. We transformed their mean velocities and velocity dispersions in the appropriate radial bins from the Galactocentric to heliocentric frame, based on the median right ascension and declination of all stars in a given field. Table 4.4 contains the parameters describing the heliocentric velocity distribution of the stellar halo component in each field.

We determined the number of components in each field by using an expectation-maximization (EM) algorithm to fit models of Gaussian mixtures to the velocity distribution of M31 RGB stars. Varying the number of components per model of each field, we utilized the Akaike information criterion (AIC) to select the best-fit Gaussian mixture, penalizing mixtures that did not significantly reduce the AIC without also decreasing the Bayesian information criterion (BIC). Based on this analysis, the number of components in S and D are 3 and 2, respectively, where one component in each field corresponds to the kinematically hot halo. The EM algorithm strongly preferred a single-component model for H based on the AIC and BIC. However, the velocity dispersion of this single-component model, 82 km $s^{-1}$, is discrepant with the velocity dispersion of M31's stellar halo between 8−14 kpc, 108 km $s^{-1}$, as measured from 525 M31 RGB stars (Gilbert et al. 2018). A two-sided Kolmogorov-Smirnov (KS) test similarly indicates that the velocity distribution of M31 RGB stars in field H is inconsistent with being solely drawn from the 8−14 kpc stellar halo model of Gilbert et al. (2018) at the 2% level. Thus, we assumed a two-component model, as opposed to a single-component model, for this field. This second component likely corresponds to the inner halo substructure known as the Southeast shelf (Section 4.6; Fardal et al. 2007; Gilbert et al. 2007), where the Southeast shelf has been identified in all shallow spectroscopic fields neighboring field H (Figure 4.1).

We sampled from the posterior distribution of the velocity model (Eq 4.1) for each field using an affine-invariant Markov chain Monte Carlo (MCMC) ensemble sampler (Foreman-Mackey et al. 2013). We enforced normal prior probability distributions for $\mu_k$ and $\sigma_k$ parameters in fields H and S based on literature measurements (Gilbert et al. 2018) for nearby fields (Figure 4.1). For H, we assumed $\mu_{1,0} = -300 \pm 20$ km $s^{-1}$ and $\sigma_{1,0} = 55 \pm 20$ km $s^{-1}$, whereas for S, we assumed $\mu_{1,0} = -490 \pm 10$ km $s^{-1}$, $\sigma_{2,0} = 25 \pm 10$ km $s^{-1}$, $\mu_{2,0} = -390 \pm 10$ km $s^{-1}$, and $\sigma_{2,0} = 20 \pm 10$ km $s^{-1}$. For field D, we assumed a flat prior, given the absence of previous modeling in the literature for the overlapping 1200G field H13d (Figure 4.1). In each case, we assumed a minimum value for all dispersion parameters, $\sigma_k$, of 10 km $s^{-1}$, based on our typical velocity uncertainty (Section 4.3). For the remainder of the bounds on each parameter, we adopted reasonable ranges that allowed for relatively unrestricted exploration of parameter space. This is intended to account for differences in the properties of our fields as compared to those of nearby fields in the literature. Additionally, we allowed $f_k$ parameters to extend down to zero for kinematically cold components.

We used 100 chains and $10^4$ steps per field, for a total of $10^6$ samples of the posterior probability distribution. We calculated the mean parameter values describing the velocity distribution model using the 50$^{\text{th}}$ percentile values of the corresponding marginalized posterior probability distributions. We constructed the marginal distributions using only the latter 50% of the MCMC chains, which are securely converged for every slitmask and model parameter in terms of stabilization of the autocorrelation time. The errors on each parameter are calculated based on the 16$^{\text{th}}$ and 84$^{\text{th}}$ percentiles of the marginal distributions.

**Probability of Substructure**

To extract the properties of the various components in each field, we assign a probability of belonging to substructure to every M31 RGB star. We computed the substructure probability under the $5\times10^5$ models from the converged portion of the MCMC chain. The total probability of belonging to substructure is,

$$p(v_i) = \frac{e^{\langle L_i \rangle}}{1 + e^{\langle L_i \rangle}}, \tag{4.2}$$

given a measurement of a star's velocity, $v_i$. $\langle L_i \rangle$ is the relative log likelihood that a M31 RGB star belongs to substructure as opposed to the stellar halo, which we express as,

$$\langle L_i \rangle = \ln \left( \frac{\sum_{k=1}^{K-1} f_k \mathcal{N}(v_i|\mu_k, \sigma_k^2)}{f_{\text{halo}} \mathcal{N}(v_i|\mu_{\text{halo}}, \sigma_{\text{halo}}^2)} \right). \tag{4.3}$$

Thus, we constructed a distribution function for the substructure probability in each field based on its full velocity model. For each M31 RGB star, we adopted the 50$^{\text{th}}$ percentile value of the probability distribution function to represent the probability of the star belonging to a particular component.

Figure 4.5 demonstrates the properties of stars likely belonging to *any* substructure component in a given field in terms of heliocentric velocity and photometric metallicity. The majority of M31 RGB stars in field D belong to M31's stellar halo as opposed to its disk, whereas field S is dominated by the GSS and the kinematically cold component. In contrast, the stars in H are approximately evenly distributed between the stellar halo and substructure. If an M31 RGB star has a probability of belonging to a particular component that exceeds 50%, i.e., it is more likely to belong to a given component than not, we associated it with the component in the subsequent abundance analysis (Section 4.5).

Table 4.4: Velocity Distribution Model Parameters for M31 Fields

| Field | $r$ (kpc) | $\mu_{halo}$ (km s$^{-1}$) | $\sigma_{halo}$ (km s$^{-1}$) | $\mu_{kcc1}$ (km s$^{-1}$) | $\sigma_{kcc1}$ (km s$^{-1}$) | $f_1$ | $\mu_{kcc2}$ (km s$^{-1}$) | $\sigma_{kcc2}$ (km s$^{-1}$) | $f_2$ |
|---|---|---|---|---|---|---|---|---|---|
| H | 12 | −315 | 108 | −295 ± 12 | $66^{+11}_{-16}$ | $0.56^{+0.23}_{-0.25}$ | ... | ... | ... |
| S | 22 | −319 | 98 | −489 ± 4 | 26 ± 3 | 0.49 ± 0.06 | −372 ± 5 | $17^{+7}_{-4}$ | $0.22^{+0.07}_{-0.06}$ |
| D | 26 | −319 | 98 | −128 ± 3 | $16^{+3}_{-2}$ | 0.43 ± 0.06 | ... | ... | ... |
| f130_2 | 23 | −317 | 98 | ... | ... | ... | ... | ... | ... |

Note. — The parameters describing the model components are mean velocity ($\mu$), velocity dispersion ($\sigma$), and normalized fractional contribution ($f$), where components are separated into the kinematically hot stellar halo and kinematically cold components (KCCs). Mean parameter values are expressed as the 50$^{th}$ percentile values of the corresponding marginalized posterior probability distributions. The errors on each parameter are calculated based on the 16$^{th}$ and 84$^{th}$ percentiles. Parameters for the halo components are adopted from Gilbert et al. (2018). In the case of the 22 kpc GSS field, the primary KCC is the GSS core.

**Resulting Velocity Distributions**

We summarize the mean velocity distribution model parameters for fields H, S, D, and f130_2 in Table 4.4 and illustrate the multiple-component models for each field in Figure 4.6. For H, we identified a relatively cold component with $\langle v \rangle = -295$ km s$^{-1}$ and $\sigma_v = 66$ km s$^{-1}$, which we attribute to the Southeast shelf (Section 4.6; Fardal et al. 2007; Gilbert et al. 2007), a tidal shell originating from the GSS progenitor. The fractional contribution of this component is uncertain, ranging from 0.3–0.8, and exhibits covariance with the velocity dispersion, where increasing (decreasing) the fractional contribution of the substructure component increases (decreases) its velocity dispersion. Substructure components are more robustly characterized in fields S and D. We find that $\langle v \rangle = -489$ km s$^{-1}$, $\sigma_v = 26$ km s$^{-1}$ for the GSS, additionally recovering the secondary kinematically cold component of unknown origin (Kalirai et al. 2006a; Gilbert et al. 2009; Gilbert et al. 2019a) separated by $\sim -120$ km s$^{-1}$ in line-of-sight velocity ($\langle v \rangle = -372$ km s$^{-1}$, $\sigma_v = 17$ km s$^{-1}$) from the primary GSS feature.[3] For M31's northeastern disk, we find $\langle v \rangle = -128$ km s$^{-1}$, $\sigma_v = 16$ km s$^{-1}$, indicating that the disk rotation velocity is 191 km s$^{-1}$ offset from M31's halo velocity in this field.[4] For a comparison of the dispersion the outer disk feature with the literature, see Appendix 4.9. The peak of our disk velocity distribution, $v = -128$ km s$^{-1}$, agrees with previous studies of disk kinematics along the northeast major axis, which measured line-of-sight velocities of $\sim -100$ km s$^{-1}$ for fields along the major axis (Ibata et al. 2005; Dorman et al. 2012). However, we note that field D ($r_{\rm maj} = 25.6$ kpc) is located beyond the maximum major axis distance probed by these studies. Although M31's disk is a prominent feature, field D is dominated by the kinematically hot stellar halo component ($f_{\rm halo} = 0.57$).

Assuming a simple model (Guhathakurta et al. 1988) for perfectly circular rotation of an inclined disk ($i = 77°$, P.A. = 38°), the line-of-sight mean velocity of the disk feature corresponds to $v_{\rm rot} = 229$–$244$ km s$^{-1}$ in the disk plane. Based on a rotation curve inferred from H I kinematics between 10–30 kpc and corrected for the inclination of M31's disk (Klypin, Zhao, and Somerville 2002; Ibata et al. 2005), the expected circular velocity at field D ($r_{\mathrm{disk}}$ = 35 kpc) is ~240 km s$^{-1}$, corresponding to a line-of-sight velocity of −119 km s$^{-1}$ (Guhathakurta et al. 1988). Thus, we computed the expected deviation from perfectly circular rotation, $v_{\mathrm{lag}}$, for the disk feature in field D. Accounting for uncertainty in the mean velocity of the disk feature resulting from the fitting procedure and the membership determination, we estimated that $v_{\mathrm{lag}} = -9^{+11}_{-3}$ km s$^{-1}$. For RGB stars in M31's disk between ~5-15 kpc, Quirk et al. (2019) found that $\langle v_{\mathrm{lag}}\rangle \sim 63$ km s$^{-1}$, although our inferred value is not inconsistent with their full $v_{\mathrm{lag}}$ distribution.

[3]Relative to previous determinations of the velocity distribution in the 22 kpc field (Gilbert et al. 2018), the KCC is offset toward lower mean heliocentric velocities by ~20 km s$^{-1}$. This may result from the reduced velocity precision of the 600ZD grating (Section 4.3), or alternatively, differences in spatial configuration of the sample.

[4]We acknowledge the possibility of bias introduced into our measurement as a result of the −100 km s$^{-1}$ velocity cut utilized in our membership determination for the disk field (Section 4.4). If we have excluded a significant fraction of M31 RGB stars redshifted to low heliocentric velocity as a consequence of the disk rotation, then our measurements for the disk would underestimate the mean velocity and velocity dispersion (Appendix 4.9).

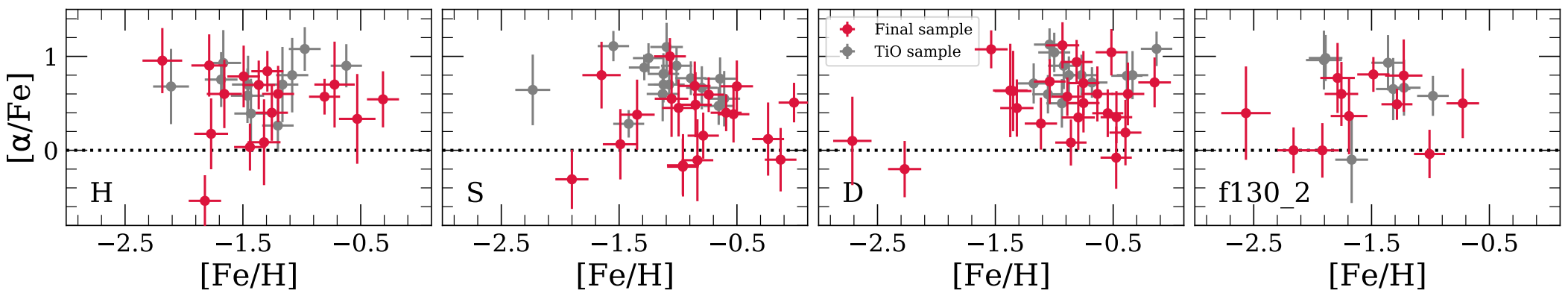

Figure 4.7: [$\alpha$/Fe] versus [Fe/H] for RGB stars in M31 (Section 4.5). We show abundance distributions for the inner stellar halo at 12 kpc (H), the GSS at 22 kpc (S), the outer disk at 26 kpc (D), and the smooth inner stellar halo at 23 kpc (f130_2). We present measurements for 70 stars comprising our final sample (*red circles*; 4.3), including 46 additional M31 RGB stars with spectroscopic evidence of TiO absorption that otherwise pass our selection criteria (*grey circles*). We find that all four fields are $\alpha$-enhanced ($\langle[\alpha/\mathrm{Fe}]\rangle \gtrsim 0.35$), and that the outer disk and GSS fields are more metal-rich on average than the 12 and 23 kpc halo fields.

## 4.5 Elemental Abundances of the M31 Fields

In Section 4.4, we modeled the velocity distributions of the 12 kpc inner halo (H), 22 kpc GSS (S), 26 kpc outer disk (D), and 23 kpc smooth halo (f130_2) fields, identifying substructure in the first three fields. Hereafter, we refer to the 12 kpc substructure as the SE shelf (Section 4.6), the primary 22 kpc substructure as the GSS core, the secondary 22 kpc substructure as the KCC, and the 26 kpc substructure as the disk, for clarity of interpretation when analyzing the abundance distributions. A catalog of stellar parameters and elemental abundances for individual M31 RGB stars across the 4 fields is contained in Appendix A.

### Full Abundance Distributions

We present [$\alpha$/Fe] versus [Fe/H] for 70 M31 RGB stars across the 12 kpc halo field, 22 kpc GSS field, 26 kpc outer disk field, and 23 kpc smooth halo field in Figure 4.7. We also show 46 M31 RGB stars with TiO absorption that otherwise pass our selection criteria (Section 4.3). Table 4.5 summarizes the [Fe/H] and

Table 4.5: Abundances in M31 Fields (H, S, D, f130_2)

| Comp.[a] | ⟨[Fe/H]⟩[b] | $\sigma$([Fe/H])[c] | ⟨[$\alpha$/Fe]⟩ | $\sigma$([$\alpha$/Fe]) |
|---|---|---|---|---|
| 12 kpc Halo Field (H) | | | | |
| Field[d] | $-1.30^{+0.12}_{-0.11}$ | $0.47 \pm 0.08$ | $0.50^{+0.10}_{-0.11}$ | $0.38^{+0.09}_{-0.13}$ |
| SE Shelf | $-1.30^{+0.13}_{-0.12}$ | $0.49^{+0.08}_{-0.09}$ | $0.53^{+0.08}_{-0.10}$ | $0.36^{+0.09}_{-0.11}$ |
| Halo | $-1.30 \pm 0.11$ | $0.45^{+0.07}_{-0.08}$ | $0.45^{+0.12}_{-0.13}$ | $0.42^{+0.09}_{-0.14}$ |
| 22 kpc GSS Field (S) | | | | |
| Field | $-0.84 \pm 0.10$ | $0.46^{+0.07}_{-0.08}$ | $0.41^{+0.08}_{-0.09}$ | $0.35^{+0.06}_{-0.05}$ |
| GSS | $-1.02^{+0.15}_{-0.14}$ | $0.45^{+0.10}_{-0.11}$ | $0.38^{+0.17}_{-0.19}$ | $0.45^{+0.07}_{-0.08}$ |
| KCC | $-0.71 \pm 0.11$ | $0.27 \pm 0.09$ | $0.35^{+0.08}_{-0.09}$ | $0.18^{+0.04}_{-0.05}$ |
| Halo | $-0.66^{+0.16}_{-0.18}$ | $0.44^{+0.07}_{-0.10}$ | $0.49^{+0.05}_{-0.06}$ | $0.21^{+0.05}_{-0.04}$ |
| 26 kpc Disk Field (D) | | | | |
| Field | $-0.92^{+0.10}_{-0.12}$ | $0.55^{+0.11}_{-0.12}$ | $0.58 \pm 0.08$ | $0.36^{+0.04}_{-0.05}$ |
| Disk | $-0.82 \pm 0.09$ | $0.28^{+0.07}_{-0.09}$ | $0.60^{+0.09}_{-0.10}$ | $0.28^{+0.05}_{-0.06}$ |
| Halo | $-1.00^{+0.17}_{-0.19}$ | $0.68^{+0.12}_{-0.14}$ | $0.55 \pm 0.13$ | $0.40^{+0.06}_{-0.08}$ |
| 23 kpc Halo Field (f130_2) | | | | |
| Field | $-1.54 \pm 0.14$ | $0.47^{+0.08}_{-0.09}$ | $0.43^{+0.11}_{-0.12}$ | $0.31 \pm 0.05$ |

Note.— All quantities are calculated from bootstrap resampling of the final sample. For a discussion of bias in the sample, see Section 4.3. (a) For the components of each field, measurements are additionally weighted by the probability of belonging to a given component (Section 4.4, 4.5) (b) Inverse-variance weighted mean. (c) Inverse-variance weighted standard deviation. (d) "Field" refers to all M31 RGB stars present in a field, regardless of association with a kinematic component.

[$\alpha$/Fe] abundances for all M31 RGB stars in our final sample (i.e., without TiO, $\delta$([Fe/H]) $< 0.5$, and $\delta$([$\alpha$/Fe]) $< 0.5$) in each field. Given that we have a finite sample subject to bias, we performed bootstrap re-sampling (with $10^4$ draws) to estimate mean abundances and abundance spreads for each field, including 68% confidence intervals on each parameter. Since the percentage of M31 RGB stars affected by TiO absorption across all four fields is similar, we anticipate that the relative metallicity differences between fields are accurate. Figure 4.8 provides a visual representation of the data in Table 4.5, where we have included equivalent measurements of ⟨[Fe/H]⟩ and ⟨[$\alpha$/Fe]⟩ in M31 RGB stars in the outer halo (Vargas et al. 2014) and a 17 kpc GSS field (G19).

On average, we find that our M31 sample is $\alpha$-enhanced ($0.40 \lesssim \langle[\alpha/\mathrm{Fe}]\rangle \lesssim 0.60$)

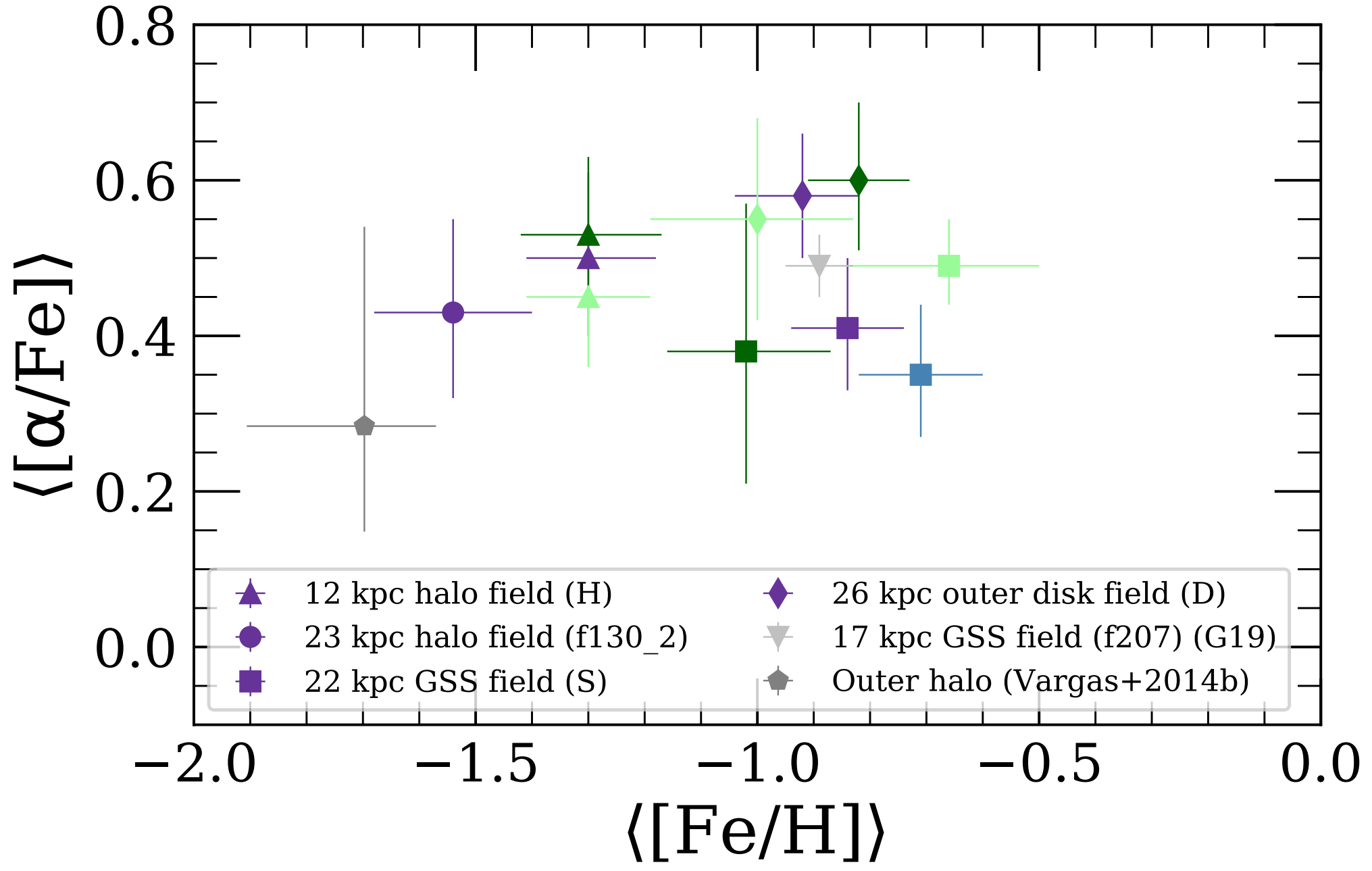

Figure 4.8: $\langle[\alpha/\mathrm{Fe}]\rangle$ versus $\langle[\mathrm{Fe/H}]\rangle$ for all M31 fields (Section 4.5). The data are presented in Table 4.5. We show the averages for the entire field (*purple*; Section 4.5), regardless of kinematic component, in addition to the probabilistic average for each kinematic component (Section 4.5): the stellar halo (*light green*), the primary KCC (*dark green*), and the secondary KCC for the GSS field (*blue*). We overplot average abundance measurements from similarly deep spectra of M31 RGB stars (*grey points*) in a 17 kpc GSS field (G19) and four outer halo stars between ~70-140 kpc (Vargas et al. 2014).

and spans a metallicity range of $-1.5 \lesssim \langle[\mathrm{Fe/H}]\rangle \lesssim -0.9$. High $\alpha$-element abundances indicate that the stellar populations in our M31 fields, regardless of the various galactic structures to which they belong, are likely characterized by rapid star formation and dominated by the yields of core-collapse supernovae. The range of $\langle[\mathrm{Fe/H}]\rangle$ indicates a range of star formation duration. Additionally, stars in all four fields possess a similar spread in [Fe/H](~0.47-0.55), supporting either extended star formation for a single origin, or a multiple-progenitor hypothesis. The GSS field and outer disk fields are the most metal-rich, suggesting either more extended or efficient SFHs compared to the 12 kpc and 23 kpc stellar halo fields. Considering simple field averages, stars in the GSS field and outer disk field are indistinguishable from one another in terms of [Fe/H]. Interestingly, the GSS field may be less $\alpha$-enhanced than the 26 kpc disk field, with a difference in $\langle[\alpha/\mathrm{Fe}]\rangle$ of $0.17^{+0.11}_{-0.12}$. If so, this suggests different relative star formation timescales between Types Ia and core-collapse supernovae, or differences in star formation efficiency, between M31's outer disk and the GSS progenitor. In accordance with expectations of stellar halo formation, the 23 kpc smooth halo field appears to be more metal-poor than the 12 kpc halo field, by 0.24 ± 0.18 dex on average. We discuss the possibility of radial abundance gradients, in both [Fe/H] and [$\alpha$/Fe], in the stellar halo of M31 in Section 4.6.

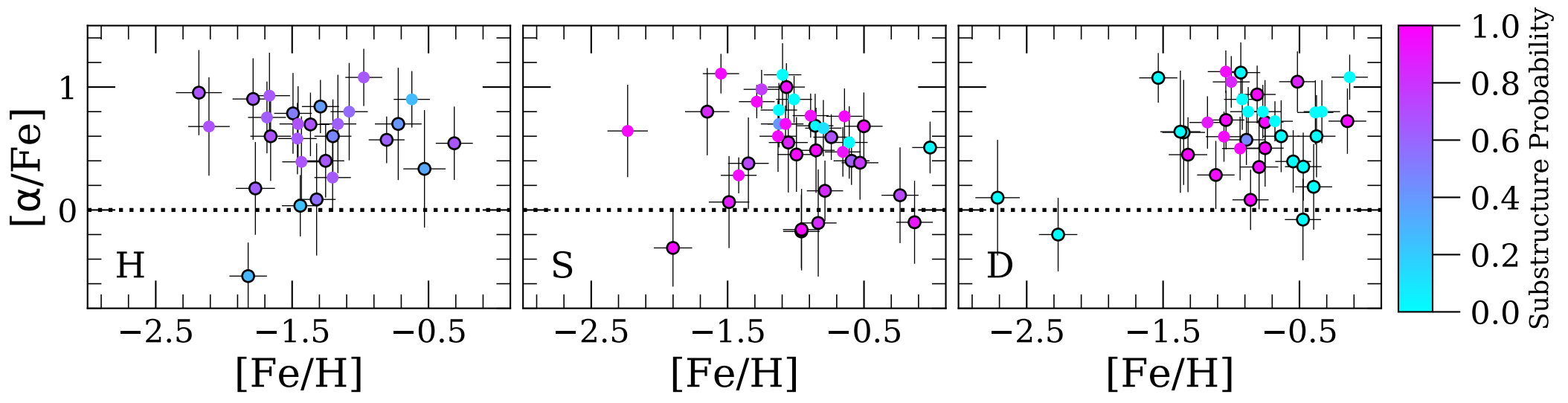

Figure 4.9: [$\alpha$/Fe] versus [Fe/H] for M31 RGB stars in fields with substructure (i.e., excluding the 23 kpc smooth halo field, f130_2), color-coded as in Figure 4.5 (Section 4.5). We show M31 RGB stars in the final sample (*black outlined circles*) and the TiO sample (Section 4.3). The abundances in the final sample of the 22 kpc GSS field (S) probe substructure almost exclusively, whereas in the 12 kpc halo field (H) and 26 kpc disk field (D), the final sample of abundances represent a mixture of the stellar halo and substructure.

**Abundance Distributions of Individual Kinematic Components**

Given that we have identified substructure (Section 4.4) in the 12 kpc halo field, 22 kpc GSS field, and 26 kpc disk field, we separate the full abundance distributions (Section 4.5) into the underlying kinematic components. Using the modeled velocity distributions, we assign each M31 RGB star in fields with substructure a probability of belonging to each individual component (Section 4.4). Figure 4.9 shows [$\alpha$/Fe] versus [Fe/H] for the 12 kpc halo, 22 kpc GSS, and 26 kpc disk fields, where we have indicated the probability that an individual M31 RGB star belongs to *any* substructure component. Our abundance measurements in the 22 kpc GSS field probe substructure almost exclusively, whereas the abundances in the 12 kpc halo and 26 kpc disk fields represent a mixture of the stellar halo and substructure. Figure 4.10 shows the probabilistic distributions of [Fe/H] and [$\alpha$/Fe] for each kinematic component, where we have plotted [$\alpha$/Fe] and [Fe/H] against heliocentric velocity. At a glance, the SE shelf is difficult to chemically distinguish from the stellar halo, where this statement also applies between the GSS core and KCC. M31's disk appears narrow in [Fe/H] relative to the stellar halo.

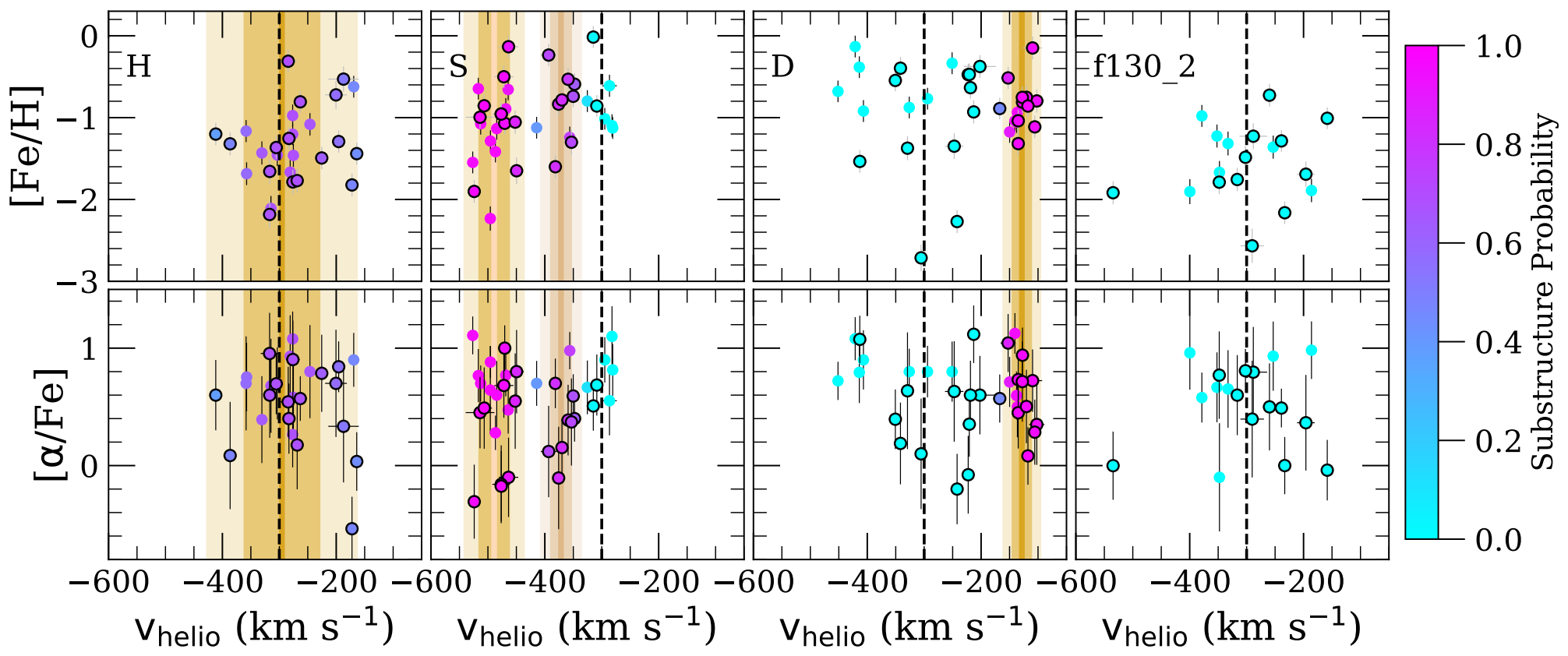

Figure 4.10: Spectroscopic [Fe/H] (*top panels*) and [$\alpha$/Fe] (*bottom panels*) versus heliocentric radial velocity for the same samples and color-coding as Figure 4.9 (Section4.5), except including the 23 kpc smooth halo field (f130_2). The dashed vertical lines and bands are the same as Figure 4.5, representing the median parameters of the velocity distributions in each field (Section4.4). The substructure in the 12 kpc halo field (H) is difficult to distinguish from the stellar halo in terms of [Fe/H] and [$\alpha$/Fe], where the same is true between the GSS core and KCC of unknown origin in the 22 kpc GSS field (S). M31's disk in the 26 kpc disk field (D) appears narrow in [Fe/H] relative to the stellar halo.

Figures 4.9 and 4.10 emphasize that the association of a M31 RGB star with any given component is not definitive. Thus, when computing $\langle$[Fe/H]$\rangle$ and $\langle$[$\alpha$/Fe]$\rangle$ for each component (Table 4.5), we weighted each abundance measurement by its probability of belonging to a particular component, in addition to weighting by the inverse variance of the measurement uncertainty. For clarity of illustration, Figures 4.11 and 4.12 show [Fe/H] and [$\alpha$/Fe] abundances for the kinematic components in each of the three fields with substructure, where we have assigned each star to the component to which it is most likely to belong (Section 4.4). The M31 RGB stars in the final abundance sample of the 12 kpc and 26 kpc fields represent the relative fraction of the stellar halo and substructure components (Table 4.4) accurately. In contrast, M31 RGB stars in the final abundance sample the 22 kpc field under-represent the estimated stellar halo fraction by ~10% and over-represent the KCC.

In addition to representing field averages, Figure 4.8 shows the average probabilistic [Fe/H] and [$\alpha$/Fe] for each kinematic component in the three M31 fields with substructure. The bias against red stars, which are presumably more metal-rich, largely incurred by the omission of TiO stars (Section 4.3) affects the final abundance distribution of the SE shelf and GSS core disproportionately relative to other kinematic components present in the 12 kpc and 22 kpc fields (Figure 4.3). We also note that there is a population of stars falling on the solar metallicity isochrone attributed to the KCC for which we were unable to measure abundances. We anticipate that the difference in ⟨[Fe/H]⟩ between the SE shelf and 12 kpc stellar halo may be larger than the quoted values (Table 4.5), whereas it is difficult to predict how these effects would impact the abundances of the GSS core compared to the KCC. An equivalent number of M31 RGB stars in both the disk and 26 kpc stellar halo were omitted from the final sample, such that the chemical composition of each component should be similarly impacted.

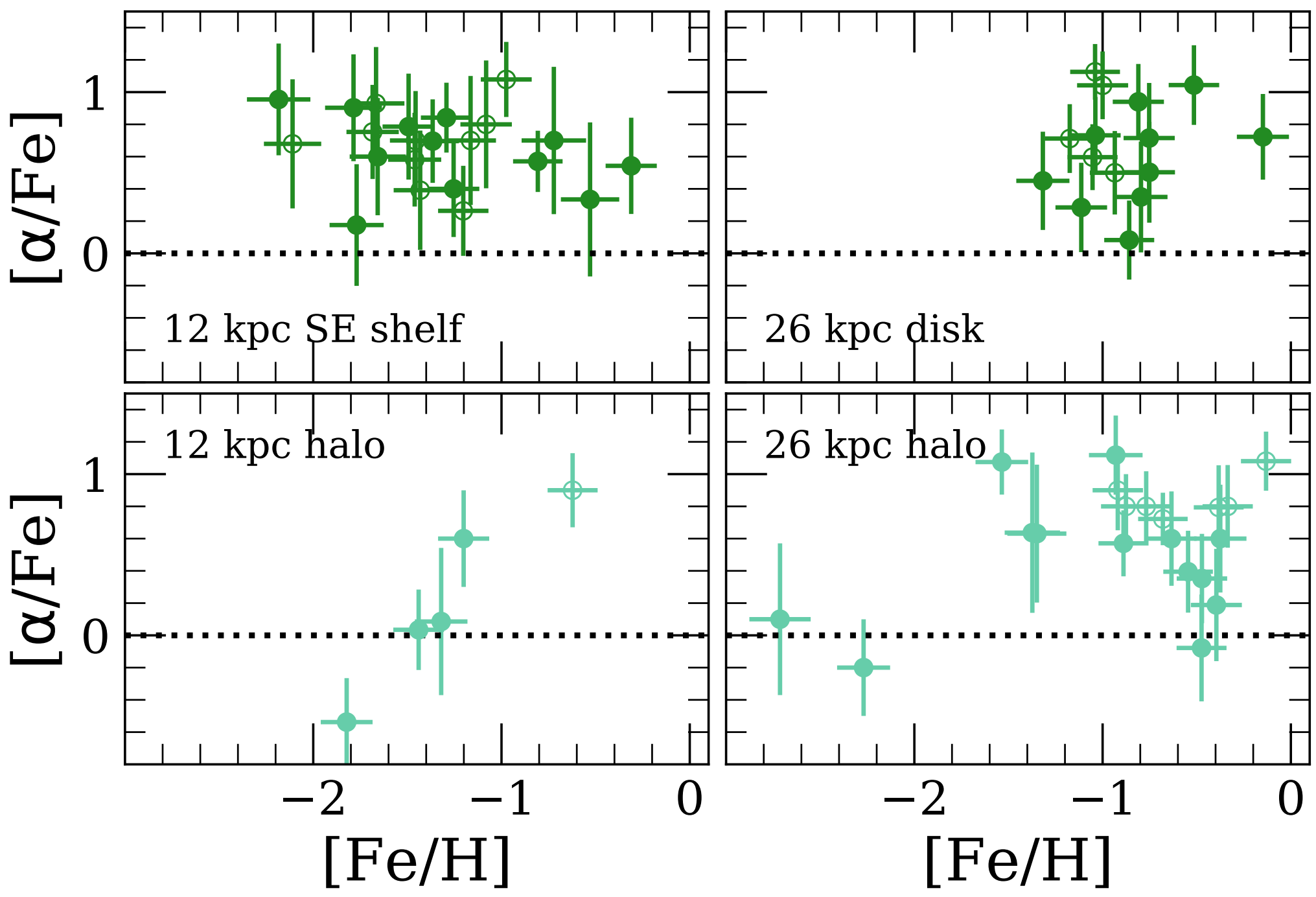

Figure 4.11: [$\alpha$/Fe] versus [Fe/H] for M31 RGB stars with $\delta([\alpha/\text{Fe}]) < 0.5$ in the 12 kpc inner halo field (H; Section 4.5) and 26 kpc outer disk field (D; Section 4.5). We separated each field into its kinematic components by assigning stars to the component to which it has the highest probability of belonging, based on its modeled velocity distribution (Section 4.4). Stars with TiO absorption (Section 4.3) are represented as open circles. We show abundances of M31 RGB stars in the SE shelf feature (*upper left*), M31's disk (*upper right*), and M31's stellar halo (*bottom panels*).

**12 kpc Halo Field**

For the 12 kpc halo field, we find that ⟨[Fe/H]⟩ and ⟨[$\alpha$/Fe]⟩ for the SE shelf cannot be statistically distinguished from the stellar halo (Table 4.5). Although we weighted our field sample by substructure probability computed from the velocity distribution, stars that are more likely to belong to the SE shelf ($p > 0.5$; Section 4.4) still have an average probability of 35% of belonging to the stellar halo. Considering that our final sample for this field does not include many of the red stars that are more likely to populate the SE shelf (Figure 4.3), it is possible that the SE shelf is more metal-rich than the halo. Given the uncertainty on ⟨[$\alpha$/Fe]⟩, the SE shelf and stellar halo may be similarly $\alpha$-enhanced, or the SE shelf may in fact be more $\alpha$-rich than the halo. We discuss the possibility that the SE shelf is related to the GSS progenitor in Section 4.6.

**22 kpc GSS Field**

When separating the GSS core from the KCC, we do not find evidence of a decline of [$\alpha$/Fe] with [Fe/H] for the GSS or the KCC. Many of the RGB stars populating the apparent gradually declining [$\alpha$/Fe] plateau of the 22 kpc GSS field when considered as a whole (Figure 4.9) have a higher probability of belonging to the KCC. We cannot identify the characteristic "knee" in the [$\alpha$/Fe] vs. [Fe/H] distribution based on our abundances for the GSS core. However, the 22 kpc GSS core abundance distribution overlaps with that of a 17 kpc GSS field (Figure 4.13), where the "knee" is located at [Fe/H] $\sim -0.9$ (G19). Taking into account observational uncertainty, computing the intrinsic dispersion (not to be confused with the standard deviation) of the [Fe/H] and [$\alpha$/Fe] distributions yields $0.46^{+0.24}_{-0.13}$ and $\leq 0.46$, respectively, for the 22 kpc GSS field and $0.28^{+0.15}_{-0.08}$ and 0, respectively, for the 17 kpc GSS field. Based on this, we can conclude that the intrinsic dispersion of the abundance distributions between the 22 kpc and 17 kpc GSS fields are marginally consistent. Thus, the GSS abundance distributions do not differ substantially in [Fe/H] and [$\alpha$/Fe] across the $\sim$16$-$23 kpc radial range probed by the two fields along the GSS core.

We find that the GSS core in the 22 kpc GSS field may be more metal-poor than the KCC by $0.31^{+0.19}_{-0.18}$ dex on average, with the caveat of bias against red stars in the GSS core. For the 17 kpc GSS field, G19 found that the KCC differed in ⟨[Fe/H]⟩ from the GSS core by $0.14^{+0.54}_{-0.59}$ based on probabilistic [Fe/H] distributions computed from their velocity model. Using a two-sample KS test, we found that the [$\alpha$/Fe] distributions of the GSS core ($p_{GSS} > 0.5$; Section 4.4) and KCC ($p_{KCC} > 0.5$) are statistically consistent in the 22 kpc GSS field, whereas the [Fe/H] distributions are inconsistent at the 2% level.

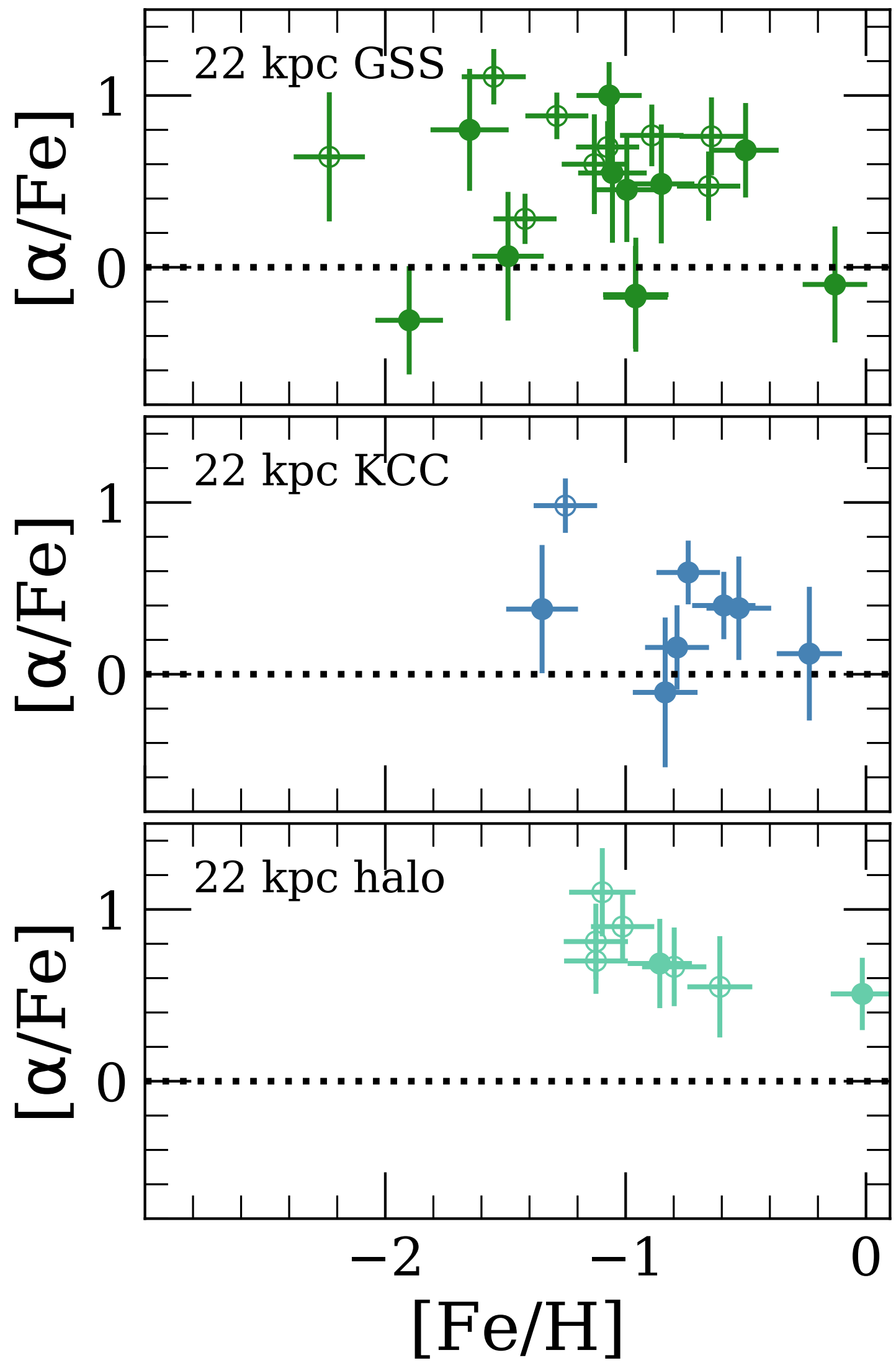

Figure 4.12: Same as Figure 4.11, except for the 22 kpc GSS field (S; Section 4.5). We show abundances of M31 RGB stars in the GSS (*top panel*), the KCC of unknown origin (*middle panel*), and M31's stellar halo (*bottom panel*).

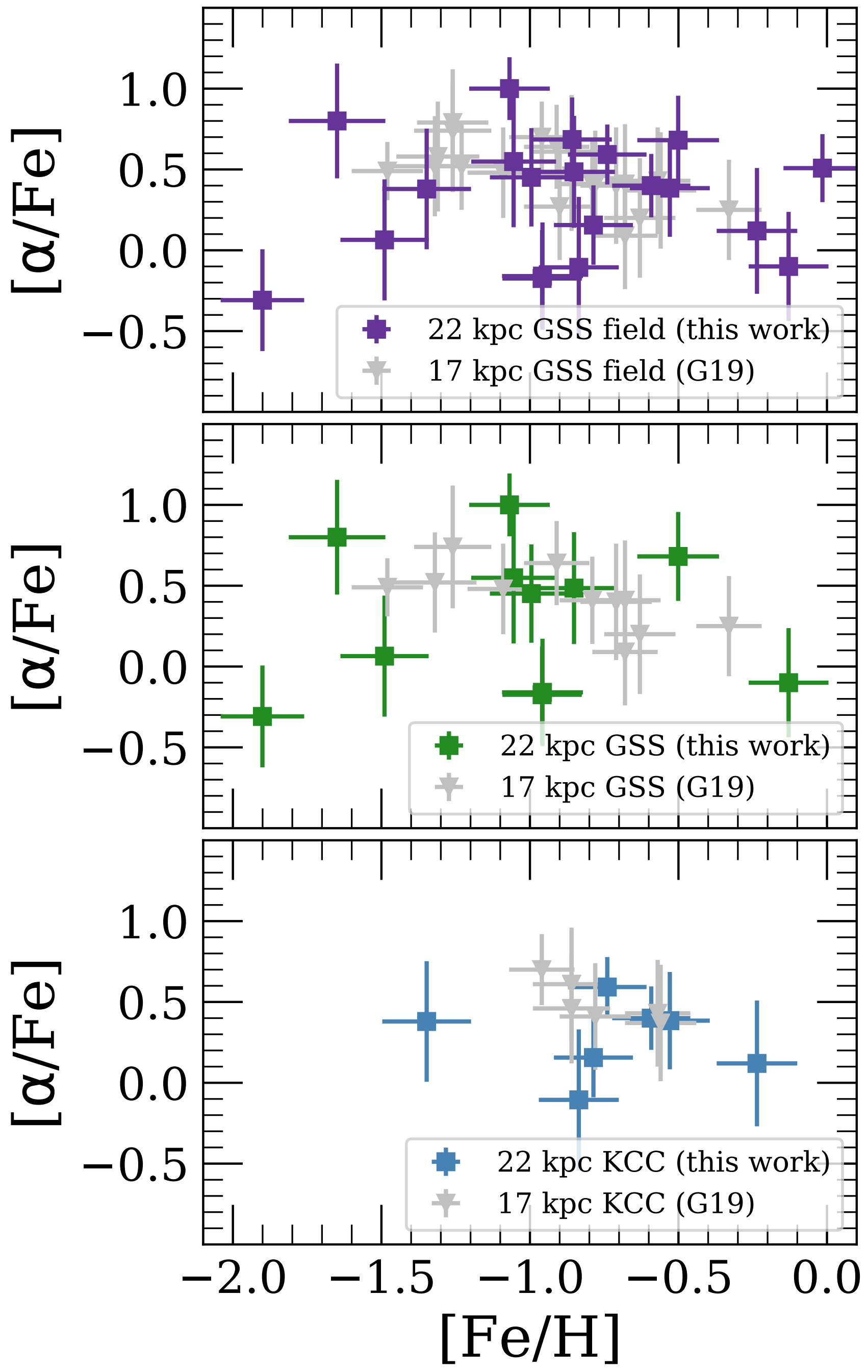

Figure 4.13: [$\alpha$/Fe] versus [Fe/H] for M31 RGB stars in the 22 kpc GSS field (S; *colored squares*, Section 4.5) compared to a 17 kpc GSS field (*grey triangles*; G19). We present abundances for all M31 RGB stars in a given field (*top panel*), the GSS core (*middle panel*), and the KCC of unknown origin (*bottom panel*). We find that the abundance distributions for the GSS between 17 and 22 kpc are consistent.

The stellar halo in the 22 kpc GSS field appears to be more metal-rich than the GSS core and KCC, although the uncertainty in $\langle$[Fe/H]$\rangle$ is large. This is because our final sample in the 22 kpc GSS field over-represents substructure and provides poor constraints on the stellar halo in this region (Figure 4.10). G19 similarly found that they could not constrain the [$\alpha$/Fe] vs. [Fe/H] abundance distribution of the stellar halo in the vicinity of the GSS at 17 kpc, owing to insufficient sample size. However, [Fe/H] for the 22 kpc stellar halo is consistent with G19's probabilistic MDF for the 17 kpc stellar halo along the GSS.

### 26 kpc Disk Field

When separating the 26 kpc disk field into the stellar halo and outer disk, we found that the disk and halo are similar in $\langle$[$\alpha$/Fe]$\rangle$ and $\langle$[Fe/H]$\rangle$, where the disk is slightly more metal-rich. However, much of this difference is driven by the two halo stars at low [Fe/H] ($\lesssim -2$). Omitting these two stars, we found that $\langle\text{[Fe/H]}\rangle_{\text{halo}} = -0.78^{+0.11}_{-0.13}$ and $\langle[\alpha/\text{Fe}]\rangle_{\text{halo}} = 0.63^{+0.12}_{-0.13}$. The metal-rich nature of the disk relative to the halo is not preserved in this case. It is unclear if the metal-poor stars are outliers or representative of a metal-poor tail of the halo distribution that was not well-sampled by our target selection. Given their M31-like velocities ($v_{\text{helio}} < -200$ km s$^{-1}$; Figure 4.10), it is unlikely that these stars are MW foreground dwarf stars. We compare our abundances to the literature for the disks of M31 and MW in Section 4.6.

## 4.6 Discussion

In this section, we placed our abundance measurements in the context of the literature on M31 and the MW. We also explored the implications of our abundance measurements for galaxy formation scenarios for M31.

### Chemical Differences Between the Inner and Outer Halo of M31 and the MW

We investigated whether the [Fe/H] and [$\alpha$/Fe] abundances in our four M31 fields, combined with data from the outer halo of M31 (Vargas et al. 2014), provides evidence for radial chemical abundance gradients in the stellar halo of M31. Previous studies have established the existence of a global radial metallicity gradient in M31's stellar halo based on spectroscopic (Kalirai et al. 2006b; Koch et al. 2008; Gilbert et al. 2014) and photometric (Ibata et al. 2014) samples of individual stars, although metallicity measurements have been primarily CMD-based with small samples of calcium-triplet based measurements. In particular, Gilbert et al. (2014) used the largest spectroscopically confirmed data set to date to analyze the CMD-based metallicity distribution of the stellar halo, with over 1500 M31 halo stars across 38 fields and detections extending beyond 100 kpc.

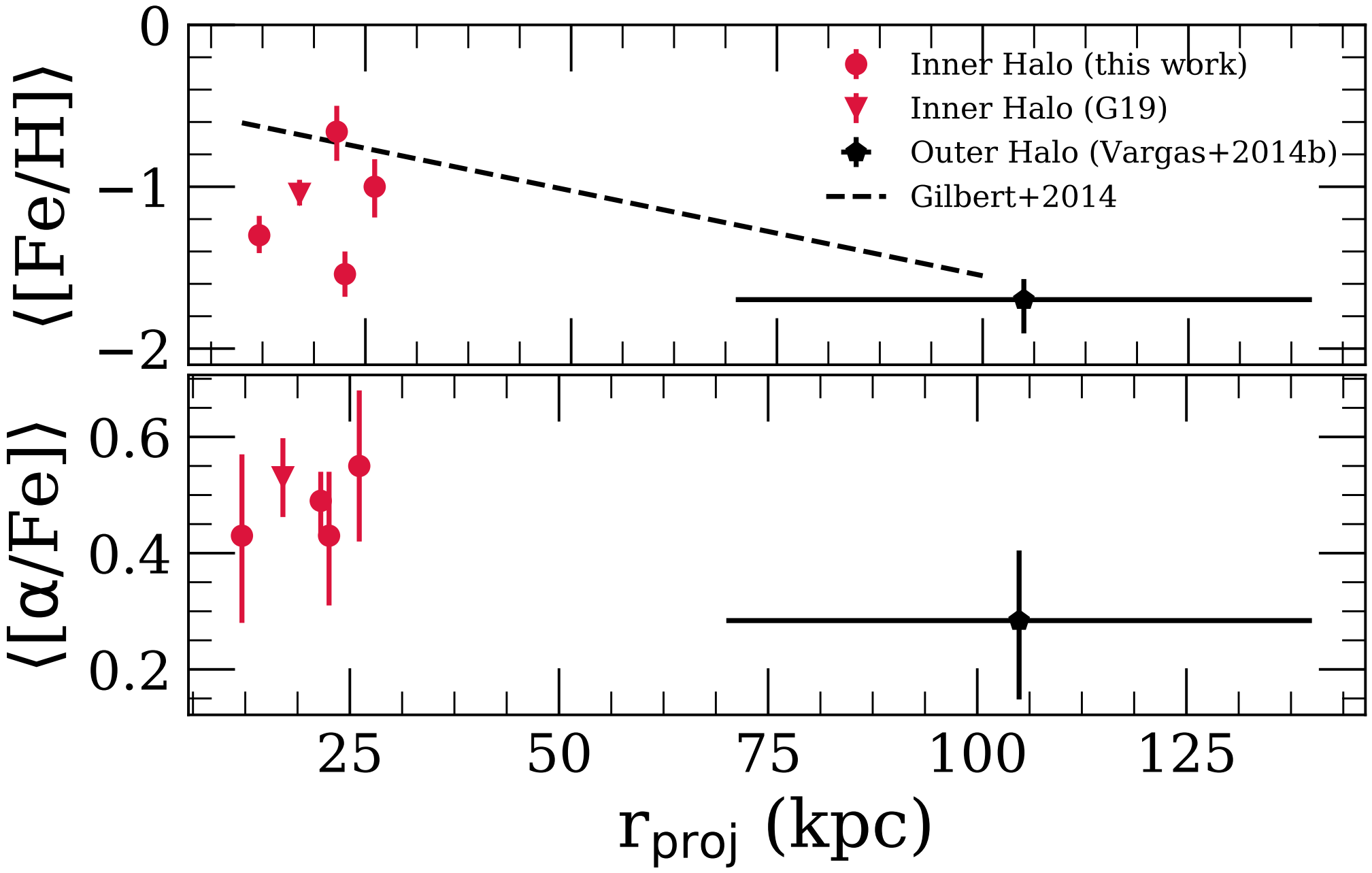

Figure 4.14: ⟨[Fe/H]⟩ ( *top*) and ⟨[$\alpha$/Fe]⟩ ( *bottom*) as a function of projected radius in the stellar halo (i.e., with substructure removed) of M31 (Section 4.6). We show the four fields presented in this chapter (*red circles*), a 17 kpc GSS field ( *red triangle*; G19), and an average for the four outer halo stars of Vargas et al. (2014) (70−140 kpc) placed at 105 kpc. The inner halo fields are biased against red stars with more metal-rich $[Fe/H]_{phot}$ (Section 4.3). The dashed black line represents the photometric radial metallicity gradient of M31's stellar halo of Gilbert et al. (2014), $-0.011$ dex kpc$^{-1}$, where substructure has been removed, assuming a normalization of $\langle[Fe/H]\rangle_{phot} = -0.5$. If the Vargas et al. (2014) halo stars are representative of the outer halo, we find tentative evidence of a negative radial [$\alpha$/Fe] gradient between the inner and outer halo of M31.

Figure 4.14 illustrates ⟨[Fe/H]⟩ and ⟨[$\alpha$/Fe]⟩ as a function of projected radius from the center of M31 for the stellar halo component (Section 4.4) in each field. We referred to the stellar halo components in each field as belonging to the "inner halo" based on their projected radius ($r_{proj} < 30$ kpc), as opposed to any definition based on structural properties of the halo (Dorman et al. 2013). We probabilistically removed substructure from each field in order to probe the properties of the "smooth" stellar halo. For comparison, we show the stellar halo (i.e., with substructure removed) radial metallicity gradient of Gilbert et al. (2014), $-0.011 \pm 0.0013$ dex kpc$^{-1}$, assuming a normalization of $\langle[\mathrm{Fe/H}]\rangle_{\mathrm{phot}} = -0.5$. Owing to the exclusion of red stars with signatures of TiO in their spectra from our final sample, the inner halo fields (including the 17 kpc GSS field; G19) are biased toward lower $[\mathrm{Fe/H}]_{\mathrm{phot}}$ (Section 4.3). Figure 4.14 also includes data for the 4 M31 outer halo stars of Vargas et al. (2014), which span a large radial range (70−140 kpc), shown at $r_{\mathrm{proj}}$ = 105 kpc. Measurements of [Fe/H] and [$\alpha$/Fe] from spectral synthesis appear to support the existence of negative radial abundance gradients in M31's stellar halo, although larger samples of data in the outer halo are necessary to confirm this possibility.

Theoretical studies of stellar halo formation (Font et al. 2011; Tissera et al. 2014; D'Souza and Bell 2018b; Monachesi et al. 2019) have shown that M31's negative radial metallicity gradient is relatively steep compared to predictions from typical simulations. Based on such comparisons, Gilbert et al. (2014) speculated that the magnitude of M31's radial metallicity gradient implies that, in addition to a population of stars having formed *in situ* in the inner regions, massive progenitors have contributed significantly to the formation of the halo. Additionally, spatial and chemical field-to-field variation in the outer halo (Gilbert et al. 2012; Gilbert et al. 2014) suggests that less massive progenitors are the dominant contributors in this region.

Comparatively few theoretical studies have explored the relationship between radial gradients in [$\alpha$/Fe] and accretion history in detail. Font et al. (2006a) found no large-scale radial [Fe/H] or [$\alpha$/Fe] gradients in their hierarchically formed stellar halos, which they attributed to their simulated stellar halos being dominated by early accretion in both the inner and outer halo. Including contributions from stellar populations formed *in situ*, Font et al. (2011) found ubiquitously negative radial [Fe/H] gradients and largely flat radial [$\alpha$/Fe] gradients in their simulated stellar halos. They ascribed the lack of a [$\alpha$/Fe] trend to the prevalence of core-collapse supernovae at all radii for both *in situ* and accreted stellar halo components, which is a consequence of the typically old stellar age ($\sim$11-12 Gyr) of the latter component. A globally $\alpha$-poor outer halo would likely be caused by progenitors accreted at late epochs (Robertson et al. 2005; Font et al. 2006a; Johnston et al. 2008). Thus, if the stellar halo of M31 possesses both negative radial [Fe/H] and [$\alpha$/Fe] gradients, it may be a consequence of the contrast between massive, $\alpha$-enhanced progenitors and/or *in situ* star formation dominating the inner halo and less massive, chemically evolved progenitors dominating the outer halo.

Similar to M31, the MW exhibits indications of negative radial metallicity and $\alpha$-element abundance gradients (c.f. Conroy et al. (2019) concerning the MW halo's radial [Fe/H] gradient). The peaks of the MDFs of the MW's inner and outer halo correspond to [Fe/H] $\sim -1.5$ and [Fe/H] $\sim -2$, respectively (Carollo et al. 2007; Carollo et al. 2010; de Jong et al. 2010; An et al. 2013; Fernández-Alvar et al. 2017). Stellar populations with distinct $\alpha$-element abundances have been identified for stars with halo-like kinematics (Fulbright 2002; Gratton et al. 2003; Roederer 2009; Ishigaki, Chiba, and Aoki 2010; Nissen and Schuster 2010; Ishigaki, Chiba, and Aoki 2012; Ishigaki, Aoki, and Chiba 2013; Hawkins et al. 2015; Hayes et al. 2018a). As opposed to relying on a kinematic decomposition, Fernández-Alvar et al. (2015) and Fernández-Alvar et al. (2017) examined the variation of [Fe/H] and [$\alpha$/Fe] as a function of galactocentric radius, confirming that the low-$\alpha$ population is associated with the outer halo ($r_{GC} > 15$ kpc) of the MW. The dichotomy in [$\alpha$/Fe] and [Fe/H], respectively, between the inner and outer halo in the MW has generally been interpreted to mean that its outer halo corresponds to an accreted population with extended SFHs, whereas its inner halo was constructed by stars formed *in situ* and/or stars accreted from chemically distinct progenitor(s).

In comparison to the MW, the metallicity of individual RGB stars attributed to the metal-poor component of M31's inner stellar halo ([Fe/H] $\sim -1.5$; E19) and the outer halo of M31 ([Fe/H] $\sim -1.7$; Vargas et al. 2014) suggest that both the "smooth" inner halo and the outer halo of M31 are more metal-rich on average at a given projected radius than the MW. The stellar halo of M31 also appears to be $\alpha$-enhanced at all radii compared to the MW, only approaching MW halo-like [$\alpha$/Fe] at large radii in M31.

**Constructing the Inner Stellar Halo of M31 from Present-Day M31 Satellite Galaxies**

Numerous simulations have investigated stellar halo formation via accretion in the context of $\Lambda$CDM cosmology, where stellar halos of massive host galaxies are predicted to form hierarchically from smaller, disrupted stellar systems (Bullock and Johnston 2005; Font et al. 2006a; Font et al. 2008; Font et al. 2011; Zolotov et al. 2009; Zolotov et al. 2010; Cooper et al. 2010; Tissera et al. 2013). The chemical abundance distributions of the stellar halo of MW and M31-like galaxies should therefore reflect the properties of the constituent progenitor galaxies. Given that the [$\alpha$/Fe] distribution at a given metallicity of the MW stellar halo disagrees with that of present-day MW dSphs (Unavane, Wyse, and Gilmore 1996; Shetrone et al. 2003; Venn et al. 2004), we investigated whether the stellar halo of M31 could have formed from a population of progenitors similar to present-day M31 satellite galaxies.

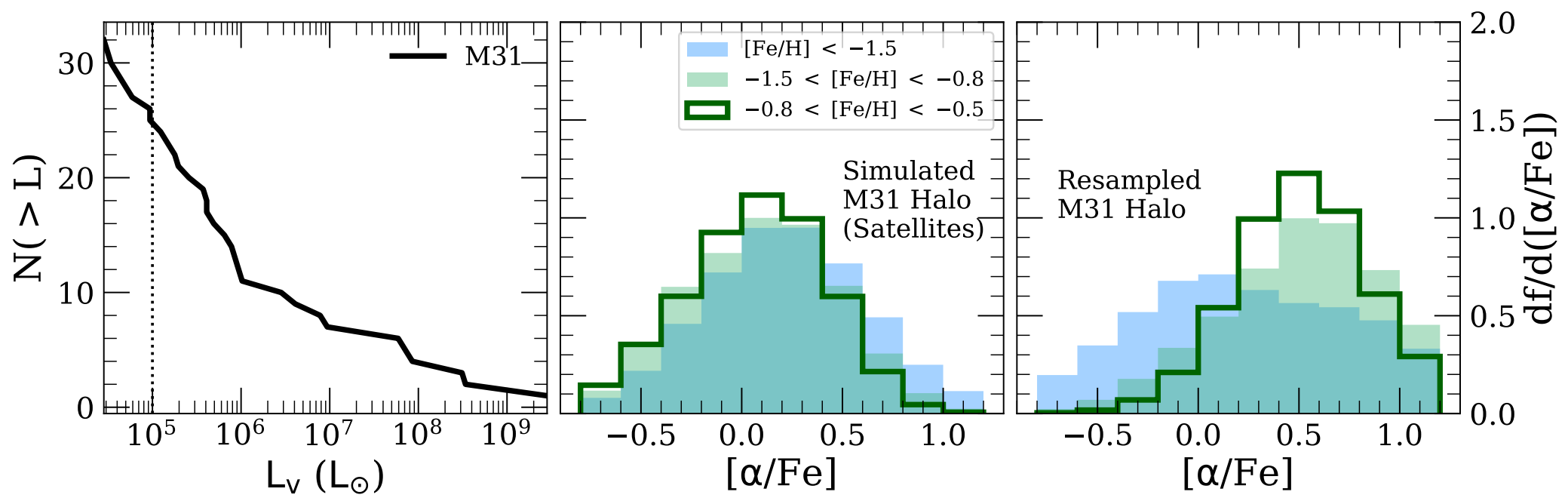

Figure 4.15: The construction of the inner stellar halo of M31 from present-day M31 satellite galaxies (Section 4.6). (*Left panel*) V-band luminosity function of satellite galaxies within 300 kpc of M31, where absolute V-band magnitudes were taken from the compilation by McConnachie (2012). The dotted line represents the luminosity above which the luminosity function is likely to be complete ($L_V > 10^5$ $L_\odot$). The luminosity function is used to assign weights to the abundance distributions of M31 satellite galaxies contributing to the simulated stellar halo of M31 (*middle panel*). The simulated stellar halo is represented by [$\alpha$/Fe] distribution functions, separated into three metallicity bins. (*Right panel*) The bootstrap re-sampled observed [$\alpha$/Fe] distribution functions, separated into metallicity bins, of the stellar halo of M31, as probed by the stellar halo components in 5 M31 fields (this chapter, G19; $r_{\rm proj} \lesssim 26$ kpc). The smooth inner stellar halo of M31 is more $\alpha$-enhanced for [Fe/H] $\gtrsim -1.5$ than would be expected for a stellar halo constructed from present-day M31 satellite galaxies.

To construct simulated abundance distributions for a M31 stellar halo formed from M31 satellite galaxies, we assumed that the progenitors in this scenario possessed a luminosity function equivalent to the luminosity function of satellite galaxies within 300 kpc of M31 (left panel of Figure 4.15), where properties for M31 satellite galaxies were taken from the compilation by McConnachie (2012). We utilized M31 satellites with existing [$\alpha$/Fe] and [Fe/H] abundance measurements ($N > 20$) from Vargas, Geha, and Tollerud (2014) (NGC 185, And II) and Kirby et al. (2020) (And VII, And I, And III, And V), spanning $M_\star \sim 10^{5-7}$ $M_\odot$, based on deep

DEIMOS 1200G spectra.[5] Each individual RGB star, $i$, with measurements of [$\alpha$/Fe] and [Fe/H] was assigned a probability, $p_{i,j}$, of contributing to the simulated stellar halo based on the stellar mass, $M_{\star,j}$, and V-band luminosity, $L_{V,j}$, of its host satellite galaxy,

$$p_{i,j} = \frac{M_{\star,j}\Phi(L_{V,j})/N_j}{\Sigma_{j=1}^{N_{gal}} M_{\star,j}\Phi(L_{V,j})}, \tag{4.4}$$

where $\Phi$ is the V-band luminosity function of present-day M31 satellite galaxies, $N_j$ is the number of RGB stars with abundance measurements in galaxy $j$, and $N_{gal}$ is the total number of M31 satellite galaxies contributing to the abundance distribution of the simulated stellar halo. We consider only the luminosity range over which the luminosity function is likely to be complete ($L_V > 10^5$ $L_\odot$), and only RGB stars with [Fe/H] $< -0.5$ (Section 4.6) and $\delta$([$\alpha$/Fe]) $< 0.5$ (Section 4.3).

To construct the abundance distributions, we drew $10^6$ random samples from the observed abundance distribution of M31 satellite galaxies ($N_{\rm tot}$ = 278) according to the probability distribution defined in Eq. 4.4. Additionally, we perturbed the observed abundance distribution during each draw by the uncertainties on the measurements, assuming Gaussian errors. Figure 4.15 (middle panel) presents [$\alpha$/Fe] distributions for the simulated stellar halo of M31 for a few metallicity bins. The [$\alpha$/Fe] distributions for the high metallicity bins ([Fe/H] $> -1.5$) are less $\alpha$-enhanced on average compared to the low metallicity bin (0.07−0.09 dex vs. 0.22 dex), reflecting the typical declining abundance pattern of [$\alpha$/Fe] vs. [Fe/H] for present-day dwarf galaxies.

Figure 4.15 also shows bootstrap re-sampled [$\alpha$/Fe] distributions of the observed abundance distribution of M31's stellar halo ($r_{\rm proj} \lesssim 26$ kpc) for various metallicity bins. We constructed the abundance distributions based on abundances from the stellar halo components ($p < 0.5$; Section 4.4) of the 5 total M31 fields presented in this chapter and G19 ($N_{\rm tot} = 29$), using the same criteria as in the case of the simulated stellar halo. The stellar halo of M31 is more $\alpha$-enhanced by 0.43−0.50 dex between $-1.5 <$ [Fe/H] $< -0.5$ than expected for a stellar halo formed from progenitors with properties similar to those of present-day M31 satellites[6]. Interestingly, $\langle[\alpha/\mathrm{Fe}]\rangle$ for the low metallicity bin ([Fe/H] $< -1.5$) of the re-sampled stellar halo is nearly identical to that of the simulated stellar halo. Using two-sample KS tests, with $10^4$ draws of $N = 29$ measurements from the parent stellar halo distributions, we find that the [$\alpha$/Fe] distributions at high metallicity ([Fe/H] $> -1.5$) are inconsistent between the re-sampled stellar halo and the simulated stellar halo at the $p < 1\%$ level, whereas the low metallicity distributions are consistent.[7]

[5] The median S/N of the Kirby et al. sample of dSphs is ~23 Å$^{-1}$, which is slightly higher than the stellar halo sample (~17 Å$^{-1}$). The S/N of the Vargas, Geha, and Tollerud sample ranges from 15−25 Å$^{-1}$. The measurement uncertainties on [Fe/H] are comparable between the combined Vargas, Geha, and Tollerud and Kirby et al. M31 satellite sample ($\delta$([Fe/H]) ~ 0.13, $\delta$([$\alpha$/Fe]) ~ 0.23) and our M31 stellar halo sample ($\delta$([Fe/H]) ~ 0.14, $\delta$([$\alpha$/Fe]) ~ 0.29). Thus, we anticipate that the bias from S/N limitations (Section 4.3) similarly affects both samples.

Thus, based on currently available abundance measurements, we conclude that the metal-rich ([Fe/H] $> -1.5$) inner stellar halo of M31 ($r_{\mathrm{proj}} \lesssim 26$ kpc) is unlikely to have formed from disrupted dwarf galaxies with properties similar to present-day M31 satellite galaxies. This is in agreement with findings that the global properties of M31's stellar halo are consistent with dominant contributions from massive progenitor(s) with $M_\star \sim 10^{8-9}\ M_\odot$ (Font et al. 2011; Deason, Mao, and Wechsler 2016; D'Souza and Bell 2018b; Monachesi et al. 2019).

**Inner Halo Substructure and Present-Day Satellite Galaxies**

The progenitor of the GSS is predicted to have been a massive dwarf galaxy of at least $M_\star \sim 10^9\ M_\odot$ (e.g., Fardal et al. 2006; Mori and Rich 2008), and therefore abundances in the GSS should in principle reflect abundance patterns characteristic of massive dwarf galaxies. If the SE shelf in fact originates from the GSS progenitor (Section 4.6), we might also expect its abundance distributions to match that of dwarf galaxies. Thus, we compare the metallicity and $\alpha$-element abundances of substructure in the 12 kpc halo and 22 kpc GSS fields to a sample of M31 satellite dwarf galaxies with measured abundances (NGC 185 and And II from Vargas, Geha, and Tollerud 2014; And VII, And I, And III, and And V, from Kirby et al. (2020). Figure 4.16 illustrates a subset of this comparison. We classified M31 RGB stars as belonging to substructure if they were more likely to be associated with substructure than the stellar halo (Section 4.4). In the case of the GSS field, we do not distinguish between the GSS core and the KCC.

Using a KS test, we find that the metallicity distribution of substructure in the 22 kpc

[6]The intermediate and high metallicity bins are statistically consistent with one another for the re-sampled stellar halo, although the high metallicity bin has a lower $\langle[\alpha/\mathrm{Fe}]\rangle$ by $\sim$0.08. The difference in the means may be a result of small sample sizes, or alternatively contamination in the stellar halo by substructure at [Fe/H] $> -0.8$, owing to limitations of our kinematic decomposition (Section 4.4)

[7]Given that we compared [$\alpha$/Fe] distributions in metallicity bins and consider only [Fe/H] $< -0.5$, the bias against red, presumably metal-rich, stars affected by TiO absorption in the M31 stellar halo sample (Section 4.3) should not alter these conclusions.

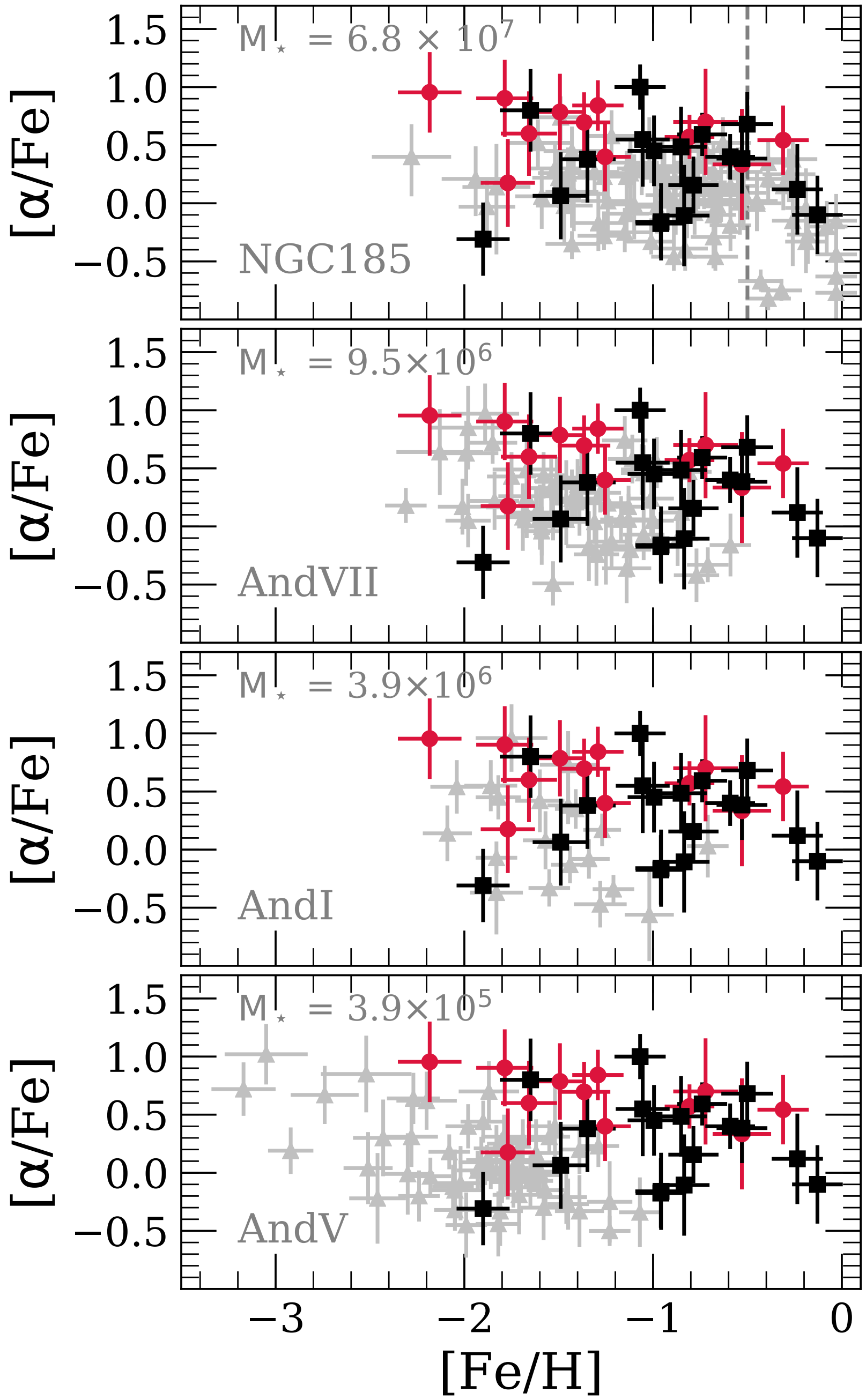

Figure 4.16: [$\alpha$/Fe] vs. [Fe/H] for M31 RGB stars that are likely to belong to 12 kpc SE shelf feature (*red circles*) and 22 kpc GSS core and KCC (*black squares*) compared to the M31 satellite galaxies (*grey triangles*; Section 4.6) NGC 185 (Vargas, Geha, and Tollerud 2014), And VII, And I, and And V (Kirby et al. 2020). From top to bottom, the satellite galaxies are ordered according to decreasing luminosity, where stellar masses are adopted from McConnachie (2012). The vertical dashed line ([Fe/H] = −0.5) delineates the metallicity above which the [$\alpha$/Fe] measurements of Vargas, Geha, and Tollerud (2014) become uncertain. The [Fe/H] distributions of substructure in the 12 kpc and 22 kpc fields resemble satellite galaxies with $M_\star \sim 10^6$ M$_\odot$ and $M_\star \gtrsim 10^7$ $M_\odot$, respectively.

GSS field is consistent with a dwarf galaxy at least as massive as NGC 185 ($M_\star$ = 6.8 $\times 10^7$ $M_\odot$; McConnachie 2012).[8] Based on the mean metallicity of GSS abundances at 17 kpc (−0.87 ± 0.10 dex), G19 used the stellar mass–metallicity relation for Local Group dwarf galaxies (Kirby et al. 2013) to estimate that the GSS progenitor had a stellar mass of at least ~0.5−2×$10^9$ $M_\odot$. Given that the mean metallicity of the GSS at 22 kpc agrees with that at 17 kpc ($\langle$[Fe/H]$\rangle_{\rm GSS,22kpc}$ − $\langle$[Fe/H]$\rangle_{\rm GSS,17kpc}$ = −0.15 ± 0.17), our results corroborate the GSS progenitor mass inferred by G19, where both samples are similarly biased against red stars (Section 4.3).

Stars in both the 17 kpc and 22 kpc GSS fields are more $\alpha$-enhanced than NGC 185. G19 found that the $\alpha$-element abundances of the GSS at 17 kpc were similarly $\alpha$-enhanced compared to Sagittarius, the Large Magellanic Cloud, and the Small Magellanic Cloud, where these conclusions also apply to the GSS at 22 kpc (Figure 4.13). The $\alpha$-element abundances of the GSS at 17 kpc and 22 kpc imply that the GSS progenitor experienced a higher star formation efficiency than NGC 185. Based on *HST* imaging, NGC 185 shows evidence for recent and extended star formation within its inner 200 pc (Butler and Martínez-Delgado 2005; Weisz et al. 2014), quenching ~3 Gyr ago. The *HST* CMD-based SFH for the GSS field (Table 4.3) implies that star formation ceased in the GSS progenitor ~4-5 Gyr ago (Brown et al. 2006), presumably when interactions with M31 began to affect its evolution. Thus, although the GSS progenitor may have quenched ~1-2 Gyr earlier than NGC 185, the galaxy had reached at least the same metallicity by that epoch, further supporting the hypothesis of a comparatively high star formation efficiency for the GSS progenitor.

Although the [$\alpha$/Fe] distributions of the GSS fields and NGC 185 differ, they have a similar metallicity spread. NGC 185 possesses a negative radial metallicity gradient out to ~2.2 kpc (Vargas, Geha, and Tollerud 2014), assuming $d_\odot$ = 617 kpc (McConnachie et al. 2005) and $r_h$ = 1.5' (De Rijcke et al. 2006), although its stellar mass is significantly lower than the inferred mass of the GSS progenitor. In accordance with expectations (e.g., Fardal et al. 2008), the GSS progenitor may have had a radial metallicity gradient. If so, the abundances of the 17 kpc and 22 kpc GSS fields may probe stellar populations from a large radial range in the progenitor (G19; Hammer et al. 2018).

Interestingly, the 22 kpc GSS field possesses an [$\alpha$/Fe] distribution that is statistically consistent with that of satellite galaxies with $M_\star \sim 0.83-9.5 \times 10^6$ $M_\odot$, although the metallicity distribution of the substructure is incompatible with that of the lower mass ($M_\star < 10^7$ $M_\odot$) dwarf galaxies. These lower mass dwarf galaxies had relatively truncated SFHs, forming at least 50% of their stellar mass as of 10 Gyr ago (Weisz et al. 2014; Skillman et al. 2017). This may indicate that stars in the GSS core, KCC, and lower mass dwarf galaxies may have similar contributions of core-collapse supernovae relative to Type Ia supernovae, with the caveat that the GSS progenitor likely experienced a higher star formation efficiency and extended SFH compared to these systems.

[8] We considered only [Fe/H] < −0.5 for the comparison between the abundances of the substructure components in H and S and NGC 185, owing to uncertainty in the abundances of NGC 185 above this metallicity (Vargas, Geha, and Tollerud 2014).

The metallicity distribution of the SE shelf resembles that of satellite galaxies with $3.9-9.5 \times 10^6$ $M_\odot$, although its $\alpha$-element distribution is inconsistent with the sample of M31 satellite galaxies across the entire analyzed mass range. The implications of this comparison are less straightforward, particularly considering the bias against red stars in the SE shelf (Section 4.3, Section 4.5) and the possibility of contamination of the SE shelf sample by halo stars. If the SE shelf abundances are representative, the SE shelf could originate from a progenitor galaxy with $M_\star \sim 10^{6-7} M_\odot$, which possessed relatively short Type Ia supernovae timescales compared to present-day satellites of similar mass, that is distinct from the GSS progenitor. Alternatively, the GSS progenitor could have possessed a significant radial metallicity gradient, such that SE shelf originates from a chemically distinct region of the GSS progenitor. We further evaluate these possibilities in Section 4.6.

**Is the SE Shelf Related to the GSS Progenitor?**

The inner halo of M31 contains abundant substructure, most of which is likely associated with the extended disk or the GSS merger event (e.g., Ferguson et al. 2005; Ibata et al. 2007; McConnachie et al. 2018). In particular, Gilbert et al. (2007) identified a kinematically cold feature at $-300$ km s$^{-1}$ using spectroscopy of $\sim$1000 M31 RGB stars between 9–30 kpc in M31's southeastern quadrant. The velocity dispersion of the feature decreased with increasing projected radial distance, from $\sigma_v = 56$ km s$^{-1}$ at 12 kpc to $\sigma_v = 10$ km s$^{-1}$ at 18 kpc, reflecting the characteristic pattern of a shell system originating from a disrupted progenitor galaxy. Based on its spatial and kinematic properties, Gilbert et al. (2007) associated the $-300$ km s$^{-1}$ cold component with the SE shelf, a predicted, faint shell corresponding to the fourth pericentric passage of GSS progenitor stars (Fardal et al. 2007).

The 12 kpc field overlaps with DEIMOS fields (Figure 4.1) in which Gilbert et al. (2007) identified the SE shelf. The velocity dispersion of the 12 kpc substructure ($\sigma_v$ = 66 km s$^{-1}$; Table 4.4) is similar to that of the SE shelf at the same radius. Figure 4.17 shows the heliocentric velocity versus the radial projected distance of the 12 kpc field compared to M31 RGB stars in DEIMOS fields with shallow 1200G spectroscopy, where Gilbert et al. (2007) identified the fields as contributing to the SE shelf. The M31 RGB stars that are most likely to belong to substructure in the 12 kpc field fall within the observed spatial and kinematical profile of the SE shelf (Gilbert et al. 2007). Thus, based on these properties alone, the 12 kpc field is likely polluted by material from the GSS progenitor.

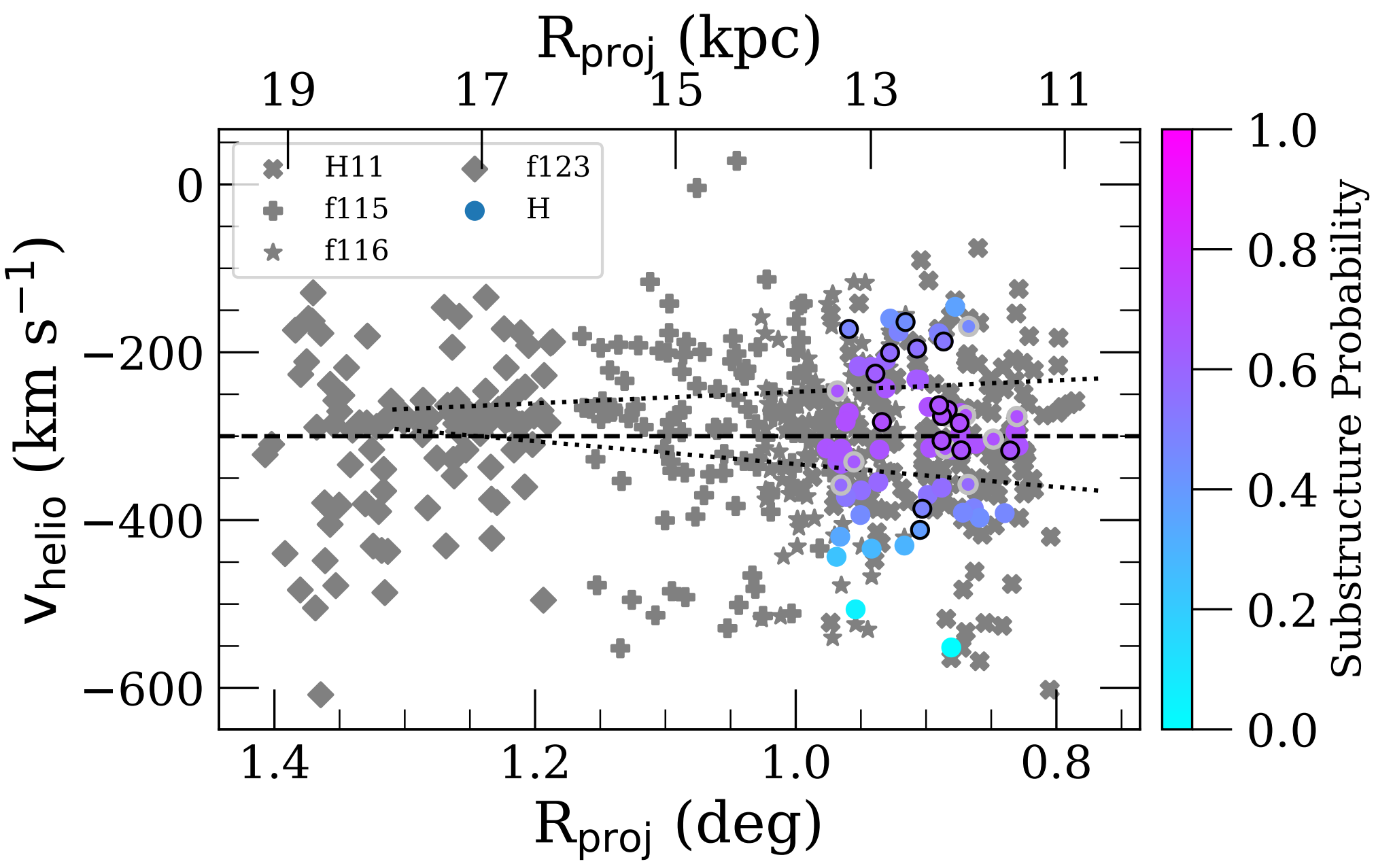

Figure 4.17: Heliocentric velocity versus projected distance of M31 RGB stars (Section 4.6). The 12 kpc field (H) corresponds to circles color-coded according to probability of belonging to substructure (Section 4.4), where M31 RGB stars in our final abundance sample (Section 4.3) are outlined in black and TiO stars are outlined in light grey. Dark grey points represent various DEIMOS fields with shallow 1200G spectroscopy that show evidence of the Southeast shelf (Figure 4.1; Gilbert et al. 2007), the predicted shell formed from GSS progenitor stars on their fourth pericentric passage (Fardal et al. 2007). The dashed horizontal line corresponds to M31's systemic velocity, whereas the dotted lines correspond to the observed boundaries of the Southeast shelf in this space (Gilbert et al. 2007). The substructure in the 12 kpc field fits within the spatial and kinematical profile of the Southeast shelf.

The properties of the stellar population in the vicinity of the 12 kpc field also argue in favor of its contamination by GSS progenitor stars. Brown et al. (2006) and Richardson et al. (2008) found that the stellar age and photometric metallicity distributions in the *HST*/ACS halo11 field (Figure 4.1, Table 4.3), which overlaps with the 12 kpc field, and the *HST*/ACS stream field were remarkably similar. Additionally, Gilbert et al. (2007) observed that $[Fe/H]_{phot}$ was similar between M31 RGB stars likely belonging to the −300 km s$^{-1}$ cold component and the GSS.

If the 12 kpc substructure corresponds to the SE shelf, it may differ from the mean metallicity of the GSS core by $-0.28^{+0.20}_{-0.18}$ dex (Table 4.5). This quoted value is weighted by the probability of belonging to kinematic substructure for all stars in the field. However, the maximum substructure probability is low (69%). In other words, M31 RGB stars with kinematic properties matching that of the SE shelf (with $p > 0.5$; Section 4.4) have a 35% chance on average of belonging to the stellar halo.

If the quoted metallicity difference between the SE shelf feature and the GSS core is accurate, this could indicate that the SE shelf originated from a chemically distinct region of the GSS progenitor. Although no metallicity gradient has been observed along the GSS, there is evidence of a gradient between the GSS core and its outer envelope (Ibata et al. 2007; Gilbert et al. 2009), such that GSS formation models have explored the possibility of the observed metallicity gradient originating from a radial gradient in the GSS progenitor (Fardal et al. 2008; Miki, Mori, and Rich 2016; Kirihara et al. 2017).

**Abundances in the Outer Disk of M31 and the MW**

Few studies of the metallicity of stars in M31's outer disk exist in the literature. Collins et al. (2011) measured Ca II triplet based [Fe/H] for 21 DEIMOS fields between 10-40 projected kpc on the sky from M31's center along the southwestern major axis of M31, finding that $\langle[Fe/H]\rangle_{CaT, thin} = -0.7$ and $\langle[Fe/H]\rangle_{CaT, thick} = -1.0$, where the thin disk has an average velocity dispersion of 36 km s$^{-1}$ vs. 51 km s$^{-1}$ for the thick disk. Thus, both the metallicity ($\langle[Fe/H]\rangle = -0.82$) and velocity dispersion ($\sigma_v$ = 16 km s$^{-1}$) of the 26 kpc disk suggest it is similar to M31's thin disk, or potentially the extended disk of M31 (Section 4.6).

In the inner disk of M31, Gregersen et al. (2015) constructed photometric stellar metallicity distributions, assuming constant stellar age and $[\alpha/Fe] = 0$, based on 7 million RGB stars across the PHAT (Dalcanton et al. 2012) footprint in M31's northeastern disk. They found a radial metallicity gradient of −0.020 dex kpc$^{-1}$ between $r_{disk}$ ∼4−20 kpc, with a $[Fe/H]_{phot}$ normalization of ∼0.11. Extrapolating this metallicity gradient, we estimated that $[Fe/H]_{phot}$ would be −0.6 at the location of 26 kpc disk field, $r_{disk}$ = 35 kpc. In comparison, we calculated $[Fe/H]_{phot}$ = −0.88 for our isolated disk feature in this field. We caution that the behavior of the radial metallicity gradient from individual RGB stars in M31's disk is unknown at large radii (Kwitter et al. 2012; Sick et al. 2014), and that differences in metallicity measurement methodology will impact the absolute metallicity normalization.

The dearth of chemical abundance data in the outer disk of M31 applies to the MW as well. However, the stellar metallicity distribution in the MW disk has been well-studied through spectroscopic surveys out to ∼15 kpc, finding a comparatively steep radial metallicity gradient of $\sim -0.06$ dex kpc$^{-1}$ (e.g., Cheng et al. 2012b; Hayden et al. 2014; Boeche et al. 2014; Mikolaitis et al. 2014). In particular, using ∼70,000 RGB stars from APOGEE, Hayden et al. (2014) found [Fe/H] = −0.43 in the MW disk plane between 13−15 kpc. If we perform the same exercise as in the case of M31's disk and extrapolate the MW's metallicity gradient to $r_{disk}$ = 35 kpc, we would obtain [Fe/H] $\sim -0.8$, which is similar to our measured mean metallicity in the 26 kpc M31 disk field.

Interestingly, spectroscopic abundances exist for the Triangulum-Andromeda (TriAnd) overdensity (Majewski et al. 2004; Rocha-Pinto et al. 2004), a distant structure ($r_{GC}$ ∼ 20 kpc) potentially associated with an extension of the MW disk (Price-Whelan et al. 2015; Xu et al. 2015; Li et al. 2017). Recently, Hayes et al. (2018a) found $\langle[Fe/H]\rangle = -0.8$ for TriAnd, in agreement with our measured metallicity for M31's 26 kpc disk. Additionally, TriAnd has chemical abundances (including $\alpha$-element abundances) similar to the most metal-poor stars in the MW's "outer disk" ($r_{GC}$ > 9 kpc) (Hayes et al. 2018a; Bergemann et al. 2018). The outer disk of M31 may be chemically the most similar to a potential third component of the MW's disk, known as the metal-weak thick disk (Carollo et al. 2019), which is metal-poor ([Fe/H] $\sim -1$) and relatively $\alpha$-enhanced ([$\alpha$/Fe] $\gtrsim$ 0.22). However, evidence for this component has thus far only been detected in the solar neighborhood, and its kinematics ($\sigma_v \sim$ 60 km s$^{-1}$) are inconsistent with those of M31's disk at 26 kpc.

Given that this chapter presents the first [$\alpha$/Fe] measurements in M31's disk, we are limited to the MW's disk for comparisons of [$\alpha$/Fe] measurements based on individual stars. The MW disk is known to possess high-$\alpha$ and low-$\alpha$ sequences at subsolar [Fe/H] (Bensby et al. 2011; Adibekyan et al. 2013; Nidever et al. 2014). High-$\alpha$ stars have been associated with the MW's thick disk (e.g., Bensby et al. 2005; Reddy, Lambert, and Allende Prieto 2006; Lee et al. 2011) and have ages exceeding ~7 Gyr, where the most $\alpha$-enhanced stars tend toward older ages (Bensby et al. 2005; Haywood et al. 2013). However, a population of young $\alpha$-rich stars has also been identified in the MW disk (Martig et al. 2015; Chiappini et al. 2015). In this instance, high-$\alpha$ refers to [$\alpha$/Fe] ~ 0.3, which is significantly lower than our measurement of [$\alpha$/Fe] ~ 0.6 for M31's disk (and M31's stellar halo) at $r_{\rm disk}$ = 35 kpc. If the mean stellar age of stars in the 26 kpc disk field is ~7 Gyr (Brown et al. 2006; Bernard et al. 2015a), with a negligible population of stars with ages $\gtrsim$10 Gyr, then it is similar in age to, if not younger than, the MW disk's high-$\alpha$ population. Assuming that the 26 kpc disk feature is representative of M31's outer disk, the expected discrepancy in [$\alpha$/Fe] between the MW and M31's outer disk is potentially even larger, considering that the low-$\alpha$ sequence is more prominent in the outer disk of the MW (Cheng et al. 2012a; Bovy et al. 2012; Nidever et al. 2014; Hayden et al. 2015).

**Disk Formation Scenarios: the MW vs. M31**

The patterns of [$\alpha$/Fe] vs. [Fe/H] in the MW disk with respect to scale length and scale height (Bensby et al. 2011; Cheng et al. 2012a; Bovy et al. 2012; Anders et al. 2014; Nidever et al. 2014; Hayden et al. 2015) provide support for scenarios in which the inner disk of the MW formed prior to the outer disk (and the chemical thick disk was formed before the chemical thin disk). In particular, the dominance of the low-$\alpha$ sequence in the outer disk and the homogeneity of the high-$\alpha$ sequence in the inner disk (where the scale length is ~2 kpc) could result from a combination of an initial stellar population that formed from a gas-rich, well-mixed, turbulent interstellar medium and multiple distinct stellar populations in the outer disk (Nidever et al. 2014). These outer disk populations could result from a transition from low- to high- star formation efficiency coupled with extended pristine gas infall (Chiappini, Matteucci, and Romano 2001) or increasing outflow rate with increasing galactocentric radius. Based on the chemical abundance patterns of the MW inner vs. outer disk (Bovy et al. 2012; Nidever et al. 2014), radial migration appears to have played a significant role in the evolution of the MW's disk (e.g., Sellwood and Binney 2002; Schönrich and Binney 2009), although its efficiency must have been limited to match the lack of observed high-$\alpha$ stars in the outer disk (Cheng et al. 2012a). In general, the abundance patterns of the MW disk seem to disfavor an external origin (e.g., Brook et al. 2012; Minchev, Chiappini, and Martig 2014), although this possibility cannot be excluded (see also Mackereth et al. 2019a).

The fact that the outer disk of M31 is $\alpha$-enhanced relative to the MW disk between

~9-15 kpc suggests that M31's outer disk may have experienced a different formation history or internal evolution. Marked differences in the structural morphology (Ibata et al. 2005) and dynamics (Dorman et al. 2015; Quirk et al. 2019) of M31's disk support this hypothesis. Perhaps the most distinguishing feature of M31's disk relative to the MW is its ubiquitous burst of star formation that occured 2−4 Gyr ago (Bernard et al. 2015a; Bernard et al. 2015b; Williams et al. 2015; Williams et al. 2017). Taking into account the unusual SFH of M31's disk, coupled with its relatively large velocity dispersion and steep age–velocity dispersion correlation (Dorman et al. 2015), Hammer et al. (2018) found that the observed properties of M31's disk are consistent with a 4:1 major merger, in which first passage occurred 7−10 Gyr ago and nuclei coalescence occured 2−3 Gyr ago.

Possible origins for the extended disk of M31 ($r \lesssim 40$ kpc) are the accretion of multiple small systems or a single secondary progenitor (Ibata et al. 2005). An episode of star formation induced by a major merger offers the advantage of explaining both the disk-like kinematics and chemical homogeneity of the extended disk ($[\mathrm{Fe/H}]_{CaT}$ = −0.9). The low velocity dispersion of the 26 kpc disk ($\sigma_v \sim 16$ km s$^{-1}$), its high $\alpha$-element abundance ($\langle[\alpha/\mathrm{Fe}]\rangle = 0.58$), and relatively young stellar age (4−8 Gyr old; Brown et al. 2006) are consistent with an extended disk that experienced rapid star formation induced by a major merger. The accretion of multiple small systems along the plane of the disk is less likely to result in such a high $\alpha$-element abundance, presuming that such systems would relatively chemically evolved, and thus more $\alpha$-poor. Based on the relatively young stellar age of the disk compared to the 23 kpc field (~7.5 Gyr in the disk field vs. 10−11 Gyr in the 23 kpc field; Brown et al. 2006; Brown et al. 2007), the expectation from the accretion of small systems would be that the 23 kpc field is more $\alpha$-enhanced than the disk, in contradiction to our abundance measurements for these fields. Furthermore, the chemical abundances of the GSS are consistent with a major merger scenario (as in Hammer et al. 2018 or D'Souza and Bell 2018a), assuming that the stars in the GSS core do not predominantly originate from the center of the progenitor and the GSS progenitor had a metallicity gradient (G19).

Internal mechanisms, such as radial migration, are problematic in terms of explaining the $\alpha$-enhancement of M31's disk at 26 kpc. This scenario requires the $\alpha$-enhanced population in the outer disk to have originated from an old, centrally concentrated stellar population, whereas the 26 kpc field contains a significant population of young stars (Brown et al. 2006). Additionally, the velocity dispersion of M31's disk is larger than that of the MW (Dorman et al. 2015), where the efficiency of radial migration is expected to decrease with increasing velocity dispersion (Solway, Sellwood, and Schönrich 2012). In light of current observations of M31's disk, we find that star formation induced by a major merger provides the simplest explanation for the chemical abundances of the 26 kpc disk.

## 4.7 Conclusions

We measured [$\alpha$/Fe] and [Fe/H] from deep, low-resolution DEIMOS spectroscopy of 70 M31 RGB stars across the inner halo, Giant Stellar Stream (Ibata et al. 2001a), and outer disk of M31. This is the largest detailed abundance sample in M31 to date, and in combination with Escala et al. (2019) and Gilbert et al. (2019a), presents the first measurements of spectroscopic [Fe/H] and [$\alpha$/Fe] in the inner halo, GSS, and disk of M31. Using a kinematic decomposition, we separated the stellar populations in our spectroscopic fields into "smooth" stellar halo and substructure. The substructure identified at 12 kpc, 22 kpc, and 26 kpc correspond to the Southeast shelf (Fardal et al. 2007; Gilbert et al. 2007), the GSS core and KCC (Kalirai et al. 2006a; Gilbert et al. 2009; Gilbert et al. 2019a) and M31's outer disk, respectively. We summarize our primary results below.

1. The inner halo, GSS, and outer disk of M31 are $\alpha$-enhanced ($\langle[\alpha/\mathrm{Fe}]\rangle \gtrsim 0.35$), where the 26 kpc disk and 22 kpc GSS fields are more metal-rich than the 12 and 23 kpc inner halo fields ($\Delta$([Fe/H])$_{26\text{-}12\mathrm{kpc}}$ = 0.38 ± 0.16, $\Delta$([Fe/H])$_{26\text{-}23\mathrm{kpc}}$ = $0.62^{+0.17}_{-0.18}$, $\Delta$([Fe/H])$_{22\text{-}12\mathrm{kpc}}$ = 0.46 ± 0.15, $\Delta$([Fe/H])$_{22\text{-}23\mathrm{kpc}}$ = 0.70 ± 0.17).

2. Measurements of [Fe/H] and [$\alpha$/Fe] between 17 kpc (G19) to 22 kpc along the GSS are fully consistent. This is in agreement with previous studies illustrating the absence of a metallicity gradient along the GSS (Ibata et al. 2007; Gilbert et al. 2009).

3. The inner halo of M31 ($r_{\mathrm{proj}} \lesssim 26$ kpc) appears to be more $\alpha$-enhanced than the MW inner halo at all radii. Additionally, we find suggestions that the outer halo of M31 (Vargas et al. 2014) is more $\alpha$-poor than the inner halo, although more data are necessary to confirm such a gradient. If a negative [$\alpha$/Fe] gradient is present, it would agree with the implications of the steep negative [Fe/H] gradient of M31 (Gilbert et al. 2014), providing support for different progenitor(s) and/or formation mechanisms contributing to the inner versus outer halo.

4. Based on currently available data, the [$\alpha$/Fe] distribution of the metal-rich ([Fe/H] > $-1.5$) inner stellar halo ($r_{\rm proj} \lesssim 26$ kpc) of M31 (i.e., with substructure removed) is inconsistent with having formed from the disruption of progenitors with chemical properties similar to present-day M31 satellite galaxies ($M_\star \sim 10^{5-7}$ $M_\odot$).

5. In agreement with G19, comparisons to the abundance distributions of M31 satellite galaxies (Vargas, Geha, and Tollerud 2014; Kirby et al. 2020) suggest that the chemical properties of the GSS are consistent with a massive progenitor ($\gtrsim 0.5$–$2\times10^9$ M$_\odot$; G19) that experienced a high star formation efficiency. Such comparisons also point to the SE shelf resembling lower mass dwarf galaxies ($M_\star \sim 10^6$ $M_\odot$), with the caveat of bias against red stars and potential stellar halo contamination in the SE shelf sample.

6. We found tentative evidence that the SE shelf is more metal poor than the GSS by $\gtrsim 0.10$ dex, taking into account observational uncertainty. If the SE shelf in fact originates from the GSS progenitor (Fardal et al. 2007; Gilbert et al. 2007), then a radial metallicity gradient in the GSS progenitor (e.g., Fardal et al. 2008) could explain the observed metallicity difference.

7. M31's disk at $r_{\rm proj}$ = 26 kpc ($r_{\rm disk}$ = 35 kpc) is consistent with nearly circular rotation (Guhathakurta et al. 1988), with $v_{\rm lag} = -9^{+11}_{-3}$ km s$^{-1}$, and is dynamically cold ($\sigma_v$ = 16 km s$^{-1}$). The disk is highly $\alpha$-enhanced ([$\alpha$/Fe] = 0.58) compared to the high-$\alpha$ population of the MW's disk ([$\alpha$/Fe] $\sim$ 0.30). The metallicities of stars in the 26 kpc disk feature ([Fe/H] = $-0.82$) agree with predictions at comparable radii in the MW (based on extrapolation of its metallicity gradient, e.g., Cheng et al. 2012b; Hayden et al. 2014) and distant, possibly disk-related structures such as TriAnd (Bergemann et al. 2018; Hayes et al. 2018a).

8. Taking into account the observed structural and dynamical properties of M31's disk (Ibata et al. 2005; Dorman et al. 2015), we find that a global episode of active star formation induced by a major merger (Hammer et al. 2018; D'Souza and Bell 2018a) is the simplest explanation for the observed chemical abundances of M31's disk at 26 kpc, .

The following chapter presents an increase in the sample size of M31 RGB stars with abundance measurements, such that we can place more stringent constraints on the accretion history of M31 and the formation of its stellar halo.

We thank the anonymous reviewer of the manuscript on which this chapter is based for a thorough reading and helpful comments. We also thank Stephen Gwyn for reducing the photometry for slitmasks H, S, and D and Jason Kalirai for the reductions of f130_2. I.E. acknowledges support from a National Science Foundation (NSF) Graduate Research Fellowship under Grant No. DGE-1745301. This material is based upon work supported by the NSF under Grants No. AST-1614081 (E.N.K.), AST-1614569 (K.M.G, J.W.), and AST-1412648 (P.G.). E.N.K gratefully acknowledges support from a Cottrell Scholar award administered by the Research Corporation for Science Advancement, as well as funding from generous donors to the California Institute of Technology. E.C.C. was supported by an NSF Graduate Research Fellowship and an ARCS Foundation Fellowship, as well as NSF Grant AST-1616540. The analysis pipeline used to reduce the DEIMOS data was developed at UC Berkeley with support from NSF grant AST- 0071048.

We are grateful to the many people who have worked to make the Keck Telescope and its instruments a reality and to operate and maintain the Keck Observatory. We wish to extend special thanks to those of Hawaiian ancestry on whose sacred mountain we are privileged to be guests. Without their generous hospitality, none of the observations presented herein would have been possible.

## 4.8 Appendix: Comparison Between DEIMOS 600ZD and 1200G Elemental Abundances

In Section 4.5, Section 4.6, and Section 4.6, we simultaneously utilized measurements of [Fe/H] and [$\alpha$/Fe] derived from the 1200G (Vargas, Geha, and Tollerud 2014; Kirby et al. 2020) and 600ZD (this chapter) gratings on DEIMOS. The spectral synthesis methods employed in each case (Kirby, Guhathakurta, and Sneden 2008; Kirby et al. 2009; Escala et al. 2019) are the same in principle, but rely on spectra of differing resolution ($R \sim 6000$ vs. 2500) and wavelength coverage (6300– vs. 4500–9100 Å).

In this section, we illustrate the general consistency of the two measurement techniques using a sample of individual giant stars in MW globular clusters (GCs; NGC 2419, NGC 1904, NGC 6864) and MW dSphs (Draco, Canes Venatici I) observed with both the 1200G and 600ZD gratings. Measurements of [Fe/H] and [$\alpha$/Fe] for the GCs were determined using 600ZD spectra by E19 and using 1200G spectra by Kirby et al. (2016), where identical slitmasks were used for each GC. E19 pre-sented measurements of a limited sample of stars from the 600ZD observations of dSphs. The measurements were restricted to those stars with previous abundance measurements in the literature. In order to build a larger sample to compare results from 600ZD with 1200G, we measured abundances for the complete sample of 600ZD dSph stars in this chapter. The 1200G dSph abundance measurements are drawn from stars identified as members in the catalog of Kirby et al. (2010). The photometry is identical for each star in common between 1200G and 600ZD measurements, where $T_{\mathrm{eff,phot}}$ and log $g$ are taken as inputs into the spectral synthesis software (Kirby, Guhathakurta, and Sneden 2008; Escala et al. 2019). We refine our sample selection according to the relevant criteria outlined in Section 4.3.

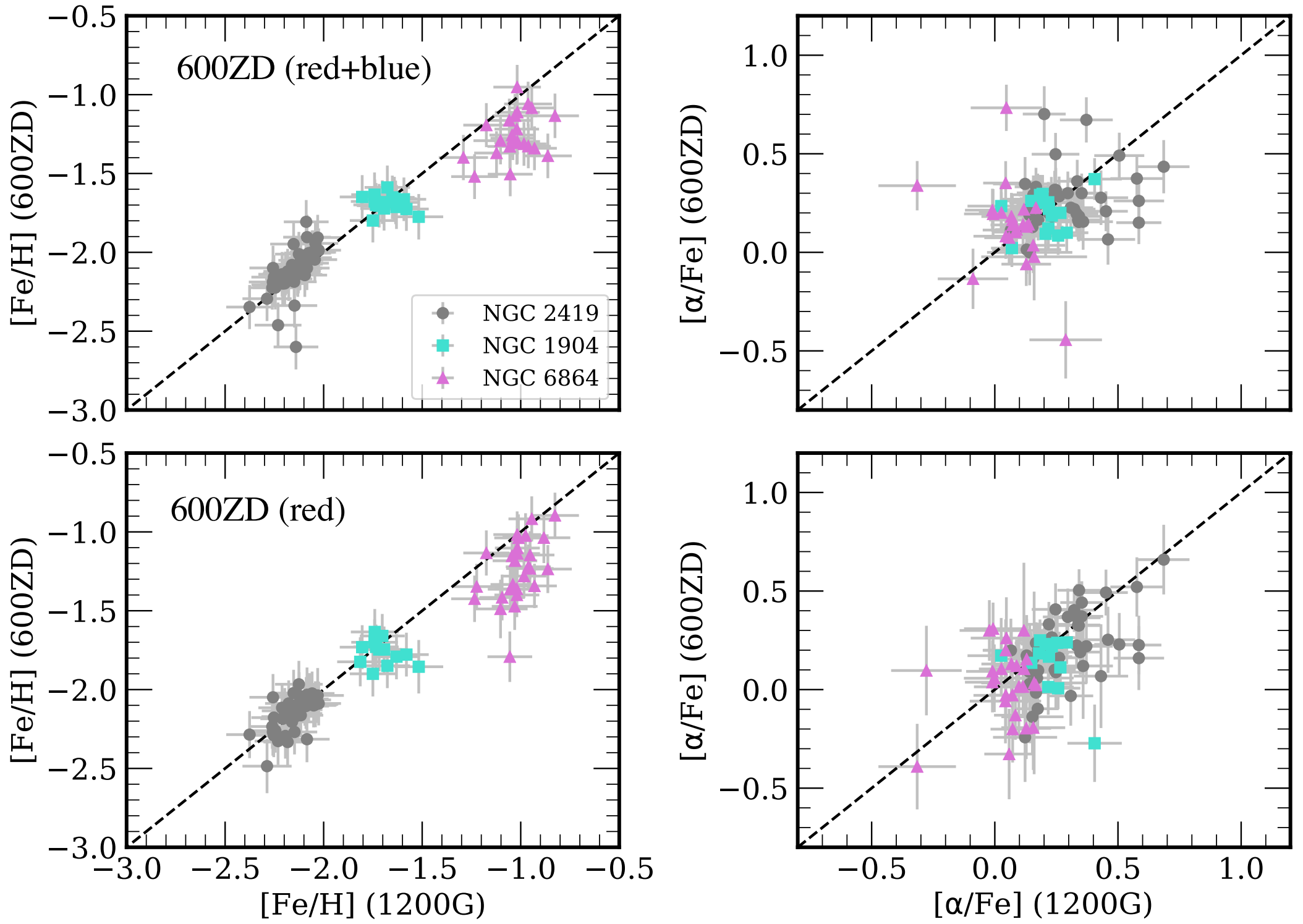

Figure 4.18: A star-by-star comparison between [Fe/H] and [$\alpha$/Fe] measurements for giant stars from a sample of MW globular clusters (Kirby et al. 2016; Escala et al. 2019) using spectra obtained with the 1200G and 600ZD gratings on DEIMOS. The dashed line represents a one-to-one relation. (*Top panels*) Standard 600ZD-based abundance measurements ($N$ = 81) using the entire usable wavelength range (4500−9100 Å) are broadly consistent (at 1.2−1.3$\sigma$) with 1200G-based abundances within the uncertainties. (*Bottom panels*) Restricting the abundance measurement for 600ZD spectra to only red wavelengths (6300−9100 Å) reduces the scatter between the two techniques (to 0.9$\sigma$, $N$ = 84), indicating that any excess scatter is owing to the inclusion of bluer wavelengths (4500-6300 Å).

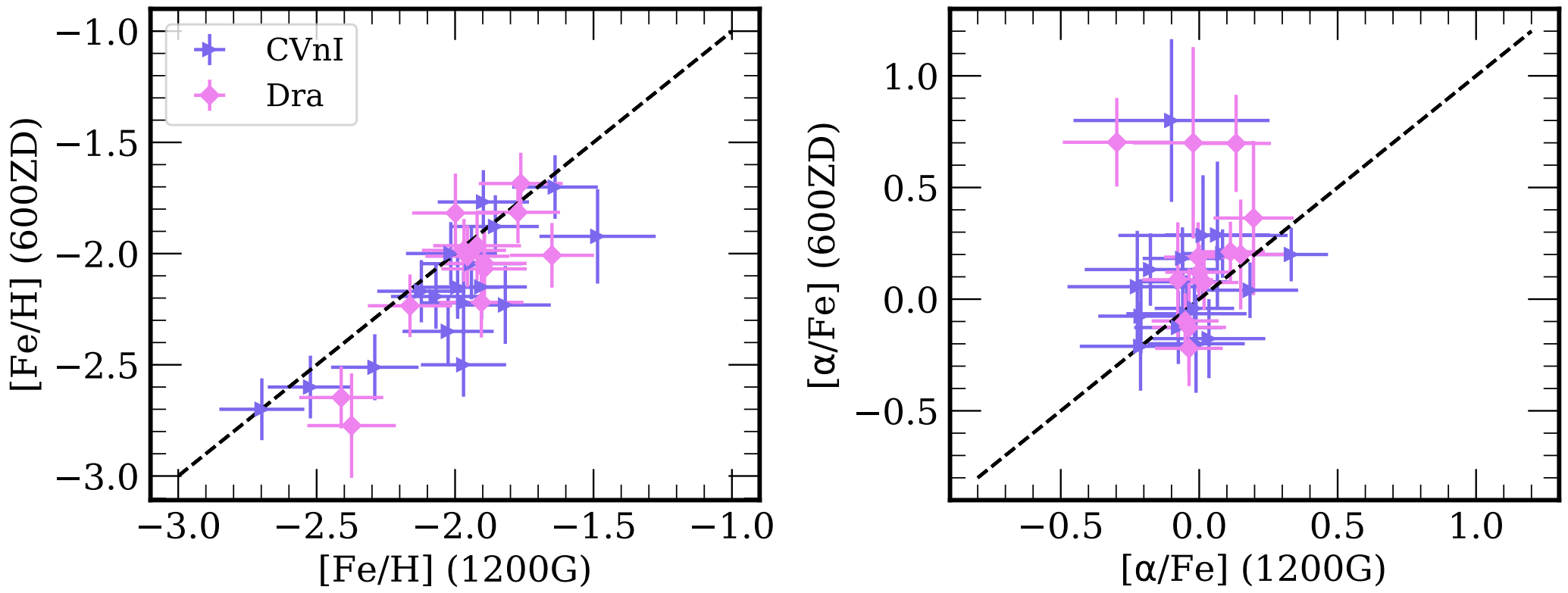

Figure 4.19: A star-by-star comparison between [Fe/H] and [$\alpha$/Fe] for 30 giant stars from a sample of MW dSphs (Kirby et al. 2010; Escala et al. 2019) using spectra obtained with the 1200G and 600ZD gratings on DEIMOS. The dashed line represents a one-to-one relation. As in the case of the GCs, the measurements are broadly consistent ($\sigma_{\text{[Fe/H]}} = 1.28$, $\sigma_{[\alpha/\text{Fe}]} = 1.14$) between the two techniques. The 600ZD [Fe/H] measurements appear to be more metal-poor than the 1200G [Fe/H] measurements for the dSph stars by ~0.1 dex, although it is unclear if this trend is representative of M31.

Figure 4.18 shows a star-by-star comparison of [Fe/H] and [$\alpha$/Fe] measured from both 1200G and 600ZD spectra for the GC sample. We do not find evidence of a statistically significant (to 3$\sigma$) error-weighted mean offset in either [Fe/H] or [$\alpha$/Fe] between the 1200G and 600ZD measurements for this sample ($\langle \Delta\text{[Fe/H]}_{\text{1200G-600ZD}} \rangle = 0.0 \pm 0.02$, $\langle \Delta[\alpha/\text{Fe}]_{\text{1200G-600ZD}} \rangle = -0.08 \pm 0.03$). We quantify the consistency between the two measurement techniques by computing the standard deviation of their error-weighted difference,

$$\sigma_\epsilon = \text{stddev}\left(\frac{\epsilon_{1200G} - \epsilon_{600ZD}}{(\delta\epsilon_{1200G})^2 + (\delta\epsilon_{600ZD})^2}\right), \tag{4.5}$$

where $\epsilon$ is a given chemical abundance measurement ([Fe/H] or [$\alpha$/Fe]) and $\delta\epsilon$ is the associated measurement uncertainty. We omitted outliers in this calculation, performing iterative 5$\sigma$ clipping on the discrepancy between the 1200G and 600ZD abundances until the comparison sample converged. We find that $\sigma_{\text{[Fe/H]}} = 1.15$ and $\sigma_{[\alpha/\text{ Fe}]} = 1.34$ for the GC sample, which indicates that the scatter between the 1200G and 600ZD measurements is not completely accounted for by the uncertainties ($\sigma_\epsilon = 1$).

To investigate the source of this excess scatter, we re-measured abundances from the 600ZD spectra for the GCs, but we restricted the chemical abundance analysis to the same wavelength range used for 1200G spectra (6300−9100 Å). Figure 4.18 also illustrates the results of a star-by-star comparison between 1200G measurements and 600ZD measurements restricted only to redder wavelengths. In this case, the excess scatter disappears ($\sigma_{\mathrm{[Fe/H]}}$ = 0.86, $\sigma_{[\alpha/\mathrm{Fe}]}$ = 0.93), where the measurements are consistent within the uncertainties. This indicates that the source of the excess scatter when utilizing the full wavelength range of the 600ZD spectra is the inclusion of bluer wavelengths (4500−6300 Å), as opposed to lower spectral resolution. Thus, the scatter between 1200G- and 600ZD-based measurements is likely driven by the additional information contained in absorption features between 4500−6300 Å. For example, the bluer spectral regions contain a different balance of $\alpha$ absorption lines, such as the Mg b triplet, than the red side. Therefore the meaning of "$\alpha$" is different depending on the spectral range. (This explanation does not apply to [Fe/H].) Alternatively, it is possible that the bluer regions of the spectrum have higher continuum normalization errors owing to the high density of absorption features at these wavelengths.

Based on our GC sample, we conclude that the 600ZD and 1200G measurements are broadly consistent (within approximately 1.3$\sigma$). We further illustrate this point using our dSph sample, which spans a larger range of [$\alpha$/Fe] than the GCs. Figure 4.19 shows a star-by-star comparison between 1200G- and 600ZD-based measurements for the dSphs, where the measurements are broadly consistent within the uncertainties ($\sigma_{\mathrm{[Fe/H]}}$ = 1.28 and $\sigma_{[\alpha/\mathrm{Fe}]}$ = 1.14). The abundance measurements clearly track one another between the two techniques, most notably in the case of [$\alpha$/Fe] as compared to the GC sample. In contrast to the GCs, we find an error-weighted mean offset between [Fe/H] measurements for the dSphs, where $\langle\Delta\mathrm{[Fe/H]}_{\mathrm{600ZD\text{-}1200G}}\rangle$ = $-0.13 \pm 0.02$. It is unclear whether this offset present in the dSph sample is representative of a general offset in [Fe/H] between 1200G- and 600ZD-based measurements, given that it is not present in our larger GC sample ($N_{\mathrm{dSphs}}$ = 30 vs. $N_{\mathrm{GCs}}$ = 81). We do not anticipate that this potential offset in [Fe/H] could significantly alter any of our conclusions as stated in Section 4.7 that rely on a comparison between 1200G- and 600ZD-based abundances, given that the quoted uncertainties in, e.g., ⟨[Fe/H]⟩, are similar in magnitude.

## 4.9 Appendix: The Velocity Dispersion of the Outer Disk of M31

The dispersion of the disk feature in field D is low ($16^{+3}_{-2}$ km s$^{-1}$; Table 4.4) compared to expectations based on previous analyses of M31's northeastern disk kinematics at smaller disk radii ($\lesssim$ 20 projected kpc; Ibata et al. 2005; Dorman et al. 2012;

Dorman et al. 2015). Collins et al. (2011) analyzed slitmasks at similar radii to field D, albeit in M31's southwestern disk. Although a less stringent velocity cut to identify M31 RGB stars (Section 4.4) would increase the velocity dispersion of the disk feature in field D, the increase is insufficient to resolve the discrepancy. Assuming that all stars with $v_{\rm helio} < -50$ km s$^{-1}$ in field D are bona fide M31 RGB stars ($N_{\rm star}$ = 73) results in the dispersion of M31's disk increasing to $\sigma_v$ ~25 km s$^{-1}$ based on estimates from the EM algorithm for fitting Gaussian mixtures to the velocity distribution. Entirely removing radial velocity as a criterion for M31 membership ($N_{\rm star}$ = 76), we instead found $\sigma_v \sim 40$ km s$^{-1}$ for the disk feature in field D, which is comparable to the values found by Dorman et al. (2012), Ibata et al. (2005), and Collins et al. (2011), although these studies accounted for MW foreground star contamination by various means. Although we used a relatively conservative velocity cut of $v_{\rm helio} < -100$ km s$^{-1}$ to identify M31 RGB field stars in field D, the MW contamination fraction appears to be low in this field based on the absence of an velocity peak at $\sim -50$ km s$^{-1}$ (Figure 4.6) corresponding to MW foreground stars.

Regardless of the details of sample selection, M31's northeastern disk exhibits intrinsic spatial variation in disk kinematics across its entire radial range, where the velocity dispersion on large scales decreases with increasing disk radius (Dorman et al. 2015). We also expect that the local velocity dispersion of a dynamically cold stellar population will be smaller when computed in individual DEIMOS fields as compared to subregions of the disk with a larger extent in position angle. Measuring the collective velocity dispersion of the disk (e.g., Dorman et al. 2015) for studies with wide spatial coverage requires assuming a disk model, which may affect measurements of the velocity dispersion. This likely explains why our measurement is more similar to studies that have averaged velocity dispersion measurements across individual DEIMOS slitmasks in M31's disk (Ibata et al. 2005; Collins et al. 2011; Dorman et al. 2012). Thus, we conclude that our measured velocity dispersion of ~15−20 km s$^{-1}$ is an accurate representation of the dynamics of the feature that we have identified as part of M31's disk.

*5*

# ELEMENTAL ABUNDANCES IN M31: PROPERTIES OF THE INNER STELLAR HALO

Ivanna Escala[1,2], Evan N. Kirby[1], Karoline M. Gilbert[3,4], Jennifer Wojno[4], Emily C. Cunningham[5], Puragra Guhathakurta[6]

[1]Department of Astronomy, California Institute of Technology, 1200 E California Blvd, Pasadena, CA, 91125, USA
[2]Department of Astrophysical Sciences, Princeton University, 4 Ivy Lane, Princeton, NJ, 08544, USA
[3]Space Telescope Science Institute, 3700 San Martin Dr., Baltimore, MD 21218 USA
[4]Department of Physics & Astronomy, Bloomberg Center for Physics and Astronomy, John Hopkins University, 3400 N. Charles St, Baltimore, MD 21218, USA
[5]Center for Computational Astrophysics, Flatiron Institute, 162 5th Ave, New York, NY, 10010, USA
[6]UCO/Lick Observatory, Department of Astronomy & Astrophysics, University of California Santa Cruz, 1156 High Street, Santa Cruz, California 95064, USA

**Abstract**

We present measurements of [Fe/H] and [$\alpha$/Fe] for 129 individual red giant branch stars (RGB) in the stellar halo of M31, including its Giant Stellar Stream (GSS), obtained using spectral synthesis of low- and medium-resolution Keck/DEIMOS spectroscopy ($R \sim 3000$ and 6000, respectively). We observed four fields in M31's stellar halo (at projected radii of 9, 18, 23, and 31 kpc), as well as two fields in the GSS (at 33 kpc). In combination with existing literature measurements, we have increased the sample size of [Fe/H] and [$\alpha$/Fe] measurements to a total of 230 individual M31 RGB stars. From this sample, we investigate the chemical abundance properties of M31's inner halo, and found $\langle$[Fe/H]$\rangle$ = −1.09 ± 0.04 and $\langle$[$\alpha$/Fe]$\rangle$ = 0.40 ± 0.03. Between 8–34 kpc, the inner halo has a steep [Fe/H] gradient (−0.024 ± 0.002 dex kpc$^{-1}$) and negligible [$\alpha$/Fe] gradient, where substructure in the inner halo is systematically more metal-rich than the smooth component of the halo at a given projected distance. Although the chemical abundances of the inner stellar halo are largely inconsistent with that of present-day M31 satellite dwarf galaxies, we identified 37 RGB stars kinematically associated with the smooth stellar halo that have chemical abundance patterns similar to M31 dwarf galaxies. This stellar population is more metal-poor ($\langle$[Fe/H]$\rangle$ = $-1.74 \pm 0.05$) and less $\alpha$-enhanced ($\langle$[$\alpha$/Fe]$\rangle$ = $0.17 \pm 0.07$) than the inner stellar halo as a whole. Comparisons to abundances of the Milky Way (MW) halo indicate that M31 and the MW may both have populations of low-$\alpha$ stars with halo-like kinematics, whereas comparisons to the Magellanic Clouds suggest M31's low-$\alpha$ population had similarly low star formation efficiency. We discuss formation scenarios for M31's halo, concluding that M31's low-$\alpha$ population has an accretion origin.

## 5.1 Introduction

In the $\Lambda$CDM cosmological paradigm, $L_\star$ galaxies like the Milky Way (MW) and Andromeda (M31) form through hierarchical assembly (e.g., Searle and Zinn 1978). Debris from mergers across cosmic time are deposited within the extended stellar halo, where they remain observationally identifiable for many dynamical times (e.g., Helmi et al. 1999b; Bullock and Johnston 2005). Simulations of stellar halo formation in MW-like galaxies have shown that the mass and accretion time distributions of progenitor dwarf galaxies can imprint strong chemical signatures in a galaxy's stellar population (e.g., Robertson et al. 2005; Font et al. 2006c; Johnston et al. 2008; Zolotov et al. 2010; Tissera, White, and Scannapieco 2012), particularly in terms of [Fe/H] and [$\alpha$/Fe]. Measurements of $\alpha$-element abundance (O, Ne, Mg, Si, S, Ar, Ca, and Ti) and iron (Fe) abundance encode information concerning the relative timescales of Type Ia and core-collapse supernovae (e.g., Gilmore and Wyse 1998), such that galactic systems with different evolutionary histories will have distinct chemical abundance patterns. In this way, stellar halos serve as fossil records of a galaxy's accretion history. This theory has been extensively put into practice in the MW, where the differing patterns of [$\alpha$/Fe] and [Fe/H] between its stellar halo and satellite dwarf galaxies have revealed their fundamentally incompatible enrichment histories (Shetrone, Côté, and Sargent 2001; Venn et al. 2004).

Studies of the kinematics and chemical composition of individual stars in the MW have provided a detailed window into the formation of its stellar halo (e.g., Carollo et al. 2007; Carollo et al. 2010; Nissen and Schuster 2010; Ishigaki, Chiba, and Aoki 2012; Haywood et al. 2018; Helmi et al. 2018; Belokurov et al. 2018; Belokurov et al. 2020). However, the MW is a single example of an $L_\star$ galaxy. Observations of MW-like stellar halos in the Local Volume have revealed a wide diversity in their properties, such as stellar halo fraction, mean photometric metallicity, and satellite galaxy demographics, which likely results from halo-to-halo variations in merger history (Merritt et al. 2016; Monachesi et al. 2016; Harmsen et al. 2017; Geha et al. 2017; Smercina et al. 2019). In these studies, both the MW and M31 have emerged as outliers at the quiescent and active ends, respectively, of the spectrum of accretion histories for nearby $L_\star$ galaxies.

Owing to its proximity (785 kpc; McConnachie et al. 2005), M31 is currently the only $L_\star$ galaxy that we can study at a level of detail approaching what is possible in the MW. M31's nearly edge-on orientation ($i$ = 77°; de Vaucouleurs 1958) provides an exquisite view of its extended, highly structured stellar halo (e.g., Ferguson et al. 2002; Guhathakurta et al. 2005; Kalirai et al. 2006b; Gilbert et al. 2007; Gilbert et al. 2009; Ibata et al. 2007; Ibata et al. 2014; McConnachie et al. 2018). Most notably, M31's stellar halo contains a prominent tidal feature known as the Giant Stellar Stream (GSS; Ibata et al. 2001a), where the debris from this event litters the inner halo (Brown et al. 2006; Gilbert et al. 2007). Since the discovery of M31's halo (Guhathakurta et al. 2005; Irwin et al. 2005; Gilbert et al. 2006), its global metallicity, and kinematical properties have been thoroughly characterized from photometric and shallow (∼1 hour) spectroscopic surveys (Kalirai et al. 2006a; Ibata et al. 2007; Koch et al. 2008; McConnachie et al. 2009; Gilbert et al. 2012; Gilbert et al. 2014; Gilbert et al. 2018; Ibata et al. 2014). However, it is only recently that Vargas, Geha, and Tollerud (2014) and Vargas et al. (2014) made the first chemical abundance measurements beyond metallicity estimates in M31's halo and dwarf galaxies.

We have undertaken a deep ($\gtrsim$6 hour) spectroscopic survey using Keck/DEIMOS to probe the formation history of M31 from the largest sample of [Fe/H] and [$\alpha$/Fe] measurements in M31 to date. This has resulted in the first [$\alpha$/Fe] measurements in the GSS (Gilbert et al. 2019a), the inner halo (Escala et al. 2019; Escala et al. 2020a), and the outer disk (Escala et al. 2020a), in addition to an expanded sample of [$\alpha$/Fe] and [Fe/H] measurements in M31 satellite galaxies (Kirby et al. 2020 for individual stars; Wojno et al. 2020 for coadded groups of spectra) and the outer halo (K.M. Gilbert et al. 2020, accepted). Some of our key results include (1) evidence for a high efficiency of star formation in the GSS progenitor (Gilbert et al. 2019a; Escala et al. 2020a) and the outer disk (Escala et al. 2020a), (2) the distinct chemical abundance patterns of the inner halo compared to M31 satellite galaxies (Escala et al. 2020a; Kirby et al. 2020), (3) support for chemical differences between the inner and outer halo (Escala et al. 2020a; K.M Gilbert et al. 2020, accepted), and (4) chemical similarity between MW and M31 satellite galaxies (Kirby et al. 2020; Wojno et al. 2020). In this contribution, we analyze the global chemical abundance properties of the kinematically smooth component of M31's inner stellar halo.

This chapter is organized as follows. In Section 5.2, we present our recently observed M31 fields. We provide a brief overview of our chemical abundance analysis in Section 5.3 and develop a statistical model to determine M31 membership in Section 5.4. We present our [$\alpha$/Fe] and [Fe/H] measurements for 129 M31 RGB stars and analyze the combined sample of inner halo abundance measurements (including the literature) in Section 5.5. Finally, we compare our measurements to the MW and the Magellanic Clouds, and place our results in the context of models of stellar halo formation in Section 5.6.

## 5.2 Data

In this section, I present the spectroscopic and photometric data utilized in this chapter and summarize the known properties of the observed M31 fields.

### Spectroscopy and Data Reduction

Table 5.1 presents previously unpublished deep ($\gtrsim$ 5 hr) observations of six spectroscopic fields in M31. Fields f109_1, f123_1, f130_1, a0_1, a3_1, and a3_2 were observed in total for 7.0, 6.25, 6.74, 6.79, 6.44, and 6.60 hours, respectively. For five of these fields, we utilized the Keck/DEIMOS (Faber et al. 2003) 600 line mm$^{-1}$ (600ZD) grating with the GG455 order blocking filter, a central wavelength of 7200 Å, and 0.8" slitwidths. We observed a single field (f123_1) with the 1200 line mm$^{-1}$ (1200G) grating with the OG550 order blocking filter, a central wavelength of 7800 Å, and 0.8" slitwidths. We observed each spectroscopic field using two separate slitmasks that are identical in design, excepting slit position angles. This difference minimizes flux losses owing to differential atmospheric refraction at blue wavelengths via tracking changes in parallatic angle. The spectral resolution of the 600ZD (1200G) grating is approximately 2.8 (1.3) Å FWHM, or R~3000 (6500) at the Ca II triplet region ($\lambda \sim$ 8500 Å). Similarly deep observations of DEIMOS fields in M31, which we further analyze in this chapter, were previously published by Escala et al. (2020a) and Escala et al. (2019) (600ZD) and Gilbert et al. (2019a) (1200G).

One-dimensional spectra were extracted from the raw, two-dimensional DEIMOS data using the spec2d pipeline (Cooper et al. 2012; Newman et al. 2013), including modifications for bright, unresolved stellar sources (Simon and Geha 2007). Kirby et al. (2020) provides a comprehensive description of the data reduction process, particularly for the case of spectra observed with the 1200G grating. For 600ZD spectra, we included additional alterations to correct for atmospheric refraction, which preferentially affects bluer optical wavelengths (Escala et al. 2019).

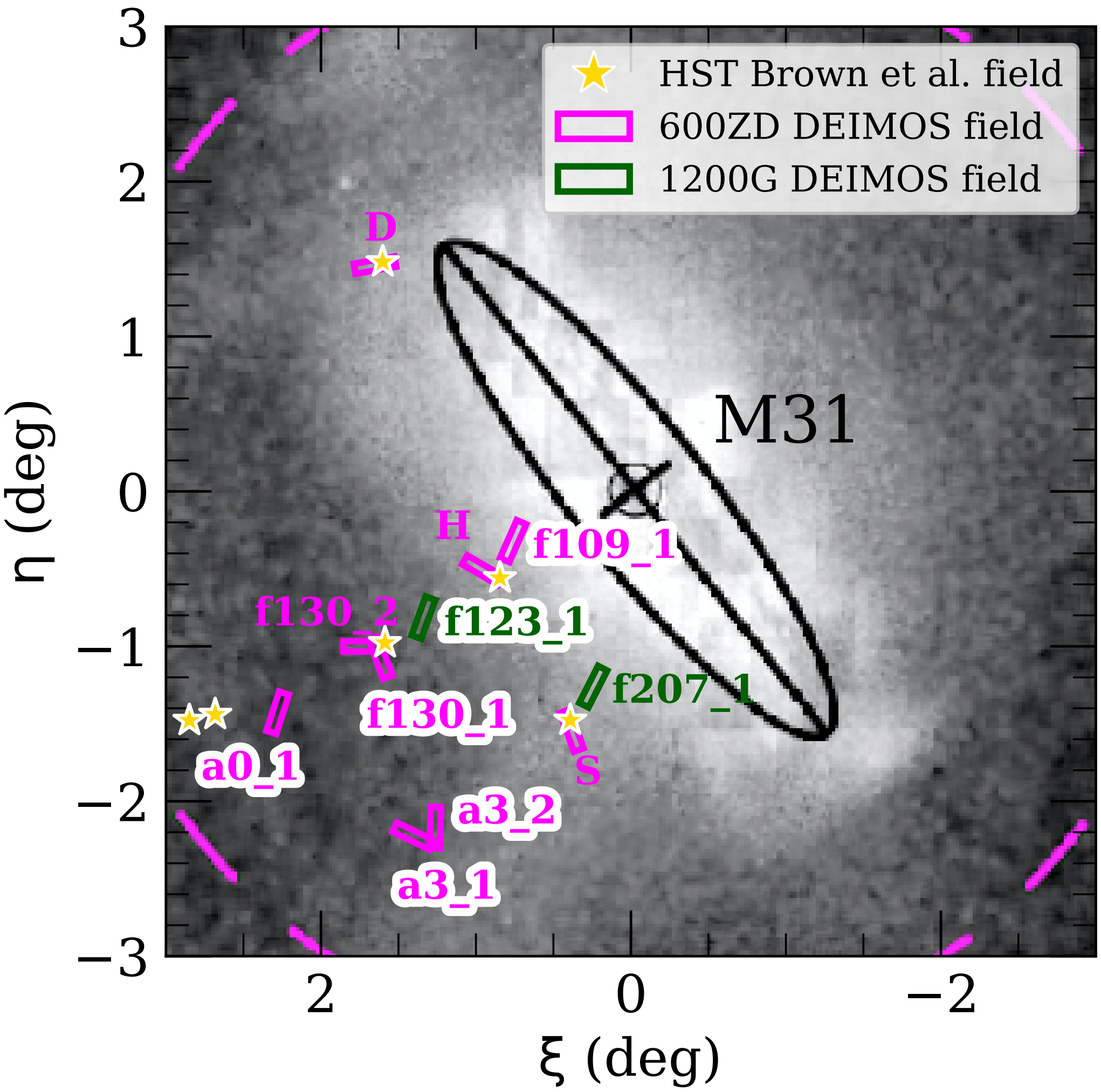

Figure 5.1: The location of deep ($\gtrsim$ 5 hr total exposure time) DEIMOS fields in M31-centric coordinates overlaid on the PAndAS red giant branch star count map (McConnachie et al. 2018). Rectangles represent the approximate size (16' x 4') and orientation of the DEIMOS fields, whereas the dashed magenta line corresponds to 50 projected kpc. We include fields observed with both the low-resolution (600ZD; magenta) and medium-resolution (1200G; green) gratings on DEIMOS. Nearby *HST*/ACS fields (202"x 202"; Brown et al. 2009) are shown as gold stars. Observations first presented in this chapter are outlined in white (Table 5.1; Section 5.2). Our additional deep fields include 9 kpc, 18 kpc, 23 kpc, and 31 kpc in the stellar halo and 33 kpc in the Giant Stellar Stream.

Table 5.1: Deep DEIMOS Observations in M31

| $\alpha_{J2000}$ | $\delta_{J2000}$ | P.A. | Grating | Slitmask[a] | Date | $\theta_s$ ($''$) | $X$ | $t_{exp}$ (hr) | $N$ |
|---|---|---|---|---|---|---|---|---|---|
| 9 kpc Halo Field (f109_1) | | | | | | | | | |
| $00^h45^m47.02^s$ | $+40°56'58.7''$ | 23.9 | 600ZD | f109_1a | 2019 Oct 24 | 0.67 | 1.35 | 2.77 | 143 |
| | | | | f109_1a | 2019 Oct 25 | 0.61 | 1.32 | 2.00 | ... |
| | | | | f109_1b | 2019 Oct 24 | 0.68 | 1.10 | 0.93 | ... |
| | | | | f109_1b | 2019 Oct 25 | 0.73 | 1.10 | 1.30 | ... |
| 18 kpc Halo Field (f123_1) | | | | | | | | | |
| $00^h48^m05.83^s$ | $+40°27'24.0''$ | −20 | 1200G | f123_1a | 2017 Oct 23 | 0.88 | 1.52 | 2.83 | 136 |
| | | | | f123_1b | 2019 Sep 25 | 0.80 | 1.34 | 3.42 | ... |
| 23 kpc Halo Field (f130_1) | | | | | | | | | |
| $00^h49^m11.90^s$ | $+40°11'50.3''$ | −20 | 600ZD | f130_1b | 2019 Sep 25 | 0.43 | 1.07 | 1.15 | 93 |
| | | | | f130_1c | 2018 Sep 10 | 0.72 | 1.29 | 0.97 | ... |
| | | | | f130_1c | 2018 Sep 11 | 0.80 | 1.25 | 2.07 | ... |
| | | | | f130_1c | 2019 Sep 25 | 0.61 | 1.25 | 2.55 | ... |
| 31 kpc Halo Field (a0_1) | | | | | | | | | |
| $00^h51^m51.31^s$ | $+39°50'26.9''$ | −17.9 | 600ZD | a0_1a | 2019 Oct 24 | 0.68 | 1.07 | 0.73 | 67 |
| | | | | a0_1a | 2019 Oct 25 | 0.57 | 1.06 | 0.43 | ... |
| | | | | a0_1b | 2019 Oct 24 | 0.66 | 1.24 | 2.84 | ... |
| | | | | a0_1b | 2019 Oct 25 | 0.70 | 1.24 | 2.79 | ... |
| 33 kpc GSS Field (a3_1) | | | | | | | | | |
| $00^h48^m22.09^s$ | $+39°02'33.1''$ | 64.2 | 600ZD | a3_1a | 2018 Sep 10 | 0.80 | 1.49 | 2.00 | 84 |
| | | | | a3_1a | 2018 Sep 11 | 0.54 | 1.50 | 2.27 | ... |
| | | | | a3_1a | 2019 Sep 26 | 0.55 | 1.44 | 1.67 | ... |
| | | | | a3_1b | 2019 Sep 26 | 0.58 | 1.17 | 0.50 | ... |
| 33 kpc GSS Field (a3_2) | | | | | | | | | |
| $00^h47^m47.22^s$ | $+39°05'50.7''$ | 178.2 | 600ZD | a3_2a | 2018 Oct 02 | 0.60 | 1.16 | 2.10 | 80 |
| | | | | a3_2a | 2019 Sep 26 | 0.64 | 1.09 | 1.00 | ... |
| | | | | a3_2b | 2019 Sep 26 | 0.61 | 1.18 | 3.50 | ... |

Note. — The columns of the table refer to right ascension, declination, position angle in degrees east of north, grating, slitmask name, date of observation (UT), average seeing, average airmass, exposure time per slitmask, and number of stars targeted per slitmask. Additional deep DEIMOS observations of M31 fields utilized in this chapter were published by Escala et al. (2019) and Escala et al. (2020a) and Gilbert et al. (2019a).

[a] Slitmasks labeled as "a", "b", etc., are identical, except that the slits are tilted according to the parallactic angle at the approximate time of observation for the slitmask.

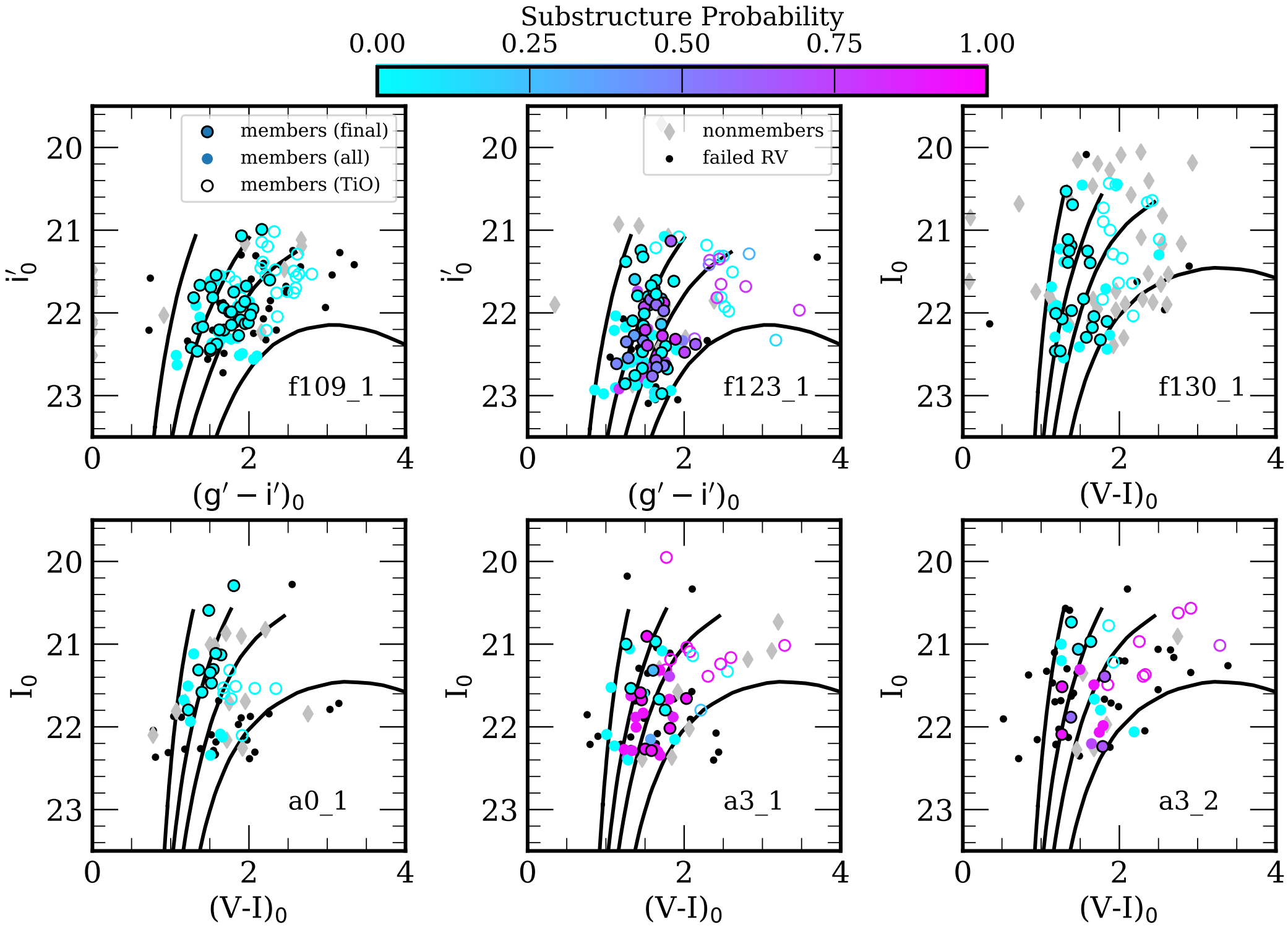

Figure 5.2: Extinction-corrected color-magnitude diagrams in the relevant photometric filters (Johnson-Cousins $V, I$ and CFHT/MegaCam $g, i$; Section 5.2) for individual stars in each spectroscopic field (Table 5.1). For reference, we overplot PARSEC isochrones (*black lines*; Marigo et al. 2017), assuming 9 Gyr ages and [$\alpha$/Fe] = 0, with [Fe/H] = −2.2, −1.0, −0.5, and 0 (from left to right). Stars without successful radial velocity measurements (Section 5.3) are represented as black points. Stars that are likely to be MW foreground dwarfs are indicated as grey diamonds, whereas likely M31 RGB stars (Section 5.4) are represented as circles color-coded according to their probability of belonging to kinematical substructure in the stellar halo (Section 5.4). Open circles indicate M31 RGB stars with strong TiO absorption, which we excluded from our final sample (Section 5.5).

**Field Properties**

Fields f109_1, f123_1, f130_1, a0_1, and a3 are located at 9, 18, 23, 31, and 33 kpc away from the galactic center of M31 in projected distance. We assumed that the M31 galactocenter is located at a right ascension of 0.71 hours and a declination of 41.3 degrees. Table 5.1 provides the mask center and mask position angle of each field on the sky. Figure 5.1 illustrates the locations of our newly observed DEIMOS fields relative to the center of M31, including previously observed fields (Escala et al. 2020a; Escala et al. 2019; Gilbert et al. 2019a) further analyzed in this chapter, overlaid on a RGB star count map from the Pan-Andromeda Archaeological Survey (PAndAS; McConnachie et al. 2018). We also show the locations of pencil-beam *HST*/ACS fields (Brown et al. 2009), some of which overlap with DEIMOS fields (Escala et al. 2020a), from which Brown et al. (2006), Brown et al. (2007), and Brown et al. (2008) derived stellar age distributions. The ACS images (202”×202”) are effectively points on the sky compared to each DEIMOS field (approximately 16’×4’).

The deep fields presented in this chapter were previously observed using shallow (∼1 hour) DEIMOS spectroscopy with the 1200G grating to obtain kinematical information for each field (Gilbert et al. 2007; Gilbert et al. 2009). Based on the shallow 1200G observations, Gilbert et al. (2007) and Gilbert et al. (2009) argued that fields f109_1, f123_1, f130_1, and a0_1 probe the properties of the stellar halo of M31, whereas a3 probes the GSS. However, multiple kinematical components can be present in a given field. For example, f123_1 contains substructure known as the Southeast shelf, which is likely associated with the GSS progenitor (Gilbert et al. 2007; Fardal et al. 2007; Escala et al. 2020a). Unlike fields along the GSS located closer to M31’s center, a3 does not show evidence for the secondary kinematically cold component (KCC) of unknown origin (Kalirai et al. 2006a; Gilbert et al. 2009; Gilbert et al. 2019a). We expect that f130_1 is associated with the “smooth”, relatively metal-poor halo of M31, based on the known properties of the overlapping DEIMOS field f130_2 (Figure 5.1; Escala et al. 2019). Similarly, the velocity distributions of a0_1 and f109_1 are fully consistent with that of M31’s stellar halo (Gilbert et al. 2007). Despite being the innermost M31 field in our sample ($r_{\rm proj}$ = 9 kpc, or $r_{\rm disk}$ = 38 kpc assuming $i$ = 77°), f109_1 does not show any kinematical evidence for significant contamination ($\gtrsim$10%) by M31’s extended disk (Gilbert et al. 2007). This also applies to the Southeast shelf, which was predicted to extend to the location of f109_1 with a large velocity dispersion (∼350 km s$^{-1}$; Fardal et al. 2007; Gilbert et al. 2007), such that any hypothetical SE shelf stars in this field could not be kinematically separated from the halo population. We refer to the spectroscopic fields by their projected M31-centric distance and predominantly traced stellar structure (as in Table 5.1; e.g., f109_1 is the 9 kpc halo field) where appropriate to emphasize their physical properties.

**Photometry**

The photometry for the majority of fields published in this chapter were obtained from MegaCam images in the $g', i'$ filters using the 3.6 m Canada-France-Hawaii Telescope (CFHT). The MegaCam images were obtained by Kalirai et al. (2006a) and reduced with the CFHT MegaPipe pipeline (Gwyn 2008). For 9 targets in field f109_1 absent from our primary catalogs, we sourced g′ and $i'$ band photometry from the Pan-Andromeda Archaeological Survey (PAndAS) point source catalog (McConnachie et al. 2018). For field f130_1, the $g'$, $i'$ magnitudes were transformed to Johnson-Cousins $V, I$ using observations of Landolt photometric standard stars (Kalirai et al. 2006a). In the case of fields a0_1 and a3, the original photometry was obtained in the Washington $M, T_2$ filters by Ostheimer (2003) using the Mosaic camera on the 4 m Kitt Peak National Observatory (KPNO) telescope and subsequently transformed to the Johnson-Cousins $V$, $I$ bands using the relations of Majewski et al. (2000).

Figure 5.2 presents the extinction-corrected color-magnitude diagrams (CMDs) for each field in the relevant photometric filters used to derive quantities based on the photometry such as the photometric effective temperature ($T_{\rm eff,phot}$), surface gravity (log $g$), and metallicity ([Fe/H]$_{\rm phot}$). We show all stars in a given field for which we extracted 1D spectra (M31 RGB stars, MW foreground dwarf stars, and stars for which we were unable to evaluate M31 membership owing to failed radial velocity measurements; Section 5.4), where each star is color-coded according to its probability of belonging to kinematically identified substructure in the stellar halo (Section 5.4). That is, cyan points are likely to be associated with the dynamically hot, "smooth" stellar halo, whereas magenta points likely belong to substructure such as the GSS (a3 fields) or Southeast Shelf (f123_1).

For fields f109_1, f130_1, a0_1, and a3, for which stellar spectra were obtained using the 600ZD grating (Section 5.2), we calculated $T_{\rm eff,phot}$, log $g$, and [Fe/H]$_{\rm phot}$ following the procedure described by Escala et al. (2020a). In summary, the color and magnitude of a star are compared to a grid of theoretical stellar isochrones to derive the above quantities. We utilized the PARSEC isochrones (Marigo et al. 2017), which include molecular TiO in their stellar evolutionary modeling, and assumed a distance modulus to M31 of $m - M = 24.63 \pm 0.20$ (Clementini et al. 2011). For the 600ZD fields, we assumed 9 Gyr isochrones based on the intermediate mean stellar ages of the stellar halo and GSS, as inferred from *HST* CMDs (Brown et al. 2006; Brown et al. 2007; Brown et al. 2008).

The photometric quantities ($T_{\rm eff,phot}$, log $g$) for the single 1200G-based field, f123_1, were derived following the procedure outlined by Kirby, Guhathakurta, and Sneden (2008), assuming an identical distance modulus and 14 Gyr isochrones from a combination of model sets (Girardi et al. 2002; Demarque et al. 2004; VandenBerg, Bergbusch, and Dowler 2006). As described in detail by Escala et al. (2019) and summarized in Section 5.5, $T_{\rm eff,phot}$ and log $g$ are used as constraints in measuring $T_{\rm eff}$, [Fe/H], and [$\alpha$/Fe] from spectra of individual stars, where these measurements are insensitive to the employed isochrone models and assumed stellar age.

## 5.3 Chemical Abundance Analysis

We use spectral synthesis of low- and medium-resolution stellar spectroscopy to measure stellar parameters ($T_{\rm eff}$) and abundances ([Fe/H]and [$\alpha$/Fe]) from our deep observations of M31 RGB stars. For a detailed description of the low- and medium-resolution spectral synthesis methods, see Escala et al. (2020a) and Escala et al. (2019) and Kirby, Guhathakurta, and Sneden (2008), respectively. The low- and medium-resolution spectral synthesis procedures are nearly identical in principle, excepting differences in the continuum normalization given the differing wavelength coverage between the low- and medium-resolution spectra (~4500−9100 vs. 6300−9100 Å). For 1200G spectra, the continuum is determined using "continuum regions" defined by Kirby, Guhathakurta, and Sneden (2008), whereas such regions would be unilaterally defined for 600ZD spectra owing to the high density of absorption features toward the blue wavelengths. Chemical abundances ([Fe/H] and [$\alpha$/Fe]) for individual stars measured using each technique are generally consistent within the uncertainties (Escala et al. 2020a).

Prior to the chemical abundance analysis, the radial velocity of each star is measured via cross-correlation with empirical templates (Simon and Geha 2007; Kirby et al. 2015; Escala et al. 2020a) observed in the relevant science configuration (Section 5.2). Systematic radial velocity errors of 5.6 km s$^{-1}$ (Collins et al. 2011) and 1.49 km s$^{-1}$ (Kirby et al. 2015) from repeat measurements of identical stars are added in quadrature to the random component of the error for observations taken with the 600ZD and 1200G gratings, respectively. The spectral resolution is empirically determined as a function of wavelength using the width of sky lines (Kirby, Guhathakurta, and Sneden 2008), and in the case of the bluer 600ZD spectra, arc lines from calibration lamps (Escala et al. 2020a; K. McKinnon et al., in preparation). The observed spectrum is then corrected for telluric absorption using a template of a hot star observed in the relevant science configuration (Kirby, Guhathakurta, and Sneden 2008; Escala et al. 2019), shifted into the rest frame based on the measured radial velocity, and an initial continuum normalization is performed.

We measured the spectroscopic effective temperature, $T_{\mathrm{eff}}$, informed by photometric constraints, and fixed the surface gravity, log $g$, to the photometric value (Section 5.2). We simultaneously measured [Fe/H] and [$\alpha$/Fe] from regions of the spectrum sensitive to Fe and $\alpha$-elements (Mg, Si, Ca – with the addition of Ti for medium-resolution spectra), respectively, by comparing to a grid of synthetic spectra degraded to the resolution of the applicable DEIMOS grating (600ZD or 1200G) using Levenberg-Marquardt $\chi^2$ minimization. The grids of synthetic spectra utilized were generated for 4100−6300 Å and 6300−9100 Å, respectively, by Escala et al. (2019) and Kirby, Guhathakurta, and Sneden (2008). The continuum determination is refined throughout this process, where $T_{\mathrm{eff}}$, [Fe/H], and [$\alpha$/Fe] are measured iteratively until $T_{\mathrm{eff}}$ changed by less than 1 K and [Fe/H] and [$\alpha$/Fe] each changed by less than 0.001. Finally, systematic errors on the abundances are added in quadrature to the random component of the error from the fitting procedure. For 600ZD-based abundance measurements, $\delta([\mathrm{Fe/H}])_{\mathrm{sys}} = 0.130$ and $\delta([\alpha/\mathrm{Fe}])_{\mathrm{sys}} = 0.107$, whereas for 1200G-based measurements, $\delta([\mathrm{Fe/H}])_{\mathrm{sys}} = 0.101$ and $\delta([\alpha/\mathrm{Fe}])_{\mathrm{sys}} = 0.084$ (Gilbert et al. 2019a).

## 5.4 Membership Determination

Separating M31 RGB stars from the intervening foreground of MW dwarf stars has served as one of the primary challenges for spectroscopic studies of individuals stars in M31. The colors and heliocentric radial velocity distributions of MW and M31 stars exhibit significant overlap (e.g., Gilbert et al. 2006), thus the difficulty in disentangling the two populations when the distances to such faint stars are unknown. Early spectroscopic studies of M31 employed simple radial velocity cuts to exclude MW stars (e.g., Reitzel and Guhathakurta 2002; Ibata et al. 2005; Chapman et al. 2006), resulting in kinematically biased populations of M31 RGB stars and relatively uncontaminated, albeit incomplete samples of M31 stars. Gilbert et al. (2006) performed the first rigorous, probabilistic membership determination in M31 using various diagnostics, including (1) heliocentric radial velocity, (2) the strength of the surface-gravity sensitive Na I $\lambda\lambda$8190 doublet, (3) CMD position, and (4) the discrepancy between photometric and calcium triplet based metallicity estimates as a distance indicator.

Given the variety of photometric filters utilized, Gilbert et al.'s method cannot be uniformly applied to all spectroscopic fields analyzed in this chapter. For inner halo fields ($r_{\rm proj} \lesssim 30$ kpc), Escala et al. (2020a) illustrated that a binary membership determination using the aforementioned diagnostics is sufficient to recover the majority of stars classified as likely M31 RGB stars by the more sophisticated method of Gilbert et al. (2006) with minimal MW contamination. However, this binary determination excludes all stars with $v_{\rm helio} > -150$ km s$^{-1}$, where some of these stars may be M31 members at the positive tail of the stellar halo velocity distribution. It also does not allow us to assign a degree of certainty to our membership determination for each star. Thus, we used Bayesian inference to assign a membership probability to each observed star with a successful radial velocity determination (Section 5.3).

**Membership Probability Model**

We evaluated the probability of M31 membership for all stars with successful velocity measurements based on (1) a parameterization of color ($X_{\rm CMD}$), (2) the strength of the Na I absorption line doublet at $\lambda\lambda$8190 (EW$_{\rm Na}$), (3) heliocentric radial velocity (Section 5.3; $v_{\rm helio}$), and (4) an estimate of the spectroscopic metallicity based on the strength of the calcium triplet ([Fe/H]$_{\rm CaT}$).

Thus, according to Bayes' theorem, the posterior probability that a star labeled by an index $j$ is an M31 member—given independent measurements $\vec{x_j}$ = (EW$_{{\rm Na},j}$, X$_{{\rm CMD},j}$, $v_{\ {\rm helio},j}$, [Fe/H]$_{{\rm CaT},j}$) with uncertainties $\delta\vec{x_j}$—is proportional to,

$$P({\rm M31}|\vec{x_j}, \delta\vec{x_j}) \propto P({\rm M31})_j \times P(\vec{x_j}, \delta\vec{x_j}|{\rm M31}), \tag{5.1}$$

where $P$(M31) is the prior probability that a star observed on a given DEIMOS slitmask is a member of M31, and $P(\vec{x_j}, \delta\vec{x_j}|{\rm M31})$ is the likelihood of measuring a given set of membership diagnostics, $(\vec{x_j}, \delta\vec{x_j})$, assuming the star is a M31 member. Analogously, we can also construct $P({\rm MW}|\vec{x_j}, \delta\vec{x_j})$, the posterior probability that a star belongs to the MW foreground population given a set of diagnostic measurements, where the sum of $P({\rm MW}|\vec{x_j}, \delta\vec{x_j})$ and $P({\rm M31}|\vec{x_j}, \delta\vec{x_j})$ is unity.

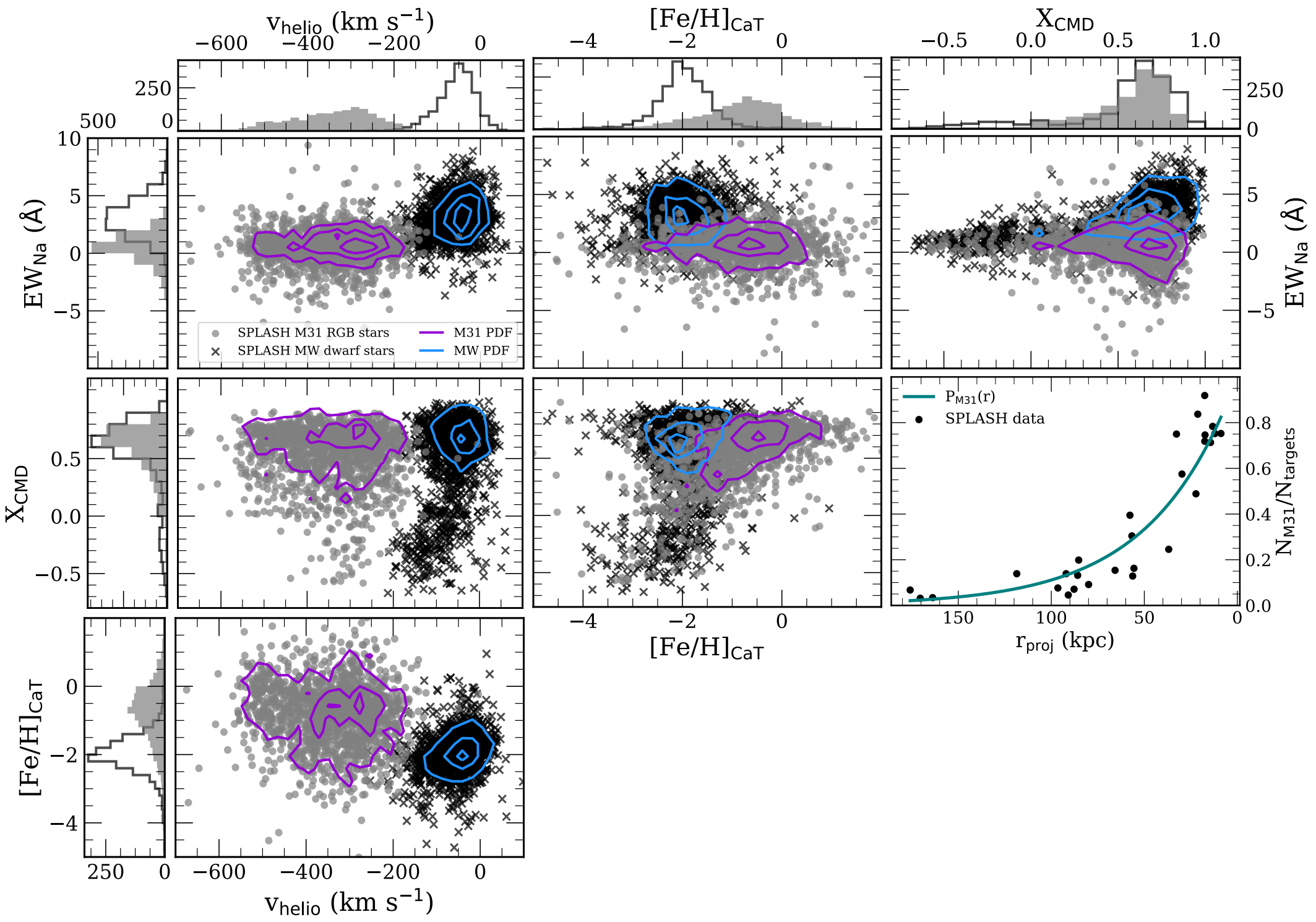

Figure 5.3: Properties of spectroscopically confirmed M31 (*grey circles, purple contours*) and MW (*black crosses, blue contours*) member stars from the SPLASH (Guhathakurta et al. 2005; Gilbert et al. 2006) survey. The contour levels correspond to $16^{th}$, $50^{th}$, and $84^{th}$ percentiles of the distributions. Histograms of star counts are also shown for both M31 and MW members. For clarity, we omit showing uncertainties in each parameter. The middle right panel illustrates the number of spectroscopically confirmed M31 members (Gilbert et al. 2006) relative to the number of targets with successful radial velocity measurements in a given field as a function of projected radius, where we have fit the relationship with an exponential distribution. We used heliocentric radial velocity ($v_{\rm helio}$), a parameterization of color ($X_{\rm CMD}$), Na I $\lambda\lambda$8190 equivalent width ($\rm EW_{Na}$), and a calcium triplet metallicity estimate ($\rm [Fe/H]_{CaT}$) as diagnostics to determine membership (Section 5.4, Section 5.4) for our sample of stars with successful radial velocity measurements.

**Measuring Membership Diagnostics**

We described the CMD position of each star by the parameter $X_{\mathrm{CMD}}$, which is analogous to photometric metallicity ($[\mathrm{Fe/H}]_{\mathrm{phot}}$). The advantage of using $X_{\mathrm{CMD}}$ rather than color and magnitude is that we can place all of our stars on the same scale, despite the diversity of the photometric filter sets (Section 5.2). Assuming 12.6 Gyr isochrones and a distance modulus of 24.47, we defined $X_{\mathrm{CMD}} = 0$ as the color of the most metal-poor PARSEC (Marigo et al. 2017) isochrone ([Fe/H] = −2.2) in the relevant photometric filter and $X_{\mathrm{CMD}} = 1$ as the most metal-rich PARSEC isochrone ([Fe/H] = +0.5). Then, we used linear interpolation to map the color and magnitude of a star to a value of $X_{\mathrm{CMD}}$. The uncertainty on $X_{\mathrm{CMD}}$ was derived from the photometric errors by using a Monte Carlo procedure. This normalization provides the advantage of easily identifying stars that are bluer than the most metal-poor isochrone at a fixed stellar age by negative values of $X_{\mathrm{CMD}}$, where stars with $X_{\mathrm{CMD}} < 0$ are $\gtrsim$10 times more likely to belong to the MW than M31 (Gilbert et al. 2006). In the sample being evaluated for membership, we classified stars with $(X_{\mathrm{CMD}} + \delta \mathrm{X}_{\mathrm{CMD}}) < 0$ as MW dwarf stars.

We measured $\mathrm{EW}_{\mathrm{Na}}$ by summing the area under the best-fit Gaussian line profiles fit to the observed spectrum between 8178−8190, 8189−8200 Å with central wavelengths of 8183, 8195 Å using least-squares minimization. This parameter is sensitive to temperature and surface gravity (Schiavon et al. 1997), thus functioning as a discriminant between MW dwarf stars and M31 RGB stars. By including both $X_{\mathrm{CMD}}$ and $[\mathrm{Fe/H}]_{\mathrm{CaT}}$ as independent diagnostics in our model, we can additionally distinguish between foreground stars at unknown distances and distant giant stars. Given that our assumed distance modulus is appropriate only for stars at the distance of M31, $[\mathrm{Fe/H}]_{\mathrm{phot}}$, and analogously, $X_{\mathrm{CMD}}$, measurements are fundamentally incorrect for MW stars. Therefore, the two stellar populations will appear distinct in $X_{\mathrm{CMD}}$ versus $[\mathrm{Fe/H}]_{\mathrm{CaT}}$ space. In order to compute $[\mathrm{Fe/H}]_{\mathrm{CaT}}$, we similarly fit Gaussian profiles to 15 Å wide windows centered on Ca II absorption lines at 8498, 8542, and 8662 Å. Then, we calculated a total equivalent width for the calcium triplet from a linear combination of the individual equivalent widths,

$$\Sigma\mathrm{Ca} = 0.5 \times \mathrm{EW}_{\lambda 8498} + 1.0 \times \mathrm{EW}_{\lambda 8542} + 0.6 \times \mathrm{EW}_{\lambda 8662}, \tag{5.2}$$

following Rutledge, Hesser, and Stetson (1997). In addition to $\Sigma$Ca, the calibration to determine $[\mathrm{Fe/H}]_{\mathrm{CaT}}$ depends on stellar luminosity,

$$[\mathrm{Fe/H}]_{\mathrm{CaT}} = -2.66 + 0.42 \times \Sigma\mathrm{Ca} + 0.27 \times (V - V_{\mathrm{HB}}), \tag{5.3}$$

where $V - V_{\mathrm{HB}}$ is the Johnson-Cousins $V$-band apparent magnitude above the horizontal branch, assuming that $V_{\mathrm{HB}} = 25.17$ for M31 (Holland, Fahlman, and Richer 1996). For fields with CFHT MegaCam $g'$ and $i'$ band photometry (Section 5.2), we approximated $V - V_{\mathrm{HB}}$ by $g' - g'_{\mathrm{HB}}$ using an empirical transformation between SDSS and Johnson-Cousins photometry (Jordi, Grebel, and Ammon 2006),

$$g - V = 0.56 \times (g - r + 0.23)/1.05 + 0.12. \tag{5.4}$$

Assuming that $g - r = 0.6$ and $g'_{\mathrm{HB}} \approx g_{\mathrm{HB}}$, we obtained $g_{\mathrm{HB}} = V_{\mathrm{HB}} + (g - V) = 25.73$ for stars in M31.

**Prior Probability of M31 Membership**

The probability of a given star observed on a DEIMOS slitmask belonging to M31, $P$(M31), increases with decreasing projected radial distance from the center of M31, owing to the increase in M31's stellar surface density. The trend of increasing probability of M31 membership with decreasing projected radius is augmented compared to expectations from M31's surface brightness profile (Courteau et al. 2011; Gilbert et al. 2012) by our photometric pre-selection of DEIMOS spectroscopic targets. These selection criteria are designed to include M31 members and exclude MW foreground stars. The photometry of the targets spans the magnitude range characteristic of M31 RGB stars ($20 < I_0 < 22.5$). We also considered narrow-band, Washington DDO51 photometry in the fields a0_1 and a3. The DDO51 filter isolates the Mg b triplet, which acts as discriminant between MW dwarf and M31 giant stars due to its sensitivity to surface gravity. We expect that an enhancement in the probability of observing an M31 RGB star owing to magnitude cuts is fairly uniform at a factor of $\lesssim 2$ within $r_{\mathrm{proj}} \lesssim 30$ kpc (Gilbert et al. 2012), where the majority of our targets are located. For fields toward the outer halo, such as a0_1 and a3, DDO51-selection bias increases the likelihood of observing an M31 RGB star by a factor of 3–4 (Gilbert et al. 2012).

Thus, we parameterized P(M31) empirically using the ratio of the number of secure M31 RGB stars (Gilbert et al. 2006), $N_{\mathrm{M31}}$, to the number of targets with successful radial velocity measurements, $N_{\mathrm{targets}}$, from the Spectroscopic and Phomtometric Landscape of Andromeda's Stellar Halo (SPLASH; Guhathakurta et al. 2005;

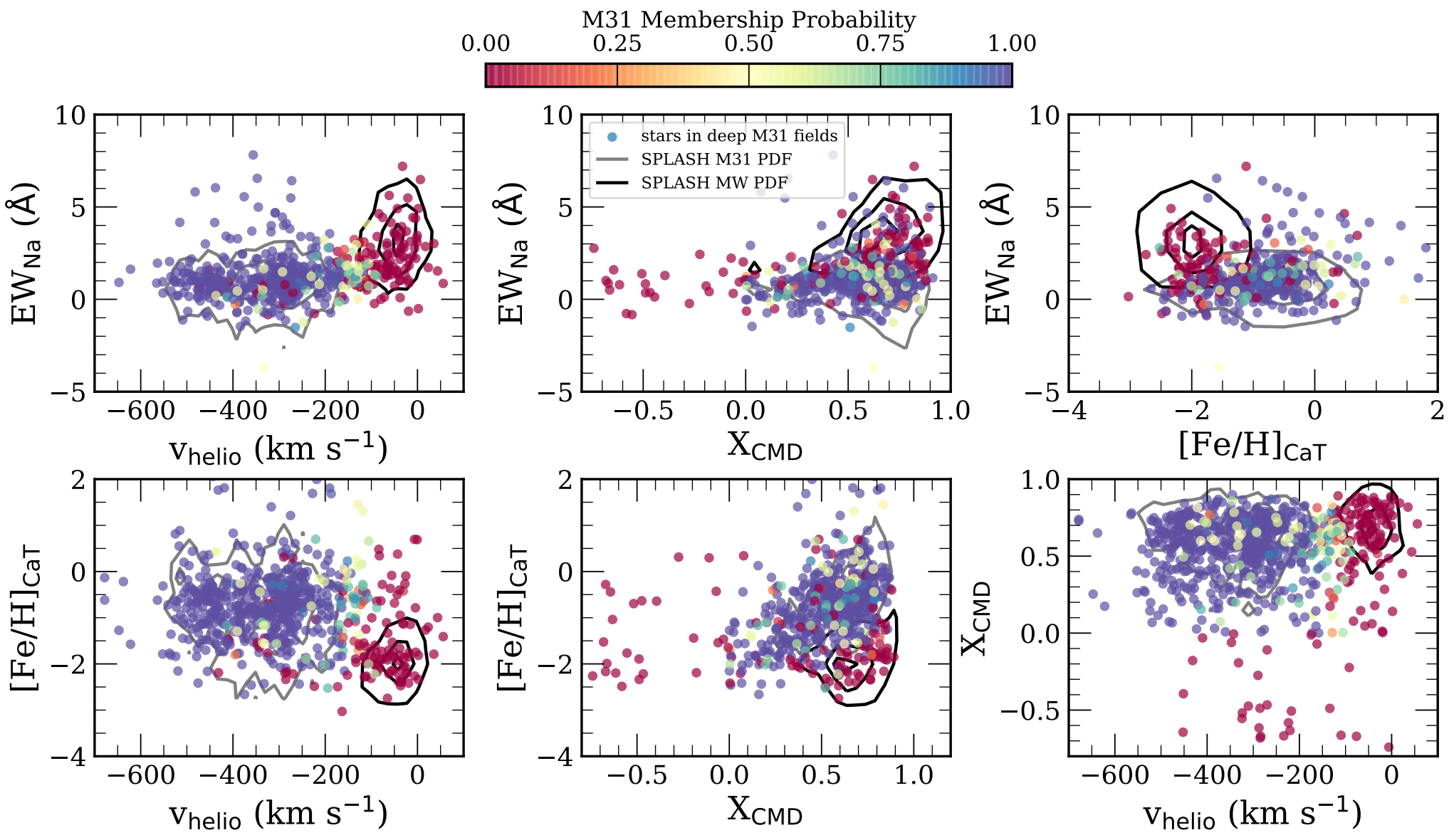

Figure 5.4: Properties used as membership diagnostics ($v_{\rm helio}$, EW$_{\rm Na}$, $X_{\rm CMD}$, [Fe/H]$_{\rm CaT}$) for stars in our spectroscopic fields (Table 5.1) with successful radial velocity measurements (Section 5.3). We also include previously observed fields H, S, D, and f130_2 (Figure 5.1; Escala et al. 2020a). For clarity, we omit showing measurement uncertainties. Each point is color-coded by its probability of belonging to M31. The SPLASH distributions utilized in our probability model (Section 5.4) are underlaid as contours. For field D, we used a distinct velocity criterion owing to the presence of M31's disk at $\sim -130$ km s$^{-1}$(Escala et al. 2020a). We identified 329 (221) secure M31 RGB stars and 91 (75) secure MW dwarf stars across the spectroscopic fields presented in this (prior) chapter.

Gilbert et al. 2006) survey. The SPLASH survey consists of tens of thousands of shallow (~1 hour exposures) DEIMOS spectra of stars along the line of sight toward M31's halo, disk, and satellite dwarf galaxies (e.g., Kalirai et al. 2010; Tollerud et al. 2012; Dorman et al. 2012; Dorman et al. 2015; Gilbert et al. 2012; Gilbert et al. 2014; Gilbert et al. 2018). In addition to $N_{\rm M31}/N_{\rm targets}$ as a function of radius, Figure 5.3 presents measurements of EW$_{\rm Na}$, $X_{\rm CMD}$, $v_{\rm helio}$, and [Fe/H]$_{\rm CaT}$ for 1,521 secure M31 members and 1,835 secure MW members across 29 spectroscopic fields in M31's stellar halo. We controlled for differences in methodology between this chapter and SPLASH by re-determining $X_{\rm CMD}$ homogeneously for the SPLASH data from the original Johnson-Cousins photometry and measuring [Fe/H]$_{\rm CaT}$ for our sample based on the same calibration (Rutledge, Hesser, and Stetson 1997) used in SPLASH.[1]

[1]Our equivalent width measurement procedure differs from that utilized in the SPLASH survey.

Based on this data, we approximated P(M31) by an exponential distribution,

$$P_{\mathrm{M31}}(r_j) = \exp(-r_j/r_p), \tag{5.5}$$

where $r_j$ is the projected radius of a star from M31's galactic center and $r_p$ = 45.5 kpc. Figure 5.3 includes $P_{\mathrm{M31}}(r)$ overlaid on the SPLASH survey data.

**Likelihood of M31 Membership**

Given the wealth of existing information on the properties of M31 RGB stars (and the MW foreground dwarf stars characteristic of our selection function) from the SPLASH survey, we assigned membership likelihoods to individual stars that were informed by this extensive data set. We described the likelihood that a star, $j$, with unknown membership belongs to M31 given its diagnostic measurements and uncertainties, $(\vec{x}_j, \delta\vec{x}_j)$, as,

$$P(\vec{x}_j, \delta\vec{x}_j|\mathrm{M31}) = \frac{1}{N_i}\sum_{i=1}^{N_i} P(\vec{x}_j|\vec{\theta}_i, \delta\vec{\theta}_i), \tag{5.6}$$

where $(\vec{\theta}_i, \delta\vec{\theta}_i)$ is a set of four diagnostic measurements and uncertainties for a star, $i$, from the SPLASH survey that is a secure M31 member. The total number of secure SPLASH member stars, $N_i$, equals 1,521 (1,835) for M31 (the MW). The likelihood that a star, $j$, with unknown membership belongs to the MW, $P(\vec{x}_j, \delta\vec{x}_j|\mathrm{MW})$, is defined analogously. Assuming normally distributed uncertainties, the log likelihood that a star, $j$, is a member of either M31 or the MW given a single set of SPLASH measurements, $i$, is a non-parametric, $N_k$-dimensional Gaussian distribution,

$$\ln P(\vec{x}_j, \delta\vec{x}_j|\vec{\theta}_i, \delta\vec{\theta}_i) = -\frac{N_k}{2}\ln 2\pi - \frac{1}{2}\sum_{k=1}^{N_k}\ln(\delta x_{j,k}^2 + \delta\theta_{i,k}^2) - \frac{1}{2}\sum_{k=1}^{N_k}\frac{(x_{j,k} - \theta_{i,k})^2}{\delta x_{j,k}^2 + \delta\theta_{i,k}^2}, \tag{5.7}$$

where $k$ corresponds to a given membership diagnostic ($\mathrm{EW_{Na}}$, $X_{\mathrm{CMD}}$, $v_{\mathrm{helio}}$, or $\mathrm{[Fe/H]_{CaT}}$) and $N_k$ is the number of diagnostics utilized for a given star. For

As opposed to summing the flux decrement in a window centered on a given absorption feature, we performed Gaussian fits. However, we do not expect this difference in methodology to significantly affect the usability of the SPLASH data to construct the likelihood (Section 5.4), given that our measured $\mathrm{EW_{Na}}$ and $\mathrm{[Fe/H]_{CaT}}$ distributions are consistent with SPLASH (Figure 5.4).

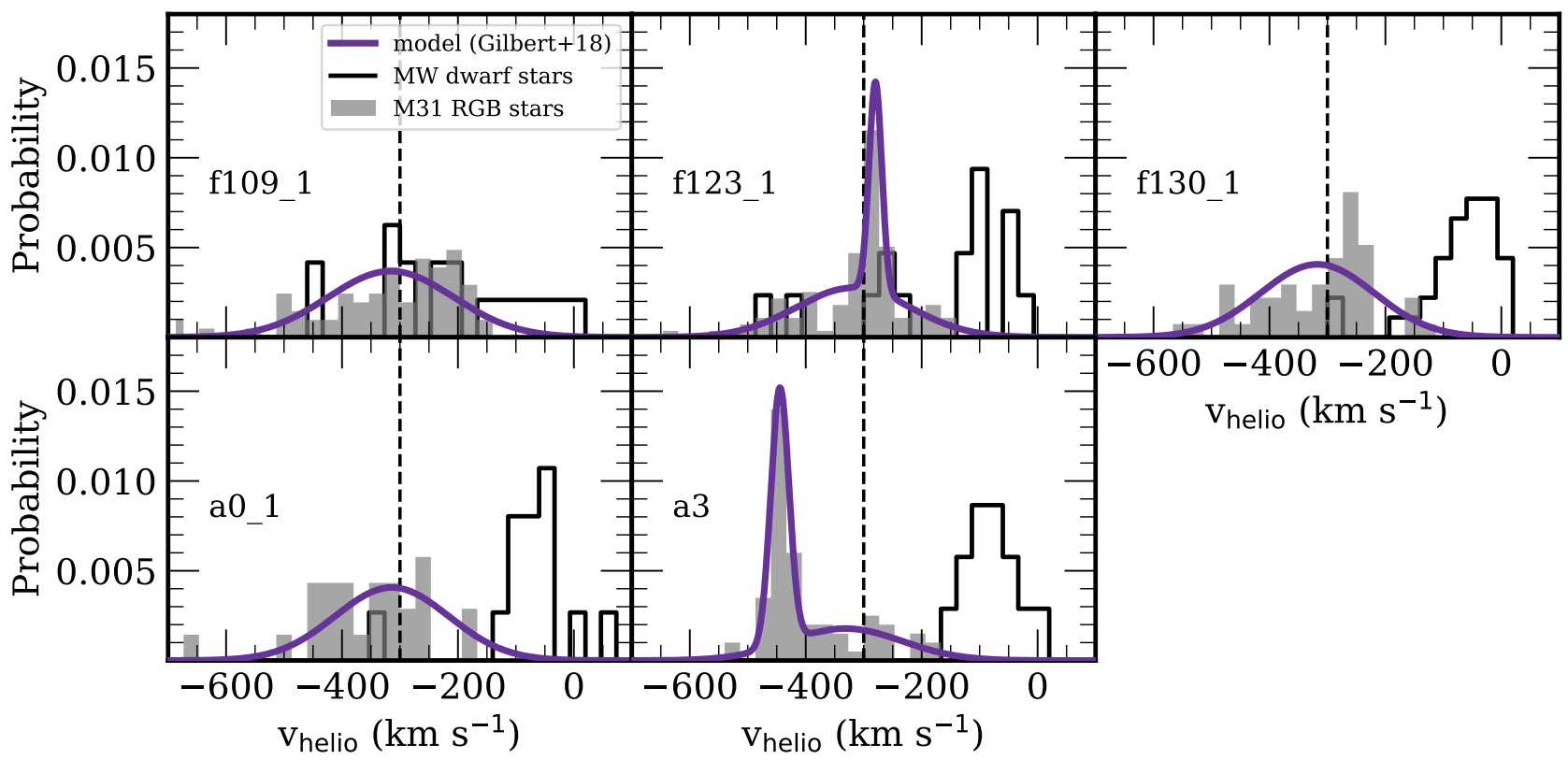

Figure 5.5: Heliocentric radial velocity distributions for stars with successful velocity measurements (Section 5.3) in each of the spectroscopic fields (Table 5.1). Grey histograms represent distributions for likely M31 RGB stars (Section 5.4), whereas black outlined histograms represent likely MW foreground dwarf stars. The systemic velocity of M31 ($v_{\rm sys}$ = −300 km s$^{-1}$) is indicated by a dashed vertical line. The adopted velocity model for M31 RGB stars in each field (Section 5.4; Gilbert et al. 2018) is overplotted as a purple curve. Halo stars are present in each field, where they are distributed in a kinematically hot component centered near the systemic velocity. The velocity distributions of the 9, 23, and 31 kpc fields (f109_1, f130_1, a0_1) are fully consistent with a "smooth" stellar halo. The Southeast shelf, a tidal feature potentially originating from the GSS progenitor (Fardal et al. 2007; Gilbert et al. 2007), is evident in the 18 kpc halo field (f123_1), whereas the substructure in the 33 kpc fields (a3_1, a3_2) correspond to the GSS.

some stars in our sample, we were unable to measure EW$_{\rm Na}$ and/or [Fe/H]$_{\rm CaT}$ as a consequence of factors such as weak absorption, low S/N, or convergence failure in the Gaussian fit. For such stars, we excluded EW$_{\rm Na}$ and/or [Fe/H]$_{\rm CaT}$ as a diagnostic, such that $N_k$ = 2−3.

Finally, we computed the probability that a star is an M31 RGB candidate, as opposed to a MW dwarf candidate, from the odds ratio of the posterior probabilities (Eq. 5.1),

$$p_{\rm M31,\mathit{j}} = \frac{P({\rm M31}|\vec{x}_j, \delta\vec{x}_j)/P({\rm MW}|\vec{x}_j, \delta\vec{x}_j)}{1 + P({\rm M31}|\vec{x}_j, \delta\vec{x}_j)/P({\rm MW}|\vec{x}_j, \delta\vec{x}_j)}, \tag{5.8}$$

where the proportionality factor in Eq. 5.1 is equivalent for $P({\rm M31}|\vec{x}_j, \delta\vec{x}_j)$ and $P({\rm MW}|\vec{x}_j, \delta\vec{x}_j)$.

**Results of Membership Determination**

Figure 5.4 summarizes our membership determination for 426 total stars with successful radial velocity measurements across the six spectroscopic fields first presented in this chapter. The probability distribution is strongly bimodal, where the majority of stars are either secure ($p_j \gtrsim 0.75$) M31 RGB ($N_{\rm M31}$ = 329) or MW dwarf ($N_{\rm MW}$ = 91) candidates, excepting 6 stars with intermediate properties (0.5 $< p_j \lesssim$ 0.75).[2]

Figure 5.4 also includes a homogeneously re-evaluated membership determination for fields H, S, D, and f130_2 (Table 5.1; Escala et al. 2020a), which we further analyze in this chapter. Across these four fields, 346 stars have successful radial velocity measurements, 221 (75) of which are classified as secure M31 (MW) stars. In total, 50 stars in these fields have intermediate properties, most of which (48) are located in field D. Owing to the presence of M31's northeastern disk at MW-like line-of-sight velocity ($v_{\rm helio} \sim -130$ km s$^{-1}$; Escala et al. 2020a), we calculated M31 membership probabilities in D without the use of radial velocity as a diagnostic. Then, we classified all stars with ($v_{\rm helio} - \delta v_{\rm helio}$) $> -100$ km s$^{-1}$ as MW contaminants, as in Escala et al. (2020a). In order to maximize our sample size across all spectroscopic fields, we considered stars that are more likely to belong to M31 than the MW ($p_j > 0.5$) to be M31 members in the following analysis. For stars in common between our dataset and SPLASH, we recovered 91.7% of stars classified as M31 RGB stars by Gilbert et al. (2006), where we used an equivalent definition of membership ($\langle L_{{\rm splash},j} \rangle > 0$). The excess MW contamination is 0.30%, in addition to the expected 2-5% from Gilbert et al.'s method. The 8.3% discrepancy results from stars at MW-like heliocentric velocities that we conservatively classified as MW stars, where these stars are considered M31 RGB stars in SPLASH.

**Kinematics of M31 RGB Stars**

Figure 5.5 illustrates the heliocentric radial velocity distributions for all stars with successful measurements (Section 5.5), including both M31 RGB stars and MW dwarf stars (Section 5.4), across the spectroscopic fields. We also show the adopted Gaussian mixture models (Gilbert et al. 2018) describing the velocity distribution for each field, which were computed using over 5,000 spectroscopically confirmed

[2]Our M31 membership yield is high, given that we had prior knowledge of the velocities of individual stars in each field from existing ~1 hour DEIMOS observations (Section 5.2). When designing our slitmasks for 5+ hour exposures, we prioritized targets known to have a high likelihood M31 membership based on this information.

M31 RGB stars across 50 fields in M31's stellar halo. Gilbert et al. omitted radial velocity as a membership diagnostic (Section 5.4) in their analysis and simultaneously fit for contributions from M31 and MW components to obtain kinematically unbiased models for each field. Table 5.2 presents the parameters characterizing the velocity model for each field, where we assumed 50$^{\text{th}}$ percentile values of Gilbert et al.'s marginalized posterior probability distribution functions. For the stellar halo components, we transformed the mean velocity from the Galactocentric to heliocentric frame using the median right ascension and declination of all stars in a given field (Gilbert et al. 2018).

We confirmed that the observed velocity distribution for each field is consistent with its velocity model using a two-sided Kolmogorov-Smirnov test. As discussed in Section 5.2, f109_1, f130_1, and a0_1 probe the "smooth" stellar halo of M31 with no detected substructure, whereas fields f123_1 and a3 show clear evidence of substructure known to be associated with the Southeast shelf (Fardal et al. 2007; Gilbert et al. 2007; Gilbert et al. 2019a; Escala et al. 2020a) and the GSS. Owing to the spatial proximity (Figure 5.1) and kinematical similarity (Gilbert et al. 2007; Gilbert et al. 2009; Gilbert et al. 2018) between fields f130_1 (this chapter) and f130_2 (Escala et al. 2020a; Escala et al. 2019) and fields a3 (this chapter), we consider them together in our subsequent abundance analysis (Section 5.5).

**Substructure Probability**

Based on the velocity model for each field, we assigned a probability of belonging to substructure in M31's stellar halo, $p_{\rm sub}$, to every star identified as a likely RGB candidate ($p_{\rm M31} > 0.5$). For smooth stellar halo fields, $p_{\rm sub} = 0$. For fields with substructure, we computed $p_{\rm sub}$ using an equation analogous to Eq. 5.8, where the Bayesian odds ratio is substituted with the relative likelihood that a M31 RGB star belongs to substructure versus the halo,

$$\mathcal{L}_{{\rm sub},j} = \frac{f_{\rm sub}\mathcal{N}(v_j|\mu_{\rm sub},\sigma^2_{\rm sub})}{f_{\rm halo}\mathcal{N}(v_j|\mu_{\rm halo},\sigma^2_{\rm halo})}, \tag{5.9}$$

where $v_j$ is the heliocentric velocity of an individual star and $\mu$, $\sigma$, and $f$ are the mean, standard deviation, and fractional contribution of the substructure or halo component in the Gaussian mixture describing the velocity distribution for a given field. In contrast, Escala et al. (2020a) calculated $p_{\rm sub}$ based on the full posterior distributions, as opposed to 50$^{\text{th}}$ percentiles alone, of their velocity models for fields with kinematical substructure. Given that we included abundance measurements of M31 RGB stars from Escala et al. (2020a) in this chapter, we re-calculated $p_{\rm sub}$ for such fields (H, S, and D; Figure 5.1) based on their 50$^{\rm th}$ percentiles, as above. In the following abundance analysis (Section 5.5), we incorporated $p_{\rm sub}$ as a weight when determining the chemical properties of the stellar halo.

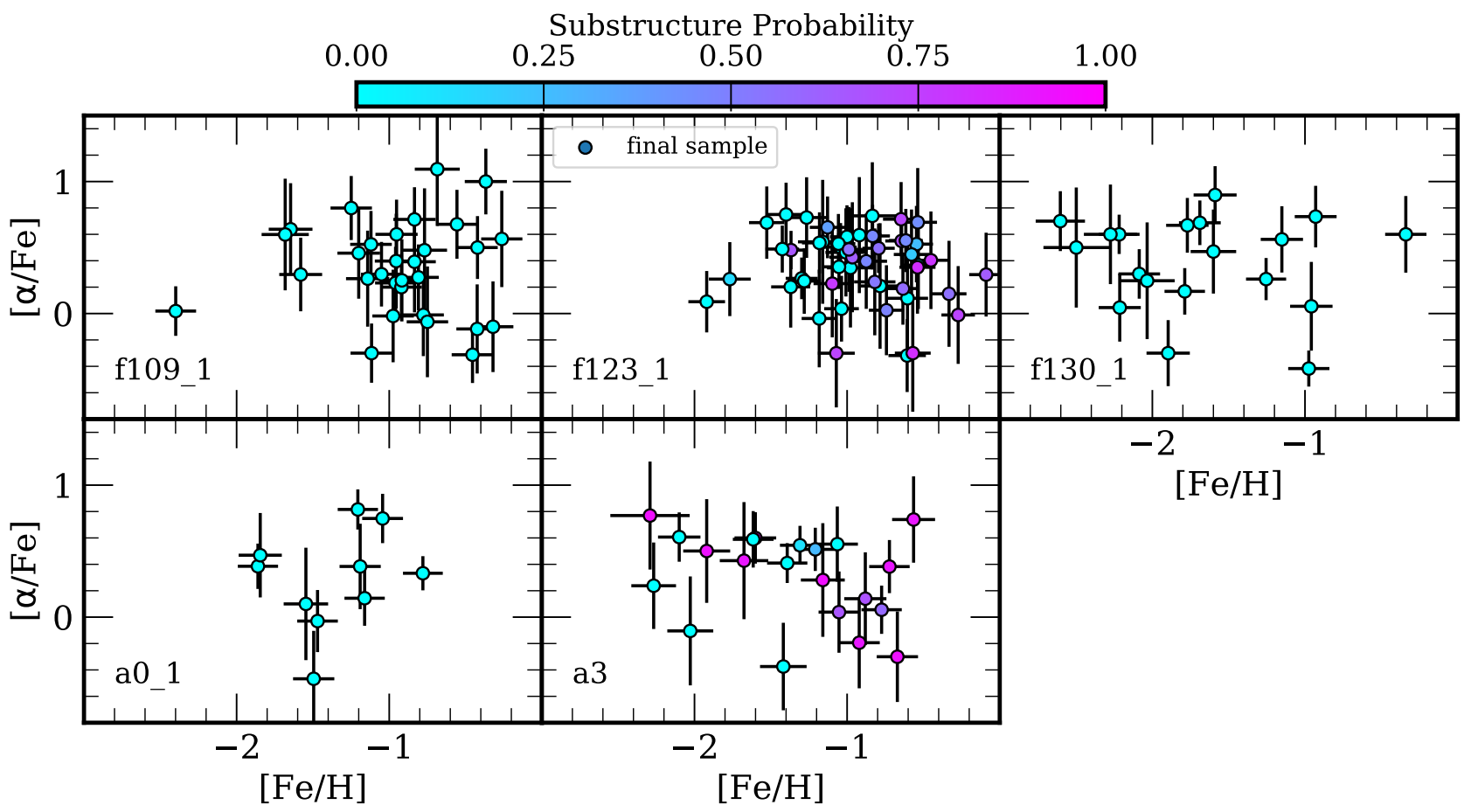

Figure 5.6: [$\alpha$/Fe] versus [Fe/H] for M31 RGB stars in the spectroscopic fields (Table 5.1), color-coded according to the probability of belonging to substructure (Eq. 5.9; Section 5.4), as in Figure 5.2. The final sample selection criteria are described in Section 5.5. Stars belonging to the stellar halo are $\alpha$-enhanced on average, but span a wide range of [$\alpha$/Fe]. Most notably, $\alpha$-poor stars ([$\alpha$/Fe] $\lesssim 0$) are consistently present in the stellar halo of M31.

## 5.5 The Chemical Properties of the Inner Stellar Halo

We measured [Fe/H] and [$\alpha$/Fe] for 129 M31 RGB stars across six spectroscopic fields spanning 9 – 33 kpc in the stellar halo. We measured [Fe/H] for 80 additional RGB stars for which we could not measure [$\alpha$/Fe]. In combination with measurements from previous chapter by our collaboration (Escala et al. 2020a; Gilbert et al. 2019a; K. M. Gilbert et al. 2020, accepted), we have increased the sample size of individual [$\alpha$/Fe] and [Fe/H] measurements in M31 to 230 RGB stars.

Figure 5.5 shows [Fe/H] and [$\alpha$/Fe] measurements for each field, including TiO stars that have been discarded from the final sample (Section 5.5), where we have color-coded each star according to its probability of belonging to substructure (Eq. 5.9; Section 5.4). Table 5.2 summarizes the chemical properties for each kinematical component present in a given field, including previously published inner halo fields with abundance measurements (Escala et al. 2020a; Escala et al. 2019; Gilbert et al. 2019a). We calculated each average chemical property from a bootstrap resampling of $10^4$ draws of the final sample for each field, including weighting by the inverse square of the measurement uncertainty and the probability that a star belongs to a given kinematical component (Section 5.4).

Hereafter, we predominantly restricted our abundance analysis to RGB stars (Section 5.4) from all six fields that are likely to be dynamically associated with the kinematically hot stellar halo of M31. Additionally, we incorporated likely M31 halo stars from inner halo fields with existing abundance measurements (Escala et al. 2020a; Gilbert et al. 2019a) into our final sample. We will analyze the chemical composition of RGB stars likely belonging to substructure, such as the Southeast shelf (f123_1) and outer GSS (a3), in a companion study (I. Escala et al., 2020, in preparation). A catalog of stellar parameters and abundance measurements for M31 RGB stars in the six spectroscopic fields is presented in Appendix A.

**Sample Selection and Potential Biases**

We vetted our final sample to consist only of *reliable* abundance measurements (Section 5.3) for M31 RGB stars. Similar to Escala et al. (2019) and Escala et al. (2020a), we restrict our analysis to M31 RGB stars with $\delta(T_{\rm eff}) < 200$ K, $\delta$([Fe/H]) $< 0.5$, $\delta([\alpha/\text{Fe}]) < 0.5$, and well-constrained parameter estimates based on the $5\sigma$ $\chi^2$ contours for all fitted parameters ($T_{\rm eff}$, [Fe/H], and [$\alpha$/Fe]). We also require that convergence is achieved in each of the measured parameters (Section 5.3). Unreliable abundance measurements often result from an insufficient signal-to-noise (S/N) ratio, translating to an effective S/N threshold of $\gtrsim 8$ Å$^{-1}$ for robust measurements of [Fe/H] and [$\alpha$/Fe]. Such S/N limitations result in a bias in our final sample against metal-poor stars with low S/N spectra, but do not affect the [$\alpha$/Fe] distributions. This bias is negligible for our innermost halo fields (f109_1, f123_1), whereas it is on the order of 0.10−0.15 dex for our remaining fields (f130_1, a0_1, a3).

We also manually inspected spectra to exclude stars with clear signatures of strong molecular TiO absorption in the wavelength range 7055−7245 Å from our final sample. It is unclear whether abundances measured from TiO stars are accurate owing to the lack of (1) the inclusion of TiO in our synthetic spectral modeling (Kirby, Guhathakurta, and Sneden 2008; Escala et al. 2019) and (2) an appropriate calibration sample. For stars with successful radial velocity measurements, 31.5%, 16.7%, 34.5%, 32.5%, 30.4%, and 31.3% of stars in fields f109_1, f123_1, f130_1, a0_1, a3_1, and a3_2 have clear evidence of TiO in their spectra. To be conservative,

Table 5.2: Kinematical[a] and Chemical[b] Properties of M31 Fields

| Field | $r_{\rm proj}$ (kpc) | Comp. | $\mu$ (km s$^{-1}$) | $\sigma$ (km s$^{-1}$) | $f$ | $\langle$[Fe/H]$\rangle$ | $\sigma$([Fe/H]) | $\langle[\alpha$/Fe]$\rangle$ | $\sigma([\alpha$/Fe]) | $N_{[\alpha/\rm Fe]}$ |
|---|---|---|---|---|---|---|---|---|---|---|
| f109_1 | 9 | Halo | −315.7 | 108.2 | 1.0 | $-0.93^{+0.08}_{-0.09}$ | 0.45 ± 0.09 | 0.32 ± 0.08 | $0.36^{+0.04}_{-0.05}$ | 30 |
| H | 12 | Halo | −315.1 | 108.2 | 0.44 | −1.30 ± 0.11 | $0.45^{+0.07}_{-0.08}$ | $0.45^{+0.12}_{-0.13}$ | $0.42^{+0.09}_{-0.14}$ | 16 |
| | | SE Shelf | −295.4 | 65.8 | 0.56 | $-1.30^{+0.13}_{-0.12}$ | $0.49^{+0.08}_{-0.09}$ | $0.53^{+0.08}_{-0.10}$ | $0.36^{+0.09}_{-0.11}$ | |
| f207_1 | 17 | Halo | −319.6 | 98.1 | 0.35 | $-1.04^{+0.09}_{-0.07}$ | 0.26 ± 0.04 | 0.53 ± 0.07 | 0.16 ± 0.04 | 21 |
| | | GSS | −529.4 | 24.5 | 0.33 | $-0.87^{+0.09}_{-0.10}$ | 0.31 ± 0.06 | $0.44^{+0.04}_{-0.05}$ | 0.16 ± 0.03 | |
| | | KCC | −427.3 | 21.0 | 0.32 | −0.79 ± 0.07 | 0.20 ± 0.04 | 0.54 ± 0.06 | 0.14 ± 0.02 | |
| f123_1 | 18 | Halo | −318.2 | 98.1 | 0.68 | −0.98 ± 0.05 | 0.33 ± 0.03 | $0.41^{+0.03}_{-0.04}$ | $0.22^{+0.04}_{-0.03}$ | 49 |
| | | SE Shelf | −279.9 | 11.0 | 0.32 | −0.71 ± 0.07 | 0.31 ± 0.05 | $0.40^{+0.04}_{-0.05}$ | 0.23 ± 0.04 | |
| S | 22 | Halo | −318.8 | 98.1 | 0.28 | $-0.66^{+0.16}_{-0.18}$ | $0.44^{+0.07}_{-0.10}$ | $0.49^{+0.05}_{-0.06}$ | $0.21^{+0.05}_{-0.04}$ | 20 |
| | | GSS | −489.0 | 26.1 | 0.49 | $-1.02^{+0.15}_{-0.14}$ | $0.45^{+0.10}_{-0.11}$ | $0.38^{+0.17}_{-0.19}$ | $0.45^{+0.07}_{-0.08}$ | |
| | | KCC | −371.6 | 17.6 | 0.22 | −0.71 ± 0.11 | 0.27 ± 0.09 | $0.35^{+0.08}_{-0.09}$ | $0.18^{+0.04}_{-0.05}$ | |
| f130[c] | 23 | Halo | −317.3 | 98.1 | 1.0 | −1.62 ± 0.10 | 0.59 ± 0.07 | $0.38^{+0.09}_{-0.10}$ | $0.37^{+0.06}_{-0.09}$ | 31 |
| D | 26 | Halo | −317.1 | 98.0 | 0.57 | $-1.00^{+0.17}_{-0.19}$ | $0.68^{+0.12}_{-0.14}$ | 0.55 ± 0.13 | $0.40^{+0.06}_{-0.08}$ | 23 |
| | | Disk | −128.4 | 16.2 | 0.43 | −0.82 ± 0.09 | $0.28^{+0.07}_{-0.09}$ | $0.60^{+0.09}_{-0.10}$ | $0.28^{+0.05}_{-0.06}$ | |
| a0_1 | 31 | Halo | −314.0 | 98.0 | 1.0 | −1.35 ± 0.10 | $0.33^{+0.06}_{-0.07}$ | 0.40 ± 0.10 | 0.30 ± 0.07 | 10 |
| a3[c] | 33 | Halo | −331.7 | 98.0 | 0.44 | $-1.48^{+0.12}_{-0.13}$ | $0.45^{+0.06}_{-0.07}$ | $0.41^{+0.05}_{-0.07}$ | 0.24 ± 0.06 | 21 |
| | | GSS | −444.6 | 15.7 | 0.56 | $-1.11^{+0.12}_{-0.13}$ | $0.46^{+0.06}_{-0.07}$ | $0.34^{+0.08}_{-0.09}$ | 0.30 ± 0.05 | |

Note. — The columns of the table correspond to field name, projected radial distance from the center of M31, kinematical component, mean heliocentric velocity, velocity dispersion, fractional contribution of the given kinematical component, mean [Fe/H], spread in [Fe/H], mean [$\alpha$/Fe], spread in [$\alpha$/Fe], and total number of RGB stars in a given field with [$\alpha$/Fe] measurements (regardless of component association). Chemical properties were calculated from a bootstrap resampling of the final sample, including weighting by the inverse variance of the measurement uncertainty and the probability that a star belongs to a given kinematical component.

[a] The parameters of the velocity model are the 50$^{\rm th}$ percentiles of the marginalized posterior probability distribution functions. These were computed by Gilbert et al. (2018) for all fields except H, S, and D. We have transformed the 50$^{\rm th}$ percentile values for the stellar halo components from the Galactocentric to heliocentric frame, based on the median right ascension and declination of all stars in a given field. For fields H, S, and D, Escala et al. (2020a) fixed the stellar halo component to the parameters derived by Gilbert et al. (2018) to independently compute the posterior distributions.

[b] Chemical abundances for fields f109_1, f123_1, f130_1, a0_1, and a3 are first presented in this chapter. We have included chemical properties of previously published M31 fields H, S, D, and f130_2 (Escala et al. 2019; Escala et al. 2020a) and f207_1 (Gilbert et al. 2019a) for reference. We further analyze the halo populations of H, S, f130_2, and f207_1 in this chapter.

[c] We combined the chemical abundance samples for fields f130_1 (this chapter) and f130_2 (Escala et al. 2019; Escala et al. 2020a) given their proximity (Figure 5.1) and the consistency of their velocity distributions (Gilbert et al. 2007; Gilbert et al. 2018). The same is true for fields a3_1 and a3_2 (this chapter), where Gilbert et al. (2009) and Gilbert et al. (2018) illustrated the similarity in kinematics between these fields.

we excluded 54 M31 RGB stars that showed TiO but otherwise would have made the final sample. In total, 129 M31 RGB stars across the six spectroscopic fields pass the above selection criteria, thereby constituting our final sample.

The final samples (Section 5.5) for fields H, S, and D are identical between Escala et al. (2020a) and this chapter despite differences in the membership determination (Section 5.4), whereas the final sample for field f130_2 contains an additional star. No stars re-classified as nonmembers in the formalism presented in this chapter were included in the final sample of Escala et al. (2020a). Three additional stars across H, S, D, and f130_2 were re-classified as M31 RGB stars with clear evidence of TiO, but otherwise passing our selection criteria (Section 5.5), and five stars were re-classified M31 RGB stars with robust measurements of [Fe/H], but not [$\alpha$/Fe].

As discussed in detail by Escala et al. (2020a), removing stars on the basis of TiO results in a bias against stars with red colors, which translates to a bias in photometric metallicity ($\mathrm{[Fe/H]_{phot}}$) against metal-rich stars. Including the sample presented in this chapter, this photometric bias ($\mathrm{[Fe/H]_{phot,mem}} - \mathrm{[Fe/H]_{phot,final}}$) ranges from 0.13−0.40 dex per field. For fields a3, more stars kinematically associated with the GSS (Section 5.4) show evident TiO absorption, whereas for field f123_1, fewer stars in the Southeast shelf substructure are affected compared to those in the halo. Therefore, the exclusion of TiO stars disproportionately impacts the $\mathrm{[Fe/H]_{phot}}$ bias depending on the given field and kinematical component. However, a bias in $\mathrm{[Fe/H]_{phot}}$ cannot be converted into a bias in spectroscopic [Fe/H], considering that $\mathrm{[Fe/H]_{phot}}$ measurements suffer from degeneracy with stellar age and [$\alpha$/Fe] from which spectroscopic [Fe/H] measurements are exempt.

**Abundance Distributions for Individual Fields**

In this section, we provide a brief overview of the chemical abundance properties of the six spectroscopic fields first presented in this work (Table 5.1; Figure 5.1). We discuss the global properties of the inner stellar halo in Section 5.5 and Section 5.5.

- *9 kpc halo field (f109_1):* The 9 kpc halo field is the innermost region of the stellar halo of M31 yet probed with chemical abundances (Gilbert et al. 2007; Gilbert et al. 2014). It does not contain any detected kinematical substructure (Section 5.2, 5.4). Excepting fields along the GSS, which have an insufficient number of smooth halo stars to constrain the abundances of the stellar halo component, the 9 kpc halo field is more metal-rich on average ($\langle\mathrm{[Fe/H]}\rangle = -0.93^{+0.08}_{-0.09}$) than the majority of halo components in fields at larger projected distance (Table 5.2; see also Section 5.5). Based on the abundances for this field, the inner halo of M31 may be potentially less $\alpha$-enhanced on average ($\langle$[$\alpha$/Fe]$\rangle$ = +0.32 ± 0.08) than the stellar halo at larger projected distances, although we did not measure a statistically significant [$\alpha$/Fe] gradient in M31's inner stellar halo (Section 5.5). Additionally, we did not find evidence for a correlation between [Fe/H] and [$\alpha$/Fe] for this field, where we computed a distribution of correlation coefficients from $10^5$ draws of the measured abundances perturbed by their (Gaussian) uncertainties.

- *18 kpc halo field (f123_1):* The 18 kpc halo field is dominated by the smooth stellar halo, but it also has a clear detection of substructure (Section 5.4; Table 5.2) known as the Southeast shelf (Section 5.2; Fardal et al. 2007; Gilbert et al. 2007). We defer further analysis of this component to future work (I. Escala et al. 2020, in preparation) owing to its likely connection to the GSS. The stellar halo in this field is metal-rich ($\langle$[Fe/H]$\rangle$ = −0.98 ± 0.05), $\alpha$-enhanced ($\langle$[$\alpha$/Fe]$\rangle$ = $+0.41^{+0.03}_{-0.04}$), and exhibits no statistically significant correlation between [$\alpha$/Fe] and [Fe/H].

- *23 kpc halo field (f130_1):* Similar to the 9 kpc halo field, the 23 kpc field does not possess detectable substructure. We combined the chemical abundance samples for this field with f130_2 (Escala et al. 2019; Escala et al. 2020a) due to their proximity (Figure 5.1) and the consistency of their velocity distributions (Gilbert et al. 2007; Gilbert et al. 2018). The average abundances for the combined sample of 31 M31 RGB stars ($\langle$[Fe/H]$\rangle$ = −1.62 ± 0.10, $\langle$[$\alpha$/Fe]$\rangle$ = $+0.38^{+0.09}_{-0.10}$; Table 5.2) agree within the uncertainties with previous determinations from an 11 star sample by Escala et al. (2019) and Escala et al. (2020a). The lack of a significant trend between [$\alpha$/Fe] and [Fe/H] is also maintained by the larger sample. The inner halo at 23 kpc appears to be more metal-poor and $\alpha$-enhanced than at 9 kpc, possibly representing a population that assembled rapidly at early times (Escala et al. 2019). The star formation history inferred for this region of M31's halo indicates that the majority of the stellar population is over 8 Gyr old (Brown et al. 2007), suggesting that an accretion origin would require the progenitor galaxies to have quenched their star formation at least 8 Gyr ago.

- *31 kpc halo field (a0_1):* The stellar population in the 31 kpc halo field is solely associated with the kinematically hot component of the stellar halo (Section 5.4). The 10 RGB star sample in this field suggests a positive trend between [$\alpha$/Fe] and [Fe/H] ($r = 0.24^{+0.12}_{-0.24}$) that is likely a consequence of small sample size. Despite its similar projected distance from the center of M31 and kinematical profile, the 33 kpc halo field appears to be more metal-rich ($\langle$[Fe/H]$\rangle = -1.35 \pm 0.10$) than the 23 kpc field, although it may be comparably $\alpha$-enhanced (Table 5.2). Nearby *HST*/ACS fields at 35 kpc (Figure 5.1) imply a mean stellar age of 10.5 Gyr (Brown et al. 2008) for the vicinity, compared to a mean stellar age of 11.0 Gyr at 21 kpc. The full age distributions suggest that the stellar populations are in fact distinct between the 21 and 35 kpc ACS fields: the star formation history of the 35 kpc field is weighted toward more dominant old stellar populations and is inconsistent with the star formation history at 21 kpc at more than $3\sigma$ significance (Brown et al. 2008). If applicable to the 31 kpc halo field, this suggests that the stellar population is both younger and more metal-rich than at 23 kpc.

- *33 kpc GSS fields (a3):* The 33 kpc field is dominated by GSS substructure (Section 5.2, Section 5.4), such that it may not provide meaningful constraints on the smooth stellar halo in this region. However, the stellar halo at 33 kpc along the GSS may be more metal-poor ($\langle$[Fe/H]$\rangle = -1.48^{+0.12}_{-0.13}$) than at 17 and 22 kpc along the GSS (Table 5.2). The abundances in this field clearly show a declining pattern of [$\alpha$/Fe] with [Fe/H], which is characteristic of dwarf galaxies. A comparison between $\langle$[Fe/H]$\rangle$ for the GSS at 17 and 22 kpc indicates that the GSS likely has a metallicity gradient, as found from photometric based metallicity estimates (Ibata et al. 2007; Gilbert et al. 2009; Conn et al. 2016; Cohen et al. 2018). We will quantify spectral synthesis based abundance gradients in the GSS and make connections to the properties of the progenitor in future work (I. Escala et al., 2020, in preparation).

In agreement with previous findings (Escala et al. 2019; Escala et al. 2020a; Gilbert et al. 2019a), the M31 fields are $\alpha$-enhanced with a significant spread in metallicity. This implies that stars in the inner stellar halo formed rapidly, such that the timescale for star formation was less than the typical delay time for Type Ia supernovae. For an accreted halo, a spread in metallicity coupled with high $\alpha$-enhancement can indicate contributions from multiple progenitor galaxies or a dominant, massive progenitor galaxy with high star formation efficiency (Robertson et al. 2005; Font et al. 2006c; Johnston et al. 2008; Font et al. 2008). We further discuss formation scenarios for M31's stellar halo in Section 5.6.

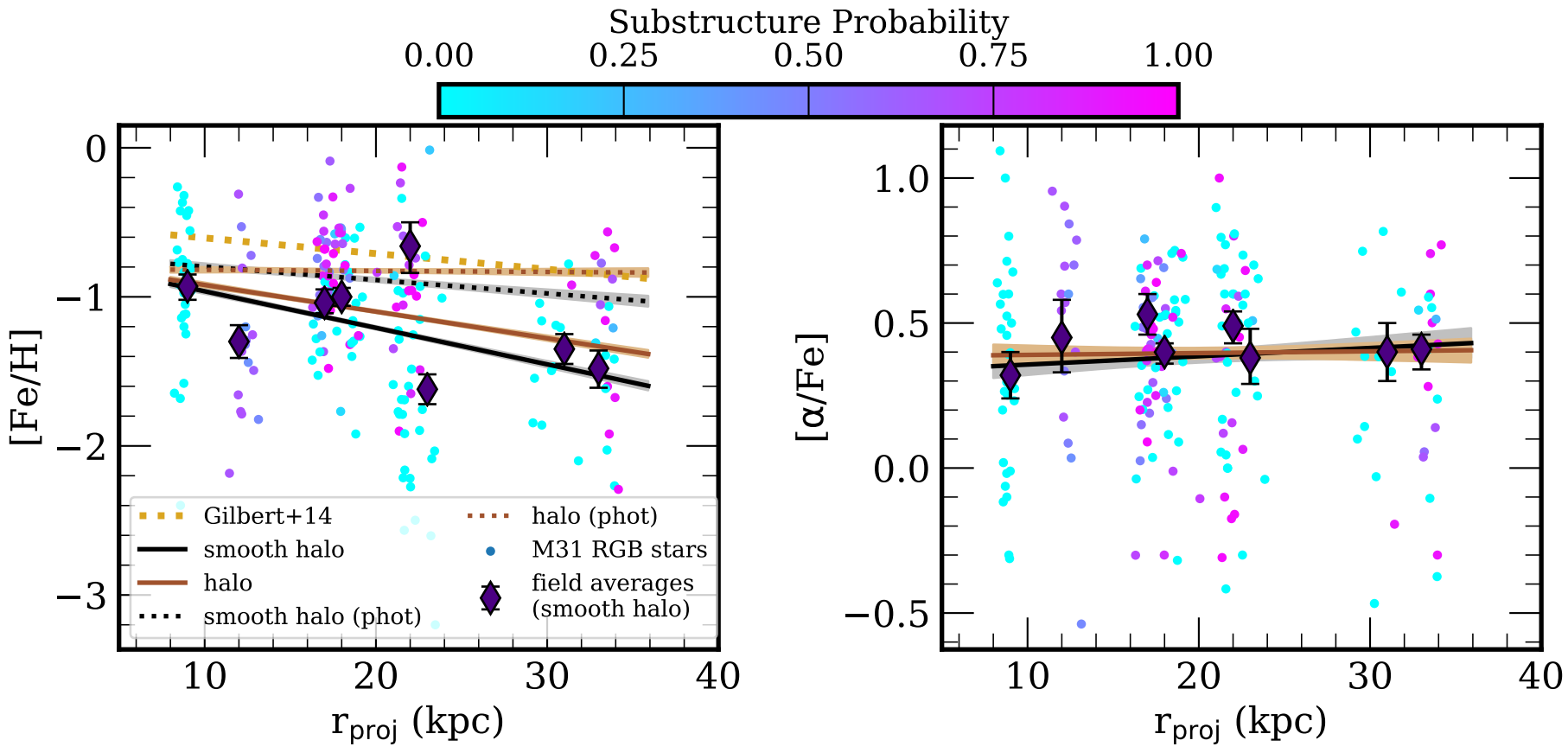

Figure 5.7: [Fe/H] (left) and [$\alpha$/Fe] (right) versus projected distance from the center of M31 for 198 M31 RGB stars (Escala et al. 2019; Escala et al. 2020a; Gilbert et al. 2019a; this work) in the inner halo ($r_{\rm proj} \lesssim 35$ kpc). Stars are color-coded according to their probability of substructure association (Section 5.4). We omit measurement uncertainties for clarity. $\langle$[Fe/H]$\rangle$ and $\langle$[$\alpha$/Fe]$\rangle$ for the halo component in each spectroscopic field (Table 5.2) are plotted as indigo diamonds. The photometric metallicity gradient of the stellar halo ($t$ = 10 Gyr, [$\alpha$/Fe] = 0) measured from spectroscopically confirmed M31 RGB stars (Gilbert et al. 2014) is indicated as a dotted gold line, assuming an intercept of $\rm [Fe/H]_{phot} = -0.5$. This slope is unchanged within the uncertainties with the inclusion/exclusion of substructure. We also display $\rm [Fe/H]_{phot}$ gradients measured from our stellar halo sample (dotted black and brown lines). The uncertainty envelopes are determined from 16$^{\rm th}$ and 84$^{\rm th}$ percentiles.

### Radial Abundance Gradients

The existence of a steep, global metallicity gradient in M31's stellar halo is well-established from both Ca triplet based (Koch et al. 2008) and photometric (Kalirai et al. 2006b; Ibata et al. 2014; Gilbert et al. 2014) metallicity estimates. Using a sample of over 1500 spectroscopically confirmed M31 RGB stars, Gilbert et al. (2014) measured a radial $\rm [Fe/H]_{phot}$ gradient of $-0.011 \pm 0.001$ dex kpc$^{-1}$ between 10−90 projected kpc in the stellar halo. The radial gradients measured with and without kinematical substructure were found to be consistent within the uncertainties.

In addition to radial [Fe/H] gradients, the possibility of radial [$\alpha$/Fe] gradients between the inner and outer halo of M31 has recently been explored. From a sample of 70 M31 RGB stars, including an additional 21 RGB stars from Gilbert et al. (2019a), with spectral synthesis based abundance measurements, Escala et al. (2020a) found tentative evidence that the inner stellar halo ($r_{\rm proj} \lesssim 26$ kpc) had higher $\langle$[$\alpha$/Fe]$\rangle$ than a sample of four outer halo stars (Vargas et al. 2014) drawn from ~70−140 kpc. K. M. Gilbert et al. 2020, accepted, increased the number of stars in the outer halo (43−165 kpc) with abundance measurements from four to nine. In combination with existing literature measurements by Vargas et al. (2014), Gilbert et al. found that $\langle[\alpha/\mathrm{Fe}]\rangle = 0.30 \pm 0.16$ for the outer halo. This value is formally consistent with the average $\alpha$-enhancement of M31's inner halo from the 91 M31 RGB star sample ($\langle[\alpha/\mathrm{Fe}]\rangle = 0.45 \pm 0.06$), indicating an absence of a gradient between the inner and outer halo, with the caveat that the sample size in the outer halo is currently limited.

With the contribution from this work of 129 M31 individual RGB stars to the existing sample of literature [Fe/H] and [$\alpha$/Fe] measurements (Vargas et al. 2014; Escala et al. 2019; Escala et al. 2020a; Gilbert et al. 2019a, K. M. Gilbert et al. 2020, accepted), we obtained $\langle[\mathrm{Fe/H}]\rangle = -1.09 \pm 0.04$ ($-1.18 \pm 0.04$) and $\langle[\alpha/\mathrm{Fe}]\rangle = 0.40 \pm 0.03$ ($0.39 \pm 0.03$) for M31's inner halo when excluding (including) substructure. Despite the formal agreement between $\langle[\alpha/\mathrm{Fe}]\rangle$ in the inner and outer halo, Gilbert et al. found that [Fe/H] and [$\alpha$/Fe] measurements of M31's outer halo are similar to those of M31 satellite dwarf galaxies (Vargas, Geha, and Tollerud 2014; Kirby et al. 2020) and the MW halo (e.g., Ishigaki, Chiba, and Aoki 2012; Hayes et al. 2018b). In comparison, [$\alpha$/Fe] is higher at fixed [Fe/H] in M31's inner halo than in its dwarf galaxies (Escala et al. 2020a). This suggests that M31's outer halo may have a more dominant population of stars with lower [$\alpha$/Fe] than the inner halo. This difference implies that the respective stellar halo populations may be, in fact, distinct. In order to better constrain $\langle[\alpha/\mathrm{Fe}]\rangle$ in the sparsely populated outer halo, abundance measurements from coadded spectra of heterogeneous samples of stars (e.g., obtained using the method of Wojno et al. 2020), will likely be necessary.

From our enlarged sample of inner halo abundance measurements, we re-assessed the presence of gradients in [Fe/H] and [$\alpha$/Fe] in the inner stellar halo of M31. Taking stars in the kinematically hot stellar halo to have $p_{\mathrm{sub}} < 0.5$ (Eq. 5.9), we identified 123 (75) stars that are likely associated with the "smooth" stellar halo (kinematically cold substructure) within a projected distance of $r_{\mathrm{proj}} \lesssim 35$ kpc of M31. We emphasize that these numbers are simply to provide an idea of the relative contribution of each component to the stellar halo–in the subsequent analysis, we *do not* employ any cuts on $p_{\mathrm{sub}}$, but rather incorporate M31 RGB stars, regardless of halo component association, by using $p_{\mathrm{sub}}$ as a weight. We have excluded all M31 RGB stars in field D (Figure 5.1) from our analysis sample owing to the presence of M31's northeastern disk. To determine the radial abundance gradients of the smooth inner halo, we fit a line using an MCMC ensemble sampler (Foreman-Mackey et al. 2013), weighting each star by $p_{\text{halo}} = 1 - p_{\text{sub}}$ and the measurement uncertainty. We also determined the radial gradients in the case of including substructure by removing $p_{\text{sub}}$ as a weight. In Figure 5.7, we refer to the stellar population including substructure simply by "halo", in contrast to "smooth halo", which excludes substructure.

Figure 5.7 shows [Fe/H] and [$\alpha$/Fe] versus projected radial distance, including our measured radial gradients and the photometric metallicity gradient of Gilbert et al. (2014) measured between 10-90 kpc. We measured a radial [Fe/H] gradient of $-0.024 \pm 0.002$ dex kpc$^{-1}$ between 8-34 kpc, with an intercept at $r_{\text{proj}} = 0$ of $-0.73 \pm 0.03$, for the smooth halo. The inclusion of substructure results in a shallower [Fe/H] gradient ($-0.018 \pm 0.001$ dex kpc$^{-1}$) with a similar intercept ($-0.75 \pm 0.03$), reflecting the preferentially metal-rich nature of substructure in M31's halo (Font et al. 2008; Gilbert et al. 2009). We did not find statistically significant radial [$\alpha$/Fe] gradients between 8-34 kpc in M31's stellar halo, both excluding ($0.0029 \pm 0.0027$ dex kpc$^{-1}$) and including ($0.00048 \pm 0.00261$ dex kpc$^{-1}$) substructure.

### Comparison to Photometric Metallicity Gradients

Our results suggest that the radial [Fe/H] gradient of the smooth halo is inconsistent with that of the halo including substructure over the probed radial range. The substructures in our inner halo fields are likely GSS progenitor debris (Fardal et al. 2007; Gilbert et al. 2007; Gilbert et al. 2009), thus the change in slope at its inclusion may reflect a convolution with the distinct metallicity gradient of the GSS progenitor (I. Escala et al. 2020, in preparation).

In order to control for differences in sample size, target selection, number and locations of spectroscopic fields utilized, and the radial extent of the measured gradient between this work and Gilbert et al. (2014), we measured a [Fe/H]$_{\text{phot}}$ (Section 5.2) gradient from our final sample, assuming $t$ = 10 Gyr and [$\alpha$/Fe] = 0 (Figure 5.7). We obtained a slope of $-0.0089 \pm 0.0018$ dex kpc$^{-1}$ ($-0.00070 \pm 0.0016$ dex kpc$^{-1}$) with an intercept of $-0.71 \pm 0.04$ ($-0.81 \pm 0.03$) when including (excluding) substructure. These gradients are inconsistent for $r_{\text{proj}} \gtrsim 20$ kpc, where the [Fe/H]$_{\text{phot}}$ gradient including substructure remains flat as the [Fe/H]$_{\text{phot}}$ gradient of the smooth halo declines. Such a difference is not detected from Gilbert et al.'s sample of over 1500 RGB stars spanning 10−90 kpc. However, the [Fe/H]$_{\text{phot}}$ gradients measured in this work provide a more direct comparison to our spectral synthesis based [Fe/H] gradients.

A possible explanation for the difference in trends with substructure between spectral synthesis and CMD-based gradients is the necessary assumption of uniform stellar age and $\alpha$-enhancement to determine $[Fe/H]_{phot}$. Although we did not measure a significant radial [$\alpha$/Fe] gradient (Figure 5.7), M31's inner stellar halo has a range of stellar ages present at a given location based on *HST* CMDs extending down to the main-sequence turn-off (Brown et al. 2006; Brown et al. 2007; Brown et al. 2008). If the smooth stellar halo is systematically older than the tidal debris toward 30 kpc, this would steepen the relative $[Fe/H]_{phot}$ gradient between populations with and without substructure. This agrees with our observation that $[Fe/H]_{phot}-[Fe/H]$ is increasingly positive on average toward larger projected distances, where the discrepancy is greater for the smooth halo than in the case of including substructure.

Additionally, a comparatively young stellar population at 10 kpc compared to 30 kpc could result in a steeper $[Fe/H]_{phot}$ gradient in better agreement with our measured [Fe/H] gradient. The difference between mean stellar age at 35 kpc and 10 kpc is 0.8 Gyr (Brown et al. 2008), though the mean stellar age of M31's halo does not appear to increase monotonically with projected radius. Assuming constant [$\alpha$/Fe], this mean age difference translates to a negligible gradient between 10–35 kpc. Thus, a more likely explanation for the discrepancy in slope between the $[Fe/H]_{phot}$ and [Fe/H] gradients are uncertainties in the stellar isochrone models at the tip of the RGB.

In general, $[Fe/H]_{phot}$ is offset toward higher metallicity, where the adopted metallicity measurement methodology can result in substantial discrepancies for a given sample (e.g., Lianou, Grebel, and Koch 2011). For example, the discrepancy between the CMD-based gradient of Gilbert et al. (2014) and literature measurements from spectral synthesis (Vargas et al. 2014; Gilbert et al. 2019a; Escala et al. 2020a) decreases when assuming an $\alpha$-enhancement ([$\alpha$/Fe] = +0.3) in better agreement with the inner and outer halo (K. M. Gilbert et al., 2020, accepted). Despite differences in the magnitude of the slope, the behavior with substructure, and intercept of the [Fe/H] gradient from different metallicity measurement techniques, our enlarged sample of spectral synthesis based [Fe/H] measurements provides further support for the existence of a large-scale metallicity gradient in the inner stellar halo, where this gradient extends out to at least 100 projected kpc in the outer halo (K. M. Gilbert et al., 2020, accepted). For a thorough consideration of the implications of steep, large-scale negative radial metallicity gradients in M31's stellar halo, we refer the reader to the discussions of Gilbert et al. (2014) and Escala et al. (2020a).

### The Effect of Potential Sources of Bias on Metallicity Gradients

Alternatively, the discrepancy between the $[Fe/H]_{phot}$ and [Fe/H] gradients could be partially driven by the bias against red, presumably metal-rich, stars incurred by the omission stars with strong TiO absorption from our final sample (Section 5.5). We investigated the impact of potential bias from both (1) the exclusion of TiO stars ($b_{\rm TiO}$) and (2) S/N limitations ($b_{\rm S/N}$; Section 5.5) on our measured radial [Fe/H] gradients by shifting each [Fe/H] measurement for an M31 RGB star in a given field by $b_{\rm TiO} + b_{\rm S/N}$. As in Escala et al. (2020a), we estimated $b_{\rm TiO}$ from the discrepancy in $\langle[\mathrm{Fe/H}]\rangle_{\rm phot}$ between all M31 RGB stars in a given field (including TiO stars) and the final sample (excluding TiO stars). We calculated $b_{\rm S/N}$ from the difference between $\langle[\mathrm{Fe/H}]\rangle$ for the sample of M31 RGB stars with [Fe/H] measurements in each field (regardless of whether an [$\alpha$/Fe] measurement was obtained for a given star) and the final sample. These sources of bias do not affect the [$\alpha$/Fe] distributions.

Incorporating these bias terms yields a radial [Fe/H] gradient of −0.022 ± 0.002 dex kpc$^{-1}$ (−0.018 ± 0.001 dex kpc$^{-1}$) and an intercept of −0.60 ± 0.03 (−0.58 ± 0.03) without (with) substructure. The primary effect of including bias estimates is a shift toward higher metallicity in the overall normalization of the gradient. The gradient slopes calculated including bias estimates are consistent with our previous measurements, which did not account for potential sources of bias. We can therefore conclude that the slopes of radial [Fe/H] gradients between 8-34 kpc in M31's stellar halo are robust against these two possible sources of bias.

### Detection of a Low-Alpha Population in the Stellar Halo

In ΛCDM cosmology, simulations of stellar halo formation for M31-like galaxies predict that the chemical abundance distributions of an accreted component should be distinct from the present-day satellite population of the host galaxy (Robertson et al. 2005; Font et al. 2006a; Tissera, White, and Scannapieco 2012), as is observed for the MW (Shetrone, Côté, and Sargent 2001; Shetrone et al. 2003; Tolstoy et al. 2003; Venn et al. 2004). This chemical distinction is driven by the early assembly of the stellar halo, where its progenitors were accreted ~8-9 Gyr ago, as opposed to ~4-5 Gyr ago for surviving satellite galaxies (Bullock and Johnston 2005; Font et al. 2006a; Fattahi et al. 2020). Accordingly, Escala et al. (2020a) showed that the [$\alpha$/Fe] distribution for the metal-rich ([Fe/H] > −1.5) component of M31's smooth, inner stellar halo ($r_{\rm proj} \lesssim 26$ kpc) is inconsistent with having formed from progenitor galaxies similar to present-day M31 satellite galaxies with measurements of [Fe/H] and [$\alpha$/Fe] ($M_* \sim 10^{5-7} M_\odot$; Vargas, Geha, and Tollerud 2014; Kirby et al. 2020). This sample of M31 dwarf galaxies consisted of NGC 185 and And II (Vargas, Geha, and Tollerud 2014) and And VII, And I, And V, and And III (Kirby et al. 2020).

However, Escala et al. were unable to statistically distinguish between the [$\alpha$/Fe] distributions for the low-metallicity ([Fe/H] $< -1.5$) component of the smooth, inner stellar halo. This could be a consequence of an insufficient sample size at low metallicity, where they identified 29 RGB stars likely belonging to the smooth stellar halo ($p_{\rm halo} > 0.5$; Eq. 5.9). Another possibility is that such low-metallicity stars, with lower average $\alpha$-enhancement ($\langle$[$\alpha$/Fe]$\rangle \sim 0.25$; Escala et al. 2020a), may represent a stellar population in the stellar halo more similar to present-day M31 satellite dwarf galaxies.

### 1-D Comparisons Between the Stellar Halo and M31's Satellite Dwarf Galaxies

To investigate whether the [$\alpha$/Fe] distributions of the stellar halo and dwarf galaxies at low-metallicity are in fact statistically indistinguishable, we repeated the analysis of Escala et al. (2020a) with our expanded sample of 198 inner halo RGB stars (excluding the 26 kpc disk field; Figure 5.1) with abundance measurements.

In contrast to Escala et al. (2020a), we applied corrections to the abundances measured by Vargas, Geha, and Tollerud (2014) to place them on the same scale as the M31 halo (Escala et al. 2019; Escala et al. 2020a; Gilbert et al. 2019a; this work) and other dwarf galaxy (Escala et al. 2020a) measurements. Kirby et al. (2020) found that the Vargas, Geha, and Tollerud measurements were systematically offset toward higher [Fe/H] by +0.3 dex compared to their measurements, based on an identical sample of spectra of M31 dwarf galaxy stars. Kirby et al. did not find evidence of a systematic offset between their [$\alpha$/Fe] measurements and those of Vargas, Geha, and Tollerud (2014). Thus, we adjusted [Fe/H] values measured by Vargas, Geha, and Tollerud (2014) by the mean systematic offset of $-0.3$ dex and did not make any changes to their [$\alpha$/Fe] values. This correction is reflected for the relevant M31 dwarf galaxies in Figures 5.8 and 5.9.

We constructed [$\alpha$/Fe] distributions for a mock stellar halo built by destroyed dwarf-galaxy-like progenitor galaxies, weighting the contribution of each M31 dwarf galaxy according to its luminosity function (McConnachie 2012).[3] When re-sampling the observed inner halo abundance distributions, we assigned each star a probability of being drawn according to its $p_{\text{halo}}$. Not only did we find that the [$\alpha$/Fe] distribution of the metal-rich, smooth inner halo is inconsistent with that of present-day M31 satellite galaxies at high significance ($p \ll 1\%$), but also that the [$\alpha$/Fe] distribution of the metal-poor component of the smooth halo disagrees with that of low-metallicity RGB stars in M31 satellite galaxies ($p < 0.3\%$).

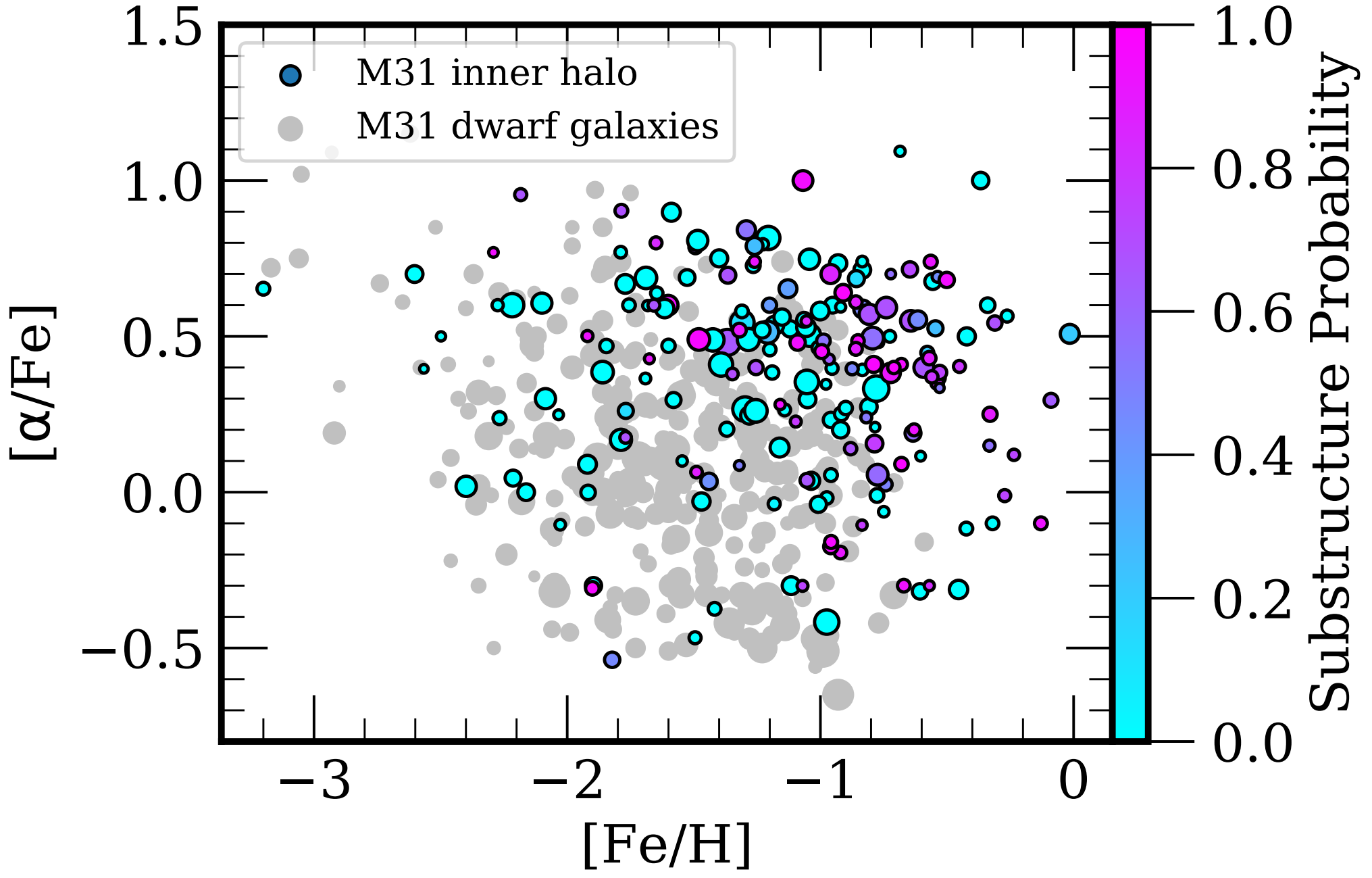

Figure 5.8: [$\alpha$/Fe] versus [Fe/H] for RGB stars in the inner halo of M31 (Gilbert et al. 2019a; Escala et al. 2020a; this work) compared to abundance trends for M31 dwarf galaxies. The combined sample of M31 dwarf galaxies consists of NGC 185, NGC 147, And II (Vargas, Geha, and Tollerud 2014) and And VII, And I, And V, And III, and And X (Kirby et al. 2020). The size of each point is proportional to the inverse variance of the measurement uncertainty, where M31 RGB stars are also color-coded by their probability of belonging to substructure (Eq. 5.9). The majority of inner halo RGB stars are inconsistent with the stellar populations of present-day M31 satellite dwarf galaxies (Escala et al. 2020a; Kirby et al. 2020), although some probable halo stars exhibit abundance patterns similar to M31 dwarf galaxies (Section 5.5).

[3]We have excluded RGB stars in NGC 185 with [Fe/H] > −0.5 (uncorrected values), owing to the uncertainty in the reliability of Vargas, Geha, and Tollerud's abundance measurements above this metallicity. Vargas, Geha, and Tollerud (2014) calibrated their measurements of bulk [$\alpha$/Fe] to approximate an $\alpha$-element abundance measured from the arithmetic mean of individual [(Mg,Si,Ca,Ti)/Fe], where their calibration is not valid for [Fe/H] > −0.5. In contrast, we have not applied such a calibration to our [$\alpha$/Fe] measurements, which are valid for [Fe/H] > −0.5. The exclusion of high-metallicity stars in NGC 185 applies to the analysis in Section 5.5 as well as Figures 5.8 and 5.9.

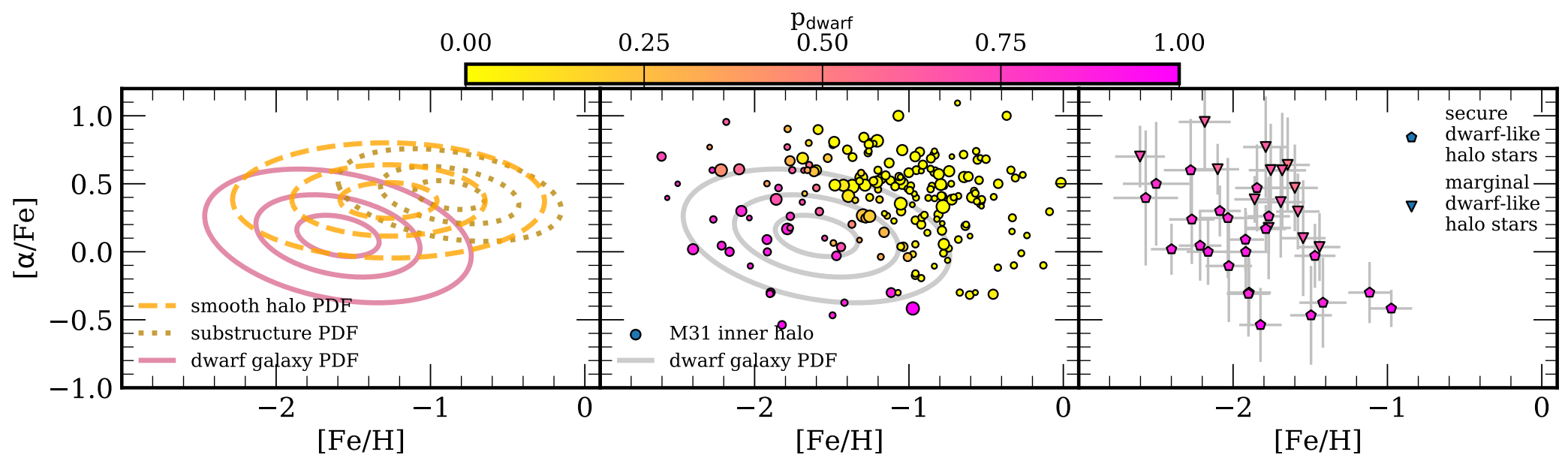

Figure 5.9: (Left) Probability density functions (PDFs) for the relationship between [$\alpha$/Fe] and [Fe/H] for the smooth component of the inner halo (dashed blue line; Gilbert et al. 2019a; Escala et al. 2020a; this work), substructure in the inner halo (dotted purple line; Gilbert et al. 2019a; Escala et al. 2020a; this work), and M31 satellite dwarf galaxies (solid green line; Vargas, Geha, and Tollerud 2014; Kirby et al. 2020), constructed from 2D bivariate normal distributions (Section 5.5; Eq. 5.10; Table 5.3). The contour levels correspond to the 16 th, 50 th, and 84 th percentiles of the 2D PDFs.

Nevertheless, constructing [$\alpha$/Fe] distributions according to metallicity bins presents a limited, 1-D view of the relationship between [$\alpha$/Fe] and [Fe/H]. Figure 5.8 displays [$\alpha$/Fe] versus [Fe/H] for the inner halo of M31 compared to M31 dwarf galaxies with abundance measurements, including NGC 147 (Vargas, Geha, and Tollerud 2014) and And X (Kirby et al. 2020). Evidently, the majority of inner halo RGB stars are inconsistent with the stellar populations of M31 dwarf galaxies, as previously found using the combined 91 RGB star sample of Escala et al. (2020a) and Gilbert et al. (2019a) and reinforced by this work. In general, [$\alpha$/Fe] tends to be higher for M31's inner halo at fixed metallicity than for the dwarf galaxies ($\langle[\alpha/\mathrm{Fe}]\rangle_{\mathrm{dwarf}} = 0.07 \pm 0.02$). Many of the high-metallicity, $\alpha$-enhanced stars in the inner halo have a high probability of being associated with kinematical substructure, such as the GSS.

In addition to these trends, Figure 5.8 suggests the existence of a stellar population preferentially associated with the dynamically hot halo that also has chemical abundance patterns similar to M31 dwarf galaxies. These stars are not contained within a well-defined metallicity bin (such as [Fe/H] $< -1.5$, the metal-poor bin utilized by Escala et al. 2020a), but rather span a region of [$\alpha$/Fe] versus [Fe/H] space coincident with the mean trend of the dwarf galaxies.

**Modeling 2-D Chemical Abundance Ratio Distributions**

In order to robustly identify M31 halo stars with abundance patterns similar to that of M31 satellite dwarf galaxies (Figure 5.8), we modeled the observed 2-D chemical abundance ratio distributions, as advocated by Lee et al. (2015). We considered the smooth component of the inner stellar halo, substructure in the inner halo, and M31 dwarf galaxies as distinct components in our modeling. We expanded upon the M31 dwarf galaxy sample analyzed in Section 5.5 by incorporating abundance measurements from galaxies that were previously excluded on the basis of their small sample sizes: And X ($N_{\rm stars}$ = 9 ;Kirby et al. 2020) and NGC 147 ($N_{\rm stars}$ = 7 ;Vargas, Geha, and Tollerud 2014).[4] Taking $\delta$([Fe/H]) < 0.5, $\delta$([$\alpha$/Fe]) < 0.5, and [Fe/H] < −0.5, these two additional galaxies result in a sample size of 293 abundance measurements for M31 dwarf galaxies.

Assuming the form of a bivariate normal distribution, the likelihood of observing a given set of abundance measurements $(x_i, y_i)$ = ([Fe/H]$_i$, [$\alpha$/Fe]$_i$) and uncertainties $(\delta x_i, \delta y_i)$ = ($\delta$[Fe/H]$_i$, $\delta$[$\alpha$/Fe]$_i$)[5] given a model described by means $\vec{\mu} = (\mu_x, \mu_y)$ and standard deviations $\vec{\sigma} = (\sigma_x, \sigma_y)$ is,

$$\mathcal{L}_i(x_i, y_i, \delta x_i, \delta y_i | \vec{\mu}, \vec{\sigma}) = \frac{1}{2\pi\sigma_{x_i}\sigma_{y_i}\sqrt{1-r^2}} \times \exp\left(-\frac{1}{2(1-r^2)}\left[\frac{(x_i-\mu_x)}{\sigma_{x_i}^2} + \frac{(y_i-\mu_y)}{\sigma_{y_i}^2} - \frac{2r(x_i-\mu_x)(y_i-\mu_y)}{\sigma_{x_i}\sigma_{y_i}}\right]\right), \quad (5.10)$$

where $i$ is an index corresponding to an individual RGB star. We incorporated the measurement uncertainties into Eq. 5.10 via the variable $\sigma_{x_i} = \sqrt{\sigma_x^2 + \delta_{x_i}^2}$, which is defined analogously for $\sigma_{y_i}$. The correlation coefficient, $r$, is an additional model parameter that accounts for covariance between $x$ and $y$.

[4]We did not include M32 in our M31 dwarf galaxy abundance sample. Vargas, Geha, and Tollerud (2014) measured abundances for 3 stars in M32 with [Fe/H] < −0.5 and concluded that they were not representative of the galaxy.

[5]We assumed independent errors $(\delta x_i, \delta y_i)$ in our model. In actuality, we expect that some amount of negative covariance, $\delta_{xy}$, exists between our measurement uncertainties. The net result of such dependent uncertainties would be a perceived decrease in the correlation coefficient, $r$ (Eq. 5.10). Thus, we acknowledge that $r$ may in fact be less negative than suggested by our model (Table 5.3).

Table 5.3: 2-D Chemical Abundance Ratio Distribution Model Parameters

| Model | $\mu_{\mathrm{[Fe/H]}}$ | $\sigma_{\mathrm{[Fe/H]}}$ | $\mu_{[\alpha/\mathrm{Fe}]}$ | $\sigma_{[\alpha/\mathrm{Fe}]}$ | $r$ |
|---|---|---|---|---|---|
| Smooth Halo | $-1.27 \pm 0.05$ | $0.53 \pm 0.04$ | $0.38 \pm 0.03$ | $0.22 \pm 0.03$ | $-0.016 \pm 0.09$ |
| Substructure | $-0.87 \pm 0.06$ | $0.38 \pm 0.05$ | $0.42 \pm 0.05$ | $0.18 \pm 0.03$ | $-0.323^{+0.17}_{-0.15}$ |
| Dwarf Galaxy | $-1.60 \pm 0.03$ | $0.45 \pm 0.02$ | $0.12 \pm 0.02$ | $0.26 \pm 0.02$ | $-0.296 \pm 0.05$ |

Note. — The columns of the table correspond to the model for the observed 2-D chemical abundance ratio distributions (Figure 5.8) for a given stellar population, and the parameters describing the bivariate normal probability density functions (Eq. 5.10). We fit for the smooth halo and substructure components simultaneously (Eq. 5.11) using a mixture model. The parameters for the 2-D chemical abundance models are the 50$^{\mathrm{th}}$ percentiles of the marginalized posterior probability distribution functions (Section 5.5), where the uncertainties on each parameter were calculated from the corresponding 16$^{\mathrm{th}}$ and 84$^{\mathrm{th}}$ percentiles.

For the full stellar halo sample, we modeled the smooth and substructure components simultaneously by combining the respective likelihood functions (Eq. 5.10) using a mixture model,

$$\mathcal{L}_i = (1 - p_i^{\mathrm{sub}})\mathcal{L}_i^{\mathrm{halo}} + p_i^{\mathrm{sub}}\mathcal{L}_i^{\mathrm{sub}}, \tag{5.11}$$

where $p_i^{\mathrm{sub}}$ is the probability that a star belongs to substructure in M31's stellar halo (Eq. 5.9). Thus, the full stellar halo abundance ratio distribution is represented by an eight parameter model ($\vec{\mu}_{\mathrm{halo}}$, $\vec{\mu}_{\mathrm{sub}}$, $\vec{\sigma}_{\mathrm{halo}}$, $\vec{\sigma}_{\mathrm{sub}}$), whereas we utilized a four parameter model ($\vec{\mu}_{\mathrm{dwarf}}$, $\vec{\sigma}_{\mathrm{dwarf}}$) for M31 dwarf galaxies. The likelihood of the entire observed data set for a given stellar population is therefore the product of the individual likelihoods,

$$\mathcal{L} = \prod_{i=1}^{N} \mathcal{L}_i. \tag{5.12}$$

Using Bayes' theorem (see also Eq. 5.1), we evaluated the posterior probability of a particular bivariate model accurately describing a given a set of abundance measurements for a stellar population,

$$P(\vec{\mu}, \vec{\sigma}|_{i=1}^{N} x_i, y_i, \delta x_i, \delta y_i) \propto P(_{i=1}^{N} x_i, y_i, \delta x_i, \delta y_i|\vec{\mu}, \vec{\sigma})P(\vec{\mu}, \vec{\sigma}), \tag{5.13}$$

where $P(_{i=1}^{N} x_i, y_i, \delta x_i, \delta y_i|\vec{\mu}, \vec{\sigma})$ is the likelihood (Eq. 5.12) and $P(\vec{\mu}, \vec{\sigma})$ represents our prior knowledge regarding constraints on $\vec{\mu}$ and $\vec{\sigma}$. We implemented noninformative priors over the allowed range for each fitted parameter, where we assumed uniform priors and inverse Gamma priors for non-dispersion and dispersion parameters, respectively. We allowed the dispersion parameters in our model to vary between $0.0 < \sigma < 1.0$, the correlation coefficients between $-1 < r < 1$, and permitted samples of $\mu$ to be drawn from $-3.0 <$ [Fe/H] $< 0.0$ and $-0.8 <$ [$\alpha$/Fe] $< 1.2$.

In the case of the variance, $\vec{\sigma}^2$, the inverse Gamma distribution is described as $\Gamma^{-1}(\alpha, \beta)$, where $\alpha$ is a shape parameter and $\beta$ is a scale parameter. This distribution is defined for $\vec{\sigma} > 0$ and can account for asymmetry in the posterior distributions for dispersion parameters, which are restricted to be positive. The bootstrap resampled standard deviations, weighted by the inverse variance of the measurement uncertainty and the probability of belonging to a given kinematical component, are $\sigma$([Fe/H]) = 0.53 ± 0.03, $\sigma$([$\alpha$/Fe]) = 0.32 ± 0.02 for the smooth halo, $\sigma$([Fe/H]) = 0.51 ± 0.03, $\sigma$([$\alpha$/Fe]) = 0.31 ± 0.02 for halo substructure, and $\sigma$([Fe/H]) = 0.49 ± 0.02, $\sigma$([$\alpha$/Fe]) = 0.34 ± 0.02 for M31 dwarf galaxies. Owing to the similarity in the dispersion for each abundance ratio between the various stellar populations, we fixed the priors on $\vec{\sigma}^2$ to $\Gamma^{-1}(3, 0.5)$ for all $\sigma^2_{\rm [Fe/H]}$ parameters and $\Gamma^{-1}(3, 0.2)$ for all $\sigma^2_{[\alpha/\rm Fe]}$ parameters. These distributions result in priors that peak at $\sigma_{\rm [Fe/H]} \sim 0.50$ and $\sigma_{[\alpha/\rm Fe]} \sim 0.30$ with a standard deviation of 0.3 dex.

With the above formulation, we sampled from the posterior distribution of each model (Eq. 5.13) using an affine-invariant Markov Chain Monte Carlo (MCMC) ensemble sampler (Foreman-Mackey et al. 2013). That is, we solved for the parameters describing the two-component stellar halo abundance distribution (Eq. 5.11), and separately for a model corresponding to M31 satellite dwarf galaxies. We evaluated each model using 100 walkers and ran the sampler for $10^4$ steps, retaining the latter 50% of the MCMC chains to form the converged posterior distribution. Table 5.3 presents the final parameters of each 2-D chemical abundance ratio distribution model. We computed the mean values from the 50$^{\rm th}$ percentiles of the marginalized posterior distributions, whereas the uncertainty on each parameter was calculated relative to the 16$^{\rm th}$ and 84$^{\rm th}$ percentiles.

Figure 5.9 displays the resulting bivariate normal distribution models for the 2-D chemical abundance ratio distributions, reflecting the general trends of the observed abundance distributions (Section 5.5) for the smooth halo, halo substructure, and dwarf galaxy populations. The models predict a larger separation between the mean metallicity of the smooth halo and substructure populations, where $\mu^{\rm sub}_{\rm [Fe/H]} - \mu^{\rm halo}_{\rm [Fe/H]} = 0.40 \pm 0.08$, compared to $\langle$[Fe/H]$\rangle_{\rm sub} - \langle$[Fe/H]$\rangle_{\rm halo} = 0.09 \pm 0.06$ from a weighted boostrap resampling (Section 5.5). However, the models predict similar mean $\alpha$- enhancements between the two halo populations.

Based on these models, we calculated the probability that a star is both kinematically associated with the smooth stellar halo has abundance patterns similar to M31 dwarf galaxies,

$$p_{\mathrm{dwarf}} = \left( \frac{\mathcal{L}_i^{\mathrm{dwarf}} / \mathcal{L}_i^{\mathrm{halo}}}{1 + \mathcal{L}_i^{\mathrm{dwarf}} / \mathcal{L}_i^{\mathrm{halo}}} \right) \times (1 - p_{\mathrm{sub,f}}), \tag{5.14}$$

where $\mathcal{L}_i$ is the likelihood from Eq. 5.12 and $p_{\mathrm{sub,f}}$ is the refined probability that a star belongs to kinematical substructure. For every sampling of the posterior distribution for each star, we calculated the odds ratio of the Bayes factor, $K$, for the substructure versus halo models, where $K = p_{\mathrm{sub}} \mathcal{L}_i^{\mathrm{sub}} / ( (1 - p_{\mathrm{sub}}) \, \mathcal{L}_i^{\mathrm{halo}})$. Then, we computed $p_{\mathrm{sub,f}}$ for each star from the 50$^{\mathrm{th}}$ percentile of these distributions. The outcome of this procedure is an updated determination of the substructure probability for each star that incorporates *both* kinematical and chemical information. For stars with nonzero $p_{\mathrm{sub}}$, the net effect is $p_{\mathrm{sub,f}} > p_{\mathrm{sub}}$ for metal-rich stars ([Fe/H] $\gtrsim -1.0$) and $p_{\mathrm{sub,f}} < p_{\mathrm{sub}}$ for metal-poor stars. This change occurs because the substructure model (Table 5.3) has a higher mean [Fe/H] than the smooth halo model, such that metal-rich stars are more likely to belong to substructure.

Figure 5.9 illustrates the distribution of smooth halo stars with M31 dwarf-galaxy-like chemical abundances, where we have identified 23 RGB stars that securely fall within this category ($p_{\mathrm{dwarf}} \geq 0.75$), in addition to 14 marginally dwarf-galaxy-like RGB stars ($0.5 < p_{\mathrm{dwarf}} < 0.75$). Based on a bootstrap re-sampling weighted by $p_{\mathrm{dwarf}}$ and the inverse measurement uncertainty, we computed $\langle$[Fe/H]$\rangle = -1.74 \pm 0.05$ and $\langle[\alpha/\mathrm{Fe}]\rangle = 0.17 \pm 0.07$ for this population of stars in the smooth stellar halo.

Hereafter, we refer to this group of stars as the "low-$\alpha$" population, owing its lower average $\alpha$-enhancement compared to the entire smooth stellar halo ($\langle[\alpha/\mathrm{Fe}]\rangle = 0.40 \pm 0.03$; Section 5.5). By definition, low-$\alpha$ stars belong to the dynamically hot component of the stellar halo in which we do not detect any kinematical substructure. These stars span the same range of parameter space in line-of-sight velocity ($-700$ km s$^{-1}$ $< v_{\mathrm{helio}} \lesssim -150$ km s$^{-1}$) and projected distance (8 kpc $< r_{\mathrm{proj}} <$ 34 kpc) as the full sample of RGB stars in M31's inner halo. Unsurprisingly, low-$\alpha$ population stars are more likely to be found in spectroscopic fields along the minor axis dominated by the smooth halo, such as the 23 kpc and 31 kpc halo fields, as opposed to fields dominated by tidal debris along the high surface-brightness core of the GSS.

Conversely, we can define a "high-$\alpha$" population of stars in the smooth halo with chemical abundances distinct from M31 dwarf galaxies by taking the complement of Eq. 5.14. This population has 66 (17) secure (marginal) smooth halo members, such that $\langle$[Fe/H]$\rangle$ = −1.05 ± 0.03 and $\langle$[$\alpha$/Fe]$\rangle$ = 0.46 ± 0.03. Similar to low-$\alpha$ stars, high-$\alpha$ stars span a wide range of heliocentric velocity and projected distance, but are more centrally concentrated along the minor-axis fields, such as the 9 kpc and 18 kpc halo fields, and are more likely to appear in the halo components of GSS fields.

### The Effect of Potential Sources of Bias on Population Detections

Potential bias from the omission of red, presumably metal-rich, stars with strong TiO absorption (Section 5.5) does not affect the robust identification of smooth halo stars with chemical abundance patterns similar to M31 dwarf galaxies. Introducing maximal bias estimates for [Fe/H] based on this source alone (Section 5.5) shifts the mean metallicity of the halo and substructure models (Table 5.3) to −1.05 ± 0.05 and −0.64 ± 0.07, respectively. TiO stars are not a significant source of bias in M31 dwarf galaxies (Kirby et al. 2020), and their abundance measurements are similarly affected by S/N limitations (Section 5.5) compared to M31 RGB stars. Thus, the net effect of the omission of TiO stars would be an increased separation between the dwarf galaxy and halo PDFs. This would result in the classification of 37 (6) secure (marginal) low-$\alpha$ stars, thereby reinforcing the detection of this population.

Although possible bias from the exclusion of TiO stars does not impact the detection of a low-$\alpha$ population in M31's stellar halo, the resulting change in $p_{\text{dwarf}}$ (Eq. 5.14) would alter the population means to approximately $\langle$[Fe/H]$\rangle$ = −1.78 ± 0.06 and $\langle$[$\alpha$/Fe]$\rangle$ = 0.24 ± 0.07. Analogously, the high-$\alpha$ population would be characterized by $\langle$[Fe/H]$\rangle$ = −0.98 ± 0.03 and $\langle$[$\alpha$/Fe]$\rangle$ = 0.46 ± 0.03, with 65 (13) secure (marginal) members. These values are formally consistent with the population means for both the low- and high-$\alpha$ populations calculated without bias estimates. None of the spatial or kinematical properties of either the low- or high-$\alpha$ populations would be altered in this bias scenario.

## 5.6 Discussion

In this section, I discuss the chemical abundance properties of M31's inner stellar halo in the context of the MW halo, the Magellanic Clouds, and halo formation models.

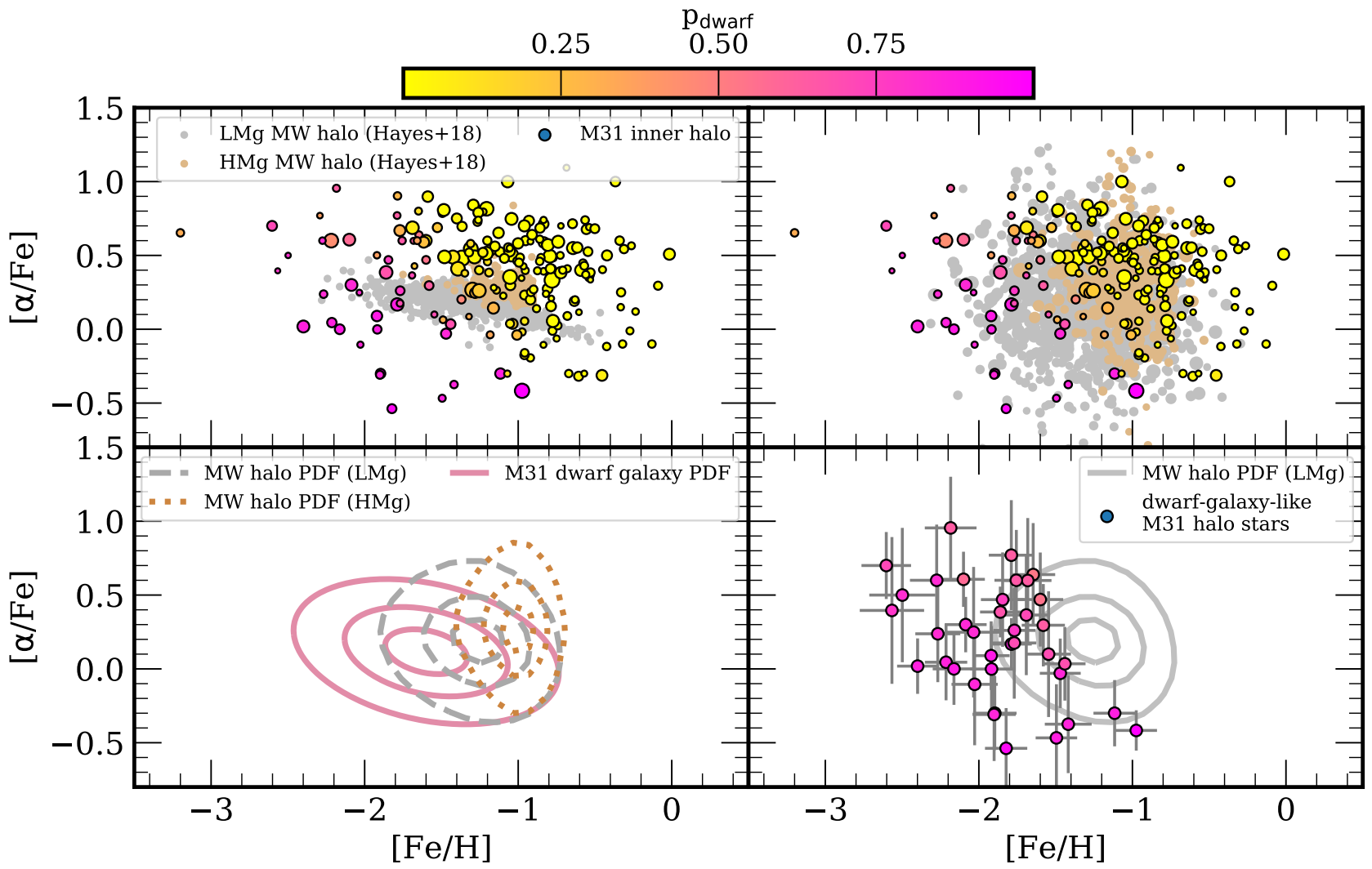

Figure 5.10: A comparison between [$\alpha$/Fe] vs. [Fe/H] for the inner halo of M31 (Escala et al. 2019; Escala et al. 2020a; Gilbert et al. 2019a; this work) and metal-poor ([Fe/H] $\lesssim -0.9$) stars in the MW halo (Hayes et al. 2018b). The MW stars are separated into low-[Mg/Fe] (LMg; grey points) and high-[Mg/Fe] (HMg; brown points) populations. (Upper left) The color-coding and point scaling for the M31 data is the same as in Figure 5.9. (Upper right) A perturbation of the MW halo abundances by our empirical error distribution ($\langle\delta_{\rm [Fe/H]}\rangle \sim 0.12$, $\langle\delta_{[\alpha/{\rm Fe}]}\rangle \sim 0.30$). (Lower left) Empirical PDFs for the LMg (grey dashed lines) and HMg (dotted brown lines) populations in the MW computed from $10^6$ perturbations of the MW abundances by the M31 error distribution. We also show our fitted model for the M31 dwarf galaxy PDF (solid pink lines; Section 5.5, Table 5.3). The contour levels correspond to $16^{\rm th}$, $50^{\rm th}$, and $84^{\rm th}$ percentiles. (Lower right) Low-$\alpha$ population stars ($p_{\rm dwarf} > 0.5$) plotted against the LMg PDF. A subset of low-$\alpha$ stars appears similar to the LMg population, but it is unclear if all low-$\alpha$ stars could originate from a LMg-like population.

### Comparison to the MW: A Similar Low-Alpha Population in the Stellar Halo of M31?

In Section 5.5, we identified populations of low- and high-$\alpha$ stars in the smooth component of M31's stellar halo. In this section, we consider whether similar populations have been identified in the stellar halo of the MW.

Indeed, Nissen and Schuster (1997) suggested the existence of distinct stellar populations with low- and high-[$\alpha$/Fe] in the MW from a small sample of 29 stars with halo-like kinematics in the solar neighborhood. In a subsequent study, Nissen and Schuster (2010) used a larger sample of 94 stars to confirm the existence of these low-$\alpha$ and high-$\alpha$ stellar populations, which were further distinguished by their abun-dances in odd-Z elements (Nissen and Schuster 2010, 2011). The high-$\alpha$ stars were preferentially bound to the Galaxy on prograde orbits, whereas the low-$\alpha$ stars were less bound with a majority on retrograde orbits. Later studies of MW stellar halo populations identified the same chemical abundance trends with respect to [$\alpha$/Fe] and [Fe/H] (Navarro et al. 2011; Ramírez, Meléndez, and Chanamé 2012; Sheffield et al. 2012; Hawkins et al. 2015; Hayes et al. 2018b; Haywood et al. 2018). Schuster et al. (2012) found that the low-$\alpha$ stars were clumped at large orbital eccentricities (>0.85) and extended out to larger projected distances (30-40 kpc) compared to the high-$\alpha$ stars, which were centrally concentrated (<16 kpc) and had more uniformly distributed eccentricities (0.4-1.0). Schuster et al. (2012) also found that the low- and high-$\alpha$ stars had ages $\gtrsim$9 Gyr (see also Hawkins et al. 2014, Gallart et al. 2019 and Das, Hawkins, and Jofré 2020). Based on this evidence, Nissen and Schuster (2010, 2011) and Schuster et al. (2012) associated the low-$\alpha$ population with the preferentially radial accretion of dwarf galaxies(s), and the high-$\alpha$ population with *in-situ* star formation, supporting a dual-origin scenario for stellar halo formation in the MW (Zolotov et al. 2009; Zolotov et al. 2010; Font et al. 2011; McCarthy et al. 2012; Tissera, White, and Scannapieco 2012; Tissera et al. 2013).

With the advent of large astrometric and spectroscopic surveys such as *Gaia* (Gaia Collaboration et al. 2016a; Gaia Collaboration et al. 2018) and APOGEE (Majewski et al. 2017), the low-$\alpha$ halo population of Nissen and Schuster has come to be associated with the last major accretion event in the MW's history, known as Gaia-Enceladus (Helmi et al. 2018), or Gaia-Sausage (Belokurov et al. 2018). Upon the discovery of a well-separated blue color sequence in the *Gaia* DR2 Hertzprung-Russel diagram of stars with high tangential velocities, Gaia Collaboration et al. (2018) suggested that this blue sequence may correspond to Nissen and Schuster's low-$\alpha$ stars. Combining kinematical and chemical information from *Gaia* DR2 and APOGEE, Helmi et al. 2018 showed that the retrograde structure corresponding to Nissen and Schuster's low-$\alpha$ population (e.g., Koppelman, Helmi, and Veljanoski 2018; Belokurov et al. 2018) has chemical abundance patterns characteristic of a massive dwarf galaxy. Using the same data sets, Haywood et al. (2018) independently concluded that the blue *Gaia* sequence was indeed accreted from a progenitor with low star formation efficiency during the Galaxy's last significant merger. Gaia-Enceladus was likely comparable to the Small Magellanic Cloud (SMC; $M_\star \sim 10^{8.5} M_\odot$) in stellar mass at infall, with $M_\star \sim 10^{8-9}\ M_\odot$ and was accreted ~10 Gyr ago (e.g., Helmi et al. 2018; Belokurov et al. 2018; Gallart et al.

2019; Mackereth et al. 2019b; Fattahi et al. 2019).[6]

To compare M31's inner halo to an analogous population in the MW, we used the low-metallicity ([Fe/H] $\lesssim -0.9$) sample from Hayes et al. (2018b) based on APOGEE (Majewski et al. 2017) data ($R \sim 22500$) presented in SDSS-III (Eisenstein et al. 2011) Data Release (DR) 13 (Albareti et al. 2017). Hayes et al. (2018b) illustrated that their data set is equivalent to the two distinct populations of Nissen and Schuster (2010, 2011), and by extension contains Gaia-Enceladus stars (Haywood et al. 2018). In contrast to earlier work (using DR12) on metal-poor MW field stars (Hawkins et al. 2015) that relied on kinematical selection, Hayes et al. (2018b) used a larger sample of stars in combination with a data-driven approach to identify stellar populations that were distinct in multi-dimensional chemical abundance space. This resulted in the identification of two kinematically and chemically distinct MW stellar halo populations characterized by low-[Mg/Fe] with negligible Galactic rotation (LMg) and high-[Mg/Fe] with significant Galactic rotation (HMg).

Based on these chemodynamical properties, Hayes et al. concluded that the origin of the LMg population is likely massive progenitors similar to the Large Magellanic Cloud (LMC; $M_\star \sim 10^{9.5} M_\odot$), accreted early in the MW's history (i.e., Gaia-Enceladus; Haywood et al. 2018). The HMg population was likely formed *in-situ*, either via dissipative collapse or disk heating. In a companion study, Fernández-Alvar et al. (2018) modeled the chemical evolution of the two populations to find that LMg stars experienced a less intense and short-lived star formation history than HMg stars.

Figure 5.10 presents a comparison between [$\alpha$/Fe] and [Fe/H] for 1,321 stars from the metal-poor halo ([Fe/H] $< -0.9$) sample of Hayes et al. (2018b) and 198 RGB stars in the inner stellar halo of M31 (Escala et al. 2019; Escala et al. 2020a; Gilbert et al. 2019a; this work). The atmospheric [$\alpha$/Fe] from Hayes et al. (2018b) is measured from spectral synthesis by the APOGEE Stellar Parameters and Chemical Abundances Pipeline (ASPCAP; García Pérez et al. 2016) based on a fit to all $\alpha$-elments (O, Mg, Si, Ca, S, and Ti; Holtzman et al. 2015) in the infrared H-band (1.514 – 1.696 $\mu$m). We show both the LMg and HMg populations for the MW, whereas the M31 inner halo stellar population is color-coded to emphasize stars that are likely members of M31's low-$\alpha$ population.

[6] Via independent estimates, Helmi et al. (2018) and Gallart et al. (2019) infer a 4:1 merger ratio, which Helmi et al. equates to $M_\star \sim 6 \times 10^8\ M_\odot$ using a derived star formation rate from Fernández-Alvar et al. (2018). Belokurov et al. (2018) only infer a virial mass. Mackereth et al. (2019b) and Fattahi et al. (2019) obtain $10^{8.5-9}$ and $10^{9-10}\ M_\odot$, respectively, based on comparisons of Gaia-Enceladus data to hydrodynamical simulations.

Although the abundance distributions of the MW and M31 halo stars clearly overlap, the spread in [$\alpha$/Fe] (and perhaps [Fe/H]) for the M31 RGB stars is much larger owing to the uncertainties on our measurements. The typical measurement uncertainties for M31 RGB stars are $\delta_{[\mathrm{Fe/H}]} \sim 0.12$ and $\delta_{[\alpha/\mathrm{Fe}]} \sim 0.30$, compared to ~0.04 for MW halo stars in the Hayes et al. (2018b) sample. In order to perform a more direct comparison between the M31 and MW chemical abundance distributions, we perturbed the MW halo abundances by uncertainties from $10^4$ draws of the empirical error distribution of our measurements.[7] During each draw, each MW halo star was perturbed by random values sourced from a normal distribution defined by the uncertainties on a single randomly selected M31 RGB star and the MW halo star[8]. The M31 RGB star is randomly selected from a metallicity bin within 0.2 dex of the MW halo star to preserve the metallicity-dependence of the M31 error distribution: M31 RGB stars with lower [Fe/H] tend to have larger [Fe/H] uncertainties. We show an example of a single perturbation of the MW halo stars in the upper right panel of Figure 5.10. The primary effect of this perturbation is that the MW halo stars now span a similar range in [$\alpha$/Fe] as the M31 halo stars.

By performing this exercise, we constructed empirical PDFs in [$\alpha$/Fe] vs. [Fe/H] space for the LMg and HMg populations (lower left panel of Figure 5.10). Despite the significantly larger uncertainties, the comparatively narrow spread in [Fe/H] and higher mean [Fe/H] and [$\alpha$/Fe] of the HMg population is preserved relative to the LMg population. Figure 5.10 also shows the M31 dwarf galaxy PDF (Section 5.5; Table 5.3), which we used to assign a probability of membership to M31's low-$\alpha$ population to each M31 RGB star in the inner halo. Both the LMg and HMg PDFs overlap the M31 dwarf galaxy PDF for [Fe/H] $\gtrsim -1.8$, although they seem to extend to higher [$\alpha$/Fe] at fixed [Fe/H]. Based on these PDFs, the M31 dwarf galaxy population appears to be chemically distinct from the accreted (LMg) and *in-situ* (HMg) populations of the MW halo. This difference between M31 dwarf galaxies and the LMg population is in accordance with expectations. Hayes et al. (2018b) concluded that the majority of the LMg population cannot be accounted for by

[7] We have assumed that our measured uncertainties in [Fe/H] and [$\alpha$/Fe] *for an individual star* are independent. The net effect of this perturbation by such independent errors is to orthogonally stretch the 2D chemical abundance distributions for the MW halo. However, as discussed in Section 5.5, we expect the errors in [$\alpha$/Fe] and [Fe/H] on a single measurement to be contravariant. Thus, the MW halo chemical abundances are not necessarily perturbed by the true underlying error distribution for the M31 inner halo stars.

[8] e.g., $\mathcal{N}(0, \sigma_{[\mathrm{Fe/H}]})$, where $\sigma_{[\mathrm{Fe/H}]} = \sqrt{(\delta^{\mathrm{M31}}_{[\mathrm{Fe/H}],i})^2 - (\delta^{\mathrm{MW}}_{[\mathrm{Fe/H}]})^2}$

the accretion of progenitors similar to present-day MW dwarf galaxies. Because M31 dwarf galaxies have similar abundance patterns as MW dwarf galaxies at fixed stellar mass (Kirby et al. 2020), this conclusion also extends to progenitors similar to present-day M31 dwarf galaxies.

In the lower right panel of Figure 5.10, we qualitatively compare M31 RGB stars that likely belong to M31's low-$\alpha$ population ($p_{\rm dwarf} > 0.5$; Eq. 5.14) to the LMg PDF. A subset of M31's low-$\alpha$ stars appear similar to the LMg population in the MW halo, but it is unclear if all of M31's low-$\alpha$ stars could originate from a LMg-like population. In particular, M31's halo may have stars that are more metal-poor ([Fe/H] $\lesssim -1.8$) than those observed in Hayes et al.'s MW halo sample.[9] We note that the APOGEE abundances attributed to Gaia-Enceladus from Helmi et al. (2018) extend down to [Fe/H] $\sim -2.5$ and up to [$\alpha$/Fe] $\sim 0.5$, potentially encapsulating the low-metallicity RGB stars in M31's low-$\alpha$ population. Regardless, the marginal similarity between the M31 low-$\alpha$ and MW LMg populations suggests that they may both originate from the accretion of dwarf galaxies $\gtrsim$8 Gyr ago (e.g., Bullock and Johnston 2005; Font et al. 2006a; Fattahi et al. 2020), such that kinematical substructure no longer remains coherent in the stellar halo. We further investigate this possibility through comparison to the Magellanic Clouds in Section 5.6 and discuss possible origin scenarios for M31's low-$\alpha$ population in Section 5.6.

Regarding the HMg population, M31's low-$\alpha$ stars do not appear to be consistent with its empirical PDF. However, M31 inner halo stars with $p_{\rm dwarf} < 0.5$ (upper right panel of Figure 5.10) do overlap in chemical abundance space with the HMg population. Some of these stars are kinematically associated with substructure, whereas the remainder are the high-$\alpha$, smooth halo population stars defined in Section 5.5. The high-$\alpha$ population in M31 appears analogous to the *in-situ* HMg population based on its [$\alpha$/Fe] and [Fe/H] measurements. However, it is unclear if high-metallicity ([Fe/H] $\gtrsim -0.5$) stars are present in the MW's HMg halo, as is the case for M31's high-$\alpha$ halo, owing to the metallicity cut ([Fe/H] $\leq -0.9$) used to define the HMg population (Hayes et al. 2018b). We further discuss formation scenarios for M31's high-$\alpha$ population in Section 5.6.

We refrain from making quantitative conclusions based on these comparisons owing to potential systematic offsets in the abundance distributions between the MW and M31 data sets. As previously discussed, the APOGEE chemical abundances are based on high-resolution spectroscopy in the near-infrared H-band, in contrast with our measurements, which we derive from low- ($R \sim 3000$) and medium- ($R \sim 6000$) resolution spectroscopy at optical wavelengths with comparatively low S/N. Additionally, the APOGEE abundances are internally calibrated against observations of metal-rich open clusters (Holtzman et al. 2015) and externally calibrated to correct for systematic offsets. Although our spectral synthesis method relies on MW globular clusters to determine the systematic uncertainty on our abundance measurements (Kirby, Guhathakurta, and Sneden 2008; Escala et al. 2019; Escala et al. 2020a), we do not calibrate the stellar parameters and elemental abundances outputted by our abundance pipelines to external observations. The selection functions also differ between Hayes et al. (2018b) and this work, where we have avoided selection criteria based on chemical abundances. Although we acknowledge these caveats, an exploration of the impact of these possible systematic offsets and selection effects between the MW and M31 data sets is beyond the scope of this chapter.

[9]We do not expect any *systematic* biases in APOGEE elemental abundances to result in the dearth of data at [Fe/H] $\lesssim -1.8$ and above the ASPCAP grid edge at [Fe/H] = $-2.5$. APOGEE is biased against *metal-rich* stars in fields that are either distant (i.e., dominated by cool giants) or enriched to supersolar metallicity (Hayden et al. 2014; Hayden et al. 2015), but there is no quantified bias for metal-poor stars in APOGEE with [Fe/H] $\gtrsim -2.5$.

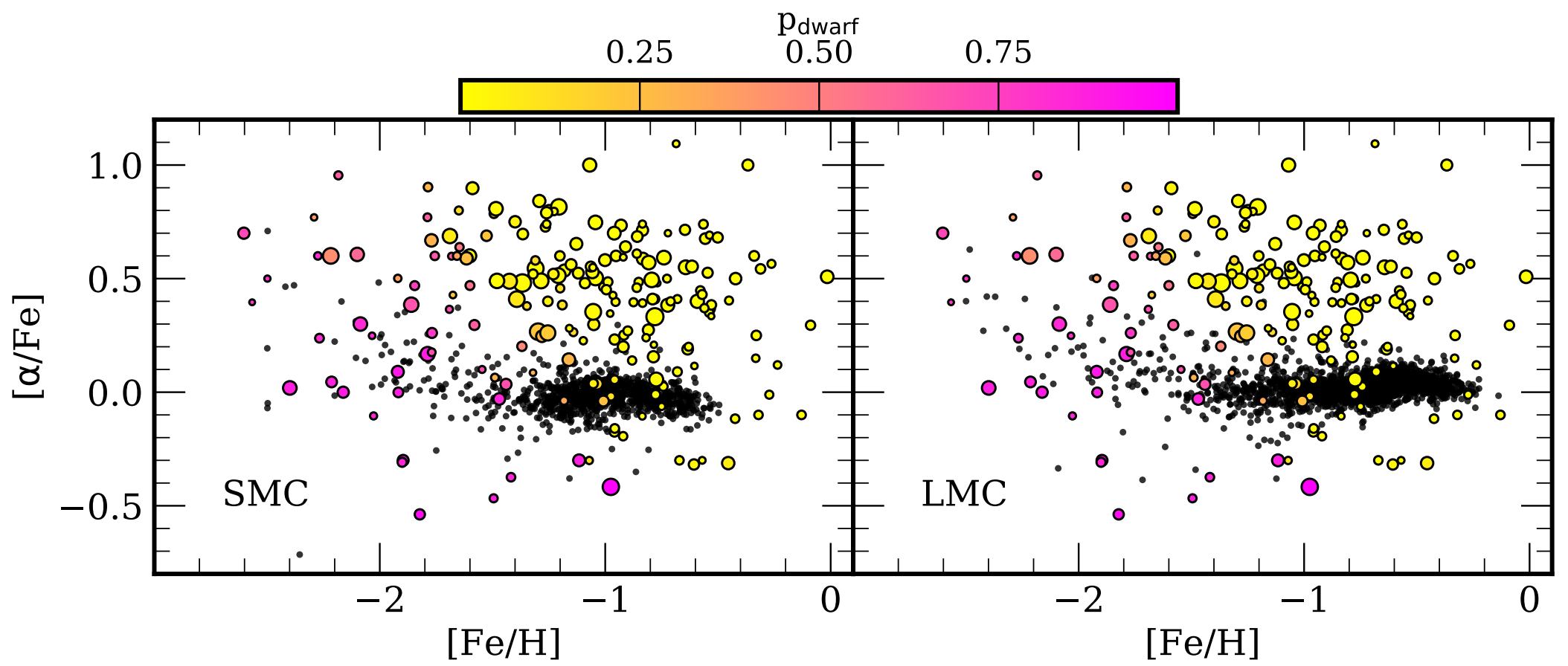

Figure 5.11: [$\alpha$/Fe] versus [Fe/H] for RGB stars in the inner halo of M31 compared to the Magellanic Clouds (black points; Nidever et al. 2019; Section 5.6): the SMC (left) and LMC (right). The color-coding and point scaling for the M31 data is the same as in Figure 5.9. We do not show the small measurement uncertainties for the MCs ($\delta_{\rm [Fe/H]} \sim 0.02$, $\delta_{[\alpha/{\rm Fe}]} \sim 0.03$). The abundance patterns of the MCs are largely inconsistent with M31's low-$\alpha$ stars ($p_{\rm dwarf} > 0.5$), except for the metal-poor tails that extend to higher [$\alpha$/Fe].

**Comparison of M31's Low-Alpha Population to the Magellanic Clouds**

In the previous section, we compared the abundances of M31's inner halo to a low-$\alpha$ population in the MW (Hayes et al. 2018b), which is likely associated with Gaia-Enceladus (Helmi et al. 2018; Belokurov et al. 2018; Haywood et al. 2018). The potential similarity between the low-$\alpha$ populations of M31 and the MW motivates the hypothesis that M31's low-$\alpha$ population has an accretion origin. In this section, we test this possibility by comparing the abundances of M31's inner halo to the Magellanic Clouds (MCs), which have stellar masses ($2.7 \times 10^9$ $M_\odot$ and $3.1 \times 10^8$ $M_\odot$ for the LMC and SMC, respectively; van der Marel et al. 2002; Stanimirović, Staveley-Smith, and Jones 2004) comparable to the inferred stellar mass of Gaia-Enceladus.

We used the MC abundance sample of Nidever et al. (2019) from APOGEE DR16 (Ahumada et al. 2019) obtained as part of SDSS-IV (Blanton et al. 2017) through the installation of a second APOGEE spectrograph in the Southern hemisphere (Wilson et al. 2019).[10] Figure 5.11 shows [$\alpha$/Fe] versus [Fe/H] for the MC samples compared to the inner halo of M31. The MCs show the same broad chemical abundance trends defined by low [$\alpha$/Fe], a relatively metal-rich body, and a metal-poor tail extending to higher [$\alpha$/Fe]. For both MCs, Nidever et al. (2019) constrained the "$\alpha$-knee" in the [$\alpha$/Fe] versus [Fe/H] relationship to [Fe/H]$\sim -2.2$, where $\langle$[$\alpha$/Fe]$\rangle$ is approximately solar. Despite these similarities, there are differences in the abundances between the MCs. The LMC is more metal-rich with $\langle$[Fe/H]$\rangle \sim -0.68$, reaching metallicities as high as $-0.2$, compared to the SMC ($\langle$[Fe/H]$\rangle \sim -0.97$), which has a maximum metallicity of [Fe/H]$\sim -0.5$. Nidever et al. (2019) also noted that the SMC abundances may be slightly biased against metal-poor stars owing to the lower S/N of the APOGEE data for the SMC relative to the LMC.

Compared to M31, the MCs have lower [$\alpha$/Fe] by ~0.4 dex at fixed [Fe/H] above [Fe/H] $\sim -1$. However, there appears to be considerable overlap between the M31 RGB stars with $p_{\rm dwarf} > 0.5$ and the metal-poor, high-$\alpha$ tails. By inspection, the decline in [$\alpha$/Fe] with [Fe/H] for the low-$\alpha$ M31 stars seems to agree with the MCs better than the MW halo LMg population / Gaia-Enceladus (Hayes et al. 2018b; Figure 5.10). If we perturb the MC abundance distributions by the measurement uncertainties of the M31 data as in the previous section, we find that the MC abundance distributions *taken as a whole* are inconsistent with M31's

[10] The discussion on how APOGEE measures abundances, and on the caveats about comparing APOGEE data to our optical spectroscopic survey in M31, in the previous section also applies to the DR16 data of the Magellanic Clouds.

low-$\alpha$ population. This is because the loci of the MCs' abundance distributions (particularly the LMC) occur at high metallicity, while M31's low-$\alpha$ population is metal-poor. However, the MC abundance distributions include metal-poor tails ([Fe/H] $\lesssim -1.5$) within the $2\sigma$ confidence limits of their 2D PDFs. In particular, an abundance distribution similar to the SMCs' metal-poor tail could potentially account for M31's low-$\alpha$ population for [Fe/H] $\gtrsim -2.0$. This implies that M31's low-$\alpha$ population may have originated from an accreted galaxy (or galaxies) that experienced chemical evolution similar to that of the MCs at low-metallicity ([Fe/H] $\lesssim -1.5$), or early in their histories.

Based on their chemical abundances, Nidever et al. (2019) concluded that the early evolution of the MCs was characterized by low star formation efficiency (gas mass converted into stellar mass). Prior studies of the $\alpha$-enhancement of the SMC (Mucciarelli 2014) and LMC (Pompéia et al. 2008; Lapenna et al. 2012; Van der Swaelmen et al. 2013) arrived at similar conclusions. This low efficiency may result from the MCs having evolved in isolation, given that they are likely on first infall into the MW's potential well (Besla et al. 2007; Besla et al. 2012). The relatively quiescent star formation history of the MCs until ~4 Gyr ago may be a consequence of such isolation (e.g., Smecker-Hane et al. 2002; Harris and Zaritsky 2009; Weisz et al. 2013).

Chemical evolution models for the LMC predict the dominance of Type Ia supernovae over core-collapse supernovae during this quiescent epoch (Bekki and Tsujimoto 2012; Nidever et al. 2019), resulting in the observed decline in [$\alpha$/Fe] with [Fe/H] in the metal-poor tails of the LMC's abundance distribution (Figure 5.11). The predicted age-metallicity relation from Nidever et al. (2019) has the LMC reaching [Fe/H]$\sim -1.5$ at 6 Gyr into its evolution (Nidever et al. 2019), suggesting that the majority of its metal-poor tail was in place by ~7 Gyr ago. The models of Bekki and Tsujimoto (2012) reach [Fe/H] $\sim -1.5$ by $\sim 9 - 11$ Gyr ago but were not well-constrained at low metallicity by the available data for the LMC at the time of their study. Thus, we can infer that the dwarf galaxy or dwarf galaxies that were accreted to form M31's low-$\alpha$ population likely had low star formation efficiency and slow chemical enrichment. We evaluate whether the multiple or single progenitor scenario is more likely for the origin of M31's low-$\alpha$ population in the following sections.

**Formation Scenarios for M31's Inner Stellar Halo**

In this section, we explore the implications of our abundance measurements in M31's stellar halo for *in-situ* and accreted formation channels. We also investigate possible origins of M31's low-$\alpha$ population in the context of the stellar mass function of accreted galaxies.

**The In-Situ vs. Accreted Halo**

Simulations predict the existence of an *in-situ* component of the stellar halo, which formed in the main progenitor of the host galaxy, in addition to an accreted component formed from external galaxies (Zolotov et al. 2009; Font et al. 2011; Tissera et al. 2013; Cooper et al. 2015). In addition to a component formed via dissipative collapse, the *in-situ* halo includes contributions from kinematically heated stars originating in the disk (Purcell, Bullock, and Kazantzidis 2010; McCarthy et al. 2012; Tissera et al. 2013; Cooper et al. 2015). The relative contributions of these two formation channels to stellar halo build-up depend on the stochastic accretion history of the host galaxy, where more active histories correspond to lower *in-situ* stellar halo mass fractions (Zolotov et al. 2009). However, the *in-situ* fraction also depends on the numerical details of a given simulation (Zolotov et al. 2009; Cooper et al. 2015). The more recent simulations of Cooper et al. (2015) have found that the *in-situ* fraction typically ranges between ~30-40%. Despite variations in the predictions of this fraction, multiple studies have found that the *in-situ* component tends to dominate within the inner ~20-30 kpc of the stellar halo (Zolotov et al. 2009; Font et al. 2011; Tissera et al. 2013; Cooper et al. 2015).

Predicted chemical signatures for the *in-situ* and accreted stellar halo can also vary depending on the simulation. In some studies (Zolotov et al. 2009; Font et al. 2011; Tissera et al. 2013; Tissera et al. 2014), *in-situ* star formation produces more metal-rich stellar halos than accretion alone. In contrast, Cooper et al. (2015) argued that the *in-situ* and accreted halo may be indistinguishable based on metallicity alone, assuming that the *in-situ* halo forms mostly from gas stripped from accreted galaxies. For galaxies that have not experienced a recent ($z \lesssim 1$) major merger, Zolotov et al. (2010) found that the *in-situ* halo has higher [$\alpha$/Fe] at fixed [Fe/H] than the accreted halo for stars with [Fe/H] $> -1$ in the solar neighborhood. Tissera, White, and Scannapieco (2012) and Tissera et al. (2013) found the opposite trend, where accreted stars tend to be more $\alpha$-enhanced (but also more metal-poor) than *in-situ* stars.

On the observational front, a bimodality in the metallicity of the MW's stellar halo with radius was interpreted as evidence in favor of a two-component halo defined by *in-situ* and accreted populations (Carollo et al. 2007; Carollo et al. 2010). After *Gaia* DR2, it has become apparent that the MW's presumably *in-situ* inner stellar halo is in actuality the remnants of Gaia-Enceladus (Helmi et al. 2018; Haywood et al. 2018; Belokurov et al. 2018; Deason et al. 2018). A veritable *in-situ* component of the inner stellar halo has recently emerged via its distinct chemical composition and kinematics in relation to Gaia-Enceladus tidal debris (Hayes et al. 2018b; Haywood et al. 2018; Di Matteo et al. 2019; Gallart et al. 2019; Conroy et al. 2019, with earlier suggestions from *Gaia* DR1 by Bonaca et al. 2017). In particular, Belokurov et al. (2020) identified this component from old, high-eccentricity, $\alpha$-rich stars on retrograde orbits[11] where with [Fe/H] $> -0.7$ in the solar neighborhood, associating it with formation from the MW's proto-disk following its last significant merger.

At this time, the presence of a significant *in-situ* component in the inner stellar halo of M31 is less clear than in the case of the MW. Along the major axis of M31's northeastern disk, there is compelling kinematical evidence for both a rotating inner spheroid (Dorman et al. 2012) and dynamically heated stars originating from the disk (Dorman et al. 2013). An extended disk-like structure possibly formed from M31's last significant merger has also been detected within the inner ~40 kpc of M31's disk plane (Ibata et al. 2005). Although these stellar structures may reasonably be various *in-situ* components, they have not been detected within the radial range spanned by our data (~8-34 kpc in M31's southeastern quadrant, along its minor axis; Figure 5.1). Even at the distance of our innermost spectroscopic field ($r_{\rm proj} \sim 9$ kpc, or $r_{\rm disk} \sim 38$ kpc assuming $i = 77°$), the contribution from the extended disk is expected to be $\lesssim$10% (Guhathakurta et al. 2005; Gilbert et al. 2007).

Interestingly, M31's global stellar halo properties appear to agree with predictions from accretion-only models for stellar halo formation (Harmsen et al. 2017), along with other MW-like, edge-on galaxies in the Local Volume from the GHOSTS survey (Radburn-Smith et al. 2011; Monachesi et al. 2016). Indeed, multiple lines of evidence indicate that accretion has played a dominant role in the formation of M31's stellar halo.

Based on this alone, both the metal-poor, low-$\alpha$ and metal-rich, high-$\alpha$ stars in M31's smooth, inner stellar halo (Section 5.5) could be accreted populations resulting from chemically distinct groups of progenitors. The similarity in the chemical abundance patterns of the low-$\alpha$ population and M31 dwarf galaxies provides strong evidence in favor of an accretion origin, in addition to its similarity to metal-poor stars associated with Gaia-Enceladus and the MCs. However, the overlap between M31's high-$\alpha$ population and the MW's *in-situ* HMg population in abundance space suggests that high-$\alpha$ stars could represent an ancient *in-situ* population instead of having an accretion origin in distinct progenitor(s). In an *in-situ* formation scenario, M31's high-$\alpha$ population may have originated from a proto-stellar disk, similar to the MW's *in-situ* halo population (Belokurov et al. 2020). Given that we are lacking multi-dimensional kinematical information or higher-dimensional chemical information, it is difficult to distinguish between accreted and *in-situ* formation scenarios for the high-$\alpha$ population without performing a detailed comparison to simulations, which is beyond the scope of this chapter.

[11] While the *in-situ* halo component of Belokurov et al. (2020) (dubbed the "Splash") most cleanly separates for metal-rich, retrograde stars, it also contains stars on prograde orbits and exhibits low-amplitude net prograde rotation.

**The Mass Function of Destroyed Galaxies**

The general consensus from simulations of MW-like galaxies is that massive dwarf galaxies ($M_\star \sim 10^{8-10}$) distinct from present-day satellites are the dominant progenitors of the accreted stellar halo (Bullock and Johnston 2005; Robertson et al. 2005; Font et al. 2006a; Cooper et al. 2010; Deason, Mao, and Wechsler 2016; D'Souza and Bell 2018b; Fattahi et al. 2020). These galaxies have a median accretion time of $\gtrsim$9 Gyr ago (Bullock and Johnston 2005; Font et al. 2006a; Fattahi et al. 2020). Furthermore, a few massive progenitors are predicted to form the bulk of the accreted halo (Cooper et al. 2010; Deason, Mao, and Wechsler 2016; D'Souza and Bell 2018b; Fattahi et al. 2020), where the secondary progenitor is on average half as massive as the primary progenitor (D'Souza and Bell 2018b). In particular, Fattahi et al. (2020) found that such massive progenitors dominate the stellar mass budget within the inner 50 kpc of the stellar halo, where contributions from progenitors with $M_\star < 10^8 M_\odot$ become non-negligible around ~100 kpc.

The details of the stellar mass function of destroyed constituent galaxies are dictated by the accretion history of the host galaxy. Deason, Mao, and Wechsler (2016) found that MW-like galaxies with more active accretion histories have higher mass accreted progenitors ($M_\star \sim 10^{9-10} M_\odot$) and larger accreted halo mass fractions than their quiescent counterparts. The wealth of substructure visible in M31's stellar halo (e.g., Ferguson et al. 2002; Ibata et al. 2007; McConnachie et al. 2018), including the GSS (Ibata et al. 2001a), and the lack of an apparent break in its density profile (Guhathakurta et al. 2005; Irwin et al. 2005; Courteau et al. 2011; Gilbert et al. 2012) suggests that M31 has likely experienced multiple and continuous contributions to its stellar halo from accreted galaxies (Deason et al. 2013; Cooper et al. 2013; Font et al. 2020). Moreover, the total accreted stellar mass of M31 ($1.5 \pm 0.5 \times 10^{10} M_{\odot}$; Harmsen et al. 2017) is on the upper end of both observed (Merritt et al. 2016; Harmsen et al. 2017) and predicted (Cooper et al. 2013; Deason, Mao, and Wechsler 2016; Monachesi et al. 2019) accreted stellar halo mass fractions for MW-like galaxies.

The large-scale negative metallicity gradient of M31's stellar halo (Gilbert et al. 2014; K.M. Gilbert et al. 2020, accepted; this work) may also lend support to it being dominated by a few massive progenitors, although this trend could also result from a significant *in-situ* component in the inner regions of the halo (Font et al. 2011; Tissera et al. 2014; Monachesi et al. 2019). Thus, M31's accreted mass function is likely weighted toward progenitor galaxies with $M_{\star} \sim 10^{9-10} M_{\odot}$. This agrees with the hypothesis that the main contributor to M31's stellar halo (including substructure) is the GSS progenitor, a massive galaxy with $M_{\star} \sim 10^{9-10} M_{\odot}$ accreted between ~1-4 Gyr ago, depending on the merger scenario (Fardal et al. 2007; Fardal et al. 2008; Hammer et al. 2018; D'Souza and Bell 2018a).

The mass spectrum of progenitor galaxies dictates the metallicity of the accreted stellar halo through the stellar mass-metallicity relation for galaxies (Gallazzi et al. 2005; Kirby et al. 2013) and its redshift evolution (Gallazzi et al. 2014; Ma et al. 2016; Leethochawalit et al. 2018; Leethochawalit et al. 2019). In particular, Leethochawalit et al. (2018) and Leethochawalit et al. (2019) found that the normalization of the stellar-mass metallicity relation evolves by $0.04 \pm 0.01$ dex Gyr$^{-1}$. Assuming that this relation extends to $M_{\star} < 10^{9}\ M_{\odot}$, this means that an SMC-mass galaxy ($M_{\star} \sim 10^{8.5}\ M_{\odot}$) accreted 10 Gyr ago would have $\langle$[Fe/H]$\rangle = -1.39 \pm 0.04$, compared to $\langle$[Fe/H]$\rangle = -0.99 \pm 0.01$ at $z = 0$ (Nidever et al. 2019). Observations of the MW-like GHOSTS galaxies (Harmsen et al. 2017) have firmly established such a relationship between stellar halo mass and metallicity, as first suggested by Mouhcine et al. (2005). The physical driving force for this relation is the fact that the most massive progenitors dominate halo assembly (D'Souza and Bell 2018b; Monachesi et al. 2019).

In addition to setting the stellar halo metallicity, the most massive progenitors largely determine the full metallicity distribution function of the stellar halo (Deason, Mao, and Wechsler 2016; Fattahi et al. 2020). For a typical progenitor mass of $10^{9-10}$ $M_{\odot}$, these galaxies contribute $\gtrsim$90% of metal-poor stars ([Fe/H] < −1) and ~20-60% of stars with [Fe/H] < −2 (Deason, Mao, and Wechsler 2016). Deason, Mao, and Wechsler also found that classical dwarf galaxies ($M_{\star} \sim 10^{5-8} M_{\odot}$) contribute the remainder of metal-poor stars, with negligible contributions from ultra-faint dwarf galaxies.

Indeed, M31's low-$\alpha$ stars most likely have an accretion origin owing to their defining similarity to the abundance patterns M31 dwarf galaxies (Section 5.5) and their distinction from the predominant abundance patterns of M31's high-$\alpha$ halo (Section 5.5). Given that (1) the chemical abundance patterns of M31's low-$\alpha$ population are broadly consistent with the low-metallicity ([Fe/H] $\lesssim -1.5$), early evolution of the MCs (Section 5.6) and (2) we do not detect any kinematical substructure for this population (Section 5.5), the galaxies(s) that built up M31's low-$\alpha$ population would need to have low star formation efficiency and have been accreted $\gtrsim$8 Gyr ago.

Given M31's halo properties and inferred accretion history (based on halo formation models), a likely scenario for the formation of the low-$\alpha$ population ($-2.5 \lesssim$ [Fe/H] $\lesssim -1.0$) is the accretion of a massive, secondary progenitor. Assuming that the GSS progenitor is the dominant progenitor, the stellar mass of this secondary progenitor should be approximately between the mass of the SMC and LMC ($M_{\star} \sim 10^{8.5-9.5} M_{\odot}$) and possibly comparable to Gaia-Enceladus. An early, massive accretion event such as this would also deposit its debris closer to the center of the host potential in the inner halo owing to dynamical friction (Cooper et al. 2010; Tissera et al. 2013; Fattahi et al. 2020). Alternatively, M31's low-$\alpha$ population could have formed exclusively from the accretion of multiple progenitors similar to classical dwarf galaxies. Although this hypothesis cannot be rejected, it is less favored by the predictions of stellar halo formation in a cosmological context (Bullock and Johnston 2005; Font et al. 2006a; Deason, Mao, and Wechsler 2016; D'Souza and Bell 2018b).

To discern between these two hypotheses, we consider the accretion of an SMC-mass galaxy onto M31 ~10 Gyr ago. Assuming a significant metallicity dispersion as observed in LG galaxies (~0.4 dex; e.g., Kirby et al. 2013; Ho et al. 2015; Kirby et al. 2020), such a galaxy would span $-1.9$ ($-2.2$) $\lesssim \langle$ [Fe/H]$\rangle \lesssim -0.9$ ($-0.6$) within $1\sigma$ ($2\sigma$). It is unclear if such a galaxy could account for the most metal-poor stars observed in M31's low-$\alpha$ population ([Fe/H] $\lesssim -2.2$). Additional progenitors with $M_\star \sim 10^{7-8}\ M_\odot$ may be necessary to explain these low-metallicity stars. On the high metallicity end, we cannot attribute stars with [Fe/H] $\gtrsim -1.0$ to the low-$\alpha$ population because the M31 dwarf galaxy PDF utilized in its detection (Section 5.5) cuts off at [Fe/H] $\gtrsim -1.0$ (Figure 5.8). Although low-$\alpha$ stars in M31 appear to exist above this metallicity, and even show overlap with the body of the SMC's abundance distribution (Figure 5.11), we cannot meaningfully isolate them from the high-$\alpha$ population with current data.

## 5.7 Summary

We have presented measurements of [$\alpha$/Fe] and [Fe/H] for 129 individual M31 RGB stars from low- ($R \sim 3000$) and medium- ($R \sim 6000$) resolution Keck/DEIMOS spectroscopy. With a combined sample of 198 M31 RGB stars with abundance measurements in inner halo fields (Escala et al. 2019; Escala et al. 2020a; Gilbert et al. 2019a; this work), we have undertaken an analysis of the chemical abundance properties of M31's kinematically smooth stellar halo between 8–34 kpc. Our primary results are the following:

1. We measured $\langle$[Fe/H]$\rangle = -1.09 \pm 0.04$ ($-1.18 \pm 0.04$) and $\langle$[$\alpha$/Fe]$\rangle = 0.40 \pm 0.03$ ($0.39 \pm 0.03$) including (excluding) substructure in M31's inner stellar halo (Section 5.5).

2. We measured a radial [Fe/H] gradient of $-0.024 \pm 0.002$ dex kpc$^{-1}$, with an intercept of [Fe/H] = $-0.73 \pm 0.03$, for the smooth halo (Section 5.5). Including substructure results in a shallower [Fe/H] gradient ($-0.018 \pm 0.001$). We did not find statistically significant radial gradients in [$\alpha$/Fe], including or excluding substructure.

3. We reaffirmed previous results based on smaller sample sizes (Escala et al. 2020a; Kirby et al. 2020) that the chemical abundance distribution of M31's inner halo is incompatible with having formed from progenitors similar to M31 present-day satellite galaxies (Vargas, Geha, and Tollerud 2014; Kirby et al. 2020) when taken as a whole (Section 5.5).

4. We robustly identified a subset of stars ($N = 37$ with $p_{\rm dwarf} > 0.5$; Eq. 5.14) belonging to the smooth halo that have [Fe/H] and [$\alpha$/Fe] measurements consistent with M31 satellite galaxies (Section 5.5). This "low-$\alpha$" population has $\langle$[Fe/H]$\rangle = -1.75 \pm 0.05$ and $\langle$[$\alpha$/Fe]$\rangle = 0.17 \pm 0.07$. The complement of the low-$\alpha$ population, M31's high-$\alpha$ population, has $\langle$[Fe/H]$\rangle = -1.05 \pm 0.03$ and $\langle[\alpha/\text{Fe}]\rangle = 0.46 \pm 0.03$ with 83 likely members. The remaining 78 M31 RGB stars are kinematically associated with substructure (to be further analyzed in I. Escala et al., in preparation).

5. We compared M31's low-$\alpha$ and high-$\alpha$ populations to analogous low-[Mg/Fe] and high-[Mg/Fe] populations in the stellar halo of the MW (Section 5.6; Hayes et al. 2018b), where the low-[Mg/Fe] population is associated with Gaia-Enceladus tidal debris (Helmi et al. 2018; Haywood et al. 2018; Belokurov et al. 2018). A subset of M31's low-$\alpha$ stars with [Fe/H] $\gtrsim -1.8$ appear similar to the MW's low-[Mg/Fe] population, but it is unclear if this applies to the most metal-poor low-$\alpha$ stars in M31.

6. We compared M31's low-$\alpha$ population to the Magellanic Clouds (Nidever et al. 2019), finding that its chemical evolution is similar to the MCs at [Fe/H] $\lesssim -1.5$ (Section 5.6). This indicates that it likely formed in an environment with low star formation efficiency.

7. We concluded that M31's low-$\alpha$ population was likely accreted (Section 5.6). We discussed potential origin scenarios, including massive ($M_\star \gtrsim 10^{8-9} M_\odot$) progenitor(s) and progenitor(s) similar to classical dwarf galaxies ($M_\star \sim 10^{7-8} M_\odot$). The high-$\alpha$ population may result from progenitor(s) distinct from those of the low-$\alpha$ population, or represent an *in-situ* stellar halo component.

The authors would like to thank Miles Cranmer and Erik Tollerud for insightful discussions regarding statistics and David Nidever for providing a catalog of APOGEE data for the Magellanic Clouds. We also thank Stephen Gwyn for reducing the photometry for slitmasks f109_1 and f123_1 and Jason Kalirai for the reduction of f130_1. I.E. acknowledges support from a National Science Foundation (NSF) Graduate Research Fellowship under Grant No. DGE-1745301. This material is based upon work supported by the NSF under Grants No. AST-1614081 (E.N.K., I.E.) AST-1614569 (K.M.G, J.W.), and AST-1412648 (P.G.). E.N.K gratefully acknowledges support from a Cottrell Scholar award administered by the Research Corporation for Science Advancement, as well as funding from generous donors to the California Institute of Technology. E.C.C is supported by a Flatiron Research Fellowship at the Flatiron Institute. The Flatiron Institute is supported by the Simons Foundation. The analysis pipeline used to reduce the DEIMOS data was developed at UC Berkeley with support from NSF grant AST- 0071048.

We are grateful to the many people who have worked to make the Keck Telescope and its instruments a reality and to operate and maintain the Keck Observatory. The authors wish to recognize and acknowledge the very significant cultural role and reverence that the summit of Maunakea has always had within the indigenous Hawaiian community. We are most fortunate to have the opportunity to conduct observations from this mountain.

# 6

# SUMMARY: A VIEW TOWARD THE FUTURE

In this thesis, I have presented both theoretical (Escala et al. 2018) and observational (Escala et al. 2019; Escala et al. 2020a; Escala et al. 2020b) studies toward a more comprehensive understanding of galaxy formation in the Local Group (LG). I have used both state-of-the art cosmological simulations and new observational techniques to probe the complex relationship between chemical abundances and merger history. This has resulted in a thorough analysis of chemical evolution predictions of simulated LG dwarf galaxies against observations and a more complete census of chemical abundance measurements in the LG.

Within the last few years, we have witnessed a factor of almost 60 increase in the number of individual stars in M31 with chemical abundance measurements beyond simple metallicity estimates (Vargas et al. 2014; Escala et al. 2019; Escala et al. 2020a; Gilbert et al. 2019a; Escala et al. 2020b; K.M. Gilbert et al. 2020, accepted). Our knowledge of the chemical composition of M31's satellite galaxy system has also been greatly augmented (Vargas, Geha, and Tollerud 2014; Kirby et al. 2020; Wojno et al. 2020). This has enabled the first systematic study of the formation history of an $L_{\star}$ galaxy other than the MW through the lens of stellar chemical abundances. This tremendous progress has been accompanied by both advances in simulating the formation and chemical evolution of Local Group galaxies (Wetzel et al. 2016; Escala et al. 2018; Garrison-Kimmel et al. 2014; Garrison-Kimmel et al. 2019; Wheeler et al. 2019; Emerick, Bryan, and Mac Low 2019; Sanderson et al. 2020) and pioneering investigations into the assembly of MW-like galaxies in the Local Volume (LV) with resolved stellar populations (Radburn-Smith et al. 2011; Geha et al. 2017; Monachesi et al. 2016; Harmsen et al. 2017; Smercina et al. 2018; Smercina et al. 2019; Bennet et al. 2019, 2020; Jang et al. 2020).

Even in these uncertain times, the unfolding story of galaxy formation in the LG will not end here. In the final chapter of this thesis, I will summarize my primary results (Section 6.1) and present a view toward the future of photometric and spectroscopic surveys of galaxies in the LG—and potentially the LV—and their anticipated implications for near-field cosmology (Sections 6.2, 6.3, and 6.4). I also briefly discuss the advances in galaxy formation simulations that will be necessary to fully utilize existing and upcoming observational data in the LG (Section 6.5).

## 6.1 Summary

In this section, I provide a brief summary of the primary results of this thesis. These conclusions are the following:

1. The inclusion of sub-grid metal mixing via the turbulent diffusion of metals in gas is necessary in Lagrangian simulations of galaxy formation and evolution to result in more realistic metallicity distribution functions and $\alpha$-element abundance patterns (Chapter 2).

2. Spectral synthesis of low-resolution spectroscopy is a viable method for measuring reliable metallicity and $\alpha$-element abundances for individual stars with low signal-to-noise spectra at the distance of M31 (Chapter 3).

3. The first measurements of [$\alpha$/Fe] in the inner halo, outer disk, and Giant Stellar Stream (Gilbert et al. 2019a) of M31 have revealed the following (Chapter 4):

    - The GSS is $\alpha$-enhanced compared to massive dwarf galaxies in the LG, implying that it experienced a truncated star formation history and was accreted onto the host halo of M31 at an early epoch (Gilbert et al. 2019a; Escala et al. 2020a). Our chemical abundance measurements could be consistent with either a major or minor merger scenario for the formation of the GSS, depending on the details of the given model.

    - The outer disk of M31 at 26 projected kpc is unusually $\alpha$-enhanced, indicating that the stellar population for this region is dominated by the yields of core-collapse supernovae. In combination with the known kinematical properties and star formation history of M31's disk, I have speculated that the chemical abundances may favor a major merger scenario for the formation of the GSS and the evolution of the disk.

    - The inner halo of M31 is inconsistent with having formed from progenitors similar to present-day M31 satellite galaxies based on their chemical abundance distributions (see also Kirby et al. 2020). There are also suggestions between differences in chemical abundances between the inner and outer halo of M31 (see also K. M. Gilbert et al. 2020, accepted).

4. An expanded sample of abundance measurements in M31's inner halo for 198 total stars has further illustrated that (Chapter 5):

- On average, the inner stellar halo of M31 appears to be both more metal-rich and $\alpha$-enhanced than that of the MW. This indicates that the build up of M31's halo is likely dominated by more massive progenitors and early accretion compared to the MW.
- Both photometric and spectroscopic metallicity measurements support the existence of global, large-scale abundance gradients in M31's stellar halo (see also K. M. Gilbert et al. 2020, accepted). This suggests that distinct formation mechanisms or dominant progenitor(s) have contributed to the formation of the inner versus outer halo.
- A subset of stars with halo-like kinematics and relatively low-$\alpha$ enhancement and low-metallicity is detectable within M31's inner stellar halo. Based on its properties, this population likely has an ancient accretion origin.

Future lines of work in M31 that will add to these numerous discoveries include (1) constraining the metallicity gradient of the GSS, (2) measuring abundances across 4–26 kpc of M31's disk, and (3) measuring abundances from spectral coaddition techniques (Wojno et al. 2020) in M31's sparse outer halo. On the simulation front, more advanced models that include realistic implementations of chemical evolution are necessary to move forward with reconstructing the formation history of M31.

## 6.2 Planned Spectroscopic Surveys of the Local Group

The current frontier of resolved stellar spectroscopy is unveiling the chemical composition and kinematics of the faintest objects in the LG. In this section, I provide an overview of future industrial-scale spectroscopic surveys of the LG, with a focus on M31.

### The Subaru Prime Focus Spectrograph

In late 2022,[1] the Subaru Prime Focus Spectrograph (PFS; Takada et al. 2014; Tamura et al. 2018) will begin to survey the LG. PFS is an optical and near-infrared multi-fiber spectrograph capable of obtaining spectra for 2394 targets within a 1.3 deg diameter wide field of view with an 8.2 m primary mirror and minimum fiber separation of 30". The spectrograph is equipped with three arms—the blue (380-

[1]The given timelines indicated in this chapter reflect pre-coronavirus estimates. Inevitably, the various spectrographs, facilities, and telescopes discussed in this chapter will be delayed as a consequence of widespread shutdowns. I have refrained from providing updated estimates owing to the uncertain nature of the global situation.

650 nm), red (630-970 nm), and near-infrared (940-1260 nm) arms—where all three arms operate in a low-resolution (LR; R$\sim$3000) mode and the red arm contains a medium-resolution mode (MR; R$\sim$5000) over a restricted wavelength range (710-885 nm). The throughput in the blue and red arms varies between $\sim$20-40%. The entire wavelength range spanned by PFS is 0.38 - 1.26 $\mu$m. Galactic archaeology is a major program goal for the instrument, targeting the MW, M31, and Local Group dwarf galaxies with PFS to obtain measurements of spectroscopic [Fe/H], [$\alpha$/Fe], and precision velocities.

The primary advantage of PFS will be an unprecedented number of stellar spectra ($\sim 10^6$ total) in these systems of sufficient quality to obtain radial velocities and homogeneously derived stellar parameters and abundances. The unifying theme of the MW, M31, and dwarf galaxy subsurveys is to provide stringent constraints on the process of hierarchical assembly on Galactic scales through the kinematics and chemical abundances of resolved, ancient stellar populations. An intimately connected goal is to probe the dark matter halos of these galaxies to test predictions of $\Lambda$CDM.

For the first time, PFS will enable spectroscopic coverage of dark-matter dominated dwarf galaxies like Sextans and NGC 6822 out to their tidal radii. Precise radial velocity measurements of their member stars will allow their dark matter profiles beyond their inner regions and their total mass to be constrained (e.g., Walker et al. 2009; Walker and Peñarrubia 2011; Hayashi and Chiba 2012). This is true of the tails of their abundance distributions (e.g., Kirby et al. 2010), where an understanding of chemical evolution in dwarf galaxies is essential for studying galaxy formation on small scales. In terms of the MW, PFS will provide a detailed view of the formation of the outer disk for the first time. With the LR mode, PFS is capable of observing stars as faint as $V = 22$, going well beyond the reach of reliable distances from *Gaia* (Gaia Collaboration et al. 2016a; Gaia Collaboration et al. 2018) and spectroscopically-derived stellar parameters and abundances from APOGEE (Majewski et al. 2017). The MW exhibits various stellar overdensities in its outer disk (Price-Whelan et al. 2015; Li et al. 2017; Sheffield et al. 2018) that are likely connected to the ongoing merger between the Sagittarius dwarf galaxy and the MW (Antoja et al. 2018; Laporte et al. 2018; Bland-Hawthorn et al. 2019; Laporte et al. 2020).

M31 particularly stands much to gain in the era of PFS. The current state of the art for deep ($\gtrsim$5 hour) spectroscopic surveys in M31 has been achieved by the Elemental

Abundances in M31 collaboration (Escala et al. 2019; Gilbert et al. 2019a; Escala et al. 2020a; Escala et al. 2020b, K. M. Gilbert et al., submmited) using the 16' × 4' DEIMOS field of view. This has yielded 230 new [$\alpha$/Fe] and [Fe/H] measurements for M31 member stars. This sample size will be dwarfed by the thousands of deep spectra obtained by PFS with an unparalleled coverage of $\gtrsim$65 deg$^2$. At least 22 nights out of the planned 300-360 nights for the entire PFS program are intended for dedicated observations of M31. At least 34 fields will be observed along the minor axis of M31 for ~5 hours per field, extensively probing the stellar halo, using both the LR and MR mode. Spectroscopic fields targeting the outer disk and nearby spiral galaxy M33 may also be planned in an expansion of the M31 halo survey. In addition, Subaru Hyper-Suprime Cam (HSC; Miyazaki et al. 2002) photometry and narrow-band (NB515) imaging have been obtained for each of the fields to identify M31 RGB stars for spectroscopic follow-up with PFS.

With the combined power of PFS and HSC, we will be able to study M31's halo formation on a scale comparable to what is currently achievable in the MW. PFS will provide more complete answers to the questions that I have partially addressed in this thesis: Are there chemical differences between M31's inner and outer halo (Escala et al. 2020a; K.M Gilbert et al. 2020, accepted)? What are the properties of the progenitors of M31's stellar halo (Gilbert et al. 2019a; Escala et al. 2020a; Escala et al. 2020b)? How do the properties of M31's halo relate to its surviving satellite galaxy population (Vargas, Geha, and Tollerud 2014; Escala et al. 2020a; Kirby et al. 2020; Wojno et al. 2020; Escala et al. 2020b)? What is the relationship between M31's halo and its extended disk (Escala et al. 2020a)? And perhaps first and foremost, how does M31 compare to the MW?

**The Maunakea Spectroscopic Explorer**

As discussed in the previous subsection, one of the primary limitations of modern multi-object spectrographs is the relatively small field-of-view (~0.02 deg$^2$) of our most powerful instruments, such as DEIMOS. Thus, spectroscopic surveys of distant galaxies that span a large area on the sky, such as M31, are observationally expensive. For example, over 120 nights on the Keck II telescope have been required over the last ~15 years to cover $\lesssim$2.5% of M31's area within 150 kpc (Gilbert et al. 2019b).

In order to ameliorate this issue, a wide field of view (1.5 deg$^2$), high throughput, massively multiplexed (3249 fibers), large aperture (11.25 m), low- to medium-resolution (R~2000-6000) optical and near-infrared (360-950 nm) spectrograph for chemodynamical studies of resolved stellar populations known as the Maunakea Spectroscopic Explorer (MSE) has been proposed as a refurbishment of the Canada-France-Hawaii-Telescope (McConnachie et al. 2016a; McConnachie et al. 2016b). The advantages of MSE relative to PFS are three-fold: it would (1) serve as a dedicated spectroscopic survey facility, (2) have a higher surveying speed, and (3) have a smaller minimum fiber separation (0.7") that would enable targeting high stellar density regions. At this time, full science operations for MSE are expected to commence in 2026. A dedicated facility such as MSE will function as an essential complement to wide-field imaging surveys such as WFIRST (Section 6.3) and next-generation telescopes such as TMT (Section 6.4).

One of MSE's primary science cases is galactic archaeology in the LG. In addition to its lower resolution modes, MSE will have a high-resolution (R∼20,000-40,0000) capability with over 1000 dedicated fibers that will be used for spectroscopic follow-up of nearby *Gaia* targets in the MW. *Gaia* will be unable to measure chemical abundances and radial velocities for stars fainter than $V \sim 12$ and $V \sim 17$, respectively, thus a ground-based spectroscopic facility such as MSE is essential to supplement the survey. Pushing further out into the Galaxy, MSE will go beyond 4-m class telescopes ($V \lesssim 16$) to perform chemical tagging in the outer components ($V \sim 20$) of the MW—the outer disk, the thick disk, and the stellar halo—with lower resolution modes, similarly to PFS. MSE additionally plans to target MW dwarf galaxies and faint streams in the Galactic halo to probe the distribution of dark matter.

A massively multiplexed and efficient spectrograph such as MSE will revolutionize our understanding of the formation history of M31 and M33 by providing unprecedented contiguous spectroscopic coverage across their halos to at least half their virial radii. In this regard, MSE's greatest contributions will be complete, magnitude-limited censuses of the dense, inner regions of M31's spheroid ($\lesssim$5 kpc), compact dwarf spheroidal galaxies embedded in its halo, and the sparse outer halo (40-150 kpc). At small M31-centric radii, multiplexing is necessary to overcome crowding effects, and at large radii, completeness is necessary to overcome significant contamination by MW foreground stars. In comparison to previous Keck/DEIMOS surveys, this would require ∼15 nights for the inner regions and ∼150 nights for the outer regions over the span of 5−10 years (McConnachie et al. 2016b). With measurements of [Fe/H] and [$\alpha$/Fe] for thousands of individual RGB stars across each structural component of M31, MSE will likely provide the definitive, holistic deconstruction of M31's merger history.

A unique component of MSE's galaxy formation science case is a spectroscopic survey of LV galaxies with $M_\star \gtrsim 10^5 M_\odot$ out to ~100 Mpc. Nearby galaxies comparable in mass to the MW, M31, and M33 within $z \lesssim 1$ could be studied in more detail than previously possible. Mapping stellar structure in these galaxies will allow the full realization of the goals of near-field cosmology by smoothly connecting high-redshift galaxies to $z = 0$ galaxies in the context of ΛCDM.

**MOONS: The Multi-Object Optical and Near Infrared Spectrograph**

MOONS is a planned multi-object optical and near-infrared (0.8-1.8 $\mu$m) spectrograph on the 8.2-m Very Large Telescope (VLT) in Cerro Paranal, Chile (Cirasuolo et al. 2014). Compared to planned spectrographs such as PFS and MSE on 8-10 m class telescopes, MOONS will have a smaller (25 arcmin diameter) field of view and less multiplexing (1000 fibers) with an intermediate minimum fiber separation (10”). Similar to MSE, MOONS will operate with both low- ($R \sim 4000-6000$) and high- ($R \sim 9000 - 18000$) resolution modes. Within its first 5 years of operation, MOONS will deliver spectra for over 3 million MW stars. MOON's key strength is high-sensitivity, high-resolution, near-infrared spectroscopy. The spectrograph is designed to have a limiting magnitude of $H \sim 17.5$ with relatively high S/N ($\gtrsim$60 per pixel) that goes beyond the current capabilities of the near-infrared APOGEE survey (Majewski et al. 2017). MOONS's lasting impact on galactic archaeology will likely be chemodynamic investigations of the relatively faint, highly obscured Galactic bulge. The bulge will be inaccessible for predominately optical spectroscopic surveys, such as WEAVE and 4MOST.

**Spectroscopic Surveys on 4-m Class Telescopes**

PFS and MSE will propel spectroscopic studies of M31 into the future. However, these are only two planned spectroscopic surveys of several targeting resolved stellar populations in the LG. Many of the remainder of these surveys rely on 4-m class telescopes with limiting magnitudes similar to *Gaia* ($V \sim 20$). Thus, these surveys will be unable to reach distant LG galaxies such as M31. However, they will produce radial velocity and chemical abundance measurements for millions of stars in the MW system. The following are planned spectroscopic surveys on 4-m class telescopes:

- **4MOST** (4-m Multi-Object Spectroscopic Telescope) is a planned dedicated spectroscopic facility on the 4-m VISTA telescope in the Southern Hemisphere with a wide (2.5 deg) diameter field of view (de Jong et al. 2014). 4MOST will be equipped with both high- ($R \gtrsim 18,500$) and moderate-resolution ($R \sim 4000-7800$) optical, fiber-fed spectrographs capable of observing 2436 targets in combination. First light is anticipated in 2022. Within 5 years, 4MOST will cover the majority of Southern sky in 2-3 epochs to obtain spectra for millions of *Gaia* targets with $V \lesssim 16$ in the MW's bulge, disk, and halo. Perhaps most interestingly, 4MOST will obtain ~500,000 spectra in the Magellanic Clouds to probe their origin and gain insight into the minor merging process.

- **WEAVE** (WHT Enhanced Area Velocity Explorer) is a planned multi-object survey spectrograph on the 4.2-m William Herschel Telescope (WHT) in the Canary Islands (Dalton et al. 2014). WEAVE's wide field of view (2 deg diameter), fiber-fed (1000 fibers) optical spectrograph has options for both moderate- ($R \sim 5000$) and high- ($R \sim 20,000$) resolution modes. Currently, on-sky operations are expected to begin in 2021. As a Northern Hemisphere facility, WEAVE is designed to be complementary to 4MOST. The highest-impact component of its galactic archaeology subsurvey is anticipated to focus on the MW's outer disk.

- **DESI** (Dark Energy Spectroscopic Instrument) is a dedicated spectroscopic survey facility on the 4-m Mayall telescope at Kitt Peak National Observatory with a 3 deg diameter field of view (DESI Collaboration et al. 2016). DESI's 10 identical 500-fiber spectrographs will have three arms in the blue, red, and near-infrared spanning spectral resolving powers of $R \sim 2000-5000$ with universally high throughput (~70%). With commissioning beginning in late 2019, full operations are anticipated in the very near future. DESI aims to probe the nature of dark energy by obtaining spectra for tens of millions of high-redshift ($z \lesssim 2$) galaxies over the course of 5 years. However, DESI has a planned stellar spectroscopy component targeting 10 million stars in the MW's disk and halo in a complement to *Gaia*. The prime advantage of DESI over surveys such as 4MOST and WEAVE is its impressively high multiplexing, wider field of view, and enhanced surveying speed.

- **SDSS-V** (Sloan Digital Sky Survey V) is the first multi-epoch, panoptic, homogeneous, all-sky spectroscopic survey. It couples novel robotic fiber positioning systems with existing 2.5-m, wide field-of-view (~2-3 deg diameter) facilities at the Apache Point Observatory and Las Campanas Observatory to obtain dual-hemisphere observations between 2020-2025 (Kollmeier et al. 2017; Kollmeier et al. 2019). The near-infrared APOGEE ($R \sim 22,000$; Majewski et al. 2017; Wilson et al. 2019) and optical BOSS ($R \sim 2000$; Smee et al. 2013; Dawson et al. 2013) spectrographs will be fiber-fed to target 300 and 500 targets simultaneously. The Milky Way Mapper (MWM) component of SDSS-V will obtain multi-object spectra of over 5 million stars in the LG. In particular, the Galactic Genesis subprogram will contiguously and densely map stellar populations at low Galactic latitudes in the MW's disk. A particularly exciting component of SDSS-V is the planned LV Mapper (LVM) with smaller aperture ($\lesssim$1.0 m) telescopes. The LVM will provide low spectral resolution ($R \sim 4000$), optical integral field unit (IFU) spectroscopy on 20 pc scales in M31 and M33 and 10 pc scales in the Magellanic Clouds. The unrivaled area coverage of the LVM will extend from the LG to the LV, a factor of $10^{3-4}$ greater than that of current IFU data sets.

## 6.3 The Power of Wide-Field Imaging Surveys

With augmented spectroscopic coverage of the LG comes the need for more expansive imaging. Whereas spectroscopy provides kinematical and chemical information, photometry of resolved stellar populations will be necessary to measure proper motions and distances to stellar structures in the outer MW halo and LG galaxies like M31. Deeper imaging surveys will also enable the discovery of ultra-faint dwarfs throughout the LG, and open the doors to obtaining globally representative, ancient star formation histories for distant galaxies. In this Section, I provide an overview of planned and proposed wide-field imaging surveys of the LG.

### The Wide Field Infrared Survey Telescope

WFIRST (The Wide Field Infrared Survey Telescope) is an infrared (0.5-2.0 $\mu$m) space-based telescope with a 2.4-m diameter mirror operated by NASA. The telescope is planned for launch in the mid-2020s with a minimum 5 year mission duration. WFIRST will have a field of view (45×23 arcmin, or 0.28 deg$^2$) on its Wide Field Imager (WFI) instrument that is 100 times larger than its predecessor while maintaining similar sensitivity and resolution. For example, the PHAT (Dalcanton et al. 2012) survey required 423 pointings of HST/WFC3 to cover M31's northeastern disk. In comparison, it will take WFIRST only 2 pointings to cover the same area. Furthermore, WFIRST will probe M31's halo and disk down to the oldest main sequence turnoff over an unprecedented area compared to previous pencil-beam deep fields (Brown et al. 2006; Brown et al. 2007; Brown et al. 2008) to derive globally representative, ancient star formation histories (Dolphin 2002; Weisz et al. 2014).

Thus, WFIRST will be transformative for photometric studies of M31. With minimal effort, WFIRST could provide the first contiguous coverage of M31's southwestern disk, of which little is known (e.g., Bernard et al. 2012). The anticipated impact of WFIRST in the southwest will be at least equivalent to that of PHAT in the northeast. The PHAT team discovered a 2-4 Gyr old global burst of star formation (Williams et al. 2015) and established that much of M31's disk was in place by ~10 Gyr ago (Williams et al. 2017). PHAT also enabled stellar spectroscopy in M31's disk by providing a catalog of resolved targets (Dorman et al. 2012), which led to discoveries of its unusually steep age-velocity dispersion relationship and rotating inner spheroid (Dorman et al. 2013; Dorman et al. 2015). Perhaps most significantly, WFIRST will provide complete coverage of M32 (e.g., Monachesi et al. 2011), a compact dwarf elliptical galaxy that is the proposed remnant of a massive galaxy in one manifestation of a major merger scenario (D'Souza and Bell 2018a).

WFIRST will also deliver the first precise, large-scale proper motion and distance measurements in M31. Proper motion measurements are essential to probe the dynamics of satellite galaxies and stellar substructure. Although bulk proper motion measurements have been measured for M31 using multi-epoch HST imaging of RGB stars (van der Marel et al. 2012; Sohn, Anderson, and van der Marel 2012) and Gaia DR2 imaging of supergiants in M31's disk (van der Marel et al. 2019), measurements for M31's halo and satellite galaxies like M32 are still lacking. The combination of WFIRST proper motion and distance measurements with heliocentric velocities measured from stellar spectroscopy will provide 6D phase space information in M31 for the first time. This will empower us to answer key questions regarding the nature of (1) the origin of the mysterious second kinematically cold component of the GSS (Gilbert et al. 2007; Fardal et al. 2007; Gilbert et al. 2009), (2) the orbit and dynamical mass of the GSS progenitor (e.g., Fardal et al. 2007; Fardal et al. 2013), (3) the currently unknown transverse motion of M32 (D'Souza and Bell 2018a), and (4) whether the inner halo contains rotation signatures characteristic of an *in-situ* stellar component (Dorman et al. 2013). In addition to addressing these important questions, WFIRST will enable the discovery of previously undetected, faint streams in M31's halo and the partial reconstruction of accretion and luminosity functions of its progenitor galaxies.

Other WFIRST initiatives in the LG include obtaining precise astrometry for MW

dwarf satellite galaxies. WFIRST will exceed both *Gaia* and LSST in its astrometric precision at a given distance (Sanderson et al. 2017), enabling usable proper motion measurements in LG dwarf galaxies beyond the current limit of ∼400 kpc (Fritz et al. 2018), particularly when combined with long-baseline HST imaging. With these measurements, WFIRST will place constraints on the nature of dark matter from the orbital and internal motions of dwarf galaxies. A complementary effort in the MW's halo will be the mapping of substructure down to the oldest main sequence turnoff out the MW's virial radius (Sanderson et al. 2017). WFIRST will also provide a detailed view of the highly-extincted, crowded Galactic bulge. Beyond the LG, WFIRST will provide resolved images of RGB stars in the stellar halos of MW-like LV galaxies within ∼10 Mpc (Williams et al. 2019).

**The James Webb Space Telescope**

JWST (The James Webb Space Telescope) is the formal successor to HST (Gardner et al. 2006). With a 6.5-m primary mirror, JWST will have ∼7 times the light collecting power of the 2.4-m HST. Unlike HST, JWST specializes in infrared observations (0.6-28 $\mu$m) to provide a new view into the faint, high-redshift universe and insight into the first galaxies. The Near Infrared Camera (NIRCam) will serve as JWST's primary imager, detecting light between 0.6-5 $\mu$m in nearby galaxies. Although NIRCam's field-of-view (129"×129") will be comparable to HST/WFC3 (160"×160"), the instrument will be complementary to the wide field-of-view of WFIRST/WFI (Section 6.3) in that it will provide higher resolution. Much like WFIRST, JWST will enable proper motion measurements for LG galaxies. The sensitivity of JWST will provide the first look at the oldest main-sequence turn-off of resolved stellar populations outside of the LG. Within 100 hours of integration, JWST can achieve this for galaxies as far as ∼3 Mpc, providing a view into the most ancient part of their star formation histories (Weisz and Boylan-Kolchin 2019). Currently, JWST is planned to launch in 2021 with a mission duration of 5-10 years.

**The Large Synoptic Survey Telescope**

LSST (The Large Synoptic Sky Telescope, or Vera C. Rubin Observatory) is a pioneering optical survey facility with an 8.4-m (6.5-m effective) primary mirror and wide field of view camera (3.5 deg) in Cerro Pachón, Chile, with science operations scheduled to commence in 2022 (Ivezić et al. 2019). Exploring the structure and formation of the MW system through its resolved stellar populations is one of the key goals of LSST. With an anticipated survey duration of 10 years,

LSST is expected to yield photometry for a total of ~20 billion resolved stars over 18,000 $deg^2$ down to $g = 27.4$ and $i = 26.8$ as part of its Wide-Fast-Deep survey. LSST will map over $10^3$ times the volume of previous optical imaging surveys.

By achieving this depth and coverage, LSST will probe the distribution of main sequence turn-off stars in the MW's halo to 300 kpc (and their photometric metallicity distributions to 100 kpc), producing a high-fidelity spatial reconstruction of the galaxy and its stellar streams out to the viral radius. This will place the first stringent constraints on outer halo predictions from hierarchical galaxy formation models (e.g., Bullock and Johnston 2005; Cooper et al. 2010). In contrast to WFIRST ($\gtrsim 25$ $\mu$as $yr^{-1}$; Section 6.3), LSST will not have the necessary precision (0.5 mas $yr^{-1}$) to measure proper motions for the MW outer halo and the LG beyond (Sanderson et al. 2019). However, LSST's optical view of the MW system will be highly synergistic with WFIRST's NIR photometry.

In addition, LSST is expected to double the number (>50) of known dwarf galaxies in the LG (Drlica-Wagner et al. 2019). In the next decade, LSST should discover ultra-faint dwarf galaxies ($M_V = -6$) throughout the LG (~1 Mpc), and classical dwarf galaxies ($M_V = -8.5$) into the LV (~5 Mpc) (Simon et al. 2019). In this way, LSST will enable detailed investigations of the nature of dark matter, addressing questions regarding the minimum dark matter halo mass (down to at least $M_h \sim 10^8 M_\odot$; Drlica-Wagner et al. 2019) for galaxy formation. Follow-up spectroscopy with next-generation ground-based telescopes (Section 6.4) will be necessary to probe dark matter microphysics in these faint galaxies.

**The Large UV / Optical / Infrared Surveyor**

LUVOIR (The Large UV/Optical/IR Surveyor) is a mission concept for a space-based observatory designed for multi-purpose astronomy proposed for launch in the mid-2030s (The LUVOIR Team 2019). As discussed in Sections 6.3, studies of resolved stellar populations for nearby galaxies is currently only possible in the LG. Currently, two different architectures are being considered for LUVOIR—an 8-m and 15-m diameter primary mirror, both of which would far surpass any other existing or planned space-based telescope. Either case would be equipped with a High Definition Imager with a wide 2×3 arcmin field of view that would enable high-precision astrometry within the LG and into the LV. With an 8-m (15-m) mirror, LUVOIR could reach the oldest main-sequence turn-off out to $\lesssim 5$ (10) Mpc for ~500 (1000) known galaxies (Weisz and Boylan-Kolchin 2019). Because SFH studies of this depth are the most observationally expensive, LUVOIR would be able to perform studies with a similar depth as PHAT (Dalcanton et al. 2012) for galaxies as far as ∼4 (7) Mpc (Gilbert et al. 2019b). In this way, LUVOIR would revolutionize our understanding of galaxy formation by enabling detailed studies (comparable to what is currently possible in M31) for a diverse sample of galaxies in the LV.

## 6.4 The Next Generation of Ground-Based Telescopes

This thesis has illustrated that we are currently pushing the limits of what is possible with the 10-m Keck telescopes, the most powerful ground-based facilities in the world, for stellar spectroscopy of resolved stellar populations. To move beyond the LG (and improve data quality within the LG), the survey and instrument facilities discussed in Section 6.2 will inevitably fall short. However, 30-m class telescopes would provide ∼2-3 extra magnitudes of depth into the universe, bringing a plethora of new galaxies at Mpc distances into our grasp. These extremely large telescopes will also facilitate statistically meaningful chemical abundance studies, which extend from the RGB down to main-sequence turn-off stars, for many galaxies in the LG. The future of near-field cosmology demands 30-m class ground-based telescopes such as TMT, GMT, and the ELT:

- **TMT** (The Thirty Meter Telescope) is a planned next-generation telescope with a 30-m diameter primary mirror (constructed from 492 segments) located on the summit of Maunakea, Hawaii, USA (Sanders 2013). TMT's first light is anticipated in 2027. With a light collecting power 9 times that of the Keck telescopes, TMT will achieve an equivalent S/N at a given magnitude 9 times faster, and will increase S/N at a given magnitude by a factor of 3 for an equivalent exposure time. As part of its first-light instrument suite, the Wide Field Optical Spectrometer (WFOS) is being developed (Pazder et al. 2006). WFOS is a near-UV and optical (500-1000 nm), low-resolution ($R \sim 1500-3500$), slit-based multi-object spectrograph with a relatively small 8×3 arcmin field of view capable of observing ∼50-80 objects simultaneously. In this way, WFOS will be similar to the low-resolution grating configuration of Keck/DEIMOS (with fewer possible targets per slitmask), except on a larger telescope. A high-resolution optical spectrograph (HROS) is being considered as a second-generation instrument for TMT.

  A significant component of TMT's science case is spectroscopic studies of the MW and nearby galaxies (see TMT Detailed Science Case 2015) to probe their formation histories and dark matter distributions. Because TMT is the only 30-m telescope planned in the Northern Hemisphere, it has the potential to revolutionize studies of resolved stellar populations in exclusive Northern Hemisphere targets like M33 and M31. These galaxies will be inaccessible to both GMT and the ELT. Thus, TMT is *essential* to advance M31 system studies post-2030, following the conclusion of industrial-scale spectroscopic surveys (Section 6.2).

- **GMT** (The Giant Magellan Telescope) is a planned next-generation, optical and near-infrared telescope located in Cerro Los Campanas, Chile (Johns et al. 2012; Bernstein et al. 2014). The primary mirror of GMT will consist of seven individual 8.4-m diameter segments that produce a combined 24.5-m effective diameter, or 6 times the light collecting power of the Keck telescopes. The GMT collaboration aims to achieve first light in 2029 and will likely be the first 30-m class telescope to be on-sky. A significant advantage of GMT is that it will be the only 30-m class telescope to include both low- and high-resolution optical spectrographs as part of its first-light instrument suite.

  GMT will be equipped with GMACS (GMT Multi-object Astronomical and Cosmological Spectrograph), an optical and near-infrared (3200 - 10000 Å), slit-based, multi-object, moderate resolution ($R \sim 1000 - 6000$), wide field of view (7.4 deg diameter) spectrograph that will be used to study resolved stellar populations in nearby galaxies (see GMT Science Book 2018). With GMACS, GMT will extend M31-like chemical abundance studies of individual stars at the tip of the red branch beyond the LG to ~2 Mpc. Additionally, GMT will have an optical echelle spectrograph (G-CLEF, the GMT-Consortium Large Earth Finder) that can be used for high-resolution ($R \sim 20,000 - 35,000$) chemical abundance studies of *Gaia* targets in the MW. Furthermore, G-CLEF can be coupled with MANIFEST (The Many-Instrument Fiber System) to achieve high multiplexing. With G-CLEF, GMT will open the door to the origins of ultra-faint dwarf galaxies via detailed chemical abundance studies for large samples of their faint stars (e.g., Ji et al. 2019).

- **ELT** (The European Extremely Large Telescope) is a planned, novel optical and near-infrared ground-based telescope located in Cerro Amazones, Chile (Gilmozzi and Spyromilio 2007). Its enormous 39-m primary mirror diameter will consist of 798 1.4-m wide segments. With a light collecting power

~15 times that of the Keck Telescopes, the ELT will revolutionize our understanding of the universe upon its scheduled first light in 2025. Although the ELT is designed primarily for detailed studies of the first galaxies and exoplanet characterization (Evans et al. 2015), its multi-object spectrograph, MOSAIC, will lead to significant progress in LV galactic archaeology with its optical, high multiplex mode (0.45-0.8$\mu$m, ~200 fibers, $R \sim 5,000, 15,000$, 32' field-of-view). The primary difference between MOSAIC and its competitors, TMT/WFOS and GMT/GMACS, is that it will provide more extensive NIR coverage in the relevant mode (0.8-1.8$\mu$m) and have higher sensitivity as a consequence of the ELT's huge diameter. The ELT will also be equipped with a high-fidelity, high-resolution ($R \sim 100,000$) optical and near-infrared (0.4-1.8 $\mu$m) echelle spectrograph (HIRES) capable of delivering high-S/N spectra for cool dwarfs in the Galactic bulge and giants in the halo and beyond (Maiolino et al. 2013). Unfortunately, MOSAIC and HIRES are second-generation instruments that will not be on-sky for first-light. However, when MOSAIC is finally on-sky, chemodynamic studies of M31-like galaxies with the ELT will reach ~1 Mpc further than GMT/GMACS.

## 6.5 Next Steps for Simulations

In this chapter, I have focused extensively on upcoming and proposed observational facilities (Sections 6.2, 6.3, 6.4). However, having the appropriate theoretical context in which to interpret these novel observations is equally meaningful for the future of near-field cosmology. For example, recent studies of simulated MW-analogs have shown that the accretion of a massive galaxies on preferentially radial orbits can reproduce observations of the MW halo while placing constraints on the nature of the Gaia-Enceladus merger remnant (Mackereth et al. 2019b; Fattahi et al. 2019; Bignone, Helmi, and Tissera 2019; Grand et al. 2020; Elias et al. 2020). Furthermore, the observed dichotomy in the [$\alpha$/Fe]-[Fe/H] plane of the MW disk may also be a consequence of this merger (Vincenzo et al. 2019; Brook et al. 2020; Grand et al. 2020). In the context of M31, simulations have brought to the forefront the hypothesis that the GSS formed in a recent major merger that significantly impacted M31's evolution by reproducing key observed properties of the stellar halo and disk (Fardal et al. 2008; Hammer et al. 2018; D'Souza and Bell 2018a).

Theoretical astrophysics has also guided the direction of observations when predictions from each sector have conflicted. For example, one of the primary science drivers for the construction of PFS (Section 6.2) is to resolve the core-cusp dis-crepancy between observations of dark-matter dominated dwarf galaxies in the LG (Battaglia et al. 2008b; Walker and Peñarrubia 2011; Amorisco and Evans 2012; Peñarrubia et al. 2012) and predictions from the CDM paradigm. Additionally, LSST (Section 6.3), in combination with spectroscopic follow-up from powerful extremely large telescopes (Section 6.4), may provide the definitive answer to whether the missing satellites problem (Klypin et al. 1999; Bullock, Kravtsov, and Weinberg 2000; Kazantzidis et al. 2004; Simon and Geha 2007; Strigari et al. 2007; Tollerud et al. 2008) is indeed a challenge for CDM. Via the extension of hierarchical assembly to small scales, simulations have predicted the existence of tiny satellites of dwarf galaxies (Sales et al. 2013; Wetzel, Deason, and Garrison-Kimmel 2015; Deason et al. 2015; Wheeler et al. 2015; Sales et al. 2017), which may in fact be the case for ultra-faint dwarf galaxies near the Magellanic Clouds (Drlica-Wagner et al. 2015; Jethwa, Erkal, and Belokurov 2016; Kallivayalil et al. 2018; Fritz et al. 2019; Patel et al. 2020). In the case of the M31 system, one of the leading major merger scenarios (D'Souza and Bell 2018a) for the formation of GSS has predicted that M32 is the remnant core of the massive progenitor, thereby triggering observational efforts to confirm or refute this scenario (HST ID15658, PI T. Sohn).

To make theoretical progress on the connection between hierarchical galaxy formation and chemical abundances, a majority of hydrodynamical simulations will need to include explicit tracking of metals beyond total metallicity, as in the FIRE project (Hopkins et al. 2014; Hopkins et al. 2018), while incorporating realistic metal mixing and injection schemes (Escala et al. 2018; Emerick et al. 2018). The lack of simulations modeling the formation of the GSS or M31 in its entirety that also track, e.g., $\alpha$-element abundances, has placed limitations on the full utilization of the recent flood of chemical abundance data in the M31 system (Gilbert et al. 2019a; Escala et al. 2019; Escala et al. 2020a; Kirby et al. 2020; Wojno et al. 2020; Escala et al. 2020b; K.M. Gilbert et al. 2020, accepted). At the same time, there is a need for flexible, yet fully cosmological, hydrodynamical models of galactic chemical evolution that allow for user post-processing to incorporate uncertainties in yield models (e.g., A. Emerick et al., in preparation), with the goal of further enabling detailed comparisons to observations.

Furthermore, hydrodynamical simulations will need to push to even higher mass and spatial resolution and be simultaneously capable of properly sampling the initial mass function to produce realistic stellar populations (e.g., Su et al. 2018; Wheeler et al. 2019; Emerick, Bryan, and Mac Low 2019). This is particularly important in the context of resolving both host galaxies and surrounding dwarf galaxies (Wetzel et al. 2016; Garrison-Kimmel et al. 2019), which are the building blocks of MW-like halos (e.g., Bullock and Johnston 2005; Font et al. 2011; Cooper et al. 2010; Deason et al. 2018; D'Souza and Bell 2018b; Monachesi et al. 2019). Additionally, large suites of low-resolution simulations of $L_\star$ galaxies with *diverse* merger histories are particularly useful for identifying candidates for re-simulation at higher-resolution to investigate specific formation scenarios (e.g., Hammer et al. 2010; Bignone, Helmi, and Tissera 2019). This is particularly relevant for M31, which may have experienced a recent ($z \lesssim 1$), significant merger (Hammer et al. 2018; D'Souza and Bell 2018a). Because some simulation suites are effectively tuned to reproduce $L_\star$ galaxies with accretion histories more similar to that inferred for the MW (e.g., Griffen et al. 2016), they do not provide suitable M31 analogs. Similarly, realistic synthetic observations constructed from hydrodynamical simulations are becoming increasingly important in the era of industrial-scale spectroscopic surveys to achieve maximally useful comparisons (e.g., Grand et al. 2018; Sanderson et al. 2020). In combination with existing and anticipated chemical abundance catalogs in M31 from PFS and MSE (Section 6.2), future simulations that incorporate many of the above features would indeed transform our understanding of the formation history of M31.

*Appendix A*

# CATALOG OF STELLAR PARAMETERS AND ABUNDANCES IN M31

Stellar parameters and elemental abundances of individual M31 RGB stars for the 9 kpc halo (f109_1), 12 kpc halo (H), 18 kpc halo (f123_1), 23 kpc halo (f130_1), 31 kpc halo (a0_1), 22 kpc GSS (S), 33 kpc GSS (a3), and 26 kpc disk (D) fields are presented in Table A.1. The table includes data for 200 M31 RGB stars with reliable [Fe/H] and [$\alpha$/Fe] measurements, in addition to 144 M31 RGB stars that only have reliable [Fe/H] measurements. Stars with signatures of TiO absorption in their spectra have been omitted from final abundance samples and were not included in this table. The errors presented for $T_{\rm eff}$ represent only the random component of the total uncertainty. However, the errors for [Fe/H] and [$\alpha$/Fe] include systematic components that account for errors propagated by inaccuracies in $T_{\rm eff}$ (Kirby, Guhathakurta, and Sneden 2008; Gilbert et al. 2019a; Escala et al. 2019; Escala et al. 2020a). Abundance measurements for the 17 kpc GSS field (f207_1) are presented in Gilbert et al. (2019a).

Table A.1: Stellar Parameters and Elemental Abundances of M31 RGB Stars

| Object | Sky Coordinates | | $v_{helio}$ | S/N | $T_{eff}$ | $\delta(T_{eff})$ | log $g$ | [Fe/H] | $\delta$([Fe/H]) | [$\alpha$/Fe] | $\delta$([$\alpha$/Fe]) |
|---|---|---|---|---|---|---|---|---|---|---|---|
| ID | RA | Dec | (km $s^{-1}$) | (Å $^{-1}$) | (K) | (K) | (dex) | (dex) | (dex) | (dex) | (dex) |
| 9 kpc Halo Field (f109_1) | | | | | | | | | | | |
| 1092408 | 00h45m44.35s | +40d54m15s | -281.6 | 15 | 3883 | 5 | 0.64 | -0.96 | 0.13 | 0.23 | 0.27 |
| 1095375 | 00h45m49.45s | +40d58m53.8s | -190.4 | 12 | 4082 | 10 | 1.06 | -0.83 | 0.14 | 0.39 | 0.38 |
| 1092785 | 00h45m44.27s | +40d54m52.2s | -288.9 | 13 | 3975 | 6 | 0.77 | -0.56 | 0.13 | 0.68 | 0.26 |
| 1094900 | 00h45m41.42s | +40d58m13s | -309.0 | 13 | 3993 | 30 | 0.97 | -0.32 | 0.17 | ... | ... |
| 1092808 | 00h45m41.94s | +40d54m51.3s | -170.1 | 14 | 4345 | 6 | 0.95 | -2.86 | 0.2 | ... | ... |
| 1098472 | 00h45m59.12s | +41d03m45.4s | -318.2 | 21 | 4127 | 8 | 0.72 | -1.12 | 0.13 | ... | ... |
| 1095109 | 00h45m39.51s | +40d58m31.4s | -380.1 | 18 | 4129 | 5 | 0.69 | -0.92 | 0.13 | 0.2 | 0.25 |
| 1092223 | 00h45m44.03s | +40d53m58s | -490.1 | 19 | 3773 | 4 | 0.54 | -0.81 | 0.13 | 0.27 | 0.25 |
| 1095905 | 00h45m43.89s | +40d59m40.6s | -189.2 | 43 | 3799 | 6 | 0.28 | -2.4 | 0.13 | 0.02 | 0.19 |
| 1093016 | 00h45m35.43s | +40d55m10.2s | -466.5 | 17 | 4329 | 5 | 0.85 | -1.58 | 0.14 | 0.3 | 0.28 |
| 1096662 | 00h45m53.44s | +41d00m41.6s | -349.5 | 19 | 3800 | 4 | 0.67 | -1.05 | 0.13 | 0.3 | 0.24 |
| 1096241 | 00h45m50.18s | +41d00m11.8s | -365.2 | 15 | 4393 | 8 | 0.94 | -0.97 | 0.14 | -0.02 | 0.35 |
| 1093749 | 00h45m35.38s | +40d56m15.4s | -190.5 | 13 | 3940 | 7 | 0.88 | -1.14 | 0.14 | 0.26 | 0.36 |
| 1097703 | 00h45m53.65s | +41d01m37s | -427.3 | 14 | 4014 | 7 | 0.99 | -1.2 | 0.14 | 0.46 | 0.35 |
| 1095318 | 00h45m41.41s | +40d58m50.6s | -275.2 | 10 | 3912 | 6 | 0.76 | -0.42 | 0.14 | -0.12 | 0.34 |
| 1098573 | 00h45m59.66s | +41d03m36.5s | -246.1 | 11 | 4259 | 13 | 0.97 | -0.88 | 0.14 | ... | ... |

*(Table A.1 Continued)*

*(Table A.1 Continued)*

| Object ID | Sky Coordinates RA | Dec | $v_{helio}$ (km $s^{-1}$) | S/N (Å $^{-1}$) | $T_{eff}$ (K) | $\delta(T_{eff})$ (K) | log $g$ (dex) | [Fe/H] (dex) | $\delta$([Fe/H]) (dex) | [$\alpha$/Fe] (dex) | $\delta$([$\alpha$/Fe]) (dex) |
|---|---|---|---|---|---|---|---|---|---|---|---|
| 1092588 | 00h45m31.45s | +40d54m29.2s | -390.0 | 16 | 4068 | 5 | 0.65 | -1.24 | 0.14 | ... | ... |
| 1097891 | 00h45m54.69s | +41d01m38.6s | -257.6 | 16 | 3789 | 5 | 0.74 | -0.92 | 0.14 | 0.25 | 0.31 |
| 1096296 | 00h45m50.53s | +41d00m14.9s | -488.2 | 18 | 3771 | 4 | 0.62 | -1.12 | 0.14 | 0.52 | 0.25 |
| 1094764 | 00h45m47.75s | +40d58m03.1s | -341.2 | 14 | 3702 | 4 | 0.62 | -1.12 | 0.14 | -0.3 | 0.22 |
| 1095722 | 00h45m51.44s | +40d59m28s | -260.9 | 18 | 4172 | 5 | 0.77 | -1.25 | 0.14 | 0.8 | 0.24 |
| 1092357 | 00h45m29.18s | +40d54m04.7s | -227.6 | 11 | 3761 | 7 | 0.74 | -0.37 | 0.14 | 1.0 | 0.25 |
| 1095484 | 00h45m52.82s | +40d59m07.8s | -276.0 | 13 | 4344 | 10 | 1.17 | -0.78 | 0.13 | -0.01 | 0.31 |
| 1097368 | 00h45m52.93s | +41d02m00.6s | -317.5 | 15 | 3819 | 3 | 0.85 | -1.3 | 0.14 | ... | ... |
| 1095248 | 00h45m53.71s | +40d58m44.2s | -248.7 | 17 | 3609 | 3 | 0.39 | -0.42 | 0.13 | 0.5 | 0.24 |
| 1098629 | 00h45m43.67s | +41d03m22.8s | -399.4 | 8 | 4072 | 11 | 1.01 | -0.2 | 0.14 | ... | ... |
| 1096986 | 00h45m48.29s | +41d03m36.1s | -637.6 | 12 | 3876 | 5 | 0.81 | -0.26 | 0.14 | 0.57 | 0.36 |
| 1094864 | 00h45m48.40s | +40d58m11.4s | -459.2 | 9 | 4009 | 58 | 0.92 | -0.49 | 0.18 | ... | ... |
| 1099333 | 00h45m55.77s | +41d02m35.7s | -193.6 | 11 | 4083 | 7 | 1.04 | -0.83 | 0.14 | 0.71 | 0.24 |
| 1092369 | 00h45m31.92s | +40d54m08.3s | -265.5 | 11 | 4265 | 8 | 1.03 | -0.32 | 0.13 | -0.1 | 0.34 |
| 1099168 | 00h45m47.62s | +41d02m39.8s | -307.8 | 10 | 3738 | 6 | 0.72 | -0.77 | 0.15 | 0.48 | 0.47 |
| 1094502 | 00h45m47.65s | +40d57m36s | -343.9 | 9 | 4250 | 8 | 1.14 | -0.45 | 0.14 | -0.31 | 0.21 |
| 1096508 | 00h45m42.46s | +41d00m33.5s | -207.5 | 12 | 3930 | 6 | 0.9 | -0.57 | 0.14 | ... | ... |
| 1092174 | 00h45m33.67s | +40d53m48.6s | -472.4 | 13 | 3995 | 6 | 0.9 | -0.95 | 0.14 | 0.6 | 0.26 |

*(Table A.1 Continued)*

*(Table A.1 Continued)*

| Object | Sky Coordinates | | $v_{helio}$ | S/N | $T_{eff}$ | $\delta(T_{eff})$ | log $g$ | [Fe/H] | $\delta$([Fe/H]) | [$\alpha$/Fe] | $\delta$([$\alpha$/Fe]) |
|---|---|---|---|---|---|---|---|---|---|---|---|
| ID | RA | Dec | (km $s^{-1}$) | (Å $^{-1}$) | (K) | (K) | (dex) | (dex) | (dex) | (dex) | (dex) |
| 1098173 | 00h45m48.79s | +41d03m57.9s | -251.2 | 8 | 3869 | 11 | 0.89 | -1.04 | 0.15 | ... | ... |
| 1094933 | 00h45m55.08s | +40d58m19.6s | -218.8 | 11 | 3835 | 6 | 0.68 | -1.63 | 0.15 | ... | ... |
| 1092257 | 00h45m36.89s | +40d53m57.1s | -237.4 | 13 | 3719 | 5 | 0.67 | -0.95 | 0.14 | 0.4 | 0.36 |
| 1095246 | 00h45m45.80s | +40d58m45s | -420.5 | 12 | 4193 | 8 | 0.75 | -0.9 | 0.14 | ... | ... |
| 1093508 | 00h45m33.48s | +40d55m54.2s | -231.6 | 10 | 3691 | 6 | 0.88 | -1.23 | 0.15 | ... | ... |
| 1096037 | 00h45m39.34s | +40d59m51.8s | -218.3 | 9 | 4199 | 11 | 0.99 | -0.69 | 0.15 | 1.09 | 0.43 |
| 1096780 | 00h45m41.01s | +41d00m48.5s | -247.4 | 10 | 3806 | 8 | 0.94 | -1.04 | 0.16 | ... | ... |
| 1097695 | 00h45m39.39s | +41d01m31.9s | -151.6 | 19 | 4121 | 5 | 0.8 | -1.65 | 0.14 | 0.64 | 0.35 |
| 1092679 | 00h45m31.13s | +40d54m36.6s | -319.1 | 13 | 3886 | 5 | 0.75 | -0.75 | 0.14 | -0.06 | 0.42 |
| 1094499 | 00h45m42.91s | +40d57m36.2s | -253.1 | 15 | 3729 | 7 | 0.6 | -1.13 | 0.16 | ... | ... |
| 1095232 | 00h45m40.89s | +40d58m41.3s | -355.9 | 11 | 3932 | 10 | 0.36 | -1.68 | 0.15 | 0.6 | 0.42 |
| 1092674 | 00h45m33.63s | +40d54m37.1s | -195.0 | 10 | 3807 | 6 | 0.83 | -1.09 | 0.15 | ... | ... |
| 12 kpc Halo Field (H) | | | | | | | | | | | |
| 1005969 | 00h46m13.39s | +40d40m10.3s | -177.4 | 13 | 4473 | 88 | 0.98 | -1.06 | 0.14 | ... | ... |
| 1007726 | 00h46m24.18s | +40d41m43s | -370.0 | 13 | 3875 | 7 | 0.81 | -2.92 | 0.27 | ... | ... |
| 1007736 | 00h46m12.93s | +40d41m43.2s | -391.1 | 6 | 4254 | 13 | 1.27 | -2.27 | 0.32 | ... | ... |
| 1009083 | 00h46m19.45s | +40d42m45.4s | -278.4 | 11 | 3569 | 106 | 0.84 | -0.67 | 0.14 | ... | ... |
| 1009202 | 00h46m21.60s | +40d42m49.3s | -551.7 | 10 | 4101 | 877 | 1.33 | -2.1 | 0.23 | ... | ... |

*(Table A.1 Continued)*

*(Table A.1 Continued)*

| Object ID | Sky Coordinates RA | Dec | $v_{helio}$ (km $s^{-1}$) | S/N (Å $^{-1}$) | $T_{eff}$ (K) | $\delta(T_{eff})$ (K) | log $g$ (dex) | [Fe/H] (dex) | $\delta$([Fe/H]) (dex) | [$\alpha$/Fe] (dex) | $\delta$([$\alpha$/Fe]) (dex) |
|---|---|---|---|---|---|---|---|---|---|---|---|
| 1009347 | 00h46m25.17s | +40d42m58.7s | -276.1 | 15 | 3871 | 5 | 0.81 | -1.79 | 0.15 | 0.9 | 0.33 |
| 1009577 | 00h46m25.90s | +40d43m07.2s | -305.4 | 17 | 3965 | 36 | 0.66 | -1.37 | 0.14 | 0.7 | 0.26 |
| 1009789 | 00h46m04.69s | +40d43m14.3s | -317.0 | 14 | 4030 | 2399 | 0.62 | -2.18 | 0.17 | 0.95 | 0.35 |
| 1010571 | 00h46m27.39s | +40d43m56.3s | -268.6 | 15 | 3873 | 5 | 0.71 | -1.77 | 0.14 | 0.18 | 0.38 |
| 1010832 | 00h46m23.89s | +40d44m05.7s | -316.8 | 13 | 3721 | 213 | 0.84 | -1.66 | 0.15 | 0.6 | 0.36 |
| 1011676 | 00h46m28.60s | +40d44m48.2s | -145.9 | 14 | 3745 | 185 | 0.91 | -1.25 | 0.15 | ... | ... |
| 1012330 | 00h46m29.11s | +40d45m22.1s | -272.4 | 10 | 3989 | 11 | 1.01 | -1.49 | 0.15 | ... | ... |
| 1012487 | 00h46m24.96s | +40d45m28.5s | -309.2 | 9 | 4502 | 509 | 1.33 | -2.5 | 0.3 | ... | ... |
| 1013217 | 00h46m33.33s | +40d46m23.6s | -284.4 | 9 | 4147 | 10 | 1.03 | -0.31 | 0.14 | 0.54 | 0.3 |
| 1152624 | 00h47m14.16s | +40d48m09s | -329.7 | 11 | 3844 | 17 | 0.89 | -2.0 | 0.22 | ... | ... |
| 1162193 | 00h46m50.17s | +40d44m25.5s | -354.7 | 16 | 4351 | 110 | 1.01 | -1.8 | 0.17 | ... | ... |
| 1162211 | 00h46m51.20s | +40d44m28.3s | -225.5 | 23 | 3881 | 42 | 0.66 | -1.49 | 0.14 | 0.79 | 0.33 |
| 1162224 | 00h46m59.42s | +40d44m35s | -272.4 | 18 | 3674 | 69 | 0.68 | -1.13 | 0.14 | ... | ... |
| 1162248 | 00h47m00.34s | +40d44m35.4s | -282.6 | 10 | 4624 | 276 | 1.36 | -2.1 | 0.27 | ... | ... |
| 1162292 | 00h46m56.95s | +40d44m42.5s | -216.8 | 9 | 4000 | 248 | 1.09 | -1.73 | 0.23 | ... | ... |
| 1162353 | 00h47m00.79s | +40d44m59.2s | -172.3 | 28 | 3919 | 183 | 0.8 | -1.82 | 0.14 | -0.54 | 0.27 |
| 1162447 | 00h46m52.84s | +40d45m16s | -316.0 | 10 | 4200 | 129 | 1.21 | -1.64 | 0.18 | ... | ... |
| 1162529 | 00h47m02.08s | +40d48m37.3s | -243.0 | 12 | 3626 | 130 | 0.95 | -1.16 | 0.14 | ... | ... |

*(Table A.1 Continued)*

*(Table A.1 Continued)*

| Object | Sky Coordinates | | $v_{helio}$ | S/N | $T_{eff}$ | $\delta(T_{eff})$ | log $g$ | [Fe/H] | $\delta$([Fe/H]) | [$\alpha$/Fe] | $\delta$([$\alpha$/Fe]) |
|---|---|---|---|---|---|---|---|---|---|---|---|
| ID | RA | Dec | (km $s^{-1}$) | (Å $^{-1}$) | (K) | (K) | (dex) | (dex) | (dex) | (dex) | (dex) |
| 1162681 | 00h46m46.53s | +40d46m35.5s | -195.7 | 18 | 3944 | 46 | 0.74 | -1.29 | 0.14 | 0.84 | 0.22 |
| 1162706 | 00h46m55.26s | +40d46m30.6s | -209.1 | 15 | 4090 | 6 | 0.98 | -1.3 | 0.14 | ... | ... |
| 1162736 | 00h47m07.88s | +40d46m22.3s | -419.8 | 13 | 4978 | 431 | 1.24 | -1.4 | 0.19 | ... | ... |
| 1162738 | 00h46m47.67s | +40d45m57.4s | -164.0 | 26 | 4466 | 3 | 0.78 | -1.44 | 0.13 | 0.03 | 0.25 |
| 1162741 | 00h47m05.25s | +40d45m39.8s | -315.1 | 10 | 3929 | 246 | 1.21 | -0.61 | 0.14 | ... | ... |
| 1162794 | 00h47m070s | +40d46m10.7s | -333.0 | 10 | 3900 | 591 | 1.17 | -2.4 | 0.3 | ... | ... |
| 1162968 | 00h46m51.86s | +40d45m52.7s | -159.9 | 16 | 3593 | 97 | 0.76 | -1.37 | 0.14 | ... | ... |
| 1163012 | 00h46m50.94s | +40d48m22.4s | -386.4 | 16 | 4153 | 162 | 0.92 | -1.32 | 0.14 | 0.09 | 0.46 |
| 1163168 | 00h47m02.97s | +40d47m37.8s | -434.0 | 9 | 3800 | 367 | 1.09 | -2.6 | 0.25 | ... | ... |
| 1163236 | 00h46m56.74s | +40d47m19s | -200.4 | 7 | 5100 | 982 | 1.38 | -0.72 | 0.17 | 0.7 | 0.46 |
| 121292 | 00h46m09.42s | +40d43m53.7s | -391.6 | 16 | 4090 | 128 | 0.65 | -1.58 | 0.14 | ... | ... |
| 225632 | 00h45m55.87s | +40d41m59s | -311.5 | 9 | 4068 | 10 | 0.94 | -2.6 | 0.28 | ... | ... |
| 237221 | 00h46m30.61s | +40d43m24.1s | -314.0 | 19 | 3819 | 3 | 0.62 | -1.18 | 0.14 | ... | ... |
| 238107 | 00h46m26.96s | +40d43m31.6s | -187.2 | 7 | 4900 | 436 | 1.51 | -0.53 | 0.15 | 0.33 | 0.48 |
| 241004 | 00h46m32.94s | +40d43m51.7s | -265.1 | 14 | 4045 | 5 | 0.93 | -1.31 | 0.14 | ... | ... |
| 241702 | 00h46m35.47s | +40d43m50.6s | -411.6 | 24 | 4450 | 4 | 0.78 | -1.2 | 0.14 | 0.6 | 0.3 |
| 247530 | 00h46m43.51s | +40d44m34.8s | -164.3 | 30 | 4687 | 154 | 0.79 | -1.91 | 0.14 | ... | ... |
| 251324 | 00h46m34.31s | +40d45m01.8s | -263.1 | 28 | 4053 | 5 | 0.59 | -0.81 | 0.13 | 0.57 | 0.19 |

*(Table A.1 Continued)*

*(Table A.1 Continued)*

| Object ID | Sky Coordinates RA | Dec | $v_{helio}$ (km s$^{-1}$) | S/N (Å$^{-1}$) | $T_{eff}$ (K) | $\delta(T_{eff})$ (K) | log $g$ (dex) | [Fe/H] (dex) | $\delta$([Fe/H]) (dex) | [$\alpha$/Fe] (dex) | $\delta$([$\alpha$/Fe]) (dex) |
|---|---|---|---|---|---|---|---|---|---|---|---|
| 284623 | 00h47m07.20s | +40d50m01.1s | -283.1 | 18 | 4276 | 9 | 0.84 | -1.25 | 0.14 | 0.4 | 0.3 |
| 285383 | 00h47m10.46s | +40d50m07.4s | -217.4 | 16 | 3876 | 136 | 0.65 | -0.96 | 0.14 | ... | ... |
| 42824 | 00h46m04.88s | +40d40m47.5s | -385.9 | 8 | 4459 | 424 | 1.19 | -1.85 | 0.21 | ... | ... |
| 18 kpc Halo Field (f123_1) | | | | | | | | | | | |
| 1230048 | 00h48m18.89s | +40d20m18.1s | -350.2 | 7 | 4825 | 51 | 1.51 | -0.81 | 0.17 | ... | ... |
| 1230053 | 00h48m05.10s | +40d20m22.8s | -158.2 | 22 | 4554 | 41 | 0.73 | -1.92 | 0.12 | 0.09 | 0.23 |
| 1230079 | 00h48m18.52s | +40d20m40.1s | -324.3 | 13 | 4198 | 39 | 0.91 | -1.0 | 0.11 | 0.58 | 0.24 |
| 1230080 | 00h48m13.83s | +40d20m36.8s | -388.6 | 11 | 3988 | 47 | 0.88 | -0.53 | 0.11 | 0.37 | 0.33 |
| 1230103 | 00h48m03.43s | +40d20m52.1s | -622.1 | 23 | 4318 | 44 | 0.58 | -1.3 | 0.1 | 0.27 | 0.16 |
| 1230106 | 00h48m01.96s | +40d20m49.8s | -392.7 | 6 | 4221 | 51 | 1.13 | -1.16 | 0.16 | 0.54 | 0.47 |
| 1230192 | 00h47m59.75s | +40d21m24s | -290.4 | 11 | 3794 | 37 | 0.81 | -0.27 | 0.11 | -0.01 | 0.37 |
| 1230207 | 00h48m19.71s | +40d21m35.6s | -173.2 | 28 | 4012 | 36 | 0.33 | -2.88 | 0.12 | ... | ... |
| 1230215 | 00h48m18.57s | +40d21m36.8s | -446.6 | 8 | 4334 | 37 | 1.22 | -1.27 | 0.14 | 0.73 | 0.31 |
| 1230250 | 00h48m04.10s | +40d21m49.3s | -389.3 | 7 | 4092 | 40 | 0.9 | -0.68 | 0.14 | ... | ... |
| 1230273 | 00h48m00.12s | +40d21m56.3s | -207.2 | 6 | 3999 | 43 | 1.15 | -0.83 | 0.16 | 0.74 | 0.41 |
| 1230302 | 00h48m14.06s | +40d22m12.2s | -190.9 | 15 | 4312 | 34 | 0.82 | -1.04 | 0.11 | 0.5 | 0.18 |
| 1230355 | 00h47m56.27s | +40d22m30.9s | -294.4 | 6 | 4165 | 63 | 1.14 | -1.26 | 0.19 | ... | ... |
| 1230452 | 00h48m16.43s | +40d23m16.7s | -515.3 | 6 | 3996 | 34 | 0.98 | -0.61 | 0.12 | -0.32 | 0.28 |

*(Table A.1 Continued)*

*(Table A.1 Continued)*

| Object ID | Sky Coordinates RA | Dec | $v_{helio}$ (km $s^{-1}$) | S/N (Å $^{-1}$) | $T_{eff}$ (K) | $\delta(T_{eff})$ (K) | log $g$ (dex) | [Fe/H] (dex) | $\delta$([Fe/H]) (dex) | [$\alpha$/Fe] (dex) | $\delta$([$\alpha$/Fe]) (dex) |
|---|---|---|---|---|---|---|---|---|---|---|---|
| 1230466 | 00h48m16.65s | +40d23m25.3s | -311.9 | 6 | 4225 | 38 | 1.09 | -1.41 | 0.16 | ... | ... |
| 1230481 | 00h48m00.83s | +40d23m31.1s | -192.4 | 7 | 4014 | 43 | 0.93 | -0.78 | 0.13 | 0.21 | 0.48 |
| 1230521 | 00h48m15.58s | +40d23m46.6s | -458.6 | 11 | 4171 | 36 | 0.78 | -1.18 | 0.11 | 0.54 | 0.23 |
| 1230554 | 00h47m56.32s | +40d23m59.6s | -304.3 | 13 | 4041 | 48 | 0.66 | -0.55 | 0.11 | 0.53 | 0.29 |
| 1230569 | 00h48m03.80s | +40d24m00.7s | -385.5 | 6 | 3962 | 36 | 0.99 | -0.6 | 0.13 | 0.12 | 0.46 |
| 1230590 | 00h48m02.43s | +40d24m09.4s | -324.5 | 8 | 4224 | 39 | 1.05 | -1.01 | 0.13 | 0.46 | 0.33 |
| 1230605 | 00h48m16.33s | +40d24m18.6s | -241.7 | 10 | 4186 | 38 | 0.74 | -1.4 | 0.13 | 0.75 | 0.24 |
| 1230662 | 00h48m13.79s | +40d24m33.9s | -259.3 | 7 | 4620 | 37 | 1.28 | -0.87 | 0.14 | 0.4 | 0.36 |
| 1230678 | 00h48m02.88s | +40d24m36.3s | -297.9 | 7 | 4007 | 40 | 0.99 | -0.82 | 0.14 | 0.24 | 0.4 |
| 1230795 | 00h48m04.27s | +40d25m15.9s | -459.1 | 6 | 4508 | 42 | 1.05 | -1.38 | 0.19 | ... | ... |
| 1230850 | 00h48m14.84s | +40d25m34.6s | -289.0 | 5 | 4168 | 40 | 0.9 | -0.91 | 0.14 | ... | ... |
| 1230918 | 00h48m05.03s | +40d25m55.1s | -276.6 | 7 | 3973 | 38 | 0.67 | -0.57 | 0.12 | -0.3 | 0.44 |
| 1230921 | 00h48m07.15s | +40d25m56s | -292.9 | 13 | 4120 | 35 | 0.65 | -0.64 | 0.11 | 0.55 | 0.2 |
| 1230936 | 00h48m10.26s | +40d26m00.9s | -477.3 | 5 | 4640 | 39 | 1.41 | -1.14 | 0.21 | ... | ... |
| 1231009 | 00h47m56.17s | +40d26m24.8s | -291.2 | 9 | 4234 | 38 | 0.82 | -0.65 | 0.12 | 0.71 | 0.28 |
| 1231010 | 00h48m06.43s | +40d26m25.8s | -258.5 | 7 | 4464 | 38 | 1.17 | -0.54 | 0.13 | 0.69 | 0.41 |
| 1231011 | 00h48m06.53s | +40d26m32.3s | -250.7 | 12 | 4424 | 40 | 0.76 | -1.77 | 0.14 | 0.26 | 0.28 |
| 1231077 | 00h48m04.94s | +40d26m53.5s | -279.6 | 9 | 3897 | 38 | 0.79 | -0.54 | 0.12 | 0.35 | 0.35 |

*(Table A.1 Continued)*

*(Table A.1 Continued)*

| Object | Sky Coordinates | | $v_{helio}$ | S/N | $T_{eff}$ | $\delta(T_{eff})$ | log $g$ | [Fe/H] | $\delta$([Fe/H]) | [$\alpha$/Fe] | $\delta$([$\alpha$/Fe]) |
|---|---|---|---|---|---|---|---|---|---|---|---|
| ID | RA | Dec | (km $s^{-1}$) | (Å $^{-1}$) | (K) | (K) | (dex) | (dex) | (dex) | (dex) | (dex) |
| 1231108 | 00h48m12.69s | +40d27m04.8s | -453.7 | 12 | 4135 | 36 | 0.6 | -1.06 | 0.11 | 0.53 | 0.23 |
| 1231121 | 00h48m11.81s | +40d27m08.2s | -375.2 | 6 | 4686 | 38 | 1.13 | -0.88 | 0.14 | ... | ... |
| 1231133 | 00h47m58.93s | +40d27m08.8s | -305.7 | 9 | 3997 | 41 | 0.79 | -0.58 | 0.12 | 0.45 | 0.34 |
| 1231138 | 00h48m11.77s | +40d27m13s | -495.3 | 5 | 4653 | 43 | 1.05 | -0.94 | 0.15 | ... | ... |
| 1231181 | 00h47m58.34s | +40d27m22.8s | -288.7 | 7 | 4290 | 38 | 1.12 | -0.92 | 0.15 | ... | ... |
| 1231219 | 00h47m54.96s | +40d27m37.7s | -142.6 | 6 | 4051 | 44 | 0.69 | -0.92 | 0.15 | 0.59 | 0.44 |
| 1231265 | 00h47m53.39s | +40d27m48.7s | -294.4 | 10 | 3742 | 39 | 0.7 | -0.09 | 0.11 | 0.29 | 0.32 |
| 1231295 | 00h47m53.96s | +40d27m58.8s | -217.0 | 10 | 3891 | 40 | 0.84 | -2.99 | 0.27 | ... | ... |
| 1231331 | 00h48m02.80s | +40d28m14.6s | -387.1 | 4 | 4422 | 47 | 1.29 | -0.87 | 0.16 | ... | ... |
| 1231349 | 00h47m57.32s | +40d28m18s | -430.6 | 7 | 4297 | 39 | 1.1 | -1.0 | 0.15 | ... | ... |
| 1231358 | 00h48m10.95s | +40d28m24.9s | -276.0 | 4 | 4049 | 41 | 0.97 | -0.99 | 0.2 | ... | ... |
| 1231379 | 00h47m55.14s | +40d28m30s | -299.4 | 13 | 4179 | 40 | 0.99 | -0.83 | 0.11 | 0.59 | 0.24 |
| 1231440 | 00h47m57.30s | +40d28m53.4s | -346.3 | 6 | 4023 | 37 | 1.01 | -1.02 | 0.17 | ... | ... |
| 1231445 | 00h48m02.84s | +40d28m55.6s | -314.9 | 6 | 4487 | 40 | 1.32 | -0.98 | 0.15 | 0.35 | 0.45 |
| 1231457 | 00h47m58.14s | +40d29m01.8s | -328.9 | 12 | 3910 | 34 | 0.49 | -1.04 | 0.11 | 0.04 | 0.25 |
| 1231462 | 00h47m48.36s | +40d29m05.9s | -261.6 | 16 | 4105 | 38 | 0.74 | -0.79 | 0.11 | 0.49 | 0.19 |
| 1231480 | 00h47m57.10s | +40d29m09.1s | -308.3 | 5 | 4097 | 38 | 1.19 | -0.78 | 0.14 | ... | ... |
| 1231564 | 00h47m56.47s | +40d29m33.4s | -315.9 | 8 | 4501 | 41 | 1.24 | -2.25 | 0.28 | ... | ... |

(Table A.1 Continued)

| Object ID | Sky Coordinates RA | Dec | $v_{helio}$ (km s$^{-1}$) | S/N (Å$^{-1}$) | $T_{eff}$ (K) | $\delta(T_{eff})$ (K) | log $g$ (dex) | [Fe/H] (dex) | $\delta$([Fe/H]) (dex) | [$\alpha$/Fe] (dex) | $\delta$([$\alpha$/Fe]) (dex) |
|---|---|---|---|---|---|---|---|---|---|---|---|
| 1231567 | 00h48m02.29s | +40d29m35.5s | -268.7 | 4 | 4071 | 41 | 1.03 | -0.63 | 0.17 | ... | ... |
| 1231588 | 00h47m55.83s | +40d29m43s | -261.9 | 11 | 4048 | 37 | 0.72 | -0.63 | 0.11 | 0.19 | 0.27 |
| 1231609 | 00h47m50.85s | +40d29m51s | -286.1 | 8 | 3974 | 40 | 0.84 | -0.45 | 0.12 | 0.4 | 0.37 |
| 1231651 | 00h47m59.53s | +40d30m03.1s | -294.6 | 7 | 4078 | 37 | 1.01 | -0.97 | 0.15 | 0.43 | 0.42 |
| 1231658 | 00h47m53.77s | +40d30m08.5s | -287.7 | 8 | 4201 | 46 | 0.92 | -1.1 | 0.12 | 0.23 | 0.41 |
| 1231664 | 00h47m51.23s | +40d30m09.9s | -293.1 | 7 | 4131 | 47 | 1.07 | -0.12 | 0.12 | ... | ... |
| 1231665 | 00h47m56.96s | +40d30m11.3s | -290.2 | 8 | 3988 | 38 | 0.96 | -0.34 | 0.12 | ... | ... |
| 1231738 | 00h47m56.93s | +40d30m43.8s | -312.0 | 4 | 4075 | 39 | 1.21 | -0.84 | 0.19 | ... | ... |
| 1231783 | 00h47m49.21s | +40d30m58.8s | -285.7 | 5 | 4578 | 40 | 1.39 | -0.89 | 0.15 | ... | ... |
| 1231849 | 00h48m00.11s | +40d32m04.2s | -293.7 | 21 | 4027 | 32 | 0.31 | -1.37 | 0.11 | 0.48 | 0.15 |
| 1231857 | 00h47m51.74s | +40d31m59.4s | -323.1 | 16 | 4119 | 33 | 0.68 | -1.05 | 0.11 | 0.35 | 0.17 |
| 1231862 | 00h47m48.97s | +40d31m55.6s | -298.8 | 10 | 4027 | 36 | 1.04 | -0.74 | 0.11 | 0.02 | 0.34 |
| 1231876 | 00h47m47.27s | +40d31m19.4s | -301.5 | 12 | 4345 | 38 | 1.02 | -1.13 | 0.12 | 0.65 | 0.23 |
| 1231891 | 00h47m51.24s | +40d34m10.4s | -244.7 | 10 | 4231 | 40 | 0.85 | -1.18 | 0.12 | -0.04 | 0.37 |
| 1231909 | 00h47m58.79s | +40d33m17.4s | -416.4 | 9 | 4387 | 36 | 0.97 | -1.37 | 0.13 | 0.2 | 0.31 |
| 1231917 | 00h47m57.27s | +40d32m58.2s | -297.2 | 9 | 4003 | 36 | 0.71 | -0.99 | 0.12 | 0.49 | 0.32 |
| 1232055 | 00h47m51.87s | +40d31m22s | -299.5 | 11 | 4483 | 35 | 1.11 | -0.62 | 0.11 | 0.55 | 0.24 |
| 1232066 | 00h47m51.58s | +40d31m12.7s | -188.2 | 6 | 4313 | 37 | 1.26 | -1.57 | 0.23 | ... | ... |

*(Table A.1 Continued)*

| Object ID | Sky Coordinates RA | Dec | $v_{helio}$ (km s$^{-1}$) | S/N (Å$^{-1}$) | $T_{eff}$ (K) | $\delta(T_{eff})$ (K) | log $g$ (dex) | [Fe/H] (dex) | $\delta$([Fe/H]) (dex) | [$\alpha$/Fe] (dex) | $\delta$([$\alpha$/Fe]) (dex) |
|---|---|---|---|---|---|---|---|---|---|---|---|
| 1232098 | 00h47m49.04s | +40d34m06s | -216.2 | 16 | 4289 | 36 | 0.6 | -1.42 | 0.11 | 0.49 | 0.18 |
| 1232109 | 00h47m49.36s | +40d33m58.9s | -290.5 | 8 | 4163 | 34 | 0.98 | -1.07 | 0.12 | -0.3 | 0.41 |
| 1232152 | 00h48m00.11s | +40d33m42.5s | -274.5 | 5 | 4043 | 42 | 1.0 | -0.85 | 0.15 | ... | ... |
| 1232196 | 00h47m47.54s | +40d33m10.6s | -552.3 | 8 | 4547 | 38 | 1.25 | -1.48 | 0.15 | ... | ... |
| 1232216 | 00h47m49.93s | +40d32m23.7s | -260.4 | 11 | 3795 | 35 | 0.75 | -0.4 | 0.11 | ... | ... |
| 1232224 | 00h47m55.13s | +40d33m08.7s | -459.9 | 14 | 4157 | 44 | 0.66 | -1.53 | 0.12 | 0.69 | 0.27 |
| 1232276 | 00h47m52.68s | +40d32m21s | -297.0 | 8 | 4074 | 36 | 1.11 | -0.33 | 0.11 | 0.15 | 0.4 |
| 1232284 | 00h47m51.14s | +40d32m56.9s | -207.9 | 15 | 4091 | 37 | 0.67 | -1.28 | 0.11 | 0.25 | 0.25 |
| 1232285 | 00h47m51.56s | +40d34m01s | -266.5 | 5 | 4346 | 55 | 0.8 | -0.69 | 0.14 | ... | ... |
| 1232295 | 00h47m57.01s | +40d32m43.3s | -234.7 | 6 | 4181 | 43 | 1.17 | -0.27 | 0.12 | ... | ... |
| | | | | | 22 kpc GSS Field (S) | | | | | | |
| 14648 | 00h44m16.87s | +39d48m55.5s | -427.7 | 10 | 4800 | 170 | 1.6 | -1.22 | 0.16 | ... | ... |
| 157934 | 00h44m00.53s | +39d35m51.8s | -314.8 | 17 | 4300 | 119 | 0.66 | -0.02 | 0.13 | 0.51 | 0.21 |
| 169191 | 00h44m02.32s | +39d37m47.1s | -472.0 | 10 | 4458 | 165 | 1.39 | -0.5 | 0.14 | 0.68 | 0.28 |
| 178993 | 00h43m56.03s | +39d39m24.1s | -202.1 | 11 | 5257 | 728 | 1.51 | -1.35 | 0.2 | ... | ... |
| 190457 | 00h44m11.76s | +39d41m14.5s | -525.0 | 9 | 3735 | 182 | 1.59 | -0.47 | 0.14 | ... | ... |
| 2000189 | 00h44m03.97s | +39d37m34s | -367.6 | 12 | 3592 | 86 | 0.93 | -1.95 | 0.16 | ... | ... |
| 2000833 | 00h44m03.16s | +39d38m26.3s | -455.2 | 21 | 3904 | 995 | 1.14 | -1.49 | 0.15 | 0.06 | 0.37 |

*(Table A.1 Continued)*

*(Table A.1 Continued)*

| Object | Sky Coordinates | | $v_{helio}$ | S/N | $T_{eff}$ | $\delta(T_{eff})$ | log $g$ | [Fe/H] | $\delta$([Fe/H]) | [$\alpha$/Fe] | $\delta$([$\alpha$/Fe]) |
|---|---|---|---|---|---|---|---|---|---|---|---|
| ID | RA | Dec | (km $s^{-1}$) | (Å $^{-1}$) | (K) | (K) | (dex) | (dex) | (dex) | (dex) | (dex) |
| 2001537 | 00h44m04.20s | +39d39m22.2s | -513.8 | 12 | 4363 | 9 | 1.3 | -1.0 | 0.14 | 0.45 | 0.3 |
| 2002001 | 00h44m13.43s | +39d39m54.3s | -350.1 | 20 | 4329 | 6 | 1.08 | -0.74 | 0.13 | 0.59 | 0.19 |
| 2002128 | 00h44m07.70s | +39d40m07.8s | -506.5 | 13 | 3990 | 39 | 1.12 | -0.85 | 0.14 | 0.49 | 0.35 |
| 2002383 | 00h44m10.21s | +39d40m26.7s | -270.2 | 13 | 3861 | 136 | 1.0 | -2.7 | 0.18 | ... | ... |
| 2002792 | 00h44m05.95s | +39d40m53.4s | -449.6 | 12 | 4655 | 8 | 1.49 | -1.65 | 0.16 | 0.8 | 0.36 |
| 2002923 | 00h44m16.86s | +39d41m01.4s | -476.6 | 12 | 4123 | 56 | 1.3 | -0.96 | 0.14 | -0.16 | 0.33 |
| 2003583 | 00h44m15.91s | +39d41m48s | -476.6 | 15 | 4343 | 77 | 1.21 | -0.96 | 0.13 | -0.17 | 0.3 |
| 2003586 | 00h44m19.62s | +39d41m48.3s | -369.9 | 20 | 3746 | 76 | 1.02 | -0.79 | 0.13 | 0.16 | 0.25 |
| 2004488 | 00h44m06.29s | +39d42m54.8s | -452.1 | 13 | 4205 | 12 | 1.33 | -1.05 | 0.14 | 0.55 | 0.41 |
| 2004641 | 00h44m13.41s | +39d43m05.8s | -347.8 | 18 | 4046 | 4 | 0.95 | -0.59 | 0.13 | 0.4 | 0.2 |
| 2004797 | 00h44m07.90s | +39d43m18.2s | -463.7 | 10 | 4158 | 36 | 1.35 | -0.13 | 0.13 | -0.1 | 0.34 |
| 2005183 | 00h44m09.84s | +39d43m43.4s | -393.3 | 11 | 3829 | 6 | 1.01 | -0.24 | 0.14 | 0.12 | 0.39 |
| 2005360 | 00h44m07.82s | +39d43m58.1s | -524.0 | 31 | 4500 | 37 | 0.89 | -1.9 | 0.14 | -0.31 | 0.32 |
| 2006139 | 00h44m14.88s | +39d44m55.6s | -470.5 | 16 | 4075 | 41 | 1.11 | -1.07 | 0.14 | 1.0 | 0.19 |
| 2006680 | 00h44m13.11s | +39d45m40.1s | -353.2 | 11 | 4400 | 113 | 1.41 | -1.35 | 0.15 | 0.38 | 0.37 |
| 2007169 | 00h44m07.63s | +39d46m19.5s | -501.7 | 11 | 4339 | 179 | 1.39 | -1.56 | 0.16 | ... | ... |
| 2008269 | 00h44m08.17s | +39d47m39.6s | -365.0 | 8 | 4485 | 40 | 1.87 | -1.29 | 0.2 | ... | ... |
| 2010061 | 00h44m11.75s | +39d49m51.8s | -375.8 | 21 | 3822 | 49 | 0.97 | -0.84 | 0.13 | -0.11 | 0.44 |

*(Table A.1 Continued)*

*(Table A.1 Continued)*

| Object | Sky Coordinates | | $v_{helio}$ | S/N | $T_{eff}$ | $\delta(T_{eff})$ | log $g$ | [Fe/H] | $\delta$([Fe/H]) | [$\alpha$/Fe] | $\delta$([$\alpha$/Fe]) |
|---|---|---|---|---|---|---|---|---|---|---|---|
| ID | RA | Dec | (km $s^{-1}$) | (Å $^{-1}$) | (K) | (K) | (dex) | (dex) | (dex) | (dex) | (dex) |
| 2010441 | 00h44m24.72s | +39d50m21.8s | -488.0 | 9 | 3987 | 21 | 1.25 | -0.85 | 0.16 | ... | ... |
| 213702 | 00h44m21.78s | +39d44m56.6s | -359.1 | 12 | 3952 | 34 | 1.06 | -0.53 | 0.13 | 0.38 | 0.3 |
| 217656 | 00h44m15.96s | +39d45m35.2s | -308.7 | 19 | 4193 | 63 | 0.92 | -0.86 | 0.13 | 0.69 | 0.26 |
| 2493 | 00h44m18.41s | +39d49m13.4s | -537.6 | 17 | 4450 | 5 | 1.09 | -1.57 | 0.14 | ... | ... |
| 252649 | 00h44m22.56s | +39d51m04.3s | -496.0 | 8 | 4100 | 1077 | 1.62 | -2.89 | 0.32 | ... | ... |
| 252859 | 00h44m27.59s | +39d51m02.6s | -422.9 | 13 | 4832 | 8 | 1.57 | -0.99 | 0.15 | ... | ... |
| 46853 | 00h44m23.91s | +39d47m16.6s | -375.0 | 13 | 4287 | 697 | 1.41 | -2.7 | 0.22 | ... | ... |
| 76769 | 00h44m18.05s | +39d46m26.3s | -374.3 | 15 | 3710 | 574 | 1.35 | -2.6 | 0.18 | ... | ... |
| 23 kpc Halo Field (f130_1) | | | | | | | | | | | |
| 1300360 | 00h49m12.01s | +40d10m00.3s | -280.2 | 14 | 3938 | 8 | 1.14 | -1.15 | 0.14 | 0.56 | 0.25 |
| 1300698 | 00h49m03.01s | +40d15m05.1s | -235.4 | 15 | 3774 | 37 | 0.98 | -0.34 | 0.14 | 0.6 | 0.29 |
| 1300024 | 00h49m06.69s | +40d04m54.2s | -360.4 | 52 | 4459 | 3 | 0.66 | -2.6 | 0.16 | 0.7 | 0.23 |
| 1300432 | 00h49m16.68s | +40d11m08.2s | -163.7 | 16 | 3956 | 7 | 0.94 | -1.9 | 0.14 | -0.3 | 0.25 |
| 1300266 | 00h49m03.58s | +40d08m42.7s | -400.5 | 13 | 3789 | 9 | 1.15 | -2.87 | 0.23 | ... | ... |
| 1300488 | 00h49m03.47s | +40d11m59s | -347.5 | 49 | 4300 | 2 | 0.63 | -2.22 | 0.13 | 0.6 | 0.15 |
| 1300466 | 00h49m18.74s | +40d11m39.8s | -232.8 | 14 | 4190 | 11 | 1.19 | -2.27 | 0.15 | ... | ... |
| 1300876 | 00h49m00.90s | +40d17m59s | -260.9 | 18 | 4300 | 15 | 1.2 | -1.59 | 0.14 | 0.9 | 0.22 |
| 1300114 | 00h49m08.36s | +40d05m53.9s | -241.8 | 22 | 4562 | 10 | 1.26 | -2.27 | 0.17 | ... | ... |

*(Table A.1 Continued)*

*(Table A.1 Continued)*

| Object | Sky Coordinates | | $v_{helio}$ | S/N | $T_{eff}$ | $\delta(T_{eff})$ | log $g$ | [Fe/H] | $\delta$([Fe/H]) | [$\alpha$/Fe] | $\delta$([$\alpha$/Fe]) |
|---|---|---|---|---|---|---|---|---|---|---|---|
| ID | RA | Dec | (km $s^{-1}$) | (Å $^{-1}$) | (K) | (K) | (dex) | (dex) | (dex) | (dex) | (dex) |
| 1300633 | 00h49m14.81s | +40d14m13.9s | -264.1 | 12 | 4222 | 10 | 0.79 | -2.27 | 0.17 | 0.6 | 0.38 |
| 1300256 | 00h49m19.02s | +40d08m36.6s | -325.0 | 56 | 4130 | 10 | 0.5 | -1.95 | 0.13 | ... | ... |
| 1300026 | 00h49m08.78s | +40d04m54.4s | -256.3 | 33 | 4477 | 4 | 0.93 | -2.9 | 0.21 | ... | ... |
| 1300502 | 00h49m11.13s | +40d12m26s | -132.2 | 29 | 3940 | 3 | 0.67 | -1.25 | 0.13 | 0.26 | 0.16 |
| 1300007 | 00h49m19.32s | +40d04m35.4s | -162.0 | 12 | 3757 | 12 | 0.79 | -2.91 | 0.35 | ... | ... |
| 1300786 | 00h49m09.28s | +40d16m36.2s | -513.6 | 24 | 3898 | 4 | 0.71 | -1.69 | 0.14 | 0.69 | 0.17 |
| 1300112 | 00h49m13.49s | +40d05m53.9s | -476.5 | 15 | 4454 | 13 | 1.2 | -2.71 | 0.24 | ... | ... |
| 1300303 | 00h49m13.49s | +40d05m53.9s | -407.4 | 32 | 4253 | 4 | 0.89 | -2.09 | 0.14 | 0.3 | 0.19 |
| 1300239 | 00h49m11.23s | +40d05m33.4s | -426.2 | 14 | 4254 | 34 | 1.38 | -2.67 | 0.29 | ... | ... |
| 1300161 | 00h49m02.94s | +40d06m49.1s | -472.3 | 18 | 4350 | 26 | 0.93 | -1.68 | 0.15 | ... | ... |
| 1300375 | 00h49m04.25s | +40d10m11.5s | -302.1 | 12 | 4441 | 16 | 1.42 | -2.5 | 0.22 | 0.5 | 0.45 |
| 1300581 | 00h48m58.36s | +40d13m32.1s | -282.4 | 39 | 4310 | 4 | 0.8 | -2.21 | 0.14 | 0.05 | 0.26 |
| 1300967 | 00h49m04.20s | +40d16m45.5s | -269.0 | 19 | 4133 | 8 | 1.08 | -1.77 | 0.14 | 0.67 | 0.21 |
| 1300799 | 00h49m06.31s | +40d16m43.2s | -294.4 | 33 | 4252 | 4 | 0.83 | -1.79 | 0.13 | 0.17 | 0.18 |
| 1300631 | 00h49m10.95s | +40d14m10.3s | -414.7 | 13 | 4315 | 16 | 1.37 | -1.6 | 0.15 | 0.47 | 0.32 |
| 1300157 | 00h49m23.11s | +40d06m47s | -370.8 | 13 | 4467 | 15 | 1.24 | -2.03 | 0.18 | 0.25 | 0.44 |
| 1300878 | 00h49m10.74s | +40d08m34.7s | -379.7 | 10 | 3766 | 9 | 1.06 | -0.7 | 0.14 | ... | ... |
| 1300621 | 00h49m00.07s | +40d13m56.8s | -237.8 | 19 | 3863 | 6 | 1.06 | -0.97 | 0.14 | -0.42 | 0.14 |

*(Table A.1 Continued)*

| Object ID | Sky Coordinates RA | Dec | $v_{helio}$ (km s$^{-1}$) | S/N (Å$^{-1}$) | $T_{eff}$ (K) | $\delta(T_{eff})$ (K) | log $g$ (dex) | [Fe/H] (dex) | $\delta$([Fe/H]) (dex) | [$\alpha$/Fe] (dex) | $\delta$([$\alpha$/Fe]) (dex) |
|---|---|---|---|---|---|---|---|---|---|---|---|
| 1300940 | 00h49m04.63s | +40d16m53.1s | -349.2 | 12 | 3825 | 8 | 1.11 | -0.96 | 0.14 | 0.05 | 0.34 |
| 1300133 | 00h49m08.81s | +40d05m07.9s | -460.4 | 9 | 4016 | 19 | 1.23 | -2.88 | 0.27 | ... | ... |
| 1300333 | 00h49m10.88s | +40d09m43s | -229.3 | 19 | 3868 | 6 | 0.99 | -0.93 | 0.14 | 0.73 | 0.23 |
| 23 kpc Halo Field (f130_2) | | | | | | | | | | | |
| 1282152 | 00h50m17.45s | +40d16m31.4s | -158.4 | 22 | 4100 | 10 | 0.63 | -1.01 | 0.13 | -0.04 | 0.26 |
| 1282178 | 00h50m13.46s | +40d18m24.3s | -475.7 | 26 | 3937 | 4 | 0.39 | -3.2 | 0.17 | 0.65 | 0.32 |
| 1282547 | 00h50m11.59s | +40d18m34.9s | -361.5 | 12 | 4034 | 9 | 0.9 | -1.27 | 0.14 | ... | ... |
| 1292468 | 00h49m56.71s | +40d18m19.1s | -259.9 | 12 | 3720 | 4 | 0.61 | -0.73 | 0.14 | 0.5 | 0.37 |
| 1292507 | 00h49m51.47s | +40d18m14.2s | -316.5 | 19 | 3912 | 8 | 0.45 | -1.76 | 0.14 | 0.6 | 0.34 |
| 1292637 | 00h49m46.52s | +40d16m53.1s | -161.5 | 6 | 3764 | 8 | 0.87 | -2.13 | 0.23 | ... | ... |
| 1292654 | 00h49m34.73s | +40d17m32.8s | -220.2 | 6 | 3864 | 12 | 1.07 | -0.15 | 0.14 | ... | ... |
| 1302581 | 00h49m09.27s | +40d15m28.7s | -290.2 | 13 | 4604 | 6 | 1.04 | -2.57 | 0.21 | 0.4 | 0.5 |
| 1302582 | 00h49m27.60s | +40d15m27.4s | -161.5 | 12 | 3634 | 7 | 0.72 | -2.73 | 0.2 | ... | ... |
| 1302618 | 00h49m16.54s | +40d18m37.7s | 399.3 | 7 | 3705 | 20 | 1.06 | -2.11 | 0.29 | ... | ... |
| 1302675 | 00h49m23.87s | +40d16m58.8s | -173.6 | 13 | 4219 | 11 | 1.07 | -2.3 | 0.17 | ... | ... |
| 1302682 | 00h49m28.12s | +40d16m56.1s | -302.1 | 19 | 4075 | 5 | 0.84 | -1.48 | 0.14 | 0.81 | 0.19 |
| 1302710 | 00h49m14.30s | +40d16m32.7s | -195.9 | 10 | 4282 | 11 | 1.07 | -1.69 | 0.16 | 0.36 | 0.41 |
| 1302825 | 00h49m12.79s | +40d17m47.4s | -169.0 | 7 | 4182 | 22 | 1.28 | -1.84 | 0.21 | ... | ... |

*(Table A.1 Continued)*

*(Table A.1 Continued)*

| Object | Sky Coordinates | | $v_{helio}$ | S/N | $T_{eff}$ | $\delta(T_{eff})$ | log $g$ | [Fe/H] | $\delta$([Fe/H]) | [$\alpha$/Fe] | $\delta$([$\alpha$/Fe]) |
|---|---|---|---|---|---|---|---|---|---|---|---|
| ID | RA | Dec | (km $s^{-1}$) | (Å $^{-1}$) | (K) | (K) | (dex) | (dex) | (dex) | (dex) | (dex) |
| 1302893 | 00h48m56.73s | +40d16m07.6s | -295.9 | 20 | 3717 | 12 | 0.39 | -2.98 | 0.27 | ... | ... |
| 1302971 | 00h49m11.94s | +40d16m47s | -348.1 | 15 | 3822 | 4 | 0.49 | -1.79 | 0.15 | 0.77 | 0.37 |
| 1303039 | 00h49m12.31s | +40d15m57.8s | -233.0 | 24 | 4242 | 4 | 0.56 | -2.16 | 0.14 | 0.0 | 0.24 |
| 1303200 | 00h49m21.67s | +40d18m13.7s | -534.6 | 26 | 4415 | 4 | 0.9 | -1.92 | 0.14 | -0.0 | 0.29 |
| 1303478 | 00h49m17.60s | +40d18m19.4s | -185.1 | 12 | 3787 | 6 | 0.82 | -3.52 | 0.29 | ... | ... |
| 1303493 | 00h49m04.20s | +40d16m45.5s | -288.3 | 8 | 4135 | 13 | 1.04 | -1.23 | 0.16 | 0.8 | 0.38 |
| 1303502 | 00h49m03.56s | +40d16m14.2s | -238.9 | 28 | 3949 | 5 | 0.4 | -1.28 | 0.13 | 0.49 | 0.17 |
| 26 kpc Disk Field (D) | | | | | | | | | | | |
| 109460 | 00h49m16.88s | +42d43m53.7s | -105.7 | 14 | 3800 | 93 | 0.96 | -1.11 | 0.14 | 0.28 | 0.28 |
| 16545 | 00h49m03.12s | +42d44m02.5s | -105.9 | 33 | 3755 | 149 | 0.74 | -2.02 | 0.13 | ... | ... |
| 21569 | 00h49m04.48s | +42d43m43.4s | -230.4 | 6 | 4515 | 41 | 1.87 | -0.75 | 0.17 | ... | ... |
| 3000373 | 00h48m53.99s | +42d45m23.6s | -120.4 | 15 | 3803 | 6 | 0.84 | -0.75 | 0.14 | 0.5 | 0.31 |
| 3000412 | 00h48m57.21s | +42d45m14.2s | -305.4 | 21 | 3730 | 229 | 0.73 | -2.71 | 0.16 | 0.1 | 0.47 |
| 3000724 | 00h49m02.02s | +42d47m38.1s | -152.6 | 16 | 4200 | 88 | 0.82 | -0.51 | 0.13 | 1.04 | 0.25 |
| 32476 | 00h49m02.30s | +42d45m14s | -413.1 | 17 | 4400 | 85 | 1.05 | -1.54 | 0.14 | 1.07 | 0.2 |
| 57165 | 00h49m06.74s | +42d44m39.4s | -127.3 | 14 | 3950 | 64 | 0.79 | -0.81 | 0.14 | 0.94 | 0.23 |
| 760462 | 00h49m57.17s | +42d41m45.5s | -167.3 | 20 | 3799 | 30 | 0.63 | -0.89 | 0.13 | 0.57 | 0.2 |
| 760533 | 00h49m56.48s | +42d41m46.4s | -117.7 | 15 | 4272 | 48 | 1.05 | -0.86 | 0.13 | 0.08 | 0.25 |

*(Table A.1 Continued)*

*(Table A.1 Continued)*

| Object ID | Sky Coordinates RA | Dec | $v_{helio}$ (km $s^{-1}$) | S/N (Å $^{-1}$) | $T_{eff}$ (K) | $\delta(T_{eff})$ (K) | log $g$ (dex) | [Fe/H] (dex) | $\delta$([Fe/H]) (dex) | [$\alpha$/Fe] (dex) | $\delta$([$\alpha$/Fe]) (dex) |
|---|---|---|---|---|---|---|---|---|---|---|---|
| 767707 | 00h49m43.61s | +42d42m11.3s | -220.7 | 11 | 4032 | 29 | 0.91 | -0.47 | 0.13 | 0.35 | 0.28 |
| 775052 | 00h49m59.73s | +42d42m35.1s | -222.7 | 14 | 4400 | 254 | 0.72 | -0.47 | 0.13 | -0.08 | 0.33 |
| 775973 | 00h49m22.16s | +42d42m38.1s | -248.1 | 13 | 3905 | 6 | 0.56 | -1.77 | 0.15 | ... | ... |
| 784724 | 00h49m45.11s | +42d43m07.8s | -130.1 | 10 | 4898 | 417 | 1.23 | -0.5 | 0.14 | ... | ... |
| 791849 | 00h50m02.98s | +42d43m31s | -350.7 | 21 | 3990 | 121 | 0.69 | -0.55 | 0.13 | 0.39 | 0.25 |
| 798409 | 00h49m50.76s | +42d43m55.6s | -118.4 | 7 | 4876 | 604 | 1.26 | -0.1 | 0.14 | ... | ... |
| 803494 | 00h49m48.61s | +42d44m14s | -109.6 | 7 | 4438 | 239 | 1.34 | -0.15 | 0.14 | 0.72 | 0.27 |
| 803662 | 00h49m40.48s | +42d44m14.3s | -202.1 | 8 | 4189 | 83 | 1.24 | -0.38 | 0.14 | 0.6 | 0.33 |
| 803996 | 00h49m25.70s | +42d44m13.4s | -139.9 | 12 | 4203 | 8 | 1.01 | -1.19 | 0.14 | ... | ... |
| 805239 | 00h49m39.22s | +42d44m17.2s | -242.2 | 29 | 4340 | 3 | 0.75 | -2.27 | 0.14 | -0.2 | 0.3 |
| 807089 | 00h49m25.06s | +42d44m26.1s | -102.1 | 9 | 4524 | 17 | 1.36 | -0.8 | 0.14 | 0.35 | 0.34 |
| 808383 | 00h50m00.46s | +42d44m28.4s | -93.1 | 22 | 3829 | 44 | 0.59 | -3.08 | 0.18 | ... | ... |
| 808773 | 00h49m40.08s | +42d44m31s | -213.0 | 12 | 3917 | 57 | 0.92 | -0.93 | 0.14 | 1.12 | 0.25 |
| 810360 | 00h48m51.97s | +42d44m36.6s | -247.2 | 9 | 3937 | 11 | 1.09 | -1.35 | 0.16 | 0.63 | 0.43 |
| 811857 | 00h49m27.53s | +42d44m41.9s | -134.8 | 17 | 3884 | 4 | 0.73 | -1.04 | 0.13 | 0.73 | 0.24 |
| 813156 | 00h49m48.54s | +42d44m46.2s | -341.7 | 10 | 4183 | 116 | 1.02 | -0.4 | 0.14 | 0.19 | 0.35 |
| 821798 | 00h48m37.33s | +42d45m17.5s | -129.6 | 9 | 4200 | 651 | 1.19 | -2.69 | 0.3 | ... | ... |
| 826857 | 00h48m40.29s | +42d45m34.2s | -146.8 | 8 | 3606 | 604 | 1.12 | -2.01 | 0.23 | ... | ... |

*(Table A.1 Continued)*

| Object | Sky Coordinates | | $v_{helio}$ | S/N | $T_{eff}$ | $\delta(T_{eff})$ | log $g$ | [Fe/H] | $\delta$([Fe/H]) | [$\alpha$/Fe] | $\delta$([$\alpha$/Fe]) |
|---|---|---|---|---|---|---|---|---|---|---|---|
| ID | RA | Dec | (km $s^{-1}$) | (Å $^{-1}$) | (K) | (K) | (dex) | (dex) | (dex) | (dex) | (dex) |
| 830877 | 00h48m42.34s | +42d45m45.5s | -200.9 | 10 | 4584 | 18 | 1.18 | -2.27 | 0.26 | ... | ... |
| 831229 | 00h48m47.41s | +42d45m50.6s | -345.9 | 4 | 3965 | 19 | 1.42 | -0.27 | 0.15 | ... | ... |
| 833457 | 00h48m38.64s | +42d45m57s | -134.8 | 18 | 3600 | 153 | 0.76 | -1.32 | 0.14 | 0.45 | 0.3 |
| 834895 | 00h48m36.56s | +42d46m00.4s | -218.7 | 17 | 4139 | 48 | 0.68 | -0.63 | 0.13 | 0.6 | 0.29 |
| 836086 | 00h49m55.87s | +42d46m06s | -108.3 | 7 | 4360 | 281 | 1.3 | -0.76 | 0.15 | ... | ... |
| 839117 | 00h49m34.92s | +42d46m17s | -329.3 | 16 | 4153 | 10 | 0.86 | -1.37 | 0.15 | 0.64 | 0.5 |
| 843412 | 00h48m51.04s | +42d46m32.9s | -371.5 | 6 | 4968 | 17 | 1.85 | -0.26 | 0.15 | ... | ... |
| 850933 | 00h48m42.26s | +42d46m58.2s | -333.5 | 15 | 3976 | 9 | 0.82 | -2.06 | 0.16 | ... | ... |
| 866741 | 00h48m58.98s | +42d47m52.3s | -127.6 | 19 | 4495 | 11 | 0.95 | -0.75 | 0.14 | 0.72 | 0.34 |
| 31 kpc Halo Field (a0_1) | | | | | | | | | | | |
| 8002454 | 00h51m59.62s | +39d46m49.7s | -317.4 | 11 | 3991 | 14 | 1.19 | -1.18 | 0.14 | ... | ... |
| 7007191 | 00h51m30.89s | +39d55m56.4s | -278.1 | 19 | 4410 | 14 | 1.13 | -1.55 | 0.15 | 0.1 | 0.43 |
| 7003904 | 00h51m49.68s | +39d51m43.1s | -268.2 | 21 | 3991 | 11 | 0.8 | -1.19 | 0.13 | 0.38 | 0.32 |
| 8004556 | 00h51m49.07s | +39d46m36.6s | -416.1 | 56 | 3922 | 3 | 0.37 | -0.78 | 0.13 | 0.33 | 0.13 |
| 7006080 | 00h51m37.18s | +39d52m15.4s | -434.4 | 17 | 4343 | 12 | 1.16 | -1.87 | 0.15 | ... | ... |
| 7004287 | 00h51m47.51s | +39d56m05s | -647.2 | 18 | 4459 | 21 | 1.03 | -2.22 | 0.17 | ... | ... |
| 7003615 | 00h51m51.90s | +39d54m00.3s | -508.9 | 22 | 4234 | 10 | 0.85 | -1.5 | 0.14 | -0.47 | 0.36 |
| 7002602 | 00h51m56.72s | +39d50m29s | -345.5 | 10 | 3870 | 32 | 1.03 | -3.03 | 0.34 | ... | ... |

*(Table A.1 Continued)*

*(Table A.1 Continued)*

| Object | Sky Coordinates | | $v_{helio}$ | S/N | $T_{eff}$ | $\delta(T_{eff})$ | log $g$ | [Fe/H] | $\delta$([Fe/H]) | [$\alpha$/Fe] | $\delta$([$\alpha$/Fe]) |
|---|---|---|---|---|---|---|---|---|---|---|---|
| ID | RA | Dec | (km $s^{-1}$) | (Å $^{-1}$) | (K) | (K) | (dex) | (dex) | (dex) | (dex) | (dex) |
| 7004904 | 00h51m44.22s | +39d57m26.3s | -290.3 | 20 | 4547 | 37 | 1.14 | -1.48 | 0.14 | ... | ... |
| 7006422 | 00h51m36.19s | +39d55m06.7s | -388.9 | 20 | 4120 | 8 | 0.91 | -1.04 | 0.13 | 0.75 | 0.19 |
| 8005126 | 00h51m45.34s | +39d48m48.2s | -339.1 | 16 | 3888 | 162 | 1.02 | -1.02 | 0.27 | ... | ... |
| 8005050 | 00h51m45.81s | +39d49m06.5s | -382.4 | 32 | 3886 | 4 | 0.59 | -1.21 | 0.13 | 0.82 | 0.15 |
| 7006200 | 00h51m36.66s | +39d54m24.7s | -272.6 | 22 | 3973 | 4 | 0.72 | -1.16 | 0.13 | 0.14 | 0.21 |
| 7007559 | 00h51m30.02s | +39d52m50.1s | -434.6 | 54 | 4170 | 4 | 0.52 | -1.86 | 0.13 | 0.39 | 0.17 |
| 7005290 | 00h51m42.06s | +39d51m02.5s | -307.5 | 28 | 3955 | 5 | 0.62 | -1.47 | 0.13 | -0.03 | 0.24 |
| 8002934 | 00h51m57.40s | +39d46m46.7s | -428.9 | 36 | 4384 | 23 | 0.84 | -1.98 | 0.15 | ... | ... |
| 7007018 | 00h51m32.14s | +39d56m52.4s | -376.6 | 24 | 4017 | 9 | 0.75 | -1.85 | 0.14 | 0.47 | 0.32 |
| 33 kpc GSS Field (a3_1) | | | | | | | | | | | |
| 7003496 | 00h03m12.68s | +39d00m28.125s | -278.5 | 11 | 4261 | 34 | 1.31 | -2.9 | 0.26 | ... | ... |
| 7003469 | 00h03m12.76s | +39d03m16.7789s | -462.2 | 9 | 3996 | 14 | 1.14 | -0.56 | 0.14 | 0.74 | 0.33 |
| 7004205 | 00h03m11.44s | +38d59m44.0424s | -355.3 | 11 | 3793 | 9 | 0.83 | -1.42 | 0.15 | -0.37 | 0.33 |
| 7002539 | 00h03m13.63s | +39d01m47.4875s | -171.5 | 9 | 4421 | 13 | 1.33 | -1.46 | 0.15 | ... | ... |
| 7003250 | 00h03m12.93s | +39d01m24.3063s | -462.5 | 10 | 3927 | 13 | 1.12 | -1.68 | 0.16 | 0.43 | 0.44 |
| 7000928 | 00h03m15.41s | +39d07m21.6779s | -357.2 | 11 | 3829 | 8 | 0.52 | -2.41 | 0.2 | ... | ... |
| 7004353 | 00h03m10.82s | +39d00m15.2161s | -406.1 | 5 | 3944 | 81 | 1.06 | -0.22 | 0.17 | ... | ... |
| 7002943 | 00h03m13.22s | +39d01m35.2377s | -251.5 | 30 | 4383 | 17 | 0.81 | -2.6 | 0.25 | ... | ... |

*(Table A.1 Continued)*

*(Table A.1 Continued)*

| Object | Sky Coordinates | | $v_{helio}$ | S/N | $T_{eff}$ | $\delta(T_{eff})$ | log $g$ | [Fe/H] | $\delta$([Fe/H]) | [$\alpha$/Fe] | $\delta$([$\alpha$/Fe]) |
|---|---|---|---|---|---|---|---|---|---|---|---|
| ID | RA | Dec | (km $s^{-1}$) | (Å $^{-1}$) | (K) | (K) | (dex) | (dex) | (dex) | (dex) | (dex) |
| 7003886 | 00h03m12.33s | +39d02m22.1768s | -270.9 | 18 | 3831 | 5 | 0.79 | -1.06 | 0.14 | 0.55 | 0.28 |
| 7003760 | 00h03m12.46s | +39d02m05.1205s | -445.1 | 24 | 3842 | 5 | 0.63 | -2.91 | 0.19 | ... | ... |
| 7005069 | 00h03m11.03s | +39d01m13.2513s | -212.8 | 24 | 4484 | 6 | 0.82 | -2.03 | 0.15 | -0.1 | 0.41 |
| 7001809 | 00h03m14.46s | +39d03m40.578s | -438.0 | 13 | 3671 | 5 | 0.7 | -0.67 | 0.14 | -0.3 | 0.34 |
| 7001891 | 00h03m14.37s | +39d05m33.5175s | -439.0 | 23 | 4048 | 7 | 0.9 | -1.6 | 0.14 | 0.6 | 0.19 |
| 7004573 | 00h03m10.95s | +39d00m18.251s | -455.4 | 12 | 3770 | 14 | 0.75 | -1.25 | 0.15 | ... | ... |
| 7005021 | 00h03m11.98s | +39d02m48.4891s | -297.2 | 38 | 3883 | 5 | 0.52 | -1.39 | 0.13 | 0.41 | 0.15 |
| 7001477 | 00h03m14.78s | +39d07m12.106s | -466.7 | 25 | 4246 | 11 | 0.98 | -2.57 | 0.2 | ... | ... |
| 7003059 | 00h03m13.12s | +39d01m48.8196s | -534.3 | 22 | 4273 | 8 | 0.95 | -2.27 | 0.15 | 0.24 | 0.33 |
| 7005383 | 00h03m12.10s | +38d59m22.7701s | -449.4 | 25 | 4066 | 24 | 0.58 | -2.29 | 0.26 | 0.77 | 0.41 |
| 7003333 | 00h03m12.87s | +39d01m08.7195s | -460.3 | 10 | 3873 | 23 | 1.1 | -3.18 | 0.38 | ... | ... |
| 7003445 | 00h03m12.75s | +39d02m47.926s | -439.6 | 22 | 4072 | 10 | 0.88 | -1.92 | 0.15 | 0.5 | 0.39 |
| 7005249 | 00h03m11.68s | +39d03m36.6367s | -430.4 | 16 | 4142 | 112 | 1.03 | -1.46 | 0.23 | ... | ... |
| 7005538 | 00h03m12.03s | +39d03m49.9026s | -411.1 | 14 | 4303 | 19 | 1.3 | -2.19 | 0.19 | ... | ... |
| 7002661 | 00h03m13.49s | +39d04m56.6995s | -443.5 | 14 | 3774 | 6 | 0.93 | -1.16 | 0.14 | 0.28 | 0.43 |
| 7002450 | 00h03m13.74s | +39d04m46.6058s | -427.3 | 14 | 3745 | 11 | 0.85 | -1.82 | 0.18 | ... | ... |
| 7001934 | 00h03m14.30s | +39d03m51.4133s | -404.1 | 24 | 3922 | 5 | 0.68 | -1.21 | 0.13 | 0.51 | 0.16 |
| 33 kpc GSS Field (a3_2) | | | | | | | | | | | |

*(Table A.1 Continued)*

*(Table A.1 Continued)*

| Object | Sky Coordinates | | $v_{helio}$ | S/N | $T_{eff}$ | $\delta(T_{eff})$ | log $g$ | [Fe/H] | $\delta$([Fe/H]) | [$\alpha$/Fe] | $\delta$([$\alpha$/Fe]) |
|---|---|---|---|---|---|---|---|---|---|---|---|
| ID | RA | Dec | (km $s^{-1}$) | ($Å^{-1}$) | (K) | (K) | (dex) | (dex) | (dex) | (dex) | (dex) |
| 6004645 | 00h03m11.76s | +39d12m03.3536s | -431.1 | 18 | 4372 | 10 | 0.98 | -0.92 | 0.13 | -0.19 | 0.35 |
| 6004424 | 00h03m12.05s | +39d10m33.0048s | -343.6 | 48 | 4314 | 5 | 0.64 | -2.1 | 0.14 | 0.61 | 0.19 |
| 7005249 | 00h03m11.68s | +39d03m36.6367s | -412.5 | 19 | 4126 | 9 | 1.03 | -0.77 | 0.13 | 0.06 | 0.18 |
| 7005151 | 00h03m11.59s | +39d05m05.1865s | -398.4 | 42 | 4118 | 3 | 0.66 | -1.31 | 0.13 | 0.54 | 0.15 |
| 7004817 | 00h03m11.12s | +39d03m09.6515s | -415.1 | 19 | 3758 | 5 | 0.62 | -1.05 | 0.13 | 0.04 | 0.31 |
| 6004687 | 00h03m11.77s | +39d12m57.6535s | -446.6 | 16 | 3786 | 12 | 0.92 | -1.98 | 0.18 | ... | ... |
| 7005234 | 00h03m11.79s | +39d05m31.3477s | -432.5 | 21 | 4299 | 9 | 1.19 | -0.72 | 0.13 | 0.38 | 0.2 |
| 6003875 | 00h03m11.47s | +39d12m36.6147s | -455.4 | 10 | 3849 | 10 | 0.72 | -2.99 | 0.27 | ... | ... |
| 7005021 | 00h03m11.98s | +39d02m48.4891s | -289.1 | 31 | 3919 | 3 | 0.52 | -1.62 | 0.13 | 0.59 | 0.21 |
| 7004990 | 00h03m11.46s | +39d00m15.4633s | -415.5 | 14 | 3804 | 8 | 1.04 | -0.88 | 0.14 | 0.14 | 0.35 |
| 6003827 | 00h03m12.11s | +39d10m38.9511s | -260.8 | 12 | 3635 | 8 | 0.87 | -1.21 | 0.16 | ... | ... |
| 6004005 | 00h03m10.85s | +39d10m38.2782s | -522.4 | 12 | 4447 | 16 | 0.89 | -0.89 | 0.15 | ... | ... |
| 7004205 | 00h03m11.44s | +38d59m44.0424s | -374.6 | 10 | 3794 | 8 | 0.83 | -1.53 | 0.15 | ... | ... |